\documentclass[manuscript,preprint,showpacs,showkeys]{revtex4-1}
\pdfoutput=1
\usepackage{graphicx,anysize,array}
\usepackage{bm,bbm,epstopdf}
\usepackage{color}
\usepackage{amsmath,amssymb,mathrsfs,amsthm}
\usepackage[perpage,para,symbol]{footmisc}
\usepackage{siunitx}
\usepackage{float}
\usepackage{makeidx}
\usepackage{epsfig}
\usepackage{subfigure,wrapfig}
\usepackage{supertabular}
\usepackage{gensymb}
\usepackage{enumerate,mdwlist,textcomp}
\usepackage{fancyhdr,lastpage,url,paralist,calc}
\usepackage{varioref,url}
\allowdisplaybreaks[1]

\newcommand{\beqn}{\begin{equation}}
\newcommand{\eeqn}{\end{equation}}

\newcommand{\be}{\begin{equation}}
\newcommand{\ee}{\end{equation}}
\newcommand{\bea}{\begin{eqnarray}}
\newcommand{\eea}{\end{eqnarray}}
\newcommand{\bdm}{\begin{displaymath}}
\newcommand{\edm}{\end{displaymath}}
\newcommand{\bit}{\begin{itemize}}
\newcommand{\eit}{\end{itemize}}
\newcommand{\ben}{\begin{enumerate}}
\newcommand{\een}{\end{enumerate}}

\usepackage{lineno}

\begin{document}

\title{Mesoscale computational studies of membrane bilayer remodeling by curvature-inducing proteins.}

\author{N. Ramakrishnan}
\email{ramn@seas.upenn.edu}
\affiliation{Department of Chemical and Biomolecular Engineering, Department of Bioengineering, Department of Biochemistry and Biophysics, University of Pennsylvania, Philadelphia, PA-19104}

\author{P. B. Sunil Kumar}
\email{sunil@physics.iitm.ac.in}
\affiliation{Department of Physics, Indian Institute of Technology Madras, Chennai, India - 600036}

\author{Ravi Radhakrishnan}
\email{rradhak@seas.upenn.edu}
\affiliation{Department of Chemical and Biomolecular Engineering, Department of Bioengineering, Department of Biochemistry and Biophysics, University of Pennsylvania, Philadelphia, PA-19104}

\begin{abstract}
Biological membranes constitute boundaries of cells and cell organelles. These membranes are soft fluid interfaces whose thermodynamic states are dictated by bending moduli, induced curvature fields, and thermal fluctuations. Recently, there has been a flood of experimental evidence highlighting active roles for these structures in many cellular processes ranging from trafficking of cargo to cell motility. It is believed that the local membrane curvature, which is continuously altered due to its interactions with myriad  proteins and other macromolecules attached to its surface, holds the key to the emergent functionality in these cellular processes. Mechanisms at the atomic scale are dictated by protein-lipid interaction strength, lipid composition, lipid distribution in the vicinity of the protein, shape and amino acid composition of the protein, and its amino acid contents. The specificity of molecular interactions together with the cooperativity of multiple proteins induce and stabilize complex membrane shapes at the mesoscale. These shapes span a wide spectrum ranging from the spherical plasma membrane to the complex cisternae of the Golgi apparatus. Mapping the relation between the protein-induced deformations at the molecular scale and the resulting mesoscale morphologies is key to bridging cellular experiments across the various length scales. In this review, we focus on the theoretical and computational methods used to understand the phenomenology underlying protein-driven membrane remodeling. Interactions at the molecular scale can be computationally probed by all atom and coarse grained molecular dynamics (MD, CGMD), as well as dissipative particle dynamics (DPD) simulations, which we only describe in passing. We choose to focus on several continuum approaches extending the Canham - Helfrich elastic energy model for membranes to include the effect of curvature-inducing proteins and explore the conformational phase space of such systems. In this description, the protein is expressed in the form of a spontaneous curvature field. The approaches include field theoretical methods limited to the small deformation regime, triangulated surfaces and particle-based computational models to investigate the large-deformation regimes observed in the natural state of many biological membranes. Applications of these methods to understand the properties of biological membranes in homogeneous and inhomogeneous environments of proteins, whose underlying curvature fields are either isotropic or anisotropic, are discussed. The diversity in the curvature fields elicits a rich variety of morphological states, including tubes, discs, branched tubes, and caveola. Mapping the thermodynamic stability of these states as a function of tuning parameters such as concentration and strength of curvature induction of the proteins is discussed. The relative stabilities of these self-organized shapes are examined through free-energy calculations. The suite of methods discussed here can be tailored to applications in specific cellular settings such as endocytosis during cargo trafficking and tubulation of filopodial structures in migrating cells, which makes these methods a powerful complement to experimental studies.
\end{abstract}


\preprint{Published as Phys. Reports  {\bf 543}, 1--60 (2014) }
\maketitle

\newpage
\tableofcontents
\newpage
\section{Introduction to membranes} \label{sec:intro}
Cell membranes are biological structures involved in a wide range of biological processes and ubiquitous in both prokaryotic and eukaryotic cells. Mouritsen, in his book ``Life as a matter of fat'' ~\cite{olemouritsen:2005}, aptly describes a membrane as functioning as a barrier, a carrier, and a host. Namely, membranes constitute a barrier that delineates the outside from the inside of a cell, separates a cell from another, and in addition, encapsulates most cell organelles in eukaryotic cells. Membranes have been long known to be involved in the trafficking of cellular cargo. They play an integral role, as a carrier, in the processes of endocytosis and exocytosis, which represent key inter- and intra-cellular transport mechanisms that aid in cellular uptake of cargo ranging from nutrients to pathogens. In its role as a host, the membrane is home to a large set of proteins, ligands, and various other macromolecules which are involved in processes that span a wide spectrum from cell signaling to cell replication. 

\begin{figure}[H]
\centering
\includegraphics[width=8.5cm,clip]{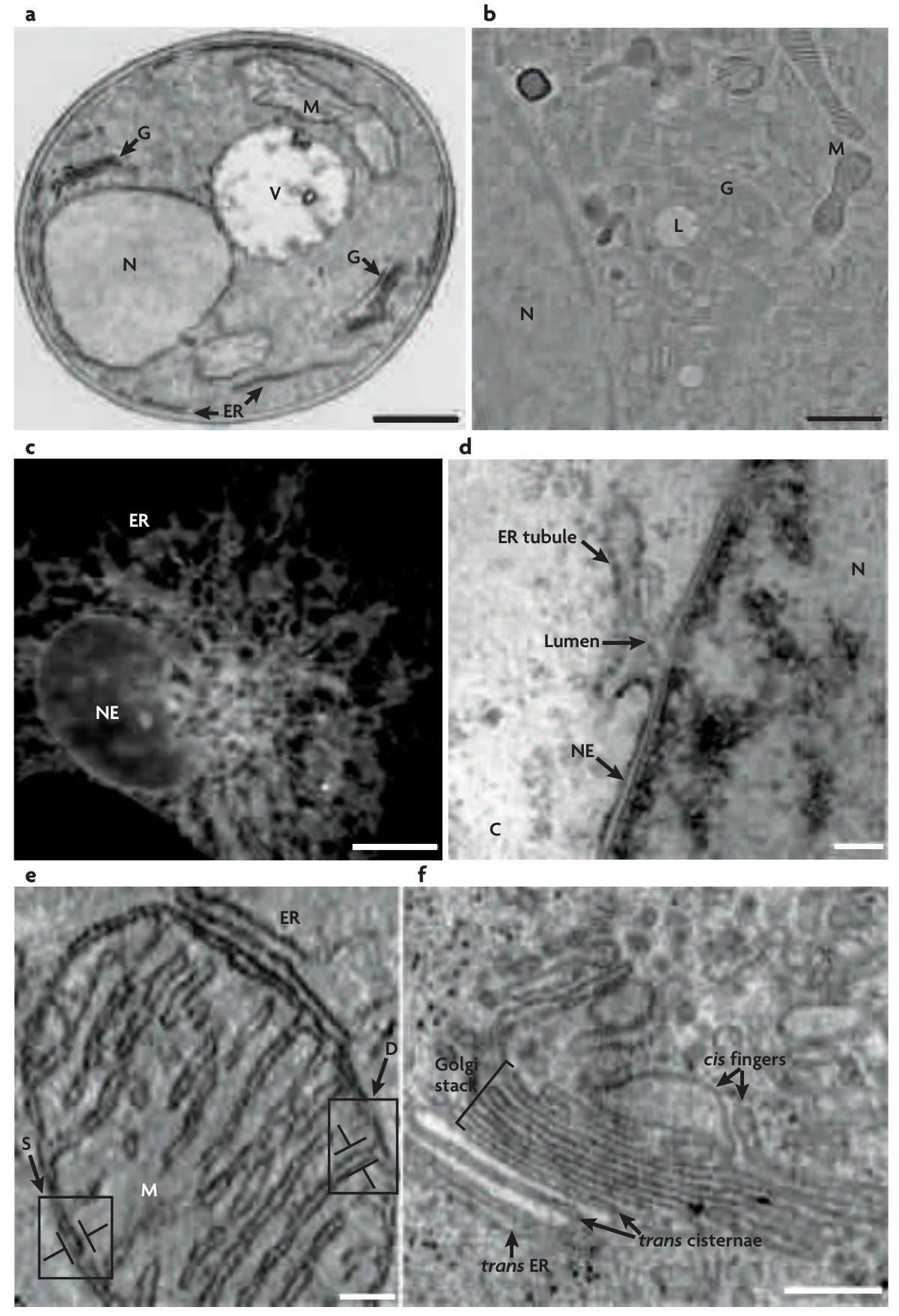}
\label{fig:organelleshapes}
\caption{Simple and complex shapes seen in cells and cell organelles. Image from  ~\cite{Voeltz:2007p1399} {\sf(Reprinted by permission from Macmillan Publishers Ltd: Nat. Rev. Mol. Cell Biol. {\bf 8}(3), 258--264  copyright (2007))}.  {\bf(a,b)} Electron microscope images showing various cell organelles in a yeast cell {\sf(a)} and in a  normal rat kidney cell {\bf (b)}. The major organelles shown\textemdash namely, the Golgi(G), the nucleus(N), the endoplasmic reticulum(ER), the vacuole(V), and the mitochondrion(M)\textemdash have identical shapes in both yeast and normal rat kidney cells,  and the shapes of these organelles are also preserved in other cells. {\bf(c)}  Fluorescence image of the nuclear envelope(NE) and ER in COS cells, wherein the bright spots denote the localization of a particular ER associated protein. {\bf(d)} The nuclear envelope that defines the boundary of a cell nucleus is a relatively smooth, double bilayer structure which merges with the rough and tubular endoplasmic reticulum. {\bf(e)} Doubled walled structure of the mitochondrion; its outer membrane has a smooth shape while its inner membrane is organized into a complex network of tubular shapes called the cristae.  {\bf(f)} Tubules and flattened sacs\textemdash the primary structures that dominate the morphology of the Golgi stacks. }
\end{figure}

A biological membrane results from the complex assembly and organization of different kinds of fatty acid molecules called lipids. In addition to the lipid molecules, a cell membrane also comprises a large number (concentration) of proteins and a relatively small number (concentration) of carbohydrates. Depending on the type of cell membrane investigated, the protein concentration varies between $18\%$ and $75\%$  with the corresponding protein to lipid ratio varying between 0.23 and 1.6  ~\cite{Guidotti:1972wd}. Carbohydrate molecules have been estimated to have concentrations in the range of $3\%-10\%$. Molecular composition and the complex interactions between the individual components are key factors that  determine the resulting macroscopic shapes of the biological membrane, which in their role as a barrier also influence the shape of the cells and cell organelles they enclose. When categorized on the basis of their complexity, membrane structures reported in the biological literature vary from simple, symmetric, primitive spherical shapes, commonly seen in the case of the plasma membrane, to the highly complex, convoluted structures displayed by organelles like the endoplasmic reticulum and the golgi.  The genesis of these shapes and the other more complex shapes like those shown in Fig.~\ref{fig:organelleshapes}, has been extensively investigated in cell biology. Despite the efforts over these years, a generic framework to explain all the observed shapes does not exist. Understanding the mechanisms governing cell membrane organization would be instructive and can help gain insight into the more complex question of ``{\it how do the cell and its organelles get their shape ?} ''. 

Currently, {\it in vivo} experimental methods have reached a degree of sophistication high enough to quantify the physical and chemical properties of the individual components of a lipid membrane. Further, computer-based algorithms and supercomputers have undergone a transformative change, which in turn allow one to model highly complex systems, such as membrane-bound macromolecules. These factors together have widened the horizons of membrane science beyond single-component and multi-component lipid membranes to also focus on the role of proteins/membrane-associated macromolecules in curvature induction, curvature detection (or sensing), and membrane remodeling. However, the problem at hand is too complex to interpret in terms of experiments alone, and the study of membranes at the cellular scale is not yet amenable to molecular simulations. 
Mechanics- and thermodynamics-based continuum modeling is a powerful alternative to investigate the behavior of membranes spanning length scales extending well above a few tens of nanometers. In this approach, the large microscopic degrees of freedom associated with the lipids and proteins in the membrane are represented in terms of a few macroscopic observables that obey well-defined mechanical and thermodynamic principles. In this article, we will focus on the various theoretical and computational methods that are widely used in the study of membranes in this continuum limit relevant to the cellular scale.
\subsection{Physiological significance of lipid membranes} \label{cellmemb}
\begin{figure}[H]
\centering
\includegraphics[width=15cm,clip]{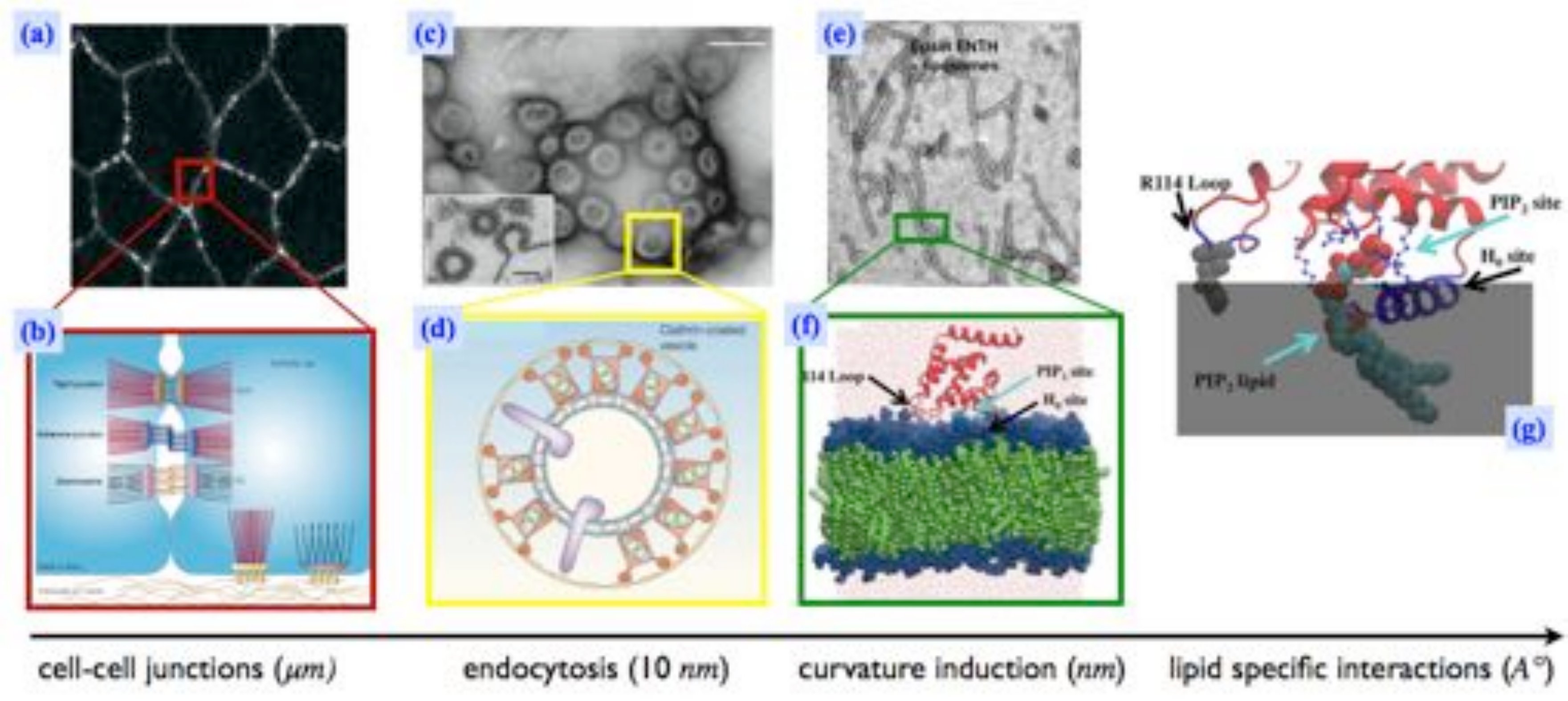}
\caption{\label{fig:multiscale-proc} Membrane protein interactions at multiple length scales. {\bf (a)} Cell junctions formed by transmembrane linker proteins; the brigt regions denote the junctions formed by Integrins and Cadherins ~\cite{Cavey:2008p751}{\sf(Reprinted by permission from Macmillan Publishers Ltd:  Nature {\bf 453}(7), 453--456  copyright (2008))}, {\bf (b)} Illustration of the various inter-cellular junctions ~\cite{Jefferson:2004cf}{(\sf Reprinted by permission from Macmillan Publishers Ltd:  Nat. Rev. Mol. Cell Biol. {\bf 5}, 542--553  copyright (2004))}, {\bf (c)} The vesicular buds formed by the process of clathrin-mediated endocytosis (CME) ~\cite{Dannhauser:2012gy} {(\sf Reprinted by permission from Macmillan Publishers Ltd:  Nat. Cell Biol. {\bf 14}(6), 634--639  copyright (2012))}, {\bf (d)} A cartoon of a clathrin coated vesicle which shows the spatial localization of various accessory-proteins involved in CME ~\cite{Wendland:2002js} {(\sf Reprinted by permission from Macmillan Publishers Ltd:  Nat. Rev. Mol. Cell Biol. {\bf 3}(12), 971--977  copyright (2002))}, {\bf (e)} Spontaneous tubulation of a liposome due to addition of epsins ~\cite{Ford:2002ifb} {(\sf Reprinted by permission from Macmillan Publishers Ltd:  Nature {\bf 419}, 361--366  copyright (2002))}, {\bf (f, g)} A molecular picture which shows the insertion of the $\alpha$-helix (helix-0), of an epsin, into a membrane leaflet and its specific interactions with a negatively charged PIP2 lipid~\cite{Lai:2012hk} {(\sf Reprinted from J. Mol. Biol., {\bf 423}(5), C-L Lai et. al, Membrane Binding and Self-Association of the Epsin N-Terminal Homology Domain, 800--817, Copyright (2012), with permission from Elsevier).} }
\end{figure}

The membrane interacts with almost all major entities of a cell\textemdash namely, proteins, nucleic acids, and polysaccharides. In addition it is an important member of many signaling pathways and hence plays a pivotal role in many crucial decisions determining cell fate such as cell motility, metabolism, proliferation, and survival. The relationship between cellular pathology and abnormality in the cellular membrane has been observed in a wide variety of diseases : for instance, sickle cell anemia is linked to enhanced phosphatidylserine levels, Duchenne muscular dystrophy results from the breakdown of cytoskeletal membrane anchoring, amyloid-related diseases like Alzheimer's show signatures of disrupted leaky membrane bilayers, and cancer metastasize in a tissue by breaking the cell-cell junction, which in turn leads to a neoplasm-promoting microenvironment in the tissue ~\cite{Joyce:2008kw,Escriba:2008hb}. 
As described above, the observed membrane anomalies can span length scales ranging  from  nanometers (molecular scale) to  microns and beyond (tissue scale). For instance, at the nanoscale, one observes local perturbations in the organization of membrane constituents,  whereas at length scales comparable to a cell, the diseased cell displays noticeable change in its structure and its organization within a tissue. In spite of being separated over large length scales, the observations at the  molecular and cellular scales can be coupled\textemdash chemical changes precede morphological changes and vice versa. The relation between the biochemistry and structure is inherently a multi-scale process, which makes membranes a complex system to deal with. In this section, using the specific examples shown in Fig.~\ref{fig:multiscale-proc}, we will re-iterate the multi-scale nature of membranes and their role in maintaining the integrity of the cell and tissue.

Cadherins are a family of transmembrane proteins found in cells. The cytoplasmic domain of a Cadherin is linked to cytoskeletal filaments while the extracellular domains of adjacent cells interact to form adherens junctions. At the scale of a tissue, cells in the tissue are kept together by these adherens junctions, and hence the integrity of the tissue is determined by the strength of these junctions. In Fig.~\ref{fig:multiscale-proc}(a) the bright regions mark the spatial location of Cadherins, and the illustration in Fig.~\ref{fig:multiscale-proc}(b) shows the representative position of the Cadherins with respect to the cell membrane. Though the  adherens junctions are formed in the extracellular region, its structure and strength are determined by the local membrane environment around the Cadherins. It has been shown that the Cadherins localize to membrane micro-domains rich in cholesterol ~\cite{Angst:2001um,Causeret:2005hb} and changes in the membrane lipid composition disrupts the structure of adherens junctions ~\cite{Marquez:2012br}. This is an obvious case of membrane biochemistry driving tissue structure, discussed at the start of this section.

At the cellular level, exocytosis and endocytosis are key processes in the bidirectional transport of inbound and outbound cargo across a membrane barrier. Interestingly, these processes also significantly influence the cadherin levels on the plasma membrane which in turn influence the stability, polarity, and motility of the cell. Fig.~\ref{fig:multiscale-proc}(c) and (d) are representative images of clathrin-mediated endocytosis (CME) that involves the formation of vesicular buds due to the cooperative action of proteins such as clathrin, adaptor protein 2 (AP2), epsin, and dynamin on a lipid bilayer environment. However, the action of just one of these proteins\textemdash namely, epsin\textemdash results in membrane tubulation (see Fig.~\ref{fig:multiscale-proc}(e)). Epsin-membrane interactions are believed to cause deformations on the membrane at length scales of nanometers due to the insertion of its helix-0 ($\alpha$-helix) into one of the leaflets of the bilayer (Fig.~\ref{fig:multiscale-proc}(f)). The nature and strength of such interactions also depend on the lipid environment such as the presence or absence of PIP$_2$ (Fig.~\ref{fig:multiscale-proc}(g)).  In order to develop an understanding of such an inherently multiscale system, there is a need to employ a scale-dependent description of the membrane. This is achieved by the technique of coarse graining, which allows for the flow of information between the different scales. As a first step of coarse graining, we will develop some insights into how the molecular picture of membranes translates into the corresponding one at the ($\sim 100$~nm) mesoscale.
\subsection{Molecular description of lipid membranes}
Lipids are one among the four building blocks of biology, with the other three being amino acids, nucleic acids, and sugars ~\cite{olemouritsen:2005}. These fatty acids, which are carboxyl-group-containing hydrocarbons, are the most abundant molecules in the cell, numbering over a thousand different types in both eukaryotes and prokaryotes ~\cite{Alberts:1994}. From a biological point of view, supramolecular organization of lipids has low functionality compared to biopolymers like proteins and DNA/RNA, which are  poly-amino acids and poly-nucleotides respectively.  A lipid molecule can either  be polar or apolar, with the former being hydrophilic and the latter being hydrophobic, depending on the chemical moieties attached to the carboxyl group. Apolar lipids in a polar solvent, aggregate into lipid droplets that are known to be the energy store of a cell ~\cite{vanMeer:2008p3294,Farese:2009p3290} and are interesting from a functional point of view.

\begin{figure}[H]
\centering
\includegraphics[height=2.0in,clip]{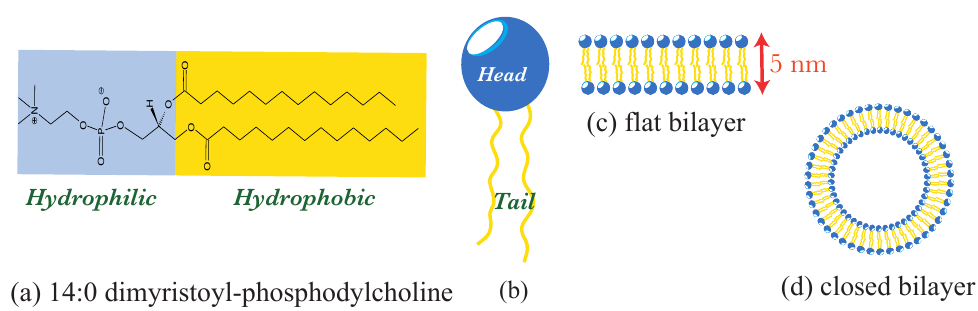}
\caption{\label{fig:lipidorg} A lipid molecule and its self organized structures\textemdash shown are, (a) the chemical composition of a DMPC lipid, with the hydrophobic and hydrophilic regions shaded differently, (b) the representative model of a lipid, with the head and tail groups marked,  (c) a cross section of a flat membrane bilayer, and (d) a cross section of a closed bilayer (generally called a vesicle).}
\end{figure}
The structurally important polar lipid molecules are characterized by a hydrophilic part called the head and a hydrophobic chain called the tail. This is illustrated in Fig.~\ref{fig:lipidorg}(a) and (b) for the case of a dimyristoyl-phosphotidylcholine or DMPC lipid. When lipid molecules are introduced into an aqueous solvent with concentration above a critical value, called the critical micelle concentration, they spontaneously partition into an interface that shields the hydrophobic tails from the solvent. The simplest realization of such an interface is a lipid bilayer, a cartoon of which is shown in Fig.~\ref{fig:lipidorg}(c) and (d). Depending on the area to volume ratio of the lipid molecule, self-assembled structures like micelles, cylindrical micelles, multi-lamellar stacks, and bi-continuous phases can also be stabilized ~\cite{Israelachvili:1977ve}. The phase diagrams of over 2000 well-characterized lipid mixtures has been compiled by Koynova and Caffrey ~\cite{Koynova:2002gb}. At the molecular scale, the organization and interaction of the lipid molecules with other biological entities is predominantly governed by the chemistry of the lipid molecules. \\

Eukaryotic and prokaryotic organisms have over 1000 types of lipid molecules, and these molecules can be broadly divided into three major classes\textemdash namely, glycerol-based lipids, cholesterol, and ceramide based sphingolipids ~\cite{Escriba:2008hb}. Even lipids belonging to the same class exhibit large chemical diversity due to variations in the hydrophilic head groups and differences in the number, length and saturation of the hydrocarbon chain (tail); see Fig.~\ref{fig:lipidorg}(a). Phospholipids for instance can have a variety of head groups, like phosphatidylcholine (PC), phosphatidylserine (PS), phosphatidylethanolamine (PE), and phospatidylglycerol (PG), and these groups can be uncharged, anionic, cationic, or zwitterionic. The organization of lipids in a multi-component lipid membrane is well described by the fluid mosaic model ~\cite{Singer:1972p2064}, which describes cell membranes as {\it ``two-dimensional solutions of lipids and other macromolecules''}. \\

Membranes in mammalian cells consist primarily of phospholipids and glycerol. Other classes of lipids that are present in smaller quantities are nevertheless essential for the cell to perform specific biological processes. Variations in lipid composition have been shown to impact a host of cellular properties like exocytosis, endocytosis, phagocytosis, sensitivity of receptor molecules to extracellular signaling molecules, and cytotoxicity ~\cite{Spector:1985vu,vanMeer:2008p3294}. Systematic studies to understand the correlation between lipid composition and membrane organization/function show that even variations in the same class of lipid across different cells can produce different effects. Hence, {\it in vitro} experiments using reconstituted cell membranes are hard to interpret due to the highly complex organizational landscape of the constituent lipid molecules. \\
 
Though it is hard to understand the morphological properties and organizational patterns even for membranes constituted from binary/ternary lipid mixtures, all biological membranes display some key microscopic properties that are key to our understanding of macroscopic models introduced later. We  given a brief summary of a few of these properties below: \\

\noindent{{\sf A two-dimensional fluid:}}
The absence of bonded interactions between the lipid molecules in a bilayer allows the lateral diffusion of lipids in the plane of the leaflet it resides in. Experimental observations based on fluorescent tagging and electron spin resonance (ESR) spin tagging of lipid molecules have estimated the diffusion constant of lipids in a bilayer membrane to be of the order of $10^{-12}$m$^{2}$s$^{-1}$ ~\cite{olemouritsen:2005,Alberts:1994}. As a result, lipid bilayers do not resist shear stresses, like a solid, but instead sustain a flow-field when sheared, like a liquid. Lipid molecules can also be translocated from one leaflet of the bilayer to another. The translocation can be a result  of thermal fluctuations  or specialized lipid translocator proteins, called flippases and floppases, that are normally found in the membranes of cells. Lipid flip flop is a slow process compared to most processes associated with a membrane. The spontaneous rate of translocation in the absence of these specialized proteins ranges from hours to days. Although they are fluid like, lipid membranes display elastic-like behavior in response to normal stresses; this topic is discussed at length in section~~\ref{sec:helfrich-model}, and also forms the basis for much of this article. In addition, lipid membranes are selectively permeable to ions ~\cite{Paula:1996p339}, poly-electrolytes ~\cite{Finkelstein:1976vo}, and many other small molecules. As a result of this semi-permeable nature, they can maintain different chemical environments in the interior and exterior regions.  It should also be noted that every cell organelle has a chemical environment different from the other, which allows them to perform a unique biological function. Semi-permeability combined with flexibility makes lipid membranes  effective barriers. As will be discussed below, many biological processes can also be controlled by modulating the curvature on the membrane surface. \\

\noindent{{\sf Domain formation in multi component membranes:}}  By the Gibbs phase rule, heterogeneity in lipid  composition can give rise to a  variety of coexisting phases in the form of lipid domains in each of the leaflets of a multi-component lipid membrane ~\cite{HonerkampSmith:2009fh,Semaru:2009p3174}. These domains are mainly formed due to mismatch in the lengths of the hydrophobic chains or due to preferential partitioning of lipids.  
\begin{figure}[H]
\centering
\includegraphics[width=12.5cm,clip]{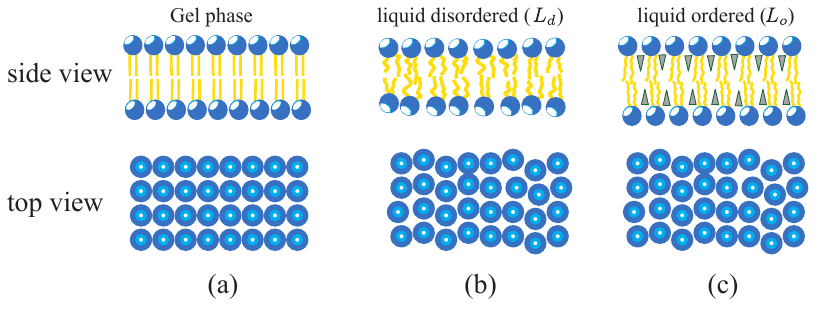}
\caption{\label{fig:lipid-phases} Representative phases in a lipid bilayer: shown are three distinct phases namely  the gel phase {\bf(a)}, the liquid-disordered, $L_{d}$ phase {\bf(b)}, and  the liquid-ordered, $L_{o}$ phase {\bf(c)}. Solid and liquid implies that the nature of translational correlations in the plane of the membrane are either solid-like or liquid-like {\sf (top view of (a) shows a lattice structure, characteristic of a solid, while (b) and (c) have no lattice structure and hence are liquid like)}. Hydrophobic chains in the ordered phase and gel phase have strong orientational correlations, whereas those in the disordered phase have random orientations and  thus have weak orientation correlations. In {\bf (c)} the shaded triangles represent smaller lipid molecules like cholesterol and sphingolipids that intercalate into the hydrophobic region.}
\end{figure}

A lipid domain can exist in three distinct phases\textemdash namely, (a) gel phase, (b) liquid-disordered ($L_{d}$), and (c) liquid-ordered ($L_{d}$), as shown in Fig.~\ref{fig:lipid-phases} ~\cite{Munro:2003p377,Edidin:2003p1606,Kaiser:2009p497}. A lipid membrane shows a gel phase at temperatures $T<T_{m}$ and makes a transition to the  $L_{d}$ or $L_{o}$ phase, depending on the concentration, when the temperature exceeds the transition temperature $T_{m}$. The gel phase is characterized by the presence of long-range translational correlations in the position of the head groups, as seen in the top view of Fig.~\ref{fig:lipid-phases}(a), whereas the absence of any such correlation is a signature of the liquid phase. On the other hand, the orientations of the lipid tails are correlated in the ordered phase and uncorrelated in the disordered phase. The gel phase of lipid domains is highly relevant to model bilayers and is not seen in biological membranes. In the case of the $L_{o}$ phase, which is commonly observed in multi-component lipid membranes, the hydrophobic chains acquire orientational order even at $T>T_{m}$ due to the intercalation of smaller lipids  into the hydrophobic region, which in turn leads to the arrest of the acyl chain degrees of freedom. The intercalating lipids are mainly cholesterol and sphingo-lipids, shown as triangles in Fig.~\ref{fig:lipid-phases}(c). The phase diagram of many two-component and three-component lipid mixtures have been well studied in the literature ~\cite{Veatch:2002ta,Bagatolli:2009SM,Veatch:2003hn,Goni:2008dt,Hamada:2011fq}. Lipid domains in a bilayer leaflet can either be correlated ~\cite{Ursell:2009p176} or uncorrelated with the lipid domains in the  other leaflet of the membrane. How these microphase separations, such as lipid rafts ~\cite{Simons:2000fa,Simons:tx,Simons:2004p187}, seen in multi-component membranes are related to (and functionally relevant to) cellular phenomena is still a matter of open debate.\\

Much of our understanding of lipid organization has been derived from the study of an {\it in vitro} membrane system called Giant Unilamellar vesicle (GUV).  These micron- to millimeter-sized vesicles are formed from lipid mixtures through processes like sonication of multi-lamellar vesicles and electroformation of dry lipid films; see ~\cite{Bagatolli:2009SM} for a detailed review of the experimental techniques. A GUV assembled from a single lipid species is called a single-component vesicle, whereas that containing multiple lipid species is called a multi-component vesicle. GUVs alleviate many complexities seen in {\it in vivo} membrane systems since their chemical heterogeneity can be precisely controlled, and the size of GUVs allows them to be observed under a microscope. These properties make GUVs the most used experimental system to investigate lipid organization in membranes. In the next two sections, based on experimental observations in GUVs, we will briefly describe how change in chemical heterogeneity  drives lipid organization and also how perturbations in the thermodynamic variables drive morphological changes.

\subsection {Chemical heterogeneity and lipid organization in multicomponent GUVs}
\begin{figure}[H]
\centering
\includegraphics[width=12.5cm,clip]{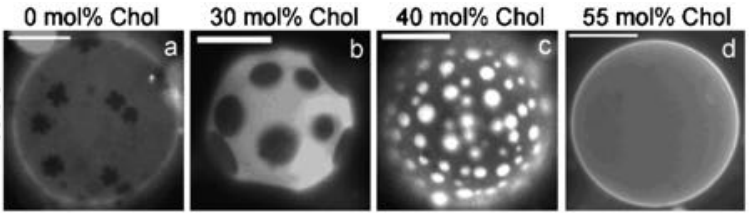}
\caption{\label{fig:threecomp-keller} Phase segregation into liquid-ordered and liquid-disordered domains in the presence of cholesterol in multi-component vesicular membranes, constituted from saturated and unsaturated DMPC lipids. The fluorescence images show the coexistence of various lipid phases at different concentrations of cholesterol; {\bf(a)} gel-$L_{d}$ at 0 mol\%; {\bf(b)} $L_{o}$-$L_{d}$ at 30 mol\%; {\bf (c)} $L_{o}$-$L_{d}$ at 40 mol\%; and {\bf(d)} no visible phase separation at 50 mol\%. Image adopted from ~\cite{Veatch:2002ta}  {\sf(Reprinted figure with permission from  Sarah L Veatch and Sarah L Keller, Phys. Rev. Lett., {\bf 89} (26), 2681011 and 2002. Copyright (2002) by the American Physical Society.)}}
\end{figure}

Fig.~\ref{fig:threecomp-keller} shows how the phases and organization of lipids are modulated by the composition of cholesterol in a  GUV formed from a ternary lipid mixture of saturated DMPC, unsaturated DMPC, and cholesterol. Composition of the 16 carbon chain long saturated and unsaturated DMPC lipid were taken at 1:1. In the absence of cholesterol (Fig.~\ref{fig:threecomp-keller}(a)), the saturated lipids organize into a non-circular solid  phase (same as the gel phase described in Fig.~\ref{fig:lipid-phases}(a)), shown as dark regions, that coexists with a liquid phase, shown by the bright regions in the micrograph. The solid domain is further characterized by decrease in lipid mobility and moves and rotates as  a rigid body. With increase in cholesterol content, for concentration in the range of $10\%-50\%$ mol, the solid phase is replaced by the $L_{o}$ phase (see Fig.~\ref{fig:lipid-phases}(c))  that coexists with a background liquid phase, leading to a liquid-liquid coexistence, as seen in Fig.~\ref{fig:threecomp-keller}(b) and (c). Phase coexistence becomes unstable beyond a cholesterol concentration of 55\% mol, and the GUV becomes uniformly bright, which is characteristic of the $L_{d}$ phase (Fig.~\ref{fig:lipid-phases}(b)).

The ternary lipid mixture discussed above resembles the lipid composition called the raft mixture.  Sphingolipids in the presence of cholesterol can assemble into a specialized cholesterol rich structure called rafts ~\cite{Simons:1997jq}, which are lipid domains in the liquid-ordered phase ($L_{o}$) ~\cite{Simons:2000fa}. Rafts are believed to be membrane micro-domains enriched in glycophosphatidylinositol and GPI-anchored proteins, which play an important role in signal transduction, membrane trafficking, cytoskeletal organization, and pathogen entry ~\cite{Munro:2003p377}. This example clearly illustrates how lipid heterogeneity affects lipid organization in membranes and how the organization can be used effectively by cells to perform biological processes. It should also be noted that the stability of lipid phases described above is a function of thermodynamic variables like temperature and pressure. The experiments described above were performed at a temperature 5\degree C \, lower than the liquid-ordered to liquid-disordered transition temperature. The cholesterol-dependent {\it gel} -$L_{d}$ and $L_{o}$-$L_{d}$ phase coexistence, described in Fig.~\ref{fig:threecomp-keller}, would show a completely different behavior or may even be unstable if the experiments are performed at temperatures above this transition temperature. 

\subsection{Morphological transitions and the role of thermodynamic variables}
\begin{figure}[H]
\centering
\centering
\includegraphics[width=8.5cm,clip]{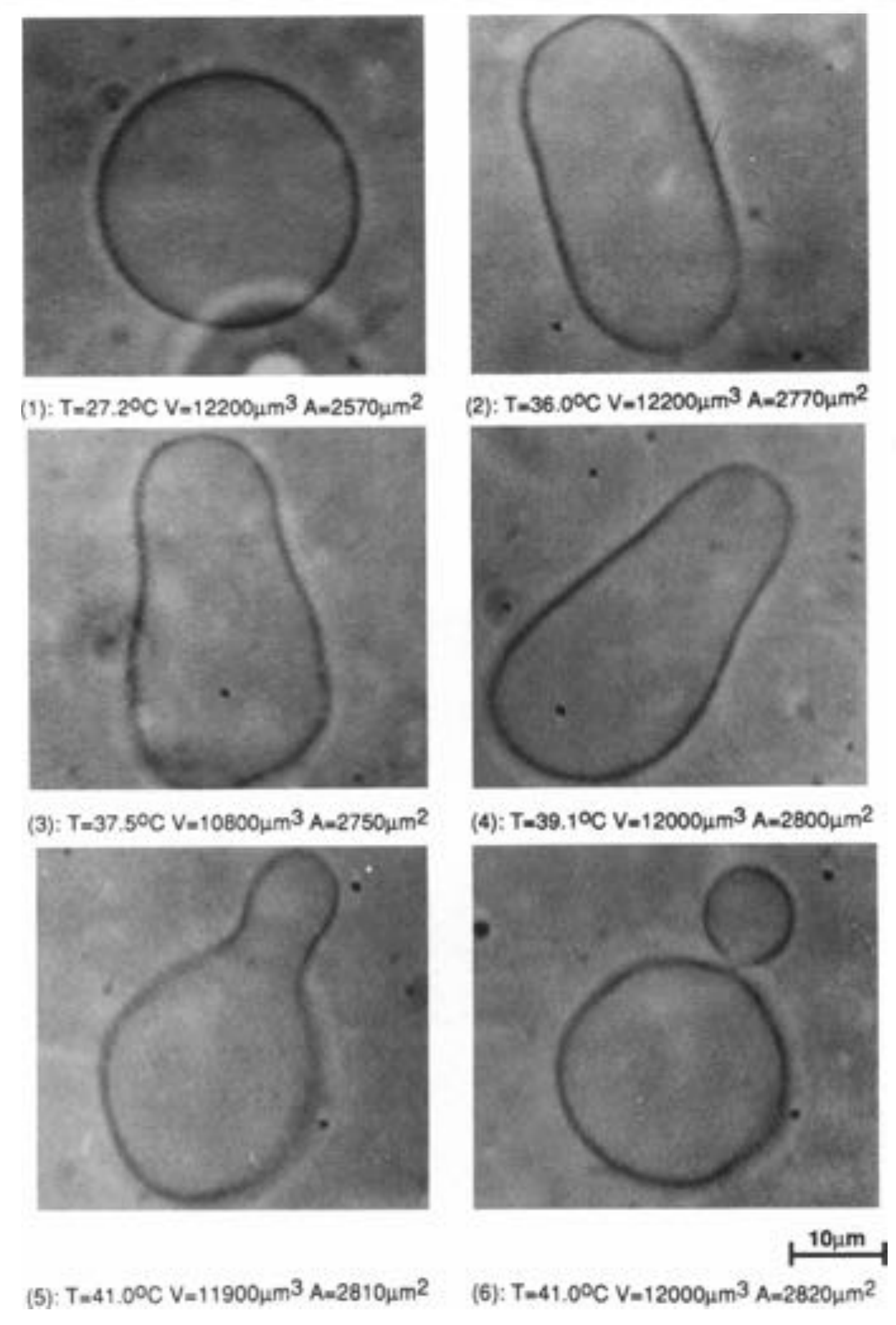}
\caption{\label{fig:shapes-expt}  Shape transformations in a DMPC vesicle in response to a change in temperature from $27.2\celsius$ - $41\celsius$. An initially spherical vesicle {\bf (1)}, at $27.2 \celsius$,  transforms into an oblate ellipsoid {\bf (2)}, at $T=36.0 \celsius$. With further increase in temperature, pear shaped vesicles {\bf (3,4,5)} are stabilized,  for $36.0 \celsius<T<41.0 \celsius$, which transforms into a budded vesicle {\bf(6)}, at $T=41 \celsius$; Image adopted from reference ~\cite{Kas:1991fk} {\sf (Reprinted from Biophys. J, {\bf 60} {\bf (4)}, J. K\"{a}s, E. Sackmann, Shape transitions and shape stability of giant phospholipid vesicles in pure water induced by area to volume changes, 825--844, Copyright (1991), with permission from Elsevier)}.}
\end{figure}

GUVs, or in general membranous structures, show noticeable morphological transitions in response to a perturbation in the thermodynamic environmental variables, like temperature and pressure. Fig.~\ref{fig:shapes-expt} shows the range of thermally undulated shapes displayed by  pure phospholipid bilayer membranes in response to temperature change. The spherical vesicle  shown in Fig.~\ref{fig:shapes-expt}(1), at $T=27.2\degree C$, transforms into the budded vesicle, shown in Fig.~\ref{fig:shapes-expt}(6), when  the temperature is changed to $T=41\degree C$. As a function of increasing temperature, the budding process proceeds through a series of conformational changes from spherical to oblate ellipsoid to prolate ellipsoid to pear shaped which discontinuously transforms into a budded membrane.  These shape changes are accompanied by an increase in the surface area of the vesicle at nearly constant volume.

\begin{figure}[H]
\centering
\centering
\includegraphics[width=15cm,clip]{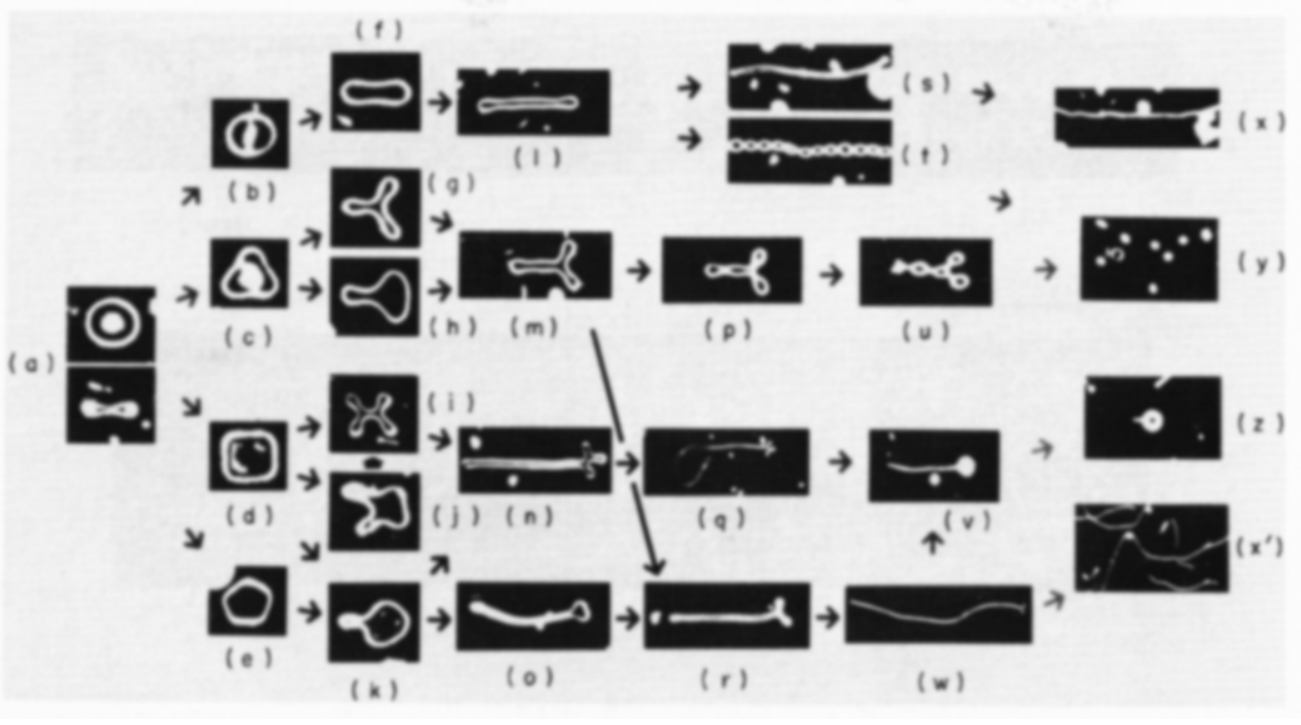}

\caption{\label{fig:osmoticshapes} Various pathways for the deformation of an initially biconcave vesicle, whose top and side views are shown in (a), to elongated filament like stable shapes, shown in frames (x), (y), (z), and (x$^{'}$), under the influence of osmotic pressure. The giant unilamellar vesicles were formed from a binary mixture of DMPC and cholesterol and imaged using phase contrast microscopy. Image adopted   from ~\cite{Hotani:1984tz} {\sf (Reprinted from  J. Mol. Bio, {\bf 178} (1), H. Hotani, Transformation pathways of liposomes, 113--120, Copyright (1984), with permission from Elsevier)}.}
\end{figure}

In addition to temperature,  in the case of closed vesicles, morphological transformations can also be driven by the  osmotic pressure difference between the inside and outside of the vesicle. An osmotic pressure difference may be a result of excess electrolyte concentration in the solvent in contact with the vesicle. Pressure-induced shape changes  may be quite drastic, as seen in Fig.~\ref{fig:osmoticshapes}, for a two component GUV formed from a binary mixture of DMPC and cholesterol ~\cite{Hotani:1984tz}. When the salt concentration is higher on its exterior, an initially biconcave vesicle, whose top and side views are shown in Fig.~\ref{fig:osmoticshapes}(a), transforms into one of the elongated filamentous shapes, seen in frames (x), (y), (z) and (x$^{'}$) of Fig.~\ref{fig:osmoticshapes}. The biconcave to elongated morphological transition has multiple pathways that are characterized by varying intermediate shapes.
 
Vesicular membranes support large morphological transformations in response to physical and chemical perturbations because they are soft systems\footnotetext{A material is said to be soft when (a) its energy density is very small (to the order that thermal energy $k_{B}T$ is relevant),  (b) it exhibits significant fluctuations, and (c) it has nanoscopic length scales. In terms of numbers, at room temperature, the thermal energy $k_{B}T$ translates to $4.1\, pN {\rm nm}$. At this scale the energy density and elastic stiffness for different classes of soft materials ranges between $10$ and $100\,k_{B}T/{\rm nm}^{3}$ and $2.5$ and $250\,k_{B}T$, respectively. The name soft arises from the fact that these values are 2 to 5 orders of magnitude smaller than those for conventional materials, like a solid crystal.}. Experiments on biological and synthetic membranes have estimated their bending rigidity to be of the order of $10-100k_{B}T$ ~\cite{Lipowsky:1991p1059,Seifert:1997p1058}, which is a clear indication of its softness. In spite of their softness, membranes can support large stresses and large shape fluctuations; this basically explains the simple to complex spectrum of vesicular shapes shown in Figs.~\ref{fig:shapes-expt} and ~\ref{fig:osmoticshapes}. \\

In summary, changes in the thermodynamic state variables can play a key role in driving shape transformation in vesicular membranes. Though these experimental observations are instructive, it should be noted that the physical and chemical environments in which these experiments have been performed are mostly non-physiological. Hence we need to look beyond simple lipid-based systems and also include the effect of other macromolecules to explain the occurrence and stability of complex cellular and cell organelle shapes described in section~~\ref{sec:intro} and shown in Fig.~\ref{fig:organelleshapes}. \\

\subsection{Membrane remodeling by curvature inducing factors}  \label{cellmemb-micro} 
A variety of factors associated with the membrane can spontaneously deform membranes and support the large curvatures associated with complex membrane shapes. A summary of the key membrane remodeling factors is given below.  \\

\noindent{\sf (a) Membrane inclusions:} \\
The membrane of a cell is home to a large set of functional macromolecules that are essential for its function. A vast set of protein machinery comprising  signaling proteins, like kinases; transport proteins, like flippases; channel proteins, like aquaporins; ion channels, like the sodium/potassium channels; pump proteins, like bacteriorhodopsins; junction proteins, like integrins; and cytoskeletal-linking proteins, like profilin,  are found in a typical membrane that encloses a cell and its organelles. These  proteins constitute $25\%-75\%$ of the weight of the membrane and this ratio depends the type of cell and on the organelle they are associated with (see Table 1 in ~\cite{Guidotti:1972wd}). Proteins can be amphipathic (contain both polar and non-polar subunits ). The degree of amphipathicity determines how the protein is associated with the membrane. In addition to proteins, other macromolecules like sugars and peptides  are also found in non-negligible quantities. In this article, we will refer to all non-lipid entities as  membrane inclusions, and the importance of these molecules in controlling membrane morphologies will be discussed in detail in section~~\ref{cellmemb-curvact}. \\

\noindent{{\sf (b) Area/compositional asymmetry:}} \\ Membranes maintain an asymmetry in the  number and composition of lipids in each monolayer for structural and functional reasons. While number asymmetry can arise due to different sizes of lipids and/or curvature in the membrane, membranes in cells also display compositional asymmetry~~\cite{Devaux:1991ti,Slochower2014} in order to orchestrate many biological processes that are specific to a particular leaflet. For example, the inner leaflet of the membrane is rich in phosphatidylinositol lipids, like PIP, PIP$_{2}$ and PIP$_{3}$, which are essential for clathrin-mediated endocytosis in the endocytic pathway. It should also be noted that many of the membrane-interacting proteins reshape cellular morphologies by generating an area asymmetry in the membrane. This mechanism will be discussed in detail in  section~~\ref{cellmemb-curvact}. \\

The aim of this review is to explore the role of these curvature-inducing factors in remodeling membranes into biologically relevant morphologies in order to gain insight into the genesis of cellular shapes. As noted above, biological membranes have widely separated length scales and time scales. They measure a few nanometers in the transverse direction and extend to a few microns laterally. Our knowledge of biological membranes has been derived both from top-down models developed from continuum theories of solid and fluid mechanics as well as from bottom-up models derived from molecular structure and packing. This review focuses on the theoretical and the associated computational methods that are prevalently used in understanding the thermodynamics and structure of lipid-based artificial and biological membranes under the influence of curvature-inducing factors with characteristic lateral size in the regime $100-1000$~nm, which includes both the meso and continuum scales. \\

The article is organized as follows. In section~~\ref{sec:therm-model}, we introduce the underlying theory for the elastic model of the membrane and discuss the role of various elastic parameters and relevant ensembles. Various analytical and computational methods employed in the study of the Canham - Helfrich elastic energy model have been reviewed in section~~\ref{sec:analyt-comp-methods} along with a brief description of the molecular methods. Starting section~~\ref{chap:sponcurvmodels} with the biological picture relevant to protein-induced deformations, we classify the  protein-induced curvature field into two categories\textemdash namely, isotropic and anisotropic. We focus on the theory and computational methods for the nematic membranes used as a model for anisotropic curvature-inducing proteins interacting with the membrane. The particle-based EM2 model is introduced in section~~\ref{sec:em2}, and the use of the model to simulate the remodeling behavior of BAR-domain-containing proteins is discussed. In section~~\ref{sec:freeener}, we describe free-energy methods based on thermodynamic integration to delineate the free-energy landscape of protein-induced remodeling and conclude in section~~\ref{sec:conclusion}.


\section{Thermodynamics-based models for membranes} \label{sec:therm-model}
Fully atomistic molecular models and coarse-grained molecular models (reviewed briefly in section~~\ref{sec:molCG}) are useful to probe the nature of the interactions at the nanometer resolution. The models closer to atomic or electronic resolution remain true to the biochemistry of interactions and can map the sensitivity of protein-lipid interactions to the underlying chemical heterogeneity (e.g., the effect of protonation or ion binding on the specific interaction of proteins with lipids). On the other hand, the models closer to the cellular length scale are adept at describing cooperative interactions and long-wavelength deformations. Conceptually, the divisive length scale between these extremes is set by a characteristic length (see section~~\ref{sec:bent-plate}): much below this length scale, representing the chemical heterogeneity is important (e.g., presence of cholesterol, PIP$_2$, Ca$^{2+}$) as much of the interactions are energy dominated. For length scales much above the characteristic length, a coarse-grained representation of the chemical interactions is often sufficient (e.g., the effect of cholesterol, PIP$_2$, Ca$^{2+}$ on bending stiffness or intrinsic curvature) as the emergent phenomena are often entropy dominated.  Owing to the limitations of computing power, models with molecular resolution have limited applicability in the investigation of structures and shapes of membrane systems, whose dimensions match cellular length scales. The limitation is primarily due to the system size needed to approach the thermodynamic limit and also due to the lower cutoff on the temporal resolution, which is essential in capturing key biochemical processes. At cellular length scales, it is apparent, however, that top-down phenomenological models can be used to gain insights into the biophysical properties of membranes. In this section, we review the Canham - Helfrich elasticity theory for biological membranes ~\cite{Canham:1970p61,Helfrich:1973p693}. In this phenomenological approach, membrane response (such as shape and undulation) to external perturbations, the associated stability, and energetics can be determined.

\subsection{Thin sheet approximation of membranes}
An elasticity-based phenomenological description of the membrane is valid if the system under consideration obeys the following constraint: the lateral extent of the membrane\textemdash for instance the diameter of a vesicle\textemdash is large ($L\sim{\cal O}(\mu m)$) compared to the bilayer thickness ($\delta \sim {\cal O}(nm)$).

\begin{figure}[H]
\centering
\includegraphics[width=9.5cm,clip]{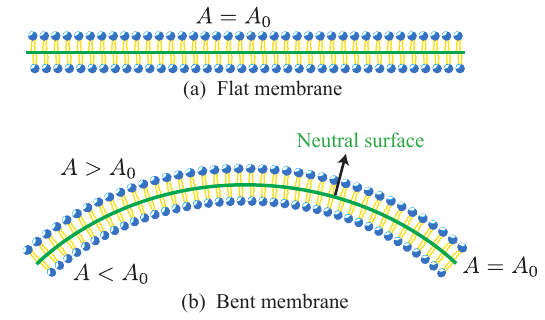}
\caption{\label{fig:neutral-surf} {\bf (a)} Cross sectional view of a flat membrane, with surface area $A_{0}$.  {\bf (b)} When bent, the flat membrane shows stretching ($A>A_{0}$) in the top monolayer and compression ($A<A_{0}$) in the bottom monolayer. The neutral surface is the plane in the bent membrane with nearly constant surface area ($A=A_{0}$). THe neutral surfaces in the flat and bent membranes are shown as solid lines.}
\end{figure}

In the above-mentioned limit,  a lipid bilayer can be represented as a thin, flexible, fluid sheet, of constant area at length scales matching its largest dimension.  The sheet represents the neutral surface, known as the unstrained plane in bent membrane, a neutral surface is shown in Fig.~\ref{fig:neutral-surf}. We will discuss the role of neutral surface in more detail in  the next section.
  
\subsection{Theory of bent plates}\label{sec:bent-plate}
\begin{figure}[H]
\centering
\includegraphics[width=15cm,clip]{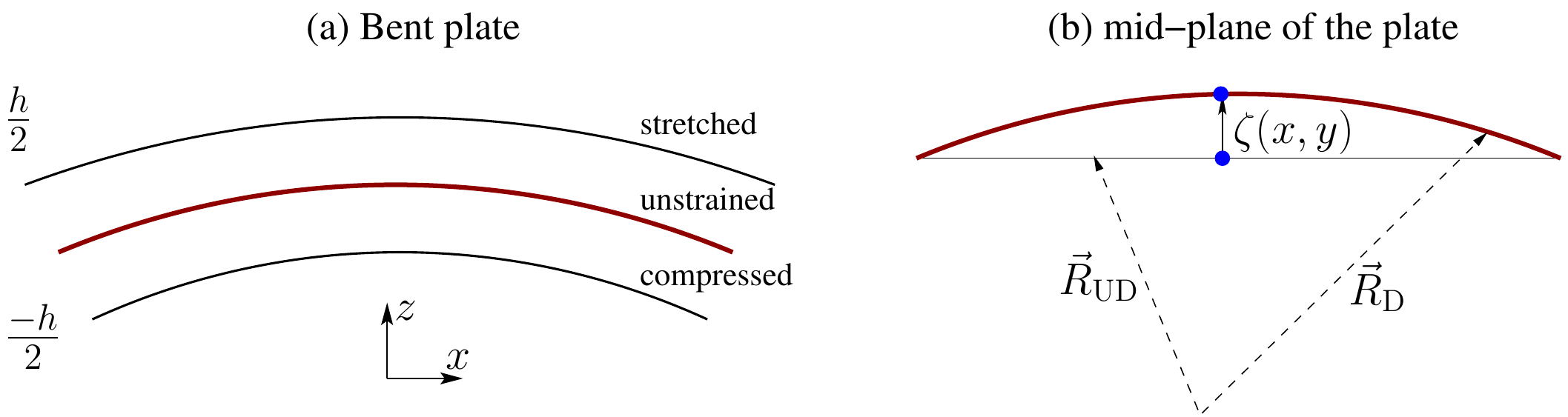}
\caption{\label{fig:bent-plate} Bending of a thin plate of thickness $h$. {\bf (a)} On bending the upper surface of the plate undergoes stretching while the lower surface gets compressed. In between these two surfaces is an unstrained surface called the {\em neutral surface}. The $xy$ plane of the cartesian coordinate system coincides with the mid-plane of the undeformed plate and the upper and lower surfaces of the plate are at $z=+h/2$ and $z=-h/2$, respectively. {\bf (b)} The position of a point (filled circle) on the neutral surface before and after deformation; the point is displaced only along the z-direction, with the $z$ component of displacement vector given by $u_{z}=\zeta(x,y)$. $R_{\rm D}$ and $R_{\rm UD}$ are respectively the coordinates of a point on the neutral surface, with respect to a global coordinate system, on the deformed and undeformed plate.}
\end{figure}

The theory of membranes originates from the elastic theory of bent plates. Consider a thin plate of thickness $h$ and lateral dimension $L$, with $h/L<<1$. The plate is bent along the $z$ direction  as shown in Fig.~\ref{fig:bent-plate}(a). On bending, the upper surface of the plate stretches while the lower surface gets compressed. The extension to compression behavior crosses over at an unstrained surface called the neutral surface of the plate. In the case of a plate with uniform thickness, the neutral surface can be identified with the mid-plane of the plate. The energy to bend the plate, can be written in terms of the stress and strain tensors as ~\cite{Landau:Elasticity}
\begin{equation}
\mathscr{H}_{\rm plate}=\int_{-h/2}^{+h/2}dz \int_{0}^{L} dy \int_{0}^{L} dx \, \frac{Y}{2(1+\sigma)} \left(u_{ik}^{2}+\frac{\sigma}{1-2\sigma}u_{ii}^{2}\right).
\label{eqn:LLfree-energy}
\end{equation}
The indices $i$ and $k$ of the strain tensor  take values $(x,y,z)$, when described in a cartesian coordinate system, and Einstein summation convention is implied. Here $Y$ and $\sigma$ are respectively the Young's modulus and Poisson's ratio of the material.  If $u_{i}$ is the $i$ th component of the displacement vector, the strain tensor is defined as $u_{ij}=\frac{1}{2}(\frac{\partial u_{i}}{\partial x_{j}}+\frac{\partial u_{j}}{\partial x_{i}})$.

A point in the undeformed plate undergoes a displacement of $\vec{u}=\{u_{x},u_{y},u_{z}\}$ to a new position on the deformed plate. When the point considered is on the neutral surface, as shown in Fig.~\ref{fig:bent-plate}(b), the in-plane displacements can be neglected ($u_{x}=u_{y}=0$) and the point moves only in the transverse direction ($u_{z}=\zeta(x,y)$). But the stretching and compressive behavior of the regions above and below the neutral surface implies that the in-plane displacements are not negligible in these regions. Hence the displacement vector can be shown to be
\begin{equation}
\vec{u}=\left \{ -z \frac{\partial \zeta}{\partial x}, -z\frac{\partial \zeta}{\partial y}, \zeta \right \}
\end{equation}
which, also satisfies the condition imposed on the neutral surface ($z=0$). The components of the strain tensor  for a thin bent plate are
\begin{eqnarray}
\label{eqn:strain-comp}
&u_{xx}=-z \dfrac{\partial^{2}\zeta}{\partial x^{2}}, & u_{yy}=-z \dfrac{\partial^{2}\zeta}{\partial y^{2}}\\ \nonumber
&u_{xy}=-z \dfrac{\partial^{2}\zeta}{\partial x\partial y}, & u_{xz}=u_{yz}=0\\  \nonumber
&{\rm and} & u_{zz}=\dfrac{\sigma}{1-\sigma}z \left( \dfrac{\partial^{2}\zeta}{\partial x^{2}}+\dfrac{\partial^{2}\zeta}{\partial y^{2}} \right).\\  \nonumber
\end{eqnarray}
  
 Using eqn.~\eqref{eqn:strain-comp} in the expression for the elastic energy (eqn.~\eqref{eqn:LLfree-energy}), we obtain
\begin{equation}
\mathscr{H}_{\rm plate}=\frac{Yh^{3}}{24(1-\sigma^{2})}\int_{0}^{L} dy \int_{0}^{L} dx \left\{ \left( \frac{\partial^{2}\zeta}{\partial x^{2}}+ \frac{\partial^{2}\zeta}{\partial y^{2}} \right)^{2}+ 2(1-\sigma) \left[ \left( \frac{\partial^{2}\zeta}{\partial x \partial y}\right)^{2}- \frac{\partial^{2}\zeta}{\partial x^{2}}\frac{\partial^{2}\zeta}{\partial y^{2}}\right] \right \}.
\label{eqn:curv-express}
\end{equation}

Rewriting the position vector $\vec{R_{\rm D}}=\vec{R}_{\rm UD}+\vec{u}$, for any point on the neutral surface, it can be shown that
\begin{eqnarray}
\frac{\partial^{2}R_{\rm D}}{\partial x^{2}}&=&\underbrace{\frac{\partial^{2}R_{\rm UD}}{\partial x^{2}}}_{0}+\frac{\partial^{2}u}{\partial x^{2}}=\frac{\partial^{2}\zeta}{\partial x^{2}} \,\, {\rm and} \nonumber \\
\frac{\partial^{2}R_{\rm D}}{\partial y^{2}}&=&\underbrace{\frac{\partial^{2}R_{\rm UD}}{\partial y^{2}}}_{0}+\frac{\partial^{2}u}{\partial y^{2}}=\frac{\partial^{2}\zeta}{\partial y^{2}},
\end{eqnarray}
 are the curvatures of the surface along the $x$ and $y$ directions, respectively. From eqn.~\eqref{eqn:curv-express}, it can be inferred that the energy of the bent plate is determined by the various  curvature measures on the neutral surface given by $\frac{\partial^{2}\zeta}{\partial x^{2}}$, $\frac{\partial^{2}\zeta}{\partial y^{2}}$ and $\frac{\partial^{2}\zeta}{\partial x \partial y}$.

\subsubsection*{Uniaxial bending}
The upper and lower surfaces of the plate show stretching and compression behavior only along one direction, say for instance the $x$ direction. The membrane has non-zero curvature only along the $x$ direction, given by $\frac{\partial^{2}\zeta}{\partial x^{2}}=\frac{1}{|\vec{R_{D}}|}$. A uniformly curved plate resembles a cylinder of radius $|R_{D}|=R$ and hence the bending energy becomes
\begin{equation}
\mathscr{H}_{\rm plate,uniaxial}=\frac{Yh^{3}}{24(1-\sigma^{2})}\int_{0}^{L} dy \int_{0}^{L} dx \left( \frac{1}{R} \right)^{2}.
\label{eqn:ener-uniaxial}
\end{equation}

\subsubsection*{Biaxial bending}
In the case of  uniform bending along both the $x$ and $y$ directions, the shape of the deformed plate is similar to the surface of a sphere. Hence $\frac{\partial^{2}\zeta}{\partial x^{2}}=\frac{\partial^{2}\zeta}{\partial y^{2}}=\frac{1}{|\vec{R_{D}}|}$ and  $\frac{\partial^{2}\zeta}{\partial x \partial y}=0$. For a uniform sphere of radius $|R_{D}|=R$, the bending cost is given by,
\begin{equation}
\mathscr{H}_{\rm plate,biaxial}=\frac{Yh^{3}}{24(1-\sigma^{2})}\int_{0}^{L} dy \int_{0}^{L} dx \left \{ \left( \frac{2}{R} \right)^{2}+2(\sigma-1)\frac{1}{R^{2}} \right \}.
\label{eqn:ener-biaxial}
\end{equation}

Eqns.~\eqref{eqn:ener-uniaxial} and ~\eqref{eqn:ener-biaxial} reveal some general features of the bending energy:
\begin{enumerate}
\item The  various material properties can be related to the bending rigidity of the plate through the relation $\kappa=\frac{Yh^{3}}{12(1-\sigma^{2})}$. $\kappa$ has the dimensions of energy.
\item The prefactor to the second term in eqn.~\eqref{eqn:ener-biaxial}, is another material property of interest called the Gaussian rigidity, $\kappa_{G}=2(\sigma-1)\kappa/2$. For materials with Poisson's ratio $\sigma=0$, the bending and Gaussian rigidity are related to each other as $\kappa_{G}=-\kappa$. 
\item The mean curvatures of the cylindrical and spherical surfaces, of radius $R$, considered in cases of uniaxial and biaxial bending are, respectively, $H_{\rm cyl}=1/2R$ and $H_{\rm sph}=1/R$. The corresponding $\kappa$-dependent part of the bending energy, in both eqns.~\eqref{eqn:ener-uniaxial} and ~\eqref{eqn:ener-biaxial}, depends on surface curvature as $\mathscr{H}_{\rm plate,uniaxial} \propto (2H_{\rm cyl})^{2}$ and $\mathscr{H}_{\rm plate,biaxial} \propto (2H_{\rm sph})^{2}$, respectively.
\end{enumerate}

These concepts are general and are applicable to any system that can be approximated as a thin plate. In the next section, we will apply these principles and derive the Canham-Helfrich Hamiltonian for biological membranes in the macroscopic length scale. 

\subsection{Canham - Helfrich phenomenological theory for membranes} \label{sec:helfrich-model}
Typical problems in membrane biophysics, which are computationally expensive to approach using molecular models, involve membranes whose lateral extension ($L$) exceeds 50~nm. If the average thickness ($h$) of a lipid membrane is taken to be 5~nm, as shown in Fig.~\ref{fig:lipidorg}, the thickness-to-length ratio ($h/L$) of these bilayer structures is less than $1/10$, which is in the regime of the thin plate theory. This heuristic estimate for $L$ can set the lower bound for the characteristic length discussed in the introduction to section~~\ref{sec:therm-model}. Later, in various sections, we will introduce and discuss other length scales that are relevant to the membrane system.

Further, there are two important properties of biological membranes that justify the use of continuum theories for their description: (a) the area per lipid molecule in the membrane is nearly constant, and (b) the fluctuation in  the thickness of a  lipid bilayer has been shown to be in the range $1 \textrm{\AA} -3.5 \textrm{\AA}$ ~\cite{Woodka:2012cf}, which is negligible when compared to its average thickness of $50 \textrm{\AA}$. In this regime, both the upper and lower surfaces in the bilayer closely follow the neutral surface and guarantee the small-deformation limit prescribed in our derivation of eqns.~\eqref{eqn:ener-uniaxial} and ~\eqref{eqn:ener-biaxial}. 

The elastic theory for membranes, known as the Canham - Helfrich Hamiltonian ~\cite{Canham:1970p61,Helfrich:1973p693}, has the form

\begin{equation}
\mathscr{H}_{\rm elastic}=\int_{0}^{L}dx\int_{0}^{L}dy \left\{\frac{\kappa}{2} (2H)^{2} +\kappa_{G} G \right\}.
\label{eqn:can-Helf}
\end{equation}

Here $G$ is the deviatoric curvature called the Gaussian curvature of the surface. Since the membrane is a self-assembled system, the relevant energies are comparable to thermal energy(${\cal O}(k_{B}T)$); this also suggests that the bending modulus $\kappa$ should be of the order of $k_{B}T$. Experimental measurements on a wide class of lipid membranes estimates the value  of $\kappa$ to be in the range of $10-100k_{B}T$. Based on these results and using the definition of $\kappa$, given in section~~\ref{sec:bent-plate}, we can estimate the Young's modulus of a  $5$nm thick lipid membrane to between $10^{7}$ and $10^{8} N/m^{2}.$

In addition to pure bending, the morphology of a membrane can also be affected by other modes that alter the membrane area. The membrane area couples to the surface tension, $\sigma$, and area elasticity modulus, ${\cal K}_{A}$. \footnotetext{\noindent Some arguments have been put forth in the literature that the surface tension of a lipid membrane in an aqueous environment is close to zero ~\cite{Jahnig:1996cv}. However, one can always impose a non-zero tension by changing other external variables, such as by applying a suction pressure in a micro-pipette aspiration experiment on a GUV or by controlling osmolarity. Another common assumption made in the literature is to drop the contribution from the area elasticity term.\medskip \\} In the case of closed vesicles, an osmotic pressure difference ($\Delta p$) between the inside and outside of a vesicle can also drive shape changes. Taking these contributions into account, eqn.~\eqref{eqn:can-Helf} can be written in a more general form as,

\begin{equation}
\mathscr{H}_{\rm sur}= \underset{S} \int d{\bf S} \left\{\frac{\kappa}{2} (2H-C_{0})^{2} +\kappa_{G} G +\sigma \right\}+  \frac{1}{2}{\cal K}_{A}(A-A_{0})^{2} + \underset{V} \int dV \Delta p.
\label{eqn:can-Helf-gen}
\end{equation}
Here, the equilibrium area is given by $A_{0}$. The geometry of the lipids can impose a preferred equilibrium curvature on the membrane, which is also captured by this energy functional through the spontaneous curvature term $C_{0}$. The integral in the first term is performed over the entire surface of the membrane,   \footnotetext{ \noindent For a parameterization $\bf{x}$, the surface area  $d{\bf S}=\sqrt{g}d{\bf x}$, where $g$ is the metric tensor defined in appendix ~\ref{app:diffgeo}} and the integration in the second term is carried out over the volume ($V$) enclosed by the surface. \\
  
\subsubsection{Gauss-Bonnet theorem and the Gaussian bending term:} \label{sec:GBtheorem} 
\begin{figure}[H]
\centering
\includegraphics[width=7.5cm]{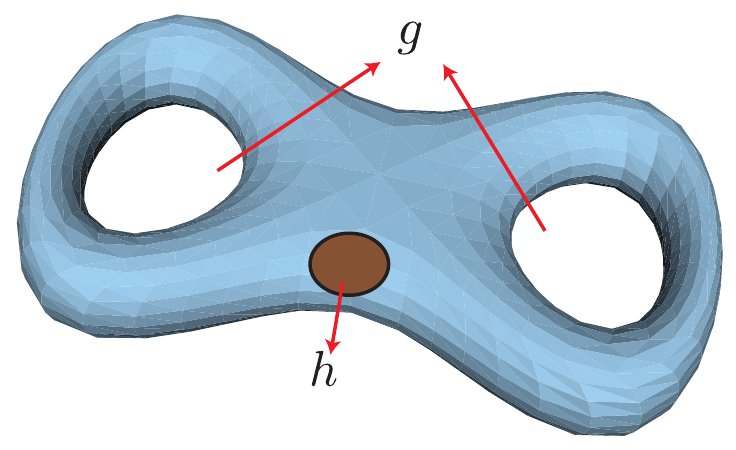}
\caption{\label{fig:genus-hole} A vesicular structure with two genus ($g=2$) and one hole ($h=1$), for which the Euler number $\chi=-3$. A hole is different from a genus in that it is constituted by a tear in the membrane surface. The membrane shape is adopted from the geometry models provided with Javaview (URL \protect\url{http://www.javaview.de}).}
\end{figure}

The topology of a membrane is described by the Euler characteristic, $\chi$, which in turn is related to the genus of the surface $g$ and  the number of holes $h$ as, $\chi=2(1-g)-h$. For instance, a membrane morphology with a spherical topology has $g=0$, $h=0$, and $\chi=2$ whereas a membrane with the topology of a torus has $g=1$, $h=0$, and $\chi=0$. As an example, a closed membrane structure with $g=2$ and $h=1$ is illustrated in Fig.~\ref{fig:genus-hole}.

The Gauss-Bonnet theorem relates the Euler number of the surface to the total Gaussian curvature of the surface as ~\cite{doCarmo:1976},
\begin{equation}
\underset{S} \int d{\bf S} \,G =2\pi \chi.
\end{equation}
For a membrane with fixed topology, by virtue of the Gauss-Bonnet theorem, it can be seen that the Gaussian energy contribution in eqn.~\eqref{eqn:can-Helf} is a constant, with values of $4\pi\kappa_{G}$ for a spherical membrane and $0$ for a toroidal membrane. For most of the studies presented here, since the topology of the membrane remains constant, the contribution from the Gaussian energy can be neglected during the analysis. We note, however, that when $\kappa_{G}$ is spatially inhomogeneous or when a planar patch of a membrane with a constant projected area is subject to periodic boundaries, in general, it will have a contribution from the Gauss curvature term.

\subsection{Thermal softening of elastic moduli} \label{sec:renormalization}
Continuum theory has been used in the renormalization studies of membrane elastic parameters ~\cite{Helfrich:1985wi,Peliti:1985p1690,Foster:1986jb,Kleinert:1986p347}, computations of the fluctuation spectrum, etc.  The elastic moduli ($\kappa$,\,$\kappa_{G}$,\,$\sigma$) introduced in the thermodynamic model for biological membranes (eqn.~\eqref{eqn:can-Helf-gen}) are conceptually different from those corresponding to a solid. In contrast to a solid, regions on a membrane away from each other by a distance $r$ display orientational decorrelations with increase in $r$. The value of $r$ at which this decorrelation occurs, characterized by the persistence length of the membrane ($\xi_{p}$), also depends on the temperature and the bending rigidity. De Gennes has suggested that $\xi_{p} \sim a\, \exp (4 \pi \kappa /3k_B T)$, which implies that for a typical bilayer, $\xi_{p} >>a$, where $a$ is the characteristic molecular dimension of an individual lipid ~\cite{Gennes:1982p240}; note here that for a typical bilayer, $\xi_{p}$ is  much larger than $L$, the characteristic length over which the elasticity model itself is valid. It was first suggested by Helfrich ~\cite{Helfrich:1985wi} that orientational decorrelations may have their origin in a scale dependent elastic modulus. Hence the elastic moduli employed in the thermodynamic description of a membrane are thermal quantities in the sense that they renormalize with system size and temperature. For example, the bending rigidity has been shown to be modulated from its bare value ($\kappa$) to a renormalized value ($\kappa_{R}$) as,
\begin{equation}
\kappa_{R}=\kappa-\alpha\,\frac{k_{B}T}{4 \pi}\log \left(\frac{L}{a_{0}}\right).
\end{equation}

There is a general consensus on the form of this relation, but the value of the prefactor depends on the statistical measure used in the calculations. Conflicting values for the prefactor\textemdash $\alpha=1$ ~\cite{Helfrich:1985wi,Helfrich:1986bx,Helfrich:1987if}, $\alpha=3$ ~\cite{Peliti:1985p1690,Foster:1986jb,Kleinert:1986p347} and $\alpha=-1$  ~\cite{Helfrich:1998dk,Pinnow:2000hr}\textemdash pointing to both thermal softening ($\alpha=1$ and $3$) and stiffening ($\alpha=-1$) have been reported. In spite of these contradictions, it should be remembered that we measure $\kappa_{R}$ from experiments and molecular simulations, whereas we impose $\kappa$ in mesoscale/continuum simulations based on the Helfrich energy functional. Similar renormalization behavior exists for other elastic moduli too ~\cite{Kleinert:1986p347,Marsh:1997p865,Marsh:2006ft}. It has been shown by Cai and Lubensky that the in-plane hydrodynamic modes arising  due to the fluid nature of the membrane can lead to renormalization of the elastic parameters ~\cite{Cai:1994p410,Cai:1995un}. 

\subsection{Ensembles for thermodynamic description} \label{sec:ensemble}
It has been stated earlier that the conformations of a membrane are susceptible to thermal undulations, and hence, it is important to understand the various thermodynamic ensembles associated with theoretical modeling of membranes. The interaction between the membrane degrees of freedom and key thermodynamic variables like temperature, $T$; chemical potential,  $\mu$; frame tension, $\tau$; and osmotic pressure difference, $\Delta p=p_{\rm out}-p_{\rm in}$; is entirely dependent on the thermodynamic ensemble used. Here we will describe two ensembles that are primarily used in the study of planar membranes and closed vesicular membranes (note that here closed refers to the fact that the membrane does not have free line boundaries). The thermodynamic formulation begins with considering the internal energy, $U$, as a function of all relevant extensive variables of the ensemble. Other free-energy functions and their independent variables can be derived from $U$ by defining suitable Legendre transformations as long as the free energy depends on at least one extensive variable ~\cite{Dill:2003vs}. \medskip

(a) {\it Constant $\sigma A_{p} \mu T$ ensemble for planar membranes:} A membrane patch is a model system for  supported membrane or represents a sub-region of a cell membrane. The internal energy of the planar membrane is a function of its extensive variables $S$, $A_{p}$, $N$ and $A$ and hence the elastic(internal) energy $U=U(S,A_{p},N,A)$. A  patch enclosed in a given frame is characterized by a fixed projected area $A_{p}$, which is conjugate to a type of   tension called the frame tension $\tau$. We also denote  the system entropy by $S$, the number of lipids  by $N$, and  the chemical potential of the lipids by $\mu$. In this ensemble,  the surface area of the membrane $A$ fluctuates while we fix the surface tension $\sigma$, such that $A \geq A_{p}$.  The appropriate free energy in this ensemble is given by $dF=dU - d(\sigma A) - d(TS) - d(\mu N)$ \medskip

(b) {\it Constant $\sigma \mu  V T$ ensemble for closed membranes:} The internal energy of a closed membrane is  $U=U(S,V,N,A)$. This ensemble closely represents a closed membrane of arbitrary topology seen in {\it ex-vivo} and {\it in-vivo} experimental systems. The osmotic pressure difference $\Delta p$ is determined by the solute concentration in the solvent inside and outside the vesicle and couples to $V$.  The appropriate free energy in this ensemble is given by $dF=dU - d(\sigma A) - d(TS) - d(\mu N)$. \medskip

\begin{figure}[H]
\centering
\includegraphics[width=15cm,clip]{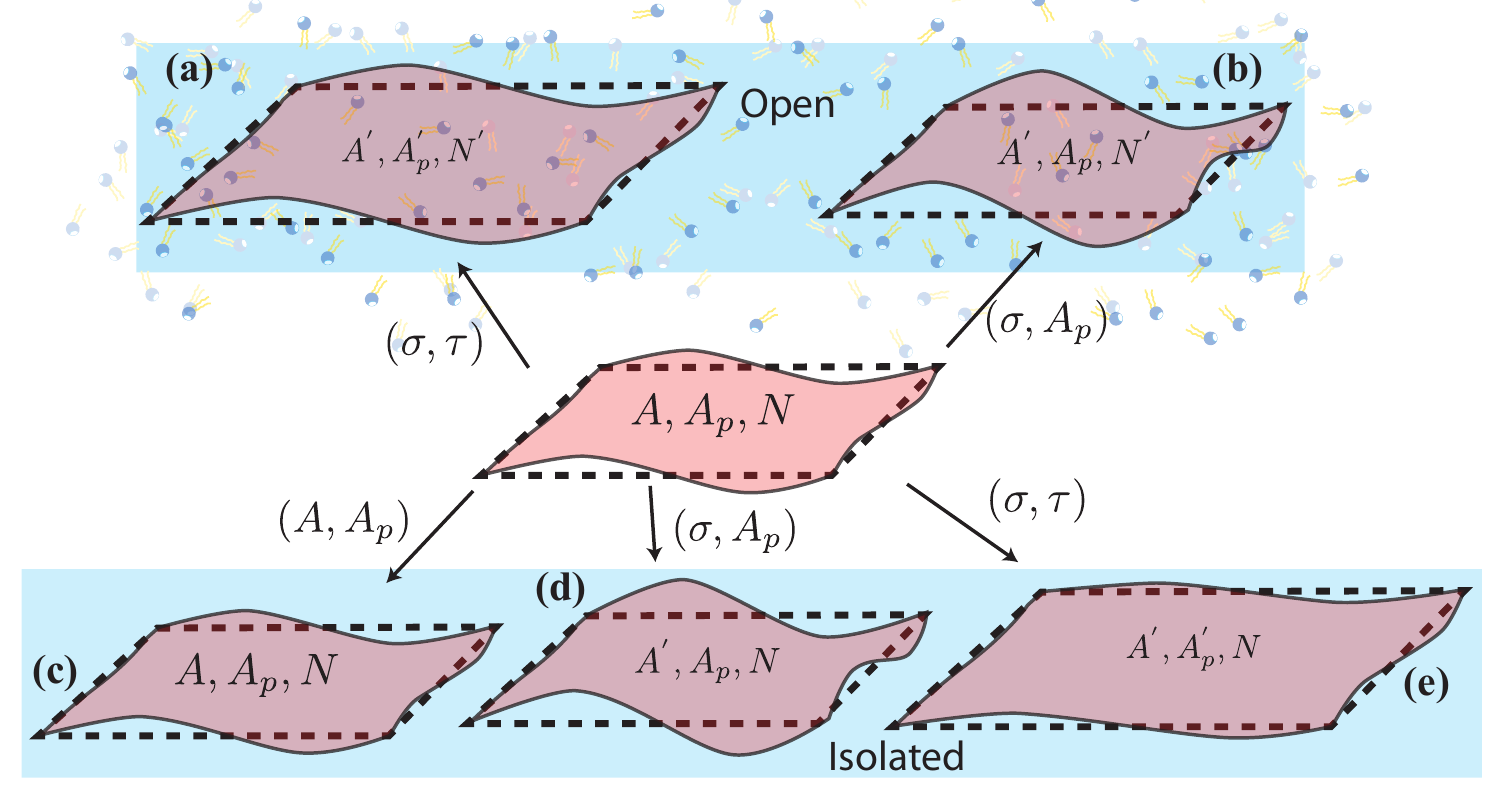}
\caption{\label{fig:planar-ensembles} {\it (center panel)} Initial state of a membrane with surface area, $A$, projected area, $A_{p}$, and number of lipids, $N$. {\it (top panel) }In the open state, the system exchanges lipid molecules with a reservoir and the corresponding variables in the final state are {\bf (a)} ($A^{'},A^{'}_{p},N^{'}$) in the ($\sigma,\tau$) ensemble and {\bf(b)} ($A^{'},A_{p},N^{'}$) in the ($\sigma,A_{p}$) ensemble. {\it (bottom panel)} In the absence of number fluctuations  the state of the membrane is given by {\bf(c)} ($A,A_{p},N$) in the ($A,A_{p}$) ensemble, {\bf(d)} ($A^{'},A_{p},N$) in the ($\sigma,A_{p}$) ensemble, and {\bf(e)} ($A^{'},A^{'}_{p},N$) in the ($\sigma,\tau$) ensemble.}
\end{figure}

The different ensembles for open and isolated systems are pictorially highlighted in Fig.~\ref{fig:planar-ensembles} ~\cite{DAVID:1991p3533,Piran:2003}. \\

Multiple variants of eqn.~\eqref{eqn:can-Helf-gen} that impose constraints on the external variables associated with the membrane, like the bilayer coupling model, and the area difference elasticity model ~\cite{Miao:1994p112} have also been used for modeling continuum membranes. For more details, we refer the reader to the detailed review on theoretical methods in membranes by Seifert ~\cite{Seifert:1997p1058}.  Any further constraints on the thermodynamic variables can be imposed in the form of a Lagrange multiplier.  Such constraints make physical sense when one accounts for the various constituents making up the cell. A case in point is the constraint on the volume enclosed by the membrane in the presence of cytoskeletal filaments or due to the incompressible nature of the cytoplasmic fluid.  The morphological transitions arising from volume constraints have long been a subject of interest, and the details can be found in an excellent review by Seifert ~\cite{Seifert:1997p1058}.


\section{Overview of analytical and computational methods}\label{sec:analyt-comp-methods}
The elastic free-energy functional can be minimized using analytical methods for membranes with well-defined geometries to gain insight into the structural and statistical properties of the system. A variety of theoretical techniques can be used in the analysis, and the choice of the method is dictated by the phenomenon investigated. The class of problems can broadly be classified into (a) determination of equilibrium properties and (b) understanding the dynamics of membranes with and without hydrodynamics. For sake of simplicity, we will use a planar membrane, parameterized using the planar Monge gauge, to illustrate each of these methods. 

In the planar Monge gauge, the surface of the membrane is described with respect to the $x-y$ plane at $z=0$ by the position vector ${\bf R}=(x,y,h(x,y))$, where $h(x,y)$ is the height of the membrane  at position $(x,y)$ on the reference plane. For simplicity, we will perform our analysis  on eqn.~\eqref{eqn:can-Helf-gen}, with $\sigma=0,\,  {\cal K}_A=0,\,$ and $\kappa_G=0$. The mean curvature of a planar membrane, which has been derived in appendix~~\ref{app:diffgeo-planar} using the differential geometry methods described in appendix~ ~\ref{app:diffgeo}, is given by,

\begin{equation}
2H=\frac{\nabla^{2}h}{\sqrt{1+\left(\nabla h \right)^{2}}}.
\end{equation}

Using this form of mean curvature in eqn.~\eqref{eqn:can-Helf-gen},  for the parameters given above,  the elastic energy of the membrane in the Monge Gauge has the form

 \begin{equation}
\label{eq:nonlinhelfMonge}
\mathscr{H}_{\rm sur}=\frac{\kappa}{2}\int d{\bf x} \,\sqrt{1+\left(\nabla h \right)^{2}} \left(\frac{\nabla^{2}h}{\sqrt{1+\left(\nabla h \right)^{2}}}-C_{0}\right)^{2}.
\end{equation}

\subsection{Energy minimization and stationary shapes}\label{sec:helf-minimize}
The planar membrane will take the shape that minimizes its bending energy. Hence the shape of the planar membrane that exhibits pure bending can be determined by minimizing the linearized form of eqn.~\eqref{eq:nonlinhelfMonge} with respect to height $h$,
\begin{equation}
\label{eq:minwrh}
\frac{\delta \mathscr{H}_{\rm sur}}{\delta h}= \frac{\kappa}{2}\,\nabla^{2} \left(\nabla^{2}h-C_{0} \right)=0.
\end{equation}
Hence the minimum energy conformations of the membrane are  those satisfying the differential equation $\nabla^{2}(\nabla^{2}h-C_{0})=0$ for the given boundary conditions. This implies that all height profiles for the membrane with curvature $\nabla^{2}h=C_{0}$ are possible solutions.

\subsection{Membrane dynamics}\label{sec:membdynamics}
Equilibrium shape analysis introduced in section~~\ref{sec:helf-minimize}, is useful in determining the long-time-scale behavior of a membrane in response to a perturbation. The short-time-scale relaxation of the membrane can be studied by analyzing the equations of motion for the membrane, which for the Monge gauge is given by the dynamical equation ~\cite{Chaikin:1995td},
\begin{equation}
\frac{\partial h({\bf x})}{\partial t}=- \int  \Gamma({\bf x},{\bf x}^{'}) \left \{\frac{\delta \mathscr{H}_{\rm sur}}{\delta h({\bf x}^{'})} +\eta({\bf x},t) \right \} d{\bf x}^{'}.
\label{eqn:eom-tdgl}
\end{equation}
The hydrodynamic kernel $\Gamma({\bf x},{\bf x}^{'})$ captures the long-range interactions between different regions of the membrane, mediated by the surrounding fluid,  and $\eta(t)$ is the noise term with $\langle \eta({\bf x},t) \rangle=0$ and $\langle \eta({\bf x},t)\eta({\bf x}^{'},t^{'}) \rangle = 2k_{B}TA_{p}\Gamma({\bf x},{\bf x}^{'})\delta({\bf x}-{\bf x}^{'}) \delta(t-t^{'})$ ~\cite{ReisterGottfried:2007ep}. For a membrane in the Monge gauge  (with $C_{0}=0$ and assuming $\sqrt{g}=1$) and using the energy given by eqn.~\eqref{eq:nonlinhelfMonge}, the dynamical equation becomes,
\begin{equation}
\frac{\partial h({\bf x},t)}{\partial t}=- \int  \Gamma({\bf x},{\bf x}^{'}) \left \{ \kappa \nabla^{4}h({\bf x}^{'},t) +\eta({\bf x},t) \right \} d{\bf x}^{'}.
\label{eqn:Monge-dynamics}
\end{equation}

The above integro-differential equation becomes amenable to theoretical analysis when represented in Fourier space. Defining $h({\bf k},t)=\int d{\bf x} \exp(-i\,{\bf k}\cdot{\bf x}) h({\bf x},t)$, eqn.~\eqref{eqn:Monge-dynamics} can be written in Fourier space as,
\begin{equation}
\frac{\partial h({\bf k},t)}{\partial t}=  \Gamma({\bf k}) \left \{ -\kappa{k}^{4} h({\bf k},t) +\eta({\bf k},t) \right \}.
\label{eqn:Monge-FSBD-dynamics}
\end{equation}
For an almost planar membrane, $\Gamma({\bf k})$ can be derived from the Stokes equation and is given by $\Gamma({\bf k})=1/{(4\eta|{\bf k}|)}$, where $\eta$ is the fluid viscosity. See reference ~\cite{Seifert:1997p1058} for a complete discussion of the method and its applications. This form of the elastic energy forms the basis for Fourier space Brownian dynamics ~\cite{Lin:2004p256001,Brown:2008cf}; see section~~\ref{sec:fsbd}. \\

The static and dynamical properties of the membrane are well represented by dynamical height correlation function given by ~\cite{Seifert:1997p1058},
\begin{equation}
{\bf S}_{\bf kk^{'}}(t,t^{'})= \left \langle h({\bf k},t)h({\bf k^{'}},t^{'}) \right \rangle =\frac{k_{B}T}{\kappa k^{4}} \exp(-\gamma_{{\bf k}} (t-t^{'})) \delta(k-k^{'}),
\end{equation}
where $k=|{\bf k}|$ and $\gamma_{k}=\kappa k^4 \Gamma(k)$. In the limit $t-t^{'} \rightarrow 0$, ${\bf S}_{\bf kk^{'}}(t,t^{'})$ reduces to the static correlation function,
\begin{equation}
 C_{\bf kk^{'}}=\frac{k_{B}T}{\kappa k^{4}} \delta(k-k^{'}),
\label{eqn:hqhmq}
 \end{equation}
 which can also be derived from eqn.~\eqref{eq:nonlinhelfMonge}. In this case, the amplitude of each undulation mode scales as $k^{-4}$, and it can be seen that the intensities of the long-wavelength modes (small $k$) dominate the fluctuation spectrum. In our description, we have considered a tensionless membrane, whereas non-zero tension gives rise to undulation modes whose spectrum scales as $k^{-2}$ for $k \leq \sqrt{\sigma /\kappa}$ and as $k^{-4}$ for $k > \sqrt{\sigma /\kappa}$. The crossover from the tension-dominated regime ($k^{-2}$ scaling)  to the bending-dominated regime ($k^{-4}$ scaling) defines a tension-dependent length scale $l_{\rm tension} = \sqrt{\kappa / \sigma}$. For deformations with radius of curvature smaller that $l_{\rm tension}$, the bending energy term dominates the Hamiltonian and for those with radius of curvature much larger than $l_{\rm tension}$, the interfacial tension dominates. The typical value of this length scale in the cellular context (assuming $\kappa= 20 k_BT$ and $\sigma = 30 \mu N/m$) is $l_{\rm tension} \sim 50 \, {\rm nm}$. At this length scale, the interfacial tension can compete with the bending energy to influence the mean shape of the membrane as well as the nature of the undulations. Additionally, we note that other length scales can also be relevant depending on the ensemble. For example, for the closed membrane ensemble discussed in section~~\ref{sec:ensemble}, one can define $l_{\rm pressure} = (\kappa / \Delta P)^{1/3}$, which in the cellular context assumes a typical value of  10~nm (assuming a typical value of $\Delta P \sim 10^6 {\rm Pa}$).
\subsection{Surface of evolution formalism}
\label{revol}
\begin{figure}[h]
\centering
\includegraphics[width=7.5cm]{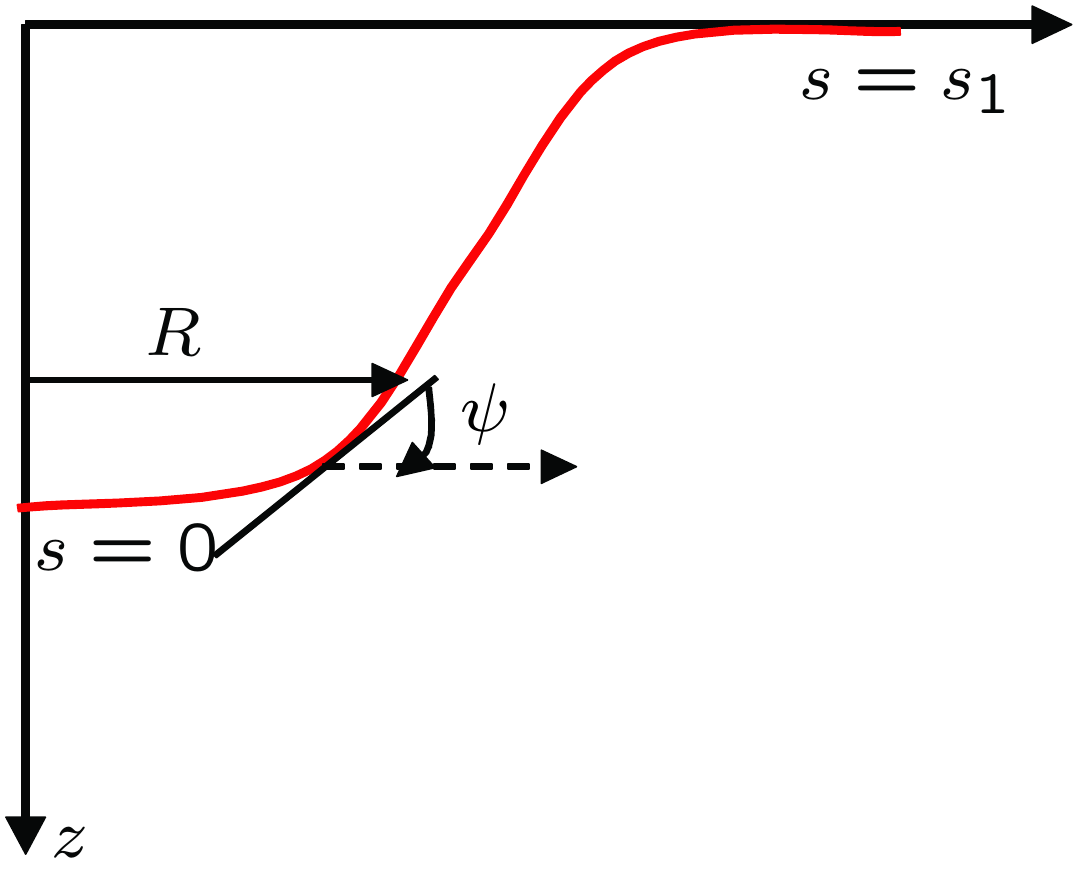}
\caption{Schematic of a membrane profile that shows the different variables used in the surface of evolution formalism.}
\label{fig:revol}
\end{figure}

The analytical techniques discussed above are extremely useful but can only handle membrane geometries that can be readily parametrized in the small-slope limit, (i.e. $| \nabla h | <<1$). 
In addition to the studies in the small-slope approximation, analytical approaches have also been used to examine the behavior of axisymmetric vesicular structures with large curvatures (i.e. curvatures beyond the small slope limit). A notable example is the study of a membrane tether pulled out from a spherical vesicle ~\cite{Evans:1990he,Evans:1996fs,Derenyi:2002kx,Powers:2002jg,Allain:2004ix}.  The analytical  approach, introduced in secs.~\ref{sec:helf-minimize} and ~\ref{sec:membdynamics}, breaks down when the membrane shape becomes nonaxi-symmetric and also when thermal fluctuations are accounted for. Here, we describe one such method in detail. The surface-of-evolution approach ~\cite{Zhongcan:1989ue,Julicher:1994bk,Julicher:1996co,Seifert91} to model the membrane at equilibrium is useful in considering large (albeit axi-symmetric) deformations, including those in which $h$ is a multivalued function of $r$. We consider a generating curve $\gamma$ parametrized by arc length $s$ lying in the $x-z$ plane.  The curve $\gamma$ is expressed as (see Fig. ~\ref{fig:revol})
 \begin{equation}
 \gamma (0,s_{1}) \rightarrow \mathbf{R}^3
 \gamma(s)=(R(s),0,z(s))
 \end{equation} 
where $s_1$ is the total arc length, which is not known a priori. This generating curve leads to a global parameterization of the membrane expressed as
 \begin{eqnarray}
 X:(0,s_{1})\times(0,2\pi) \rightarrow \mathbf{R}^3\\
 X(s,u)=(R(s)\cos(u),R(s)\sin(u),z(s))
 \end{eqnarray}
where $u$ is the angle of rotation about the z-axis.  With this parameterization, the mean curvature $H$ and the Gaussian curvature $G$ are given as follows;
\begin{eqnarray}
2H&=&-\frac{z'+R(z'R''-z''R')}{R}\quad {\rm and}\\
G&=& -\frac{R''}{R},
\end{eqnarray}
where the prime indicates differential with respect to arc-length $s$. The expressions obtained above for the mean curvature and the Gaussian curvature are cumbersome. To simplify them, an extra variable $\psi$, where $\psi(s)$ is the angle between the tangent to the curve and the horizontal direction, is introduced which leads to the following two geometric constraints:
\begin{eqnarray}
R'&=&\cos(\psi(s))\\
z'&=&-\sin(\psi(s))
\end{eqnarray}
These two constraints lead to the following simplified expressions for the mean curvature and the Gaussian curvature:
\begin{eqnarray}
2H&=&\psi'+\frac{\sin(\psi(s))}{R(s)}\\
G&=&\psi'\frac{\sin(\psi(s))}{R(s)}
\end{eqnarray}
Starting with the membrane energy $\mathscr{H}_{\rm sur}$ defined by
\begin{equation}
\mathscr{H}_{\rm sur} = \int_0^{2\pi}\int_0^{s_1}\left\{\frac{\kappa}{2}(2H-C_{0})^2+ \kappa_G G +\sigma\right\} dA
\end{equation}
where $dA$ is the area element given by $Rdsdu$, and substituting for $H,G$, we obtain the following expression for $\mathscr{H}_{\rm sur}$:
\begin{equation}
\mathscr{H}_{\rm sur} = \int_0^{2\pi}\int_0^{s_1}\left\{\frac{\kappa}{2}\left(\psi'+\frac{\sin(\psi(s))}{R(s)}-C_{0}\right)^2+ \kappa_G \psi'\frac{\sin(\psi(s))}{R(s)} +\sigma\right \}R\,ds\,du.
\end{equation}
We now proceed to determine the minimum-energy shape of the membrane. The condition that specifies the minimum-energy profile is that the first variation of the energy should be zero. That is:
\begin{equation}
\delta \mathscr{H}_{\rm sur}=0,
\end{equation}
subject to the geometric constraints $R'=\cos(\psi(s)),z'=-\sin(\psi(s))$.  These constraints can be re-expressed in an integral form as follows:
\begin{eqnarray}
	\int_{0}^{s_1}\left \{ R'-\cos(\psi(s)) \right \}ds=0\\
	\int_{0}^{s_1} \left \{ z'+\sin(\psi(s)) \right \} ds=0
\end{eqnarray}
Introducing Lagrange multipliers, we solve our constrained optimization problem as follows. We introduce the Lagrange function  $\nu,\eta$ and minimize the quantity $F$:
\begin{align}
F=&\int_0^{2\pi}\int_0^{s_1}\left\{\frac{\kappa}{2}\left(\psi'+\frac{\sin(\psi(s))}{R(s)}-C_{0}\right)^2+ \kappa_G \psi'\frac{\sin(\psi(s))}{R(s)} \sigma \right\}R\,ds\,du\notag\\ 
&+ \nu \int_{0}^{s_1}R'-\cos(\psi(s))ds+ \eta \int_{0}^{s_1}z'+\sin(\psi(s))ds.
\end{align}
Since the integrand of the double integral is independent of $u$, $F$ simplifies to:
\begin{align}
\label{eq:eqno1}
F=&2\pi \int_{0}^{s_1}\Biggl\{ \frac{\kappa R}{2}\left(\psi'+\frac{\sin(\psi(s))}{R(s)}-C_{0}\right)^2+ \kappa_G \psi'\sin(\psi(s))+\sigma R \notag\\
&+\nu \left(R'-\cos(\psi(s))\right)+\eta\left(z'+\sin(\psi(s))\right)\Biggr\} ds
\end{align}
The minimization problem is then expressed as:
\begin{equation}
\delta F=0.
\end{equation}
The resulting Euler-Lagrange equations and applications of this approach in studying membrane tethers in GUV experiments and nucleation of vesicles mediated by curvature-inducing and force-mediating proteins on cell membranes can be found in the published literature ~\cite{Derenyi:2002kx,Gao05, Liu06, Liu:2006fc, Agrawal:2010eu}.

\subsection{Direct numerical Minimization}

Complex shapes that are both  axisymmetric and non-axisymmetric can be studied using direct numerical simulations that use well-known numerical methods to compute the elastic energy and curvature forces in the membrane.

Here the continuous membrane is discretised into a computational mesh, and the energy and forces are computed from the conformation of the mesh, which evolves with time. A key requirement for the applicability of this method is that the membrane shape should be continuous and derivatives can be computed everywhere. Using the computed values of $\mathscr{H}_{\rm sur}$ and $-\nabla \mathscr{H}_{\rm sur}$, the state of the membrane can be evolved using a suitable minimization technique or by integrating the equation of motion given in eqn.~\eqref{eqn:eom-tdgl}.

Surface evolver ~\cite{Brakke:1992tn}  is a powerful numerical minimization software package that can be used to determine the zero temperature shapes of symmetric and non-axisymmetric vesicular membranes. This tool can simultaneously minimize energy contributions from surface tension, curvature energy, and gravitational energy. Further, in this approach, the total energy can be minimized subject to constraint on the external variables $A$, the surface area, and $V$, the enclosed volume, and in effect, one determines the membrane shape minimizing the energy given in eqn.~~\eqref{eqn:can-Helf-gen}.

The applications of this approach in studying the interaction of the membrane with nanoparticles of different shapes can be found in the published literature ~\cite{Dasgupta:2013iz,Dasgupta:2014hr}. Extending these analyses of non-axi-symmetric shapes to thermally undulating systems requires the use of other numerical methods, which are discussed next. We do this in two stages: first, we discuss one numerical method to incorporate thermal undulations in small slopes, and then, we proceed to describe methods handling undulations in the non-small-slope limit.

\subsection{Fourier space Brownian dynamics (FSBD)}
\label{sec:fsbd}
For a planar membrane with periodic boundaries, the starting point for FSBD is the discretization and forward integration of eqn.~\eqref{eqn:Monge-FSBD-dynamics} as:
\begin{equation}
 {\vec h}({\bf k},t+\Delta t) = \vec{h}({\bf k},t) +  \left \{ -\Lambda({\bf k}) \cdot \vec{h}({\bf k}) + \Gamma(k) \cdot \vec{\eta}({\bf k},t)\right \} \Delta t,
\label{eqn:Monge-FSBD-dynamics-disc}
\end{equation}
with $\Lambda({\bf k})=\kappa k^{4} \Gamma(k)$ being a diagonal matrix. The shape profile of the periodic membrane at every time can be obtained by an inverse Fourier transform of $h(\bf{k},t)$. In the Fourier space, the noise $\eta({\bf k})$, which is a complex number,  obeys
\begin{equation}
\langle \eta({\bf k,t}) \rangle=0 \quad {\rm and} \quad \langle \eta({\bf k,t}) \eta({\bf k^{'},t^{'}}) \rangle= 2k_{B}TA_{p}\Gamma^{-1}({\bf k})\delta({\bf k}-{\bf k}^{'}) \delta(t-t^{'}),
\end{equation}
and is drawn from a Gaussian distribution with zero mean and variances\textemdash $2k_{B}TA_{p}\Gamma({\bf k})$ and $k_{B}TA_{p}\Gamma({\bf k})$ for the real and imaginary parts, respectively. At every time step, the forces are calculated in real space, and the shape profile is computed in the Fourier space and inverted to get the final shape of the membrane. In addition to planar membranes, this model has also been used to study protein mobility on fluctuating surfaces, and the effect of cytoskeletal pinning on membrane dynamics ~\cite{Lin:2004eg,Lin:2005iv,Lin:2006ic,Brown:2007ip,Sigurdsson:2013cx}.
 
\subsection{Dynamically triangulated Monte Carlo methods.} \label{sec:dtmc}
The numerical approach described above deals with smooth geometries and is hence a zeroth-order approximation to the natural state of bilayer configurations. Membrane shapes are naturally irregular since they are susceptible to thermal fluctuations, and a rigorous analysis of these shapes should also involve  higher order terms. Monte Carlo based  methods account for the effect of thermal undulations and hence are extremely useful in studying membranes in their canonical ensemble.  Triangulated random surface models have been extensively used in high-energy physics, mainly in Euclidean string theory ~\cite{Polyakov:1981p207,David:1985p303,Kazakov:1985p295,Stella:1987p561}. In the context of membranes, these techniques were first used to study the crumpling of self-avoiding tethered membranes ~\cite{Kantor:1986p3356,Kantor:1987p3357}. Fluid membranes were first studied by Ho and Baumg{\"a}rtner ~\cite{Ho:1990p295,Ho:1990p5747} using the method of dynamically triangulated Monte Carlo (DTMC), wherein the triangulation map of the membrane is dynamic. In this representation, the membrane surface is discretized into a set of $N$ vertices that form $T$ triangles and $L$ links, with surface topology defined by the Euler number $\chi=N+T-L$. The elastic energy of the membrane $\mathscr{H}_{\rm sur}$ can estimated from the orientation of the triangles as,
\begin{figure}[H]
\centering
\includegraphics[height=2.5in]{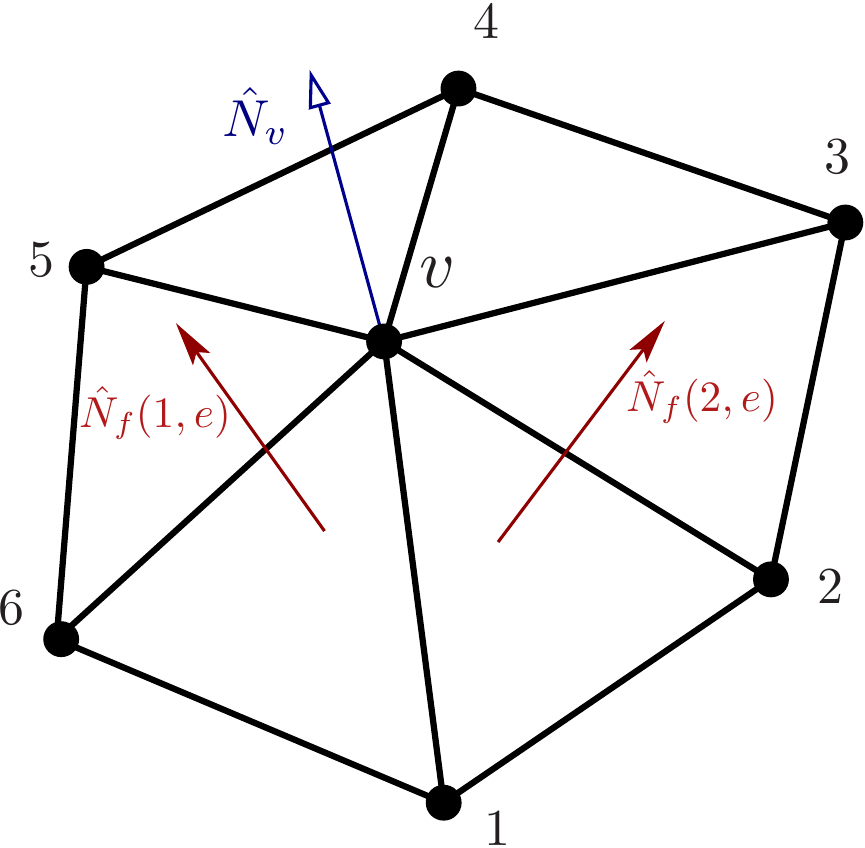}
\caption{\label{fig:patch} A patch of a triangulated membrane depicting the one-ring neighbourhood around a vertex $v$. The edge $e$ connects vertices $v$ and $1$, and is shared by the faces $f_{1}(e)$ and $f_{2}(e)$. The normal to the surface at vertex $v$ is given by $\hat{N}_{v}$. } 
\end{figure}

\begin{equation}
\label{eq:Gompdisc}
\mathscr{H}_{\rm sur}=\lambda_{b} \sum_{v} \sum_{\{e(v)\}} \left \{1-\hat{N}_{f}(1,e) \cdot \hat{N}_{f}(2,e)\right\}-\Delta p \ V.
\end{equation}
As illustrated in Fig.~\ref{fig:patch}, $\{e(v)\}$ denotes the set of links vertex $v$ makes in its one-ring neighbourhood. In general, $\hat{N}_{f}$ denotes the normal to any face $f$, whereas $\hat{N}_{f}(1,e)$ and $\hat{N}_{f}(2,e)$ explicitly represent the normals to faces $f_{1}$ and $f_{2}$ sharing an edge $e$. $V$ is the volume enclosed by the surface. $\lambda_{b}$ in eqn.~\eqref{eq:Gompdisc} and $\kappa$ in ~\eqref{eqn:can-Helf-gen} are related to each other in a geometry-dependent manner. For example, $\lambda_{b}=\sqrt{3}\kappa$ for a sphere and   $\lambda_{b}=2 \kappa/\sqrt{3}$ for a cylinder ~\cite{Piran:2003}. %

 In this discretization technique, the squared mean curvature at a vertex $v$ is approximated in an implicit manner, as given in eqn.~\eqref{eq:Gompdisc}. This calculation does not involve the computation of the principal curvatures $c_{1}(v)$ and $c_{2}(v)$ . Recently, an alternate technique to compute principal curvatures on triangulated surfaces was introduced by Ramakrishnan et al. ~\cite{Ramakrishnan:2010hk} (see Appendix ~\ref{geom-quant} for details), which allows one to represent the elastic energy as,
\begin{equation}
\mathscr{H}_{\rm sur}=\frac{\kappa}{2}\sum_{v=1}^{N} \left\{{c_{1}(v)+c_{2}(v)} \right\}^{2}A_{v}-\Delta p V.
\label{eqn:dirdisc-Helf}
\end{equation}

 The equilibrium properties of a self-avoiding membrane are determined by analyzing the total partition function,
\begin{equation}
Z(N,\kappa,\Delta p)=\frac{1}{N!} \sum_{\{\mathscr{T}\}} \prod_{v=1}^{N} \int d\vec{x}(v) \exp \left \{-\beta \left[\mathscr{H}_{\rm sur}\left (\{\vec{X}\},\{\mathscr{T}\}\right )+ V_{SA}\right ]\right\} \,\,\,\,\,
\label{eqn:Zfluidsurf}
\end{equation}
$V_{SA}$ is the self-avoidance potential, usually chosen to be the hard sphere potential. $\vec{X}$ is the position vector of all vertices in a triangulated surface, and $\mathscr{T}$ is the corresponding triangulation map.   The temperature of the system is expressed in units of $\beta=1/k_{B}T$, and the integral is carried over all vertex positions and summed over all possible triangulations. A tuple, $\eta=[ \{\vec{X}\},\{\mathscr{T}\}]$, represents a state of the membrane in its phase space. In the case of Monte Carlo (MC) studies, a change in state, $\eta \rightarrow \eta^{'}$, is effected by means of Monte Carlo moves, the rules of which correspond to importance sampling~~\cite{Frenkel:2001}. The time in MC simulations is expressed in units of Monte Carlo steps(MCS).

\begin{figure}[H]
\centering
\includegraphics[height=3.0in,clip]{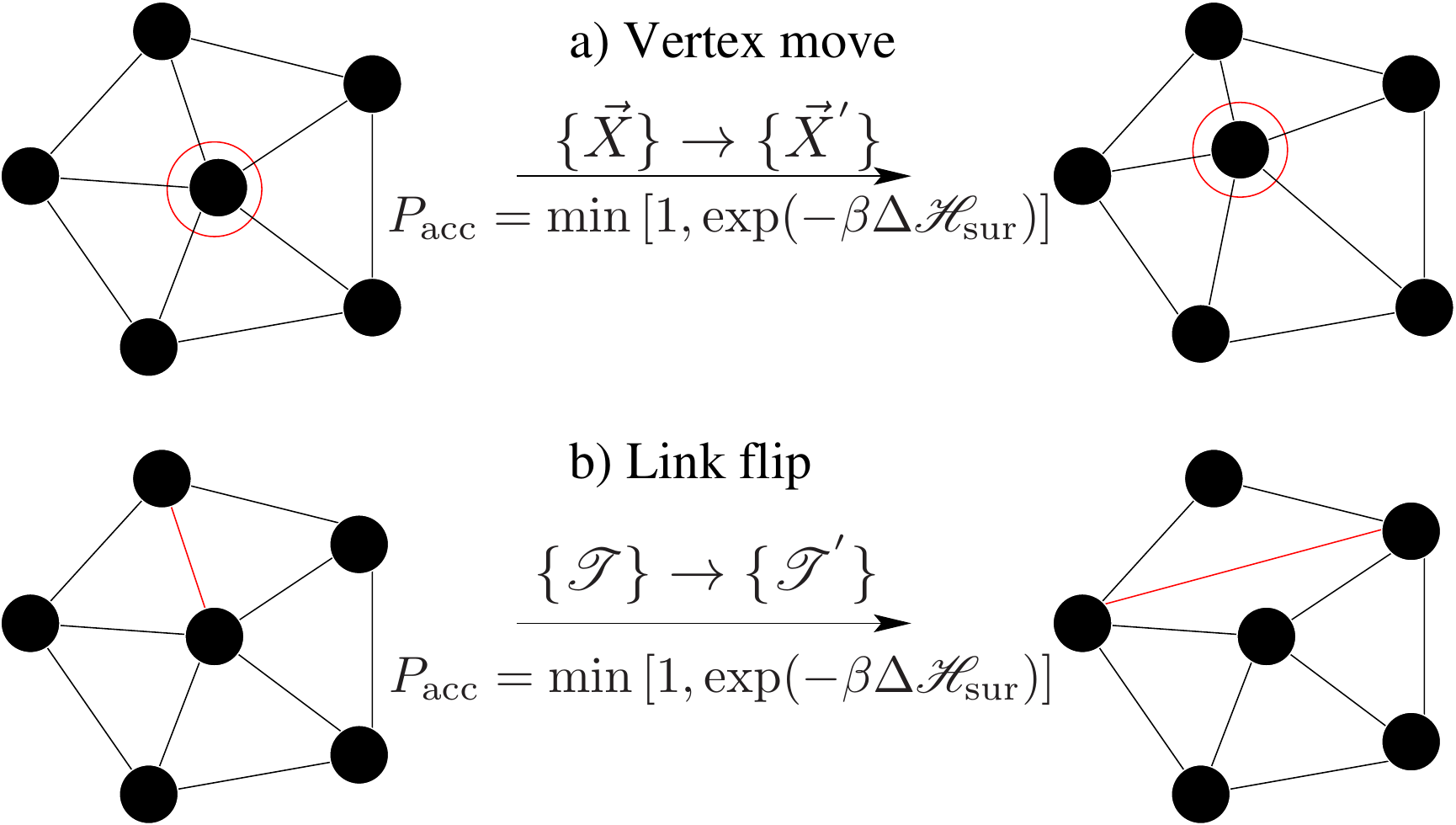}
\caption{\label{fig:ransurf-mcsmoves} Monte Carlo moves involved in the equilibrium simulations of a fluid random surface. (a) A vertex move emulates thermal fluctuations in the membrane, and (b) a link flip simulates fluidity in the membrane.}
\end{figure}

A membrane quenched to a particular microstate ($\eta$) relaxes to its equilibrium conformation mainly through thermal fluctuations and  in-plane diffusion. A Monte Carlo step captures these modes by performing $N$ attempts to displace a randomly chosen vertex and $L$ attempts to flip a randomly chosen link, as in Fig.~\ref{fig:ransurf-mcsmoves}. The rules of importance sampling and details of each move are as follows:
 
{\it (a) Vertex move :} The vertex positions of the surface are updated, $\{\vec{X}\} \rightarrow \{\vec{X}^{'}\}$, by displacing a randomly chosen vertex within a cube of side $2\sigma$ around it, with fixed triangulation $\{\mathscr{T}\}$. As a result, the old configuration of the membrane $\eta=[ \{\vec{X}\},\{\mathscr{T}\}]$ is updated to a new configuration $\eta^{'}=[ \{\vec{X}^{'}\},\{\mathscr{T}\}]$. The total probability of this MC move  obeys the  detailed balance condition given by,
\begin{equation}
\omega(\eta \rightarrow \eta^{'}) P_{\rm acc}(\eta \rightarrow \eta^{'})=\omega(\eta^{'} \rightarrow \eta) P_{\rm acc}(\eta^{'} \rightarrow \eta).
\end{equation}
Choosing the attempt probability $\omega$  for the forward transition ($\eta \rightarrow \eta^{'}$) and backward transition ($\eta^{'} \rightarrow \eta$) to be equal, $\omega(\eta \rightarrow \eta^{'})=\omega(\eta^{'} \rightarrow \eta)=(8\sigma^{3}N)^{-1}$, we get the probability of acceptance as
\begin{equation}
\label{eq:metropolis}
 P_{\rm acc}(\eta \rightarrow \eta^{'})={\rm min}\left\{ 1,\exp \left [-\beta \Delta \mathscr{H}_{\rm sur}(\eta \rightarrow \eta^{'}) \right ] \right \},
\end{equation}
which is the well-known Metropolis scheme~~\cite{metropolis:1953p1087}. $\sigma$ defines the maximum displacement of the vertex and is chosen appropriately, so that the acceptance of vertex moves is close to $50\%$.

{\it (b) Link flips :} An edge shared between two triangles is flipped to link the previously unconnected vertices of the triangles. Such a move changes the triangulation map from $\{\mathscr{T}\} \rightarrow \{\mathscr{T}^{'}\}$, in the process of which it changes the neighborhood of some vertices, which is effectively a diffusion. With fixed vertex positions, the old and new configurations in this case are $\eta=[ \{\vec{X}\},\{\mathscr{T}\}]$ and $\eta^{'}=[ \{\vec{X}\},\{\mathscr{T}^{'}\}]$, respectively. The attempt probability for flipping a link is given by $\omega(\eta \rightarrow \eta^{'})=\omega(\eta^{'} \rightarrow \eta)=(L)^{-1}$, and the acceptance probability is as in eqn.~\eqref{eq:metropolis}. 

 The thermodynamic properties of the fluid membrane computed using the DTMC approach can be found in published studies ~\cite{Kroll:1992jb,Gompper:1995fga,Ramakrishnan:2010hk,Paulose:2012dl}.

\subsection{Particle-based models} \label{sec:membmodel-particlebased}
Field-based continuum models, described above, assume the presence of a self-assembled membrane, and hence are extremely useful in understanding the membrane response to an imposed curvature field or fluctuations in its environment. This model becomes too complicated to handle when one needs to investigate the role of many other physical parameters like topological fluctuations and hydrodynamics. Meshless, particle-based mesoscopic models are powerful alternatives in this regime. In this approach, the microscopic structure of the membrane is coarse grained into particles, whose lengths are in the mesoscale, that interact via well-defined interparticle potentials. In this framework the membrane structure is formed due to the self-assembly of the coarse grained particles and alleviates many of the limitations seen in the mesh representation of the membrane. However, the formulation of interaction potentials that can reproduce both the elastic and transport coefficients of the membrane at the mesoscale is still less understood. 

The earliest particle-based mesoscopic model was proposed by Drouffe et. al. ~\cite{Drouffe:1991}, in which the spherical coarse grained particle has a hydrophobic region surrounded by two hydrophilic regions. The repulsive hard sphere potential between the particles ensures incompressibility, whereas an orientation-dependent potential ensures flexibility and bending.  A variety of mesoscale particle models\textemdash namely, the EM-DPD model ~\cite{Ayton:2002p3357}, where the bending modulus is related to the imposed bulk strain modulus; the rigid and flexible rod approximation of lipids ~\cite{Noguchi:2001bn,Noguchi:2002iv,Farago:2003jf,Cooke:2005p4710}; the spherocylinder representation ~\cite{Brannigan:2004fr}; EM2 model based on both the bending rigidity and stretching modules ~\cite{Ayton:2006ht}; and variants of the Drouffe model ~\cite{Noguchi:2006pre,Kohyama:2009p3334,Yuan:2010ww}\textemdash have been proposed. In this section, we will focus on the EM2 model based on eqn.~\eqref{eqn:can-Helf}  introduced by Ayton et al. ~\cite{Ayton:2006ht}, which has been extended to include the presence of curvature-inducing proteins in reshaping membranes ~\cite{Ayton:2007p3485,Ayton:2009gw,Cui:2009p2746,Cui:2011p1271,Lyman:2011p10430}. 

\begin{figure}[H]
\centering
\includegraphics[width=15cm,clip]{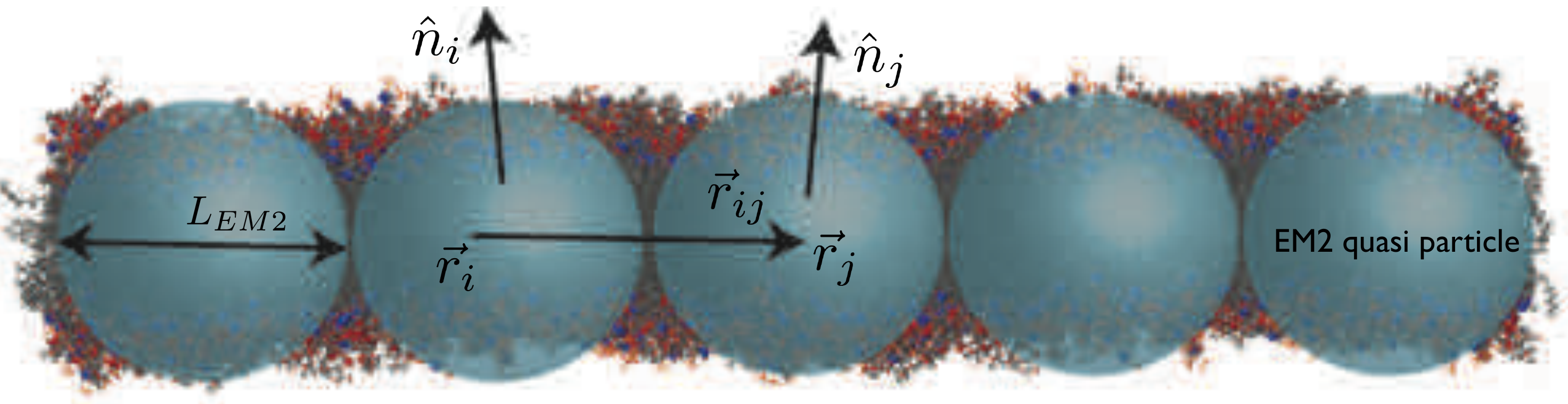}
\caption{\label{fig:EM2model} Illustration of the EM2 model where each quasiparticle is of size $L_{EM2}$. Two particles $i$ and $j$, with respective positions $\vec{r}_i$ and $\vec{r}_{j}$, interact via  pair potentials that depend on their separation, $\vec{r}_{ij}$, and normal orientations, $\hat{n}_{i}$ and $\hat{n}_{j}$. ({\em Bilayer image is courtesy of Ryan Bradley.})}
\end{figure}

In the EM2 model, a bilayer membrane is coarse grained into a set of quasi particles of length $L_{EM2}$, as shown in Fig.~\ref{fig:EM2model}, with area density $\rho_{A}=h\rho_{0}$; here $h$ and $\rho_{0}$ are, respectively, the thickness and volume density of the membrane.
The effective interaction of two EM2 particles ~\cite{Ayton:2006ht}, positioned at $\vec{r}_{i}$ and $\vec{r}_{j}$ and separated by $\vec{r}_{ij}$, has contributions from a bending potential ($\cal{H}_{\rm bend}$) and a stretching potential ($\cal{H}_{\rm stretch}$). These potentials have the form,
\begin{eqnarray}
{\cal H}_{\rm bend}&=&\frac{1}{2} \sum_{i=1}^{N} \sum_{j\in i}^{N_{c,i}} \frac{8\kappa}{\rho_{A}N_{c,i}} \frac{\left\{ (\hat{n}_{i}.\hat{r}_{ij})^{2}+(\hat{n}_{j}.\hat{r}_{ij})^{2} \right\}}{|\vec{r}_{ij}|^{2}} \\
{\cal H}_{\rm stretch}&=&\frac{1}{2} \sum_{i=1}^{N} \sum_{j\in i}^{N_{c,i}} \frac{8 \pi (r_{ij}^{0})^{2} h\lambda }{N_{c,i}^{2}} \left\{\frac{r_{ij}}{r_{ij}^{0}}-1 \right \}^{2}.
\end{eqnarray}
The unit normal vectors corresponding to particles $i$ and $j$ are $\hat{n}_{i}$ and $\hat{n}_{j}$, respectively. Summation over $j$ runs over all the $N_{c,i}$ particles found with a cutoff distance $r_{c}$ around particle $i$. $\lambda$ is the stretch modulus of the membrane, computed from all atom non-equilibrium MD simulations, and $r_{ij}^{0}$ is the distance between the particles in the undeformed state of the membrane. The properties of the EM2 model and its coupling to mesoscopic solvents like WCA and BLOBS ~\cite{Ayton:2004jr} can be found in the original article by Ayton  et al. ~\cite{Ayton:2006ht}. Details of the model and its use in studying phenomena related to protein-induced membrane remodeling  will be discussed in detail in  Sec.~~\ref{sec:em2}.


\subsection{Molecular and coarse grained approach for modeling membranes} {\label{sec:molCG}}

\begin{figure}[H]
\centering
\includegraphics[width=6in]{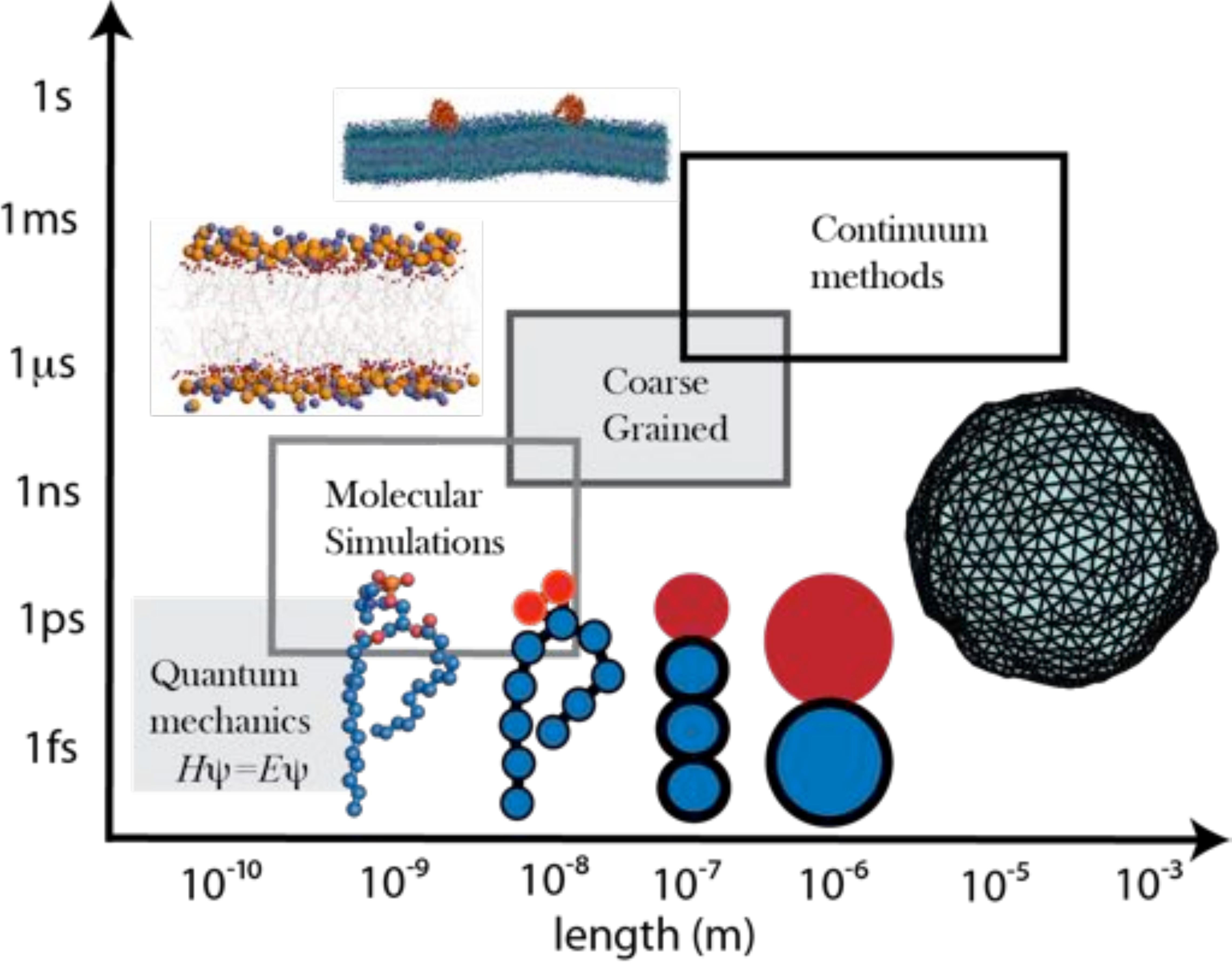}
\caption{\label{fig:multi-scale} DMPC lipid molecules described at multiple resolutions, ranging from the nanoscale to the macroscopic scale. All atoms in a lipid are explicitly represented  at the nanoscale, whereas in the mesoscale, where coarse grained simulations are used, only a reduced number of atomic coordinates are used to represent the lipid molecule. At length and time scales that fall into the continuum regime the relevant parameters are determined from the structural and physical properties of the membrane.} 
\end{figure}

The roles of lipids and protein-lipid interactions can be described/quantified by theoretical models at multiple resolutions of length and timescales, as depicted in Fig.~~\ref{fig:multi-scale}. 
In the previous sections, we have focused on the continuum scale, which is the major emphasis in this article. However, we note that models closer to atomic or electronic resolution remain true to the biochemistry of interactions and can map the sensitivity of protein-lipid interactions to the underlying chemical heterogeneity (e.g., the effect of protonation or ion binding on the specificity interaction of proteins with lipids). Hence, we conclude this section by providing a brief overview of molecular simulation methods.

Molecular dynamics (MD), where a molecule can be resolved to the level of an atom, is the most popular simulation technique and is widely used in the study of material systems. In this approach, the atoms in a molecule interact among themselves and other atoms through a set of bonded and non-bonded potentials. These potentials are in turn derived from extensive quantum calculations of the molecule involved. MD simulations are extremely useful in understanding various physical and biological processes in membranes that span over nanoscopic length and time scales, and they can also be used to understand many chemical events with average life times of the order of femto seconds. Technical details of MD simulations of lipid membranes can be found in  several reviews on this topic ~\cite{Tobias:1997gw,Tieleman:1997ve,Feller:2000bi,Marrink:2009uo}. The spectrum of membrane-related problems studied using molecular  simulations ranges from properties of self-assembled, single- and multi-component lipid bilayers ~\cite{vanderPloeg:1982by,Heller:1993wk,Tu:1996wi,Tieleman:1996dz,Berger:1997ut,Marrink:1998fi,PasenkiewiczGierula:2000jc,Mashl:2001jm,KMurzyn:2001uj,Marrink:2003iu,ChristoferHofsass:2003ub,deVries:2004vc,Leontiadou:2004ba}, fusion of vesicles ~\cite{Marrink:2003wn}, and interactions of membranes with sugars ~\cite{Sum:2003bn}, proteins ~\cite{Edholm:1995ty,Pitman:2005gf,Sansom:2005en,Lindahl:2008fc,Meyer:2008p1851}, and peptides ~\cite{Lague:2005ti}. 

The advent of faster computer processors and efficient algorithms has scaled membrane system sizes amenable to molecular simulations, with explicit atomistic representation, as shown in Fig.~\ref{fig:multi-scale} for a DMPC molecule, from 128 lipids ~\cite{vanderPloeg:1982by} to hundreds of thousands at present ~\cite{Marrink:2009uo}. However, a simple estimate shows that the total number of atoms, inclusive of both lipids and water, involved in a MD simulation of a micron-sized vesicular membrane is of the order of $10^{11}$. The number of degrees of freedom involved is an over representation of the membrane  if one is only interested in studying its physical properties at length and time scales far separated from the atomistic scales. As an alternative, coarse graining techniques can be used  to retain only the degrees of freedom that are relevant to the time and length scales under investigation, which in turn can reduce the number of atomic units used in the simulations.

The concept of coarse graining has been used extensively used in constructing a macroscopic theory of material systems from their microscopic degrees of freedom ~\cite{Chaikin:1995td}. An example of coarse graining in soft-matter systems is the blob representation of polymers, where a segment of the polymer is represented as an independent entity of a given size ~\cite{Doi:1988ug}. Using a similar strategy, the multiple microscopic variables associated with a set of atoms in a lipid molecule can be replaced by fewer macroscopic variables that capture the essential physics of these atoms: for instance, the spatial position of the atoms in the head group of a lipid molecule can be substituted with a coarse grained (CG) particle that has an average position and a characteristic size. The various coarse graining strategies differ in how these macroscopic variables are handled: is the CG particle placed at the center of mass or around the position of the largest atom? What macroscopic properties does the choice of interaction potential between the CG particles intend to capture? Coarse grained lipid models have evolved rapidly from  simple bead-spring  representations ~\cite{Goetz:1998p7397,Steven:2004p11942} to more sophisticated models with the choice of interaction potentials representing functions that bridge certain membrane properties between the atomistic and coarse grained picture ~\cite{Lyubartsev:2005de,Marrink:2009uo}.

There are three well established coarse graining strategies for simulations of soft-matter systems\textemdash namely, {\it structure based}, {\it force based} and {\it potential-energy based}. The key differences between these models lies in the  choice of the interaction potential between the CG beads, which are chosen to reproduce some specific physical properties observed in experiments or computed through detailed molecular simulations. In the case of {\it structure-based} models, the CG  potentials are designed to minimize the difference between the coarse- and fine-grained radial distribution functions. The resulting model retains the chemical structure of the atomistic system and can be an excellent choice if the study aims to investigate  the structural properties of the molecule. Such a model has also been used in scale hopping and adaptive resolution simulations ~\cite{Praprotnik:2008p646,Peter:2009tx,Peter:2010p65}. Structure-based CG methods have been employed in simulating lipid membranes ~\cite{Murtola:2004ky,Murtola:2007ja} and their interactions with proteins ~\cite{Shih:2006bi,Arkhipov:2008p3007,Yin:2009p255}. On the other hand,  the CG potentials for the {\it force-based} ~\cite{Izvekov:2005iy,Noid:2008dc} method are chosen to preserve the total force as one moves up from the atomistic description to a coarse grained description of the membrane.

The conservation of the underlying structure and forces in the CG methods described above, do not necessarily imply conservation of thermodynamic averages of various observables. Alternatively, the {\it free-energy-based} MARTINI force field by Marrink and co-workers for lipids ~\cite{Marrink:2004p750,Marrink:2007bw} and proteins ~\cite{Monticelli:2008ia} is another widely used parameterization to semiquantitatively reproduce a wide distribution of thermodynamic data. The resulting force field is thus maximally transferable to novel systems with different temperatures, pressures, and compositions. It has found wide application in modeling vesiculation ~\cite{Marrink:2003iu}, membrane fusion  ~\cite{Marrink:2003wn} and pore formation ~\cite{Leontiadou:2004ba,Lindahl:2008fc}, peptide-lipid interactions, protein-gated ion channels, and lipid raft formation. Several mechanisms of curvature interactions on membranes have been carried out using particle-based simulation methods~~\cite{Reynwar07,Shillcock:2008ko,Shillcock:2013fu,Vacha:2012bd}.

CGMD models can also be used in combination with MD models in bottom-up approaches of systematically coarse-graining the atomistic description. Bridging techniques that seamlessly integrate two distinct length  scales in this category include the integrated molecular mechanics/coarse-grained or MM/CG approaches ~\cite{klein-etal-08}. Another goal of molecular/coarse-grained modeling can also be to rigorously bridge the molecular scale with the continuum scale, so that information is effectively passed between scales in a self-consistent manner. 

The Canham - Helfrich model in eqn.~\eqref{eqn:can-Helf} is only an idealization of the free-energy of the interface. The true bilayer system can only be accurately described by invoking the statistical mechanics of inhomogeneous phases ~\cite{Chaikin:1995td}.  The MD and CGMD models can in principle capture the nature of such inhomogeneities when a computational approach to modeling is sought. However, a theoretical approach for handling the density inhomogenieties can also be pursued within the purview of classical density functional theory, wherein the free-energy can be described as a unique functional of the spatially varying density~~\cite{Henderson:1992vp}. In this context, the membrane thickness is finite with an inhomogeneous density profile, and the conjugate variables to the strain or deformation fields are defined through the pressure tensor. For an inhomogeneous system, the pressure tensor $\mathbf{P}$ at position $\mathbf{r}$ can be expressed in a tensorial form of the virial equation, and it can be split into a kinetic part, $\mathbf{P_K}$, derived from the kinetic energy and a potential part, $\mathbf{P_U}$, derived from the potential energy $U(\{\mathbf{r_i}\})$ ~\cite{schofield-82,Walton:1985to},
\begin{equation}\label{p-tensor1}
\mathbf{P}(\mathbf{r})= \mathbf{P_K}(\mathbf{r}) + \mathbf{P_U}(\mathbf{r}), \, \textrm{where}  \, \mathbf{P_K} (\mathbf{r}) = k_B T \rho(\mathbf{r})  \mathbf{I}.                                      
\end{equation}
Here, $U$ is the potential energy function, $\{\mathbf{r_i}\}$ represents the vectorial coordinate of atom $i$, $\rho$ is the density, $k_B$ is the Boltzmann constant, and $\mathbf{I}$ is the identity matrix. For pair-wise additive potentials $U=(1/2) {\sum_{i}\sum_{j,\,{i \ne\j}}} u(ij)$, where $u(ij)=u(\mathbf{r_{ij}})$ with $\mathbf{r_{ij}}=\mathbf{r_i}-\mathbf{r_j}$, $\mathbf{P_U}(\mathbf{r})$ is given by:
\begin{equation}\label{p-tensor2}
\mathbf{P_U} (\mathbf{r}) = - \frac{1}{2}      \left \langle {\sum_i \sum_{j, \, i \ne j}} \frac{\partial u(ij)}{\partial \mathbf{r_{ij}}} \, {\int}_{C_{ij}} \, d\mathbf{l} \, \delta(  \mathbf{r} - \mathbf{l} ) \, \right \rangle.                       
\end{equation}
Here, $C_{ij}$ is a contour from $\mathbf{r_i}$ to $\mathbf{r_j}$. Different conventions to choose the contour yield different (non-equivalent, see however ~\cite{Rossi:2009wh}) expressions for the pressure tensor~~\cite{irving-50,Walton:1985to,schofield-82}. For a review comparing the different methods, see references ~\cite{Varnik:2000bd,Venturoli:2006p2740}. Applications of these methods for the characterization of interfacial stresses in lipid bilayers are available ~\cite{Ollila-2009, Goetz:1998p7397}. \footnotetext{The equilibrium properties of a lipid-bilayer system investigated using the dissipative particle dynamics (DPD) method, which preserves hydrodynamic correlations (rather than regular CGMD), have been analyzed by Mouritsen ~\cite{olemouritsen:2005}, including the investigations of the anisotropic stress distributions; see eqns.~\eqref{p-tensor1} and ~\eqref{p-tensor2}, which are defined by following the Irving-Kirkwood convention~~\cite{irving-50} leading to the evaluation of parallel and perpendicular stresses in the bilayer ~\cite{schofield-82,Goetz:1998p7397,ask-besold-05}. Several applications of DPD in the study of the dynamics of bilayers, interactions with small and large molecules with bilayers, interactions between particles mediated by bilayers, and vesicle fusion have been reviewed in the literature~~\cite{Venturoli:2006tv, Grafmuller07,Shillcock:2008ko,Shillcock:2013fu}.}

The correspondence between equation for $\mathscr{H}_{\rm sur}$ (eqn.~\eqref{eqn:can-Helf-gen}), which describes the membrane as an infinitesimally thin interface, and the Hamiltonian ${\cal H}=\sum_i \frac{1}{2} m_i | {\mathbf{\dot{r}_i}}^2 | +U(\{\mathbf{r_i}\})$ governing $\mathbf{P(r)}$, which provides a molecular view of the membrane interface of finite width, can be derived by equating the corresponding free-energy changes or work done in deforming the interface in the two descriptions~~\cite{Safran:1999ty, Szleifer:1990we}. The expression for the surface tension is given by (here, we choose the direction normal to the bilayer as the $z-$axes and  calculate the pressure averages over the  lateral plane~~\cite{schofield-82,goetz-98,ask-besold-05}):   
\be
\gamma = \int_0^{L_z} dz [P_N(z)-P_L(z)]\:,
\ee
where $P_N(z)=\langle P_{zz}(z)\rangle$ and $P_L=\frac{1}{2}(\langle P_{xx}(z)+P_{yy}\rangle)$ are the average normal and lateral pressure components of the tensor $\mathbf{P}$, respectively. The expressions for bending stiffness constants are given by~~\cite{Safran:1999ty, Szleifer:1990we}:
\be
\kappa C_0 = \int_0^{L_z} dz (z-z_{\textrm{ref}}) [P_N(z)-P_L(z)]\:,
\ee
\be
\kappa_G = \int_0^{L_z} dz (z-z_{\textrm{ref}})^2 [P_N(z)-P_L(z)]\:.
\ee
These expressions were derived assuming that deforming the interface conserves the total volume. In general, the expressions relating the elastic constants and the pressure tensor are ensemble dependent; for example, analogous expressions in the constant projected area ensemble are derived by Farago~~\cite{Farago:2004vj}. Alternatively, the equivalence between $\mathscr{H}_{\rm sur}$ and ${\cal H}$ can be established by requiring that the height-height fluctuation spectrum in both models are in agreement. This equivalence provides a powerful technique to estimate membrane elastic constants in both microscopic and macroscopic simulations.  Such an approach has been used recently to determine the macroscopic bending modulus $\kappa$ and Gaussian modulus $\kappa_{G}$ from molecular simulations ~\cite{Hu:2012et,Hu:2012bi,Hu:2013gj}.


\section{Modeling membrane proteins as spontaneous curvature fields} \label{chap:sponcurvmodels}


\subsection{Curvature induction in membranes\textemdash intrinsic and extrinsic curvatures}\label{cellmemb-curvact}

Based on eqn.~~\eqref{eqn:hqhmq}, the natural thermal undulations are pronounced at large wavelengths and supressed at small wavelengths due to equipartition of energy. Thus, it is highly unlikely that curvature at the nanoscale can arise spontaneously due to bending-mediated thermal undulations. The regular but complex membrane morphologies observed in cellular systems is expected to be influenced by additional interactions defining an energy landscape associated with curvature.  Experimental observations suggest that the process of systematic deformation of cellular membranes may be local, and the macroscopic reorganization of membranes is likely driven by cooperative interactions.  

In a bilayer membrane, both the number and composition of lipid molecules in each monolayer can vary. This difference is an inherent source of spontaneous curvature in membranes. Consider a bilayer membrane, with its monolayers uniformly separated by a  distance $\delta$ and an equilibrium area difference equal to $\Delta A_{0}$ ( in the case of a planar membrane $\Delta A_{0}=0$). When the area difference between the monolayers deviates from its equilibrium value to $\Delta A$, the membrane develops a spontaneous curvature,  which can be expressed in terms of their material properties, using the area difference elasticity (ADE) model, as ~\cite{Miao:1994p112},

\begin{equation}
\label{eqn:expr-czero}
C_{0}=\delta \frac{\partial G}{\partial (\Delta A)}+\frac{\kappa_{\rm nl}}{\kappa_{\rm l}} \frac{\pi}{A_{\rm mid}\delta} \left(\Delta A-\Delta A_{0} \right).
\end{equation}
 
Here $\kappa_{\rm nl}$ and $\kappa_{\rm l}$ are the non-local and local bending rigidities, with $\kappa_{\rm nl} \approx \kappa_{\rm l}$ for most phospholipid bilayers ~\cite{Miao:1994p112}. $G=\int dA\, H^{2}$ is a dimensionless measure of the squared curvature of the surface, and $A_{\rm mid}=\int dA$ is the area of the reference surface. In the presence of an area asymmetry between the monolayers,  which implies non-zero spontaneous curvature ($C_{0} \ne 0$), the equilibrium morphology of the membrane would assume shapes with $2H=C_{0}$. 

\begin{figure}[H]
\subfigure[]{
\centering
\begin{minipage}{0.4\textwidth}
\includegraphics[height=1.9in]{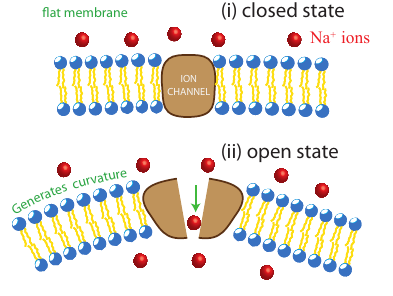}
\label{fig:channel-states}
\end{minipage}
}
\subfigure[]{
\begin{minipage}{0.4\textwidth}
\centering
\includegraphics[height=1.75in]{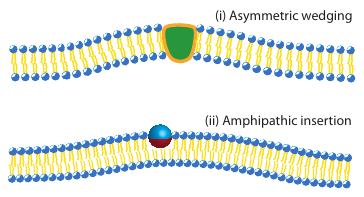}
\label{fig:wedge}
\end{minipage}
}
\caption{\label{fig:wedging} {\bf (a)} States of a symmetric transmembrane ion channel embedded inside a membrane bilayer. The membrane is flat and has no induced-curvature when the channel is in the closed state (i) and when the channel transitions to the open state the membrane is spontaneously curved under the influence of the induced-curvature (ii). {\bf (b)} Curvature induction due to the insertion of an asymmetric transmembrane protein (i) and due to the partial insertion of an amphipathic entity as shown in (ii).}

\end{figure}

A variety of mechanisms can induce an area asymmetry in the monolayers of the membrane. Common examples include  (a) flip-flop of lipids between the monolayers, (b) mixing of lipids with the solvent, (c) presence of lipids whose geometry differs from that of the bulk membrane, and (d) insertion of a non-lipid molecule into the bilayer\footnotetext{if the inserted molecule is symmetric, the insertion should span only one monolayer to induce curvature; see Fig.~\ref{fig:channel-states}}. Such an induced spontaneous curvature can either be local or be felt over the entire membrane domain. In case of (c) and (d) the strength and magnitude of $C_{0}$ is dependent on the geometry and composition of the lipids and on the macromolecules in the membrane. Uncatalyzed inter leaflet lipid transduction is an extremely slow process, and lipid molecules have diminishingly small solubility in aqueous solvents and hence can be considered to be insoluble for all practical purposes. For these reasons, the spontaneous curvature contributions (a) and (b) are negligible in our time scales of observation. It should also be noted that in experiments on model membranes, large deformations in membrane shapes are observed by changing environmental conditions like temperature and salt concentration.  However, the role of these variables as drivers of membrane shape change can be neglected since they fluctuate weakly in physiological conditions.

Under physiological conditions, the spontaneous curvature induced by area asymmetry is considered too weak ~\cite{Shibata:2009p643,Kozlov:2010p301} to induce large deformations in the membrane shapes. Zimmerberg and Kozlov ~\cite{Zimmerberg:2006p510} have shown that it takes large area asymmetry to generate vesicular shapes observed in cellular systems; e.g. their calculations show that the area asymmetry required to bud off vesicles of size $100-200$nm,  which matches  transport vesicles  in biological cells, from a cellular size GUV is of the order of $\Delta A=0.1A_{\rm mid}$. Known lipid compositions of biological membranes do not support such a large difference in the area of the monolayers, which in turns calls for other external mechanisms of curvature induction. In the past decade, a vast body of experimental evidence has highlighted the role of membrane-associated macromolecules, mainly proteins, in remodeling the curvature of membranes. Proteins induce membrane curvature through the two key mechanisms discussed below.

\noindent{\em 1. Curvature generation by wedging:  } The simplest example for protein induced spontaneous curvature can be seen in the case of membrane-spanning proteins generally called transmembrane proteins. This family of proteins represents a large class of macromolecules like pore proteins aquaporin, ion channels, ion pumps,  enzymes, and light harvesting complexes. The membrane-deforming properties of the transmembrane proteins is demonstrated in Fig. ~\ref{fig:channel-states} using a sodium ion channel. The functionality of the ion channel is dependent on its conformation, which switches between a closed and open state, as shown in Fig. ~\ref{fig:channel-states}, (i) and (ii), respectively.  Mismatch in the size and shape of the proteins and lipids tends to distort the membrane around a transmembrane protein to induce an area asymmetry between the upper and lower leaflet of the bilayer. This distortion is minimal when the  channel is in the inactive state (closed) resulting in a flat membrane. When the channel opens to allow the diffusion of sodium ions, the surrounding membrane is more distorted, and as a result, the vicinity of the channel becomes curved, as shown in Fig. ~\ref{fig:channel-states}(b). In the event of cell signaling, when all the channels open up in unison, the induced area asymmetry can be large enough to induce spontaneous curvatures required for systemic membrane deformations ~\cite{Wiggins:2004ir}. A recent experimental report on curvature sensing and partitioning ability of potassium channels reinforces the above notion~~\cite{Aimon:2014if}. \smallskip

\noindent{\em 2. Curvature induction by scaffolding:  } Proteins that localize and interact with the surface of the membrane are called peripheral proteins. Many of these proteins, in addition to their definite functional role, also interact with the membrane. These interactions result in membrane curvature induction as observed in BAR-domain-containing proteins, dynamin superfamily proteins ~\cite{Praefcke:2004p3309}, nexins, ENTH-domain-containing epsins, reticulon-containing Dp1/Yop1 ~\cite{Shibata:2008p544,Hu:2008p3289}, C2-domain-containing synaptotagmin, and clathrin complexes ~~\cite{Campelo:2008p3288}. The underlying cytoskeletons are also known to tether to the membrane and pull out long tethers, whose effects will not be considered here. Some of the peripheral proteins can also induce curvature through the wedging mechanism described earlier. 
\begin{figure}[H]
\centering
\includegraphics[height=3in,clip]{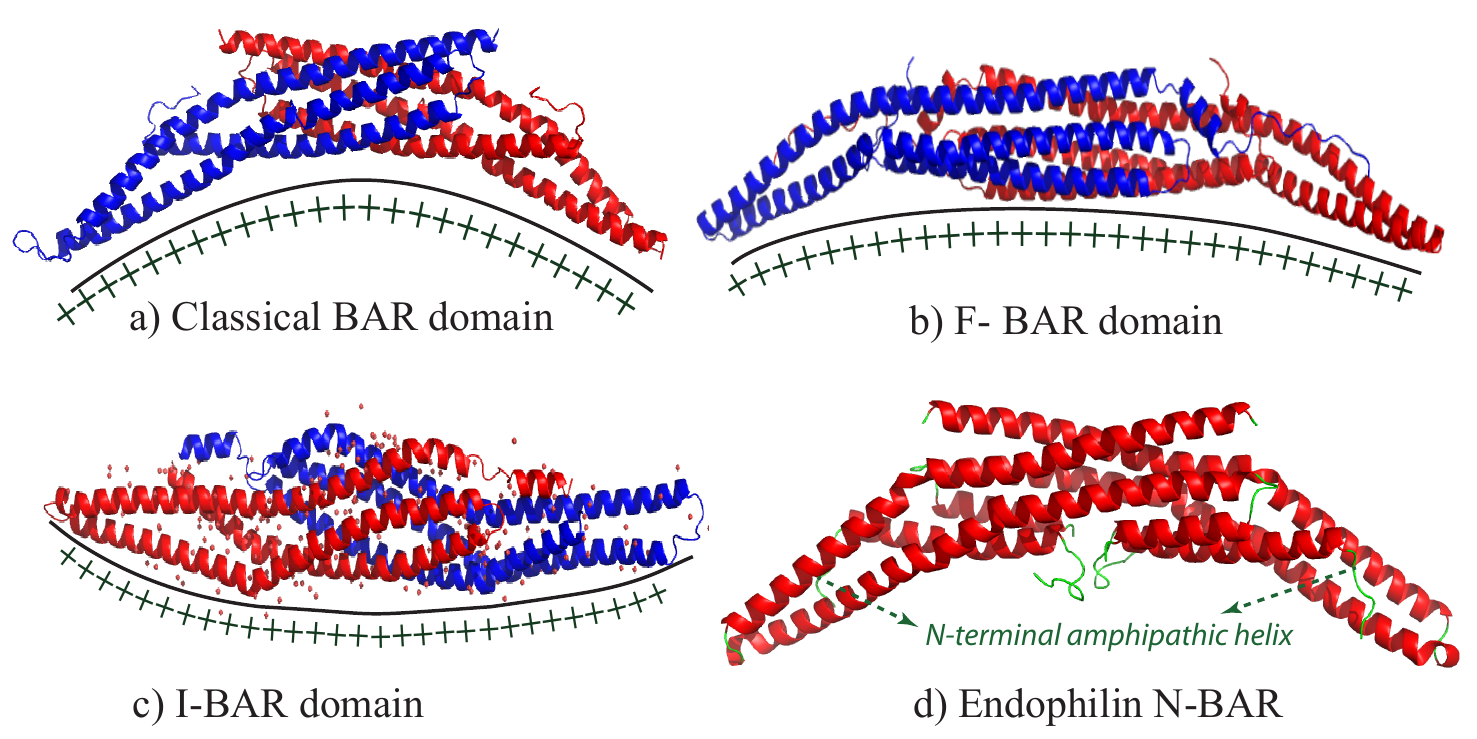}
\caption{\label{fig:BARClass} 
Classes of the BAR domains. Shown are the crystal structures for a classical BAR domain {\bf (a)}, a F-BAR domain {\bf (b)}, and  an I-BAR domain {\bf (c)}: the monomeric units that constitute the dimeric molecule are shown in blue and red, and the positive charges on each protein are localized to the curved surface; this surface being concave for the classical- and F-BAR and convex for an I-BAR. {\bf (d)} The N-terminal amphipathic helices in an N-BAR domain, whose dimeric unit is shown in red.
}
\end{figure}
The study of BAR-domain-containing proteins and their role in membrane reshaping is currently an active area of research. The BAR (Bin/Amphiphysin/Rvs) domain, a dimeric molecule made from $260-280$ amino acids arranged primarily as $\alpha$-helices, is conserved in many proteins of the endocytic pathway ~\cite{Dawson:2006p113} and conserved in most organisms ranging from yeast to humans ~\cite{Farsad:2001p3596,Peter:2004p3597,Habermann:2004p3595}. 
These domains are known to generate high curvature due to their interactions with the membrane ~\cite{Frost:2009p157} and also by recruiting other proteins like Dynamin, Arp 2/3 ~\cite{Gallop:2006p306}. The monomeric units are arranged such that the resulting BAR domain has a curved shape, as in Fig.~\ref{fig:BARClass}, that interacts with the underlying lipid membrane. Based  on the motif of the interacting regions, the BAR domains are further classified into classical-BAR, F-BAR,  and I-BAR ~\cite{Frost:2009p157}. Both the classical and F-BAR domains are crescent (or) banana shaped, with the former being more curved than the latter. The shallowness of F-BAR proteins, which is $\sim$ 3 fold smaller than the classical BAR, arises due to a slightly different arrangement of the monomeric units ~\cite{Henne:2007p94,Frost:2007p225,Frost:2008p6}. In the conventional sense, the membrane bending into the extracellular space is said to be positively curved. I-BAR domains, on the other hand, are negatively curved ~\cite{Ahmed:2010p3565}. A special class of the positively curved dimers is the N-BAR domains, characterized by the presence of an amphipathic helix at the  N-terminal of each monomer unit ~\cite{Weissenhorn:2005p3594}, which in shown in Fig.~\ref{fig:BARClass}(d). The presence of the helix enhances the domain membrane interaction. For example, the N-BAR containing Endophilin tubulates liposomes even at lower concentrations where the classical BAR fails to ~\cite{Huttner:2002p1214}. The extra curvature is induced by the amphipathic helix that embeds into a monolayer, to the depth of the glycerol groups, serving as a wedge that generates a spontaneous curvature ~\cite{Gallop:2006p306,Masuda:2006p363} due to area asymmetry in the bilayer.

\begin{figure}[H]
\centering
\includegraphics[height=1.25in,clip]{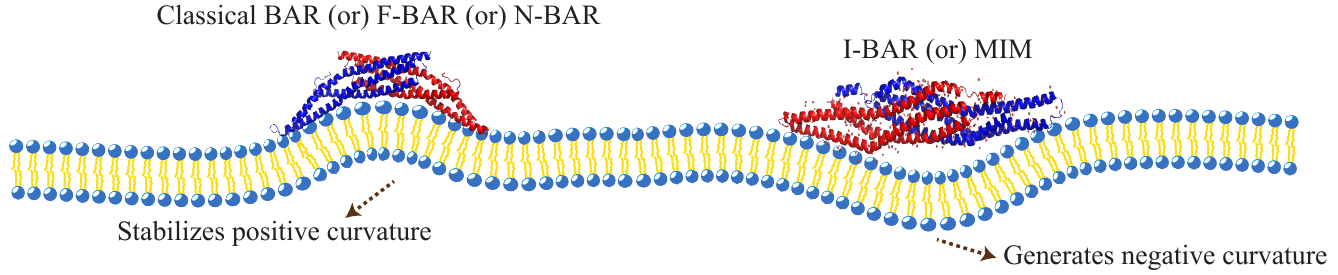}
\caption{\label{fig:bilayerBAR} An illustration of the membrane reshaping behavior of positively and negatively curving BAR domains. A BAR domain inducing positive curvature, like the classical-, F- and N- BAR, induces a positive curvature along its orientation ({\it left panel}), whereas an I-BAR stabilizes negative curvature in the membrane ({\it right panel}).}
\end{figure}

Interaction of the BAR domains with the underlying membrane is primarily electrostatic in nature. The positive charges of the protein are concentrated on the membrane proximal surface of the BAR domains (see Fig.~\ref{fig:BARClass}), which is in turn strongly attracted by the negatively charged lipid heads.  As the protein is  fairly stiff, the membrane responds to the above-mentioned attractive interactions by curving itself to match the curved surface of the BAR domain, but at the cost of the bending energy. The strong electrostatic interactions dominate over the soft elastic energy resulting in a spontaneous deformation of the membrane ~\cite{Zimmerberg:2004p129}. As shown in Fig.~\ref{fig:bilayerBAR}, the presence of a crescent like BAR domain induces a spontaneous positive curvature on the membrane. The degree of deformation depends on the type of BAR domain, the distribution of the cationic residues on the domain, and also on the nature of the lipid molecules making up the membrane.

It has further been observed that the membrane deforms preferentially in particular directions. This is a   key feature that differentiates a scaffolding protein from a transmembrane protein, whose locally induced membrane deformations tend to be isotropic. The directional bending anisotropy of the scaffolding proteins can be mapped to a $n$-atic tangent plane order, which is invariant under rotations by $2\pi/n$. The deformations induced by the amphipathic helices of the N-BAR domain are also directional in nature, hence fitting naturally into this scheme. Computational investigations into the BAR domain-membrane interactions have also established the directional nature of the BAR-domain containing proteins ~\cite{Blood:2006uc,Arkhipov:2008p257,Arkhipov:2009p3009}. More details on the various classes of membrane remodeling proteins, their mechanisms of membrane bending, and the associated biochemical network can be found in various reviews on this topic  ~\cite{Voeltz:2007p1399,Shibata:2009p643,Zimmerberg:2006p510,Zimmerberg:2004p129,Collins:2006p464,McMahon:2005p274,Shibata:2006p311,Shibata:2010p2363,Graham:2010p2433}. \smallskip

The spatial pattern of the spontaneous curvature induced in the membrane through the mechanisms listed above can have entirely different forms. For the purpose of identifying the suitable continuum model for a given system, we classify the spontaneous curvature($C_{0}$) induced by membrane-curving macromolecules (henceforth we refer to them as  curvactants) into two major classes based on the principal radii of curvature $R_{1}$ and $R_{2}$. \medskip

\noindent{\em Isotropic curvactants:} \label{sec:iso-curvactant}  Local membrane deformations induced by this class of proteins/protein complexes and the associated curvature profile have a radial symmetry ($R_{1}=R_{2}=R$). The curvature induced by symmetric transmembrane proteins/inclusions, illustrated in Fig.~\ref{fig:wedging}(a) and (b,i), falls into this category. As will be seen later, the presence of these proteins preserves the rotational symmetry  of the membrane and hence does not involves additional energy contributions due to symmetry breaking ~\cite{Chaikin:1995td}  (i.e. the energy of the membrane in a given conformation remains invariant under change in orientation of the proteins). Adhesive functionalized nanoparticles, like those used in targeted drug delivery applications, can strongly bind to membranes, which in turn indents the membrane in their vicinity leading to an isotropic spontaneous curvature ~\cite{Bahrami:2012gb,Saric:2012hb,Zhang:2012ko}\smallskip

\begin{figure}[H]
\centering
\subfigure[]{
\centering
\begin{minipage}{1.0\textwidth}
\includegraphics[height=2.0in,clip]{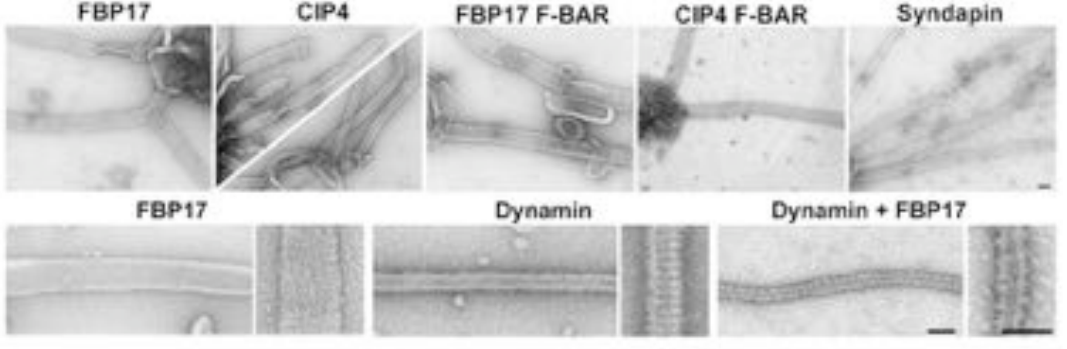}
\label{fig:lipos-tube}
\end{minipage}
}
\subfigure[]{
\centering
\begin{minipage}{0.2\textwidth}
\includegraphics[height=1.5in,clip]{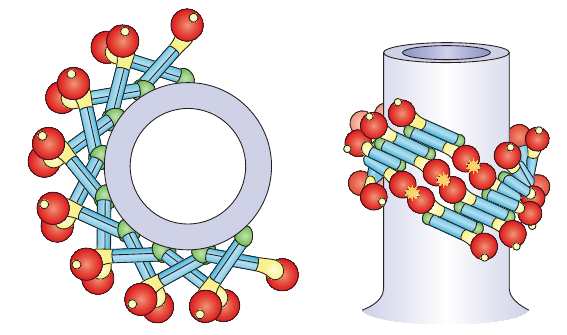}
\label{fig:cryoEM-recon}
\end{minipage}
}
\subfigure[]{
\begin{minipage}{0.35\textwidth}
\centering
\vspace*{10pt}
\includegraphics[height=1.6in,clip]{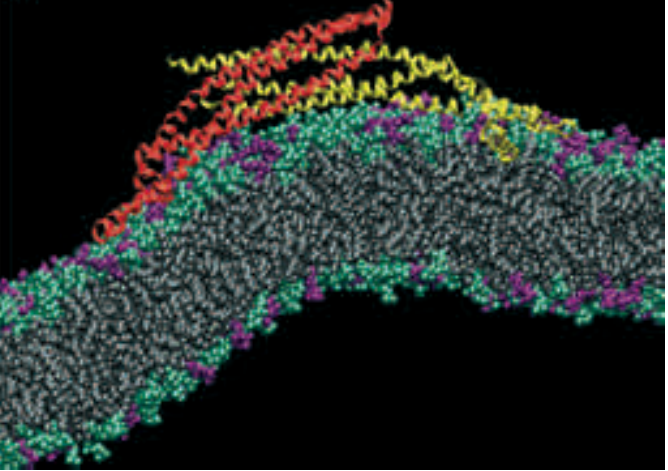}
\label{fig:NBAR}
\end{minipage}
}
\hspace*{10pt}
\subfigure[]{
\begin{minipage}{0.35\textwidth}
\centering
\includegraphics[height=1.7in,clip]{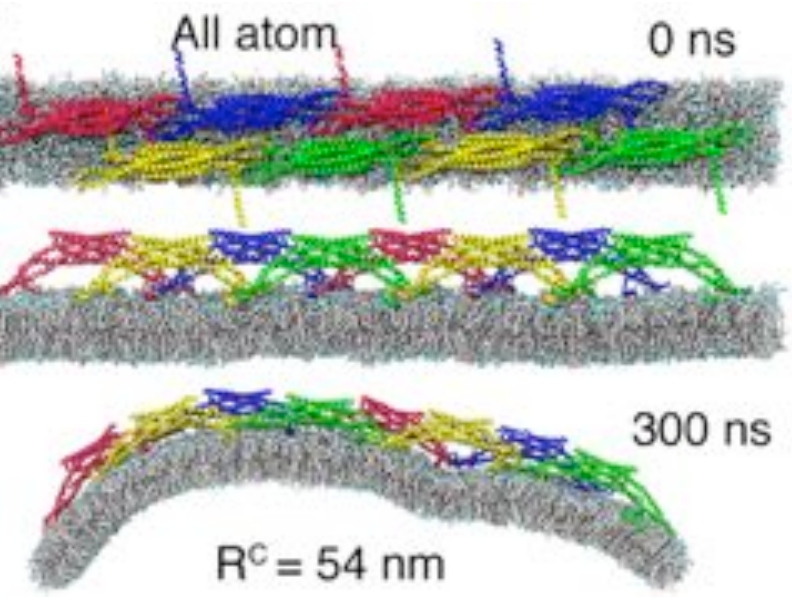}
\label{fig:msBAR}
\end{minipage}
}
\caption{\label{fig:anisotropic-curv} {\bf (a)} Liposome tubulation by F-BAR proteins, Synapdin, and Dynamin {\sf(Reprinted from Dev. Cell, {\bf 9} (6), Itoh et. al., Dynamin and the Actin Cytoskeleton Cooperatively Regulate Plasma Membrane Invagination by BAR and F-BAR Proteins, 791--804, Copyright (2005), with permission from Elsevier)}, {\bf (b)} A cartoon of the helical organization of Dynamin proteins on a tube they constrict. {\bf (c)} Snapshot from a 27 ns molecular dynamics simulation of an N-BAR domain interacting with a DOPC/DOPS membrane {\sf(Image from P. Blood and G. Voth, Direct observation of Bin-Amphiphysin-Rvs(BAR) domain-induced membrane curvature by means of molecular dynamics simulations, Proc. Nat. Acad. Sci. {\bf 103}, 15068-15072 ~\cite{Blood:2006uc} - Copyright (2006) National Academy of Sciences, U.S.A)}  {\bf (d)} All atom simulations of Amphiphysin N-BAR domains\textemdash in its dimeric form and arranged in a staggered conformation\textemdash shows the generation of spontaneous directional curvature {\sf(Reprinted from Structure, {\bf 17} (6), Ying Yin, Anton Arkhipov, Klaus Schulten, Simulations of Membrane Tubulation by Lattices of Amphiphysin N-BAR Domains, 882--892, Copyright (2009), with permission from Elsevier) }. }\end{figure}

\noindent{\em Anisotropic curvactants:} \label{sec:aniso-curvactant}   The curvature profile induced by BAR-domain-containing proteins, dynamin, and exo70-domain-containing proteins are inherently anisotropic ($R_{1} \ne R_{2}$). This anisotropy arises due to a variety of reasons like the secondary structure of the protein, the charge distribution in the membrane-facing domain of the protein, and distribution of lipids around the proteins. Direct experimental/computational evidence, for the anisotropic nature of the curvature profile is hard to obtain since these deformations at the single molecule level are not large enough to be distinguished from those resulting from thermal undulations in the membrane. Alternately, the anisotropic form of the curvature field can be observed in the macroscopic structures stabilized when these proteins interact with a liposome. For instance, liposomes tubulate in the presence of  different F-BARs, syndapin, and dynamin proteins, as shown in Fig.~\ref{fig:lipos-tube}. It is hypothesized that the spontaneous curvature induced by a protein is then related to the radius of the tube it stabilizes. The macroscopic organization of membrane-remodeling proteins on  tubular liposomes are known only for a few proteins like dynamin~\cite{Zhang:2001p3767}\textemdash Fig.~\ref{fig:cryoEM-recon} shows the helical arrangement of dynamin on a tube of radius 25nm, with the pitch of the helix measuring 15 nm ~\cite{Roux:2010p4141,Morlot:2010hr}. Computational studies have also been supportive of the anisotropic nature of many curvature-remodeling proteins. All atom simulations of N-BAR domains~\cite{Blood:2006uc} and BAR domains~\cite{Yin:2009p255}, shown respectively in Figs.~\ref{fig:NBAR} and ~\ref{fig:msBAR}, confirm the presence of directional curvatures in these systems. The latter study~\cite{Yin:2009p255} and others ~\cite{Arkhipov:2008p3007,Blood:2008p1866} also point to the role of cooperativity amongst proteins in reshaping membranes. For a given protein density, it has been shown that the radius of curvature of the deformed membrane depends on the relative orientation between the  proteins. Furthermore, the proteins were required to be on a lattice to facilitate extreme membrane deformations, as seen when a planar membrane folds to a tube ~\cite{Lai:2012hk}.  We characterize the class of anisotropic curvature-inducing proteins by two values of spontaneous curvature\textemdash $C_{0}^{\parallel}=1/R_{1}$ and  $C_{0}^{\perp}=1/R_{2}$\textemdash which are respectively the spontaneous curvatures induced along and normal to the orientation of the long axis of the protein.

\begin{figure}[H]
\subfigure[ Flat bilayer]{
\begin{minipage}{0.32\textwidth}
\centering
\includegraphics[height=1.1in]{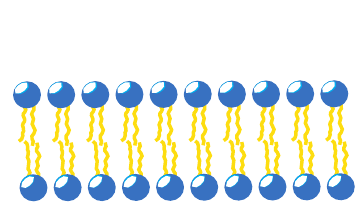}
\label{fig:flatbilayer}
\end{minipage}
}
\subfigure[Tilted phase of bilayer]{
\centering
\begin{minipage}{0.3\textwidth}
\centering
\includegraphics[height=1.1in]{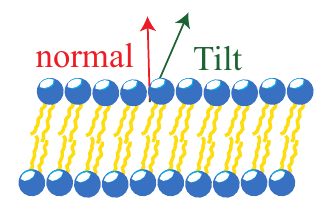}
\label{fig:tilted-phase}
\end{minipage}
}
\subfigure[Dimeric head Gemini surfactant]{
\centering
\begin{minipage}{0.3\textwidth}
\centering
\includegraphics[height=1.1in]{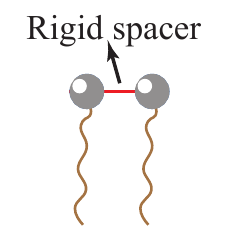}
\hspace*{1cm}
\label{fig:gemini}
\end{minipage}
}
\caption{\label{fig:lipidtilt} Shown in {\bf (a)} is a flat bilayer with negatively charged head group and its hydrophobic tails oriented along the surface normal, in {\bf (b)}  is the tilted phase of a membrane, where the orientation of its tail molecules subtends an angle with the surface normal, and in {\bf (c)} is a Gemini surfactant molecule, with its dimeric heads linked by a rigid spacer.}
\end{figure}

\noindent{\em Intrinsic anisotropy:} Anisotropic behavior in membranes can also be intrinsic, originating mainly from the geometry of the lipid molecules. The hydrophobic part of the lipid is a long molecule and hence can have a multitude of orientations. The effect of chain orientations on the conformation and physical properties of membranes have been known for some time. Of interest is the tilted bilayer phase, where the hydrocarbon chain orients at an angle with respect to the normal defining the vesicle surface, as shown in Fig.~\ref{fig:tilted-phase}. The lipid tilt  normally seen in the liquid ordered $L_{\beta^{'}}$\footnotetext{ The liquid-ordered $L_{o}$ phase shown in fig.~\ref{fig:lipidorg}(c) can further be subdivided into two phases\textemdash namely, $L_{\beta}$ and $L_{\beta^{'}}$ ~\cite{Kranenburg:2005kv}. In the $L_{\beta}$ phase, the lipid tails are oriented along  the membrane normal, whereas they show a finite tilt in the $L_{\beta^{'}}$ phase. \\ }  ~~\cite{Smith:1988p3535,Smith:1990p3537} and  $P_{\beta^{'}}$ phases \footnotetext{ \noindent The liquid-ordered to liquid-disordered transition proceeds through an intermediate phase called the rippled phase ($P_{\beta^{'}}$).} preferentially selects a direction in the membrane that constitutes an in-plane vector field on its surface~~\cite{Tardieu:1973p711}. Membranes with tilted lipids have been observed to stabilize highly curved tubular shapes.

Surface  vector order  is also seen in membranes constituted from lipid molecules with a chiral center, like  1,2-bis-(tricosa-10,12-diynoyl)-sn-glycero-3-phosphocholine ~\cite{Selinger:2001p284}, which is a diacteylene-containing lipid. The presence of chirality induces a spontaneous twist ~\cite{Sarasij:2007p3509,Harris:1999p3510} in the self-assembled structures\textemdash chiral membranes are also commonly associated with tubular shapes. The role of lipid tilt and chirality has been previously recognized, and mean field models that include their contributions have predicted the stability of tubular membranes and helical ribbons~\cite{Helfrich:1988p3004,Nelson:1992p9,Selinger:1993p108,Schnur:1993p1277}.
Signatures of in-plane order have also been observed in membrane systems such as  POPC + water ~\cite{KraljIglic:2002p1533}. It has been observed that the membrane self-assembles  into long tubes,  and the authors have theorized that this arises from the inherent anisotropy of the lipid molecules with respect to the surface normal.
In addition to the hydrophobic chains, the polar head of the lipid can also induce an in-plane order. For instance, Gemini surfactants with dimeric heads group are connected by a rigid spacer, as in  Fig.~\ref{fig:gemini}, whose orientations constitute an in-plane order. The morphologies of the vesicles with random spacer orientations are different from those with spacers in the ordered phase. In contrast to other pure lipid systems, bilayer vesicles of Gemini surfactants have been observed to stabilize tubular structures ~\cite{Oda:1997p124}. 

At this point, it should be noted that the mentioned intrinsic mechanisms can only promote weakly curved regions and hence their role in stabilizing highly curved organelles, like the golgi, and endoplasmic reticulum or in the formation of endocytic vesicles is limited ~\cite{Shibata:2009p643}; the latter would require specialized proteins to perform the job as discussed above.


Different modes by which proteins modulate the curvature of membranes have also been discussed by McMahon et al. ~\cite{McMahon:2005p274}. On the experimental front, direct measurements of bending-mediated force transduction and molecular organization in lipid membranes based on interferometry and fluorescence measurements have been reviewed ~\cite{Groves07,Neto06}.  The dynamics of molecules and the mosaic organization of the plasma membrane and their implications in cellular physiology, primarily focused on studies involving fluorescent labeling and imaging, have also been recently reviewed  ~\cite{Marguet06}. 
Several experimental studies have been carried out to investigate curvature generation and sensing. Sorre and coworkers ~\cite{Sorre:2009do,Sorre:2012if} conducted experimental investigation of the sorting of lipids on a lipid membrane tube (tether) drawn from a giant unilamellar vesicle (GUV) using an optical trap. Curvature sorting of lipids and its influence on the bending stiffness of the bilayer membrane was studied by Tian et al ~\cite{Tian:2009fu,Tian:2009hx}. Dynamic sorting of lipids and proteins has been studied by Heinrich and coworkers ~\cite{Heinrich:2010ej}. These authors observed that nucleation of disordered membrane domains occurs at the junction between the tether and giant unilamellar vesicles. Several theoretical and experimental studies have helped shed light on the curvature-mediated sorting phenomenon ~\cite{Julicher:1996co, Heinrich:2010ej,Seifert:1993bz,Capraro:2010jo,Aimon:2014if}. 

In cell membranes, protein-induced radius of curvature ranges from few a nanometers to a few tens of nanometers depending on the protein and lipid composition of the membrane. For example N-BAR domains stabilize tubular membranes with radius in the range $25-32$ nm~~\cite{Mim:2012je}, whereas dynamin-induced tubes have a radius of $25$~nm ~\cite{Marino:2005gl}. In vitro experiments have reported epsin-induced tubulation of liposomes with a tubule radius of $10$nm~~\cite{Lai:2012hk}. In the following sections, we will discuss how the triangulated surface method and particle-based EM2 method  can be readily extended to include the effect of curvature-inducing proteins. The effect of protein-induced membrane remodeling can be captured in field theoretic models, based on eqn.~\eqref{eqn:can-Helf-gen}, by substituting the protein with a suitable spontaneous curvature field whose magnitude and extent matches the deformation profile of the membrane in the vicinity of the protein. In this section, we will describe two curvature field based models that can be used to study the effect of proteins on the properties of membranes. 

\subsection{Isotropic curvature models}
\label{sec:isotropic}
Most theoretical studies on protein binding focus on protein adsorption to planar lipid bilayers (reviewed in ~\cite{Baumgart:2011en}); these studies were mainly concerned with planar membranes, and curvature effects were not discussed. Reynwar et al. ~\cite{Reynwar07} performed coarse-grained molecular dynamics simulations, which show that once adsorbed onto lipid bilayers, curvature-inducing proteins experience effective curvature-mediated attractive interactions. Jiang and Powers ~\cite{Jiang:2008je} investigated lipid sorting induced by curvature for a binary lipid mixture using a phase-field model. Das and co-workers have investigated the effect of protein sorting on tubular membranes using theoretical techniques ~\cite{Singh:2012gb,Zhu:2012eg}. Using Monte Carlo simulations, Sunil Kumar and coworkers ~\cite{SunilKumar:1999bf,SunilKumar:2001cr,Ramakrishnan:2010hk,Ramakrishnan:2013gl,Ramakrishnan:2012dk} have studied the effects  protein-induced isotropic and anisotropic curvatures have on the shape of vesicles. Using the Monte Carlo method, Liu et al. and Ramanan et al. have investigated  spatial segregation, curvature sensing, and vesiculation in bilayers with curvature-inducing proteins ~\cite{Liu:2012es,Ramanan:2011ds}. 

Models for protein diffusion in ruffled surfaces ~\cite{Gov06} and the simultaneous diffusion of protein and membrane dynamics ~\cite{Divet05, Naji07,Reister07, Atilgan07} have also been reported. In such models, there is only a one-way coupling between the membrane dynamics and the protein dynamics\textemdash i.e., a change in membrane morphology affects the diffusion of the proteins and not the reverse. These studies have been extended to simultaneous protein diffusion and membrane motion models to treat the case of curvature-inducing proteins diffusing on the membrane ~\cite{Weinstein06,Agrawal:2008ff,Sigurdsson:2013cx}. The new aspect introduced in these latter models  is the two-way coupling between the protein and membrane motion. In this case, the membrane morphology not only influences the protein diffusion by presenting a curvilinear manifold, but also presents an energy landscape for protein diffusion. The protein diffusion in turn affects membrane dynamics because the spatial location of the proteins determine the intrinsic curvature functions and hence the elastic energy of the membrane ~\cite{Weinstein06}. This methodology has been utilized in exploring the equilibrium behavior of bilayer membranes under the influence of cooperative effects induced by the diffusion of curvature-inducing proteins ~\cite{Agrawal:2008ff}.

 Isotropic curvature models have been utilized to study the energetics of curvature-inducing protein interactions on a membrane ~\cite{Chou:2001bm,Grabe03, Kim98, Dan94,Wallace05, ArandaEspinoza:1996ux, Berman:1994uv, Dan:1993uv, Goulian:2007wk, Golestanian:1996wj, Goulian:1996td}. Experimental methods to probe such interactions have also been discussed~~\cite{Goulian:1996wu}. By including the effect of protein-membrane interaction as a curvature field, the assumption is that the equilibrium behavior of the system is dominated by the membrane-mediated protein-protein interaction dictated by the strength and range of the curvature field and that small-length-scale interactions (i.e. at the atomic level) are smoothed out. Justification for this assumption has recently been presented by directly parameterizing such a curvature field from molecular dynamics simulations ~\cite{Zhao:2013hi}. In ~\cite{ArandaEspinoza:1996ux}, Aranda-Espinoza et al. employed a combination of integral equation theory to describe the spatial distribution of the membrane-bound proteins and the linearized elastic free-energy model and reported that the interaction (in the absence of thermal undulations) between two membrane-bound curvature-inducing proteins is dominated by a repulsive interaction. Consistent with these published reports, the calculated binding energy (again without thermal undulations) between two membrane-bound proteins interacting through the curvature fields show dominant repulsive interactions governed by the range of the curvature field ~\cite{Agrawal:2008ff}.  Thus, purely based on energetic grounds, the previous analyses have suggested that membrane-deformation-mediated energies tend to be repulsive and should prevent, rather than promote, the formation of protein dimers or clusters.

Kozlov has discussed how the effect of fluctuations can change the repulsive nature of the interactions ~\cite{Kozlov07}. The author's discussion is based on the premise that any membrane protein locally restrains thermal undulations of the lipid bilayer. Such undulations are favored entropically, and so this increases the overall free energy of the bilayer. Neighboring proteins collaborate in restricting the membrane undulations and reduce the total free-energy costs, yielding an effective (membrane-mediated) protein-protein attraction. Indeed, for the linearized free-energy model, computing the second variation of energy (note that at equilibrium, the first variation is zero, whereas the second variation governs the stiffness of the system against fluctuations), we see that the presence of a protein (or equivalently a curvature inducing function) leads to a localized suppression of membrane fluctuations ~\cite{Agrawal:2008ff,Agrawal:2009bt}. As will be discussed later in section~~\ref{sec:freeener}, this calculation has been further verified by using a free-energy method to compute the change in Helmholtz free energy upon the introduction of a curvature field ~\cite{Agrawal:2009bt}. This provides for the possibility of  an entropically mediated protein-protein attraction. The outcome of the interplay between the attractive entropic forces and the repulsive energetic forces is context specific because both have the same dependence on the protein-protein distance and their absolute values differ only by coefficients with similar values. This has been demonstrated by examining the protein-protein pair correlation (spatial and bond-orientational) and through the effect on membrane morphology ~\cite{Agrawal:2008ff}. Indeed, the model predicts that the cooperative effects of membrane-mediated interactions between multiple proteins can drive different morphological transitions in membranes ~\cite{Agrawal:2008ff,Reynwar07,ArandaEspinoza:1996ux,Kozlov07}. This notion of cooperativity is also consistent with the analysis of Kim et al. ~\cite{Chou:2001bm}, who have shown using an energetic analysis that in the zero-temperature limit, clusters of larger than five membrane-bound curvature-inducing proteins can be arranged in energetically stable configurations. It is also worth mentioning for completeness that Chou et al. ~\cite{Chou:2001bm} have extended the energetic analysis to membrane-bound proteins that have a noncircular cross-sectional shape and to local membrane deformations that are saddle shaped (negative Gaussian curvature) and have shown that, in such cases, the interactions can be attractive even without considering fluctuations. Hence, based on the afore-mentioned simulations and analysis of the continuum models, it is hypothesized that attractive interactions between curvature-inducing proteins can result from entropy of membrane undulations ~\cite{Weikl01,Ramanan:2011ds,Agrawal:2009bt,Kozlov07} (the same phenomenon was investigated using particle-based simulations by Reynwar et al.~~\cite{Reynwar07}). Using related continuum methods, preserving bi-directional coupling of protein-induced curvature migration and membrane undulations, the role of adhesive forces as well as anisotropy of curvature fields on membrane-mediated protein interactions have also been reported ~\cite{Lee02,Liu:2012es,Liu:2012ww,Iglic:2007vq,Bohinc:2006vr}. The isotropic curvactants, described in section~~\ref{sec:iso-curvactant}, can be represented by the spontaneous curvature field $C_{0}$, defined in the continuum model described by eqn.~\eqref{eqn:can-Helf-gen}. 

The role of the spontaneous curvature $C_{0}$ in determining the conformational and thermodynamic properties of the membrane has also been studied in a variety of contexts\textemdash curvature-induced instabilities ~\cite{Leibler:1987kg,Leibler:1986kq}, intramembrane- and crystalline-domain-induced budding in multi-component lipid membranes ~\cite{Lipowsky:1992fh,Julicher:1993hg,Julicher:1996co,Kumar:1998jt,Kohyama:2003dg,SunilKumar:2001cr}, effects of lipid packing ~\cite{Hui:1989fh}, and effects of trans-bilayer sugar asymmetry ~\cite{Dobereiner:1999tm}; see chapter by  Gompper and Kroll ~\cite{Piran:2003} for  further details.   

\subsection{Membranes with in-plane order: Model for anisotropic curvature-inducing proteins }\label{chap:nemmemb}
The organelles of a biological cell  have membranes with highly curved edges and tubes, as  seen in the endoplasmic reticulum, the golgi,  and the inner membrane of mitochondria. Tubulation has also been observed, {\it in vitro}, in self-assembled systems of pure lipids~~\cite{Markowitz:1991p3276}. It has been shown  that macromolecules,  which  constitute and decorate the membrane surface, strongly influence the morphology of membranes. For instance,  proteins from the dynamin superfamily  are known to pull out membrane tubes  while  oligomerizing themselves  into a helical coat along the tube~~\cite{Praefcke:2004p3309}. The BAR domain containing proteins in general can  induce a wide spectrum of membrane shapes, ranging from protrusions to invaginations, depending on the geometry and interaction strength of the BAR domain~~\cite{Zimmerberg:2006p510,Voeltz:2007p1399,Shibata:2009p643}.

 The models for membranes with isotropic curvature fields (such as those discussed in section~~\ref{sec:isotropic}) cannot explain the emergence and stability of such highly curved structures. We will show in the remainder of this section (and also in section~~\ref{sec:em2}) that
 a source for anisotropic bending energy will be the minimal requirement to explain the emergence and stability of tubular shapes, which could arise from an in-plane orientational field on the membrane~~\cite{Fournier:1996p488,Fournier:1998p2999}.  In general, the different types of intrinsic and extrinsic in-plane order, discussed in section~~\ref{sec:aniso-curvactant}, can be represented as a $p$-atic in-plane field. This in-plane order, tangential to the surface of the membrane, can capture curvature modulations in the membrane arising from anisotropic membrane inclusions. The $p$-atic field has a rotation symmetry of $2\pi/p$ ; for instance, a nematic field ($p=2$) has a $\pi$ rotational symmetry, and a hexatic field  ($p=6$) has a $\pi/3$ rotational symmetry. In this section, we will only be dealing with nematic in-plane order since this closely resembles the curvature profile induced by a protein interacting with a membrane. As we will discuss below, a nematic field is sufficient to demonstrate the applicability of this model to explain a variety of vesicular morphologies observed in {\it in vitro} experiments involving membrane remodeling proteins like Dynamin~\cite{Hinshaw:2000fi,Praefcke:2004bi}, Epsin~\cite{Voeltz:2006ca}, BAR domains~\cite{Dawson:2006fn}, and Exo70~\cite{Zhao:2013hi}, which are all believed to induce an anisotropic curvature field on the membrane. \footnotetext{As an aside, we note that in addition to the membrane system, this model can also be used to study liquid crystal shells~~\cite{LopezLeon:2011p3971} and nematic elastomers, which are curved soft matter systems with inherent nematic symmetry.} The anisotropic contributions to the energy  can be accommodated by extending the isotropic elastic energy with additional anisotropic terms, obeying all relevant symmetries, as shown below.

\subsubsection{Membrane with in-plane field} \label{sec:inplane-meth}
\begin{figure}[H]
\centering
\subfigure[In-plane field at vertex $v$]{
\begin{minipage}{0.4\textwidth}
\centering
\includegraphics[height=1.5in]{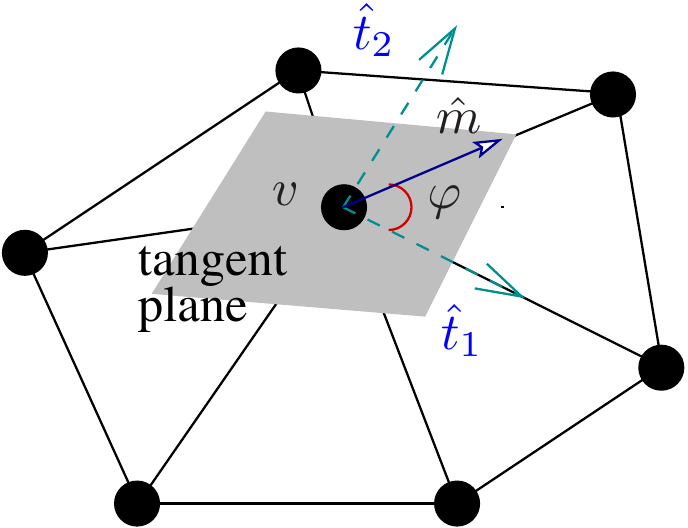}
\label{fig:inplane-field}
\vspace*{10pt}
\end{minipage}
}
\subfigure[Membrane surface with decorated in-plane field.]{
\begin{minipage}{0.5\textwidth}
\centering
\includegraphics[height=1.5in]{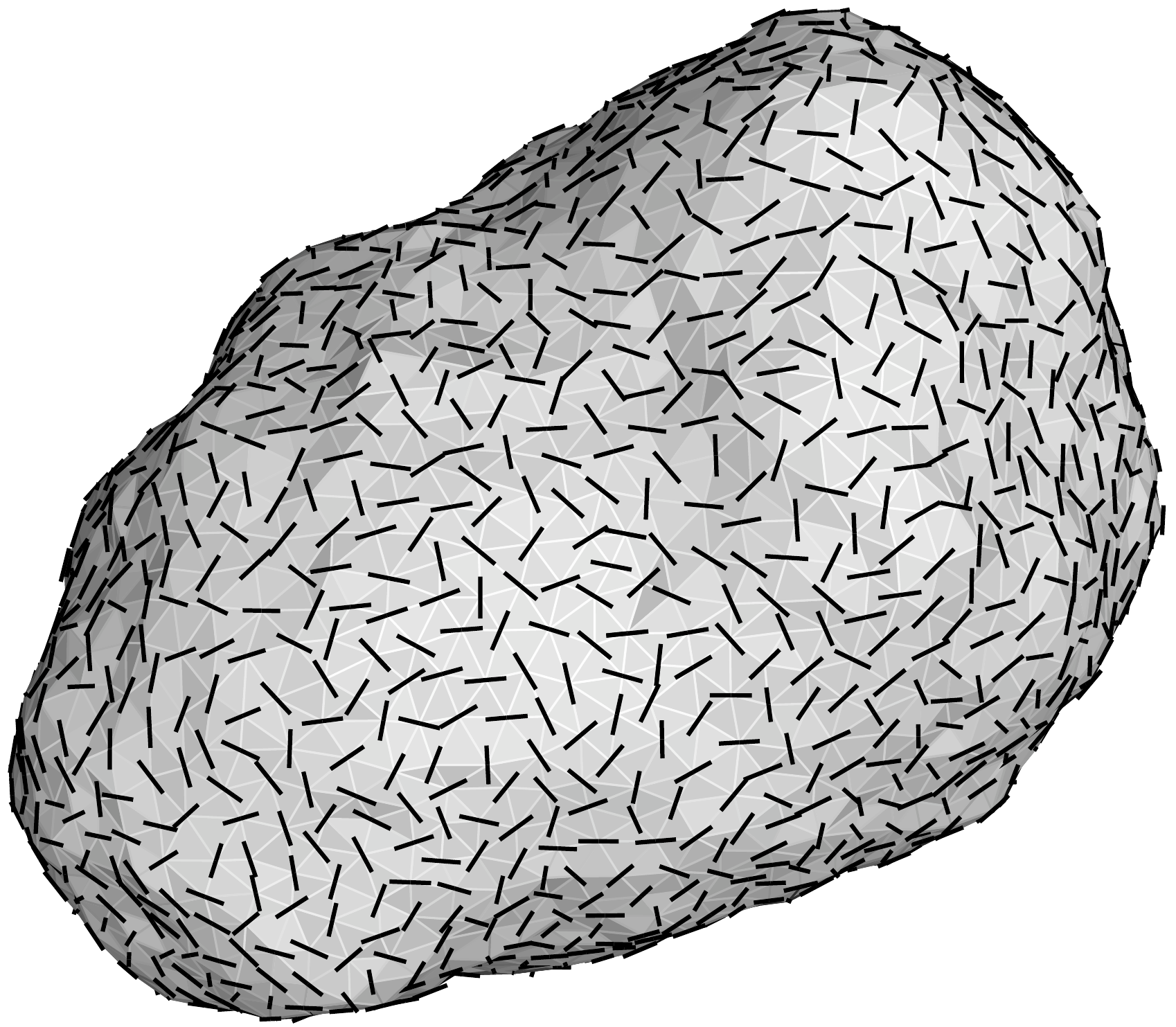}
\label{fig:model-inplane}
\vspace*{25pt}
\end{minipage}
}
\caption{{\bf (a)} An in-plane polar field, $\hat{m}(v)=a\,\hat{t}_{1}+b\,\hat{t}_{2}$, defined on the tangent plane of vertex $v$. {\bf (b)} A discretized membrane, of spherical topology, decorated with an in-plane nematic director field.}
\end{figure}

 Ramakrishnan et al.~~\cite{Ramakrishnan:2011cc,Ramakrishnan:2010hk} considered the vertices of the triangulated surface (introduced in section~~\ref{sec:dtmc}) by additionally decorating them with a nematic in-plane field $\hat{m}$, of unit length, defined on the tangent plane at each vertex. Fig.~\ref{fig:inplane-field} shows the neighbourhood of a vertex with an in-plane field and fig.~\ref{fig:model-inplane} shows a vesicular membrane with in-plane order defined at all vertices; the surface coverage can be set to the desired concentration. The field $\hat{m}$ is defined on the tangent plane of vertex $v$ ({\sl the Darboux frame to be precise}, see Appendix~~\ref{geom-quant}),  as $\hat{m}(v)=a\,\hat{t}_{1}(v)+b\,\hat{t}_{2}(v)$, and this representation is illustrated in Fig.~\ref{fig:inplane-field}. If $\hat{m}$ subtends an angle  $\varphi$ with the maximum principal direction $\hat{t}_{1}$, then $a=\cos\varphi$ and $b=\sin\varphi$.  

The self-interaction between the in-plane field is given by $\mathscr{H}_{p{\rm-atic}}$. The form of this interaction potential differs with the value of $p$. For example, the self-interaction between the in-plane field with polar symmetry ($p=1$) can be represented by the standard $XY$-like interactions ~\cite{Chaikin:1995td},
 \begin{equation}
 \label{eq:discXY}
 \mathscr{H}_{\rm 1-atic}=-J_{\rm 1}\sum_{\langle v,v^{'}\rangle}\cos(\theta_{vv^{'}}),
 \end{equation}
 For a nematic field, the Lebwohl-Lasher interaction potential ~\cite{Lebwohl:1972p426},
 \begin{equation}
 \label{eq:discLL}
 \mathscr{H}_{\rm 2-atic}=-\frac{J_{2}}{2}\sum_{\langle v,v^{'}\rangle}\left \{3\cos^{2} (\theta_{vv^{'}})-1\right\},
 \end{equation}
 has been used by Ramakrishnan et al.~~\cite{Ramakrishnan:2011cc,Ramakrishnan:2010hk} to model the self-interaction of the nematic in-plane field defined on the vertices of the membrane. $J_{1}$ and $J_{2}$ are the interaction strengths of the polar and nematic fields, respectively. Note that $\theta_{vv^{'}}$ is the curvature-dependent angle between the orientations of the in-plane field at vertices $v$ and $v^{'}$, computed using the parallel transport technique defined in appendix ~\ref{sec:partpt}. This dependence implicitly couples the in-plane field to the membrane, which in turn considerably affects the morphology of the fluid membrane.

\subsubsection{Monte Carlo procedure for nematic membranes} \label{sec:field-mcs}
 As described by Ramakrishnan et al.~~\cite{Ramakrishnan:2011cc,Ramakrishnan:2010hk}, the Monte Carlo techniques for a fluid random surface can be extended to a field-decorated membrane when the contributions arising from field orientations are accounted for in the phase space integral. The total partition function of the membrane with the in-plane field has the form,
\begin{equation}
Z(N,\kappa,\Delta p,J_{p})=\frac{1}{N!} \sum_{\{\mathscr{T}\}} \prod_{v=1}^{N} \int d\varphi(v) \int d\vec{x}(v) \exp \left \{-\beta \mathscr{H}_{\rm tot}\right\},
\label{eq:Zin-plane}
\end{equation} 

\begin{figure}[H]
\centering
\includegraphics[height=3.0in,clip]{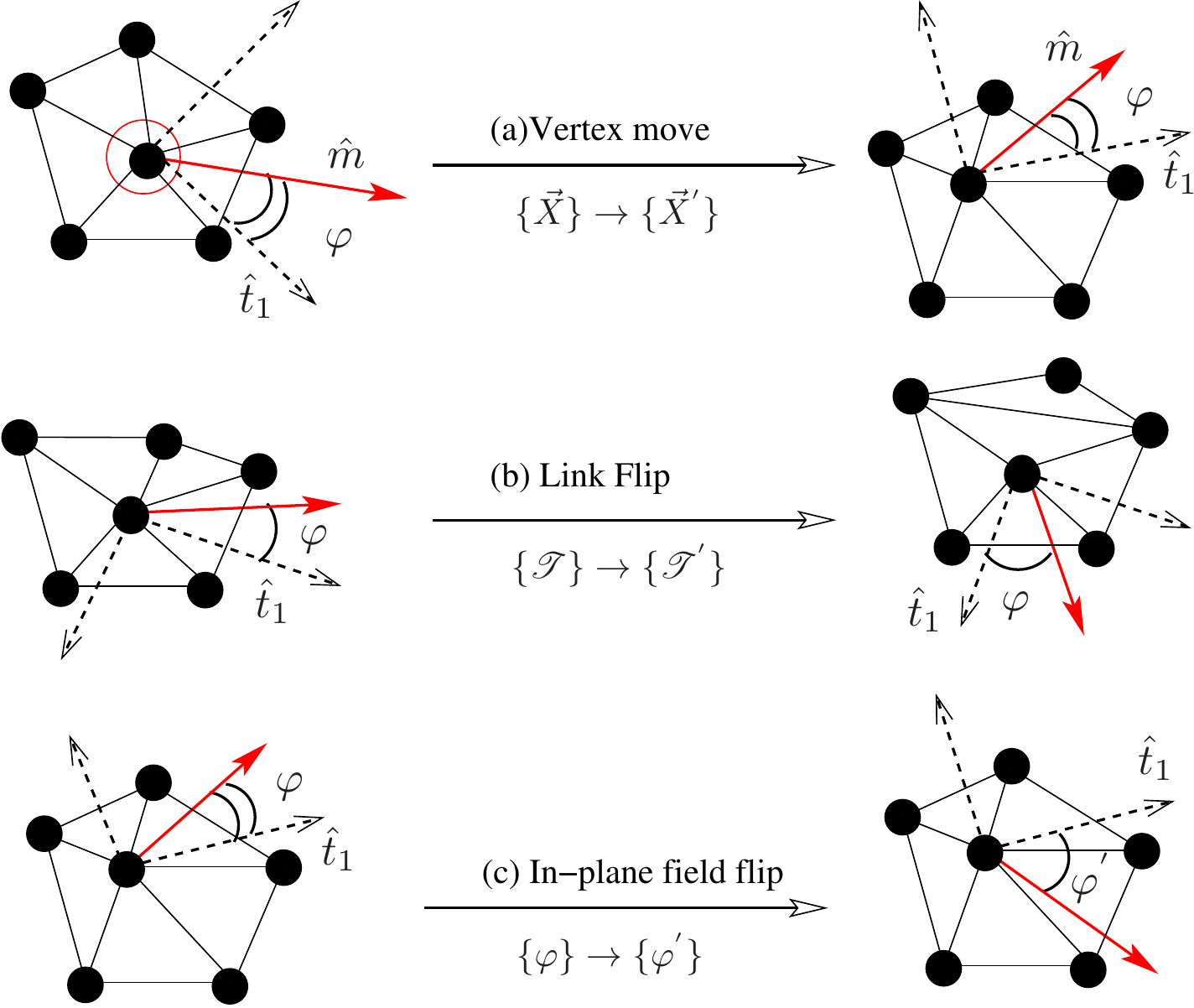}
\caption{\label{fig:inplane-MCS} Monte Carlo moves for a membrane with an in-plane field.}
\end{figure}

\noindent where $\mathscr{H}_{\rm tot}=\mathscr{H}_{\rm sur}+\mathscr{H}_{\rm 2-atic}+V_{SA}$, and the integral is performed over all possible in-plane orientations, in addition to the phase space of the random surface in eqn.~\eqref{eqn:Zfluidsurf}. The membrane explores the accessible states in its phase space, now represented by $\eta=[ \{\vec{X}\},\{\mathscr{T}\}, \{\varphi\}]$, through the set of Monte Carlo moves shown in Fig.~\ref{fig:inplane-MCS}. A vertex move and a link flip, shown, respectively, in Fig.~\ref{fig:inplane-MCS}(a) and (b), are exactly performed as described in section~~\ref{sec:dtmc}, except that in these moves, the field orientation $\{\varphi\}$ remains unchanged before and after the move. The third class of move aims at changing the orientations of the in-plane nematic field, $\{\varphi\} \rightarrow \{\varphi^{'}\}$. As shown in Fig.~\ref{fig:inplane-MCS}(c) the in-plane field $\hat{m}(v)$, at a randomly chosen vertex $v$, is updated to a new orientation $\hat{m}^{'}(v)$. The move is accepted using the Metropolis scheme with a probability given by,
\begin{equation}
P_{\rm acc}(\{\varphi\}\rightarrow \{\varphi^{'}\})={\rm min}\left \{1,\exp \left (-\beta \Delta \mathscr{H}(\eta \rightarrow \eta^{'} )\right) \right\},
\end{equation}
with $\eta=[ \{\vec{X}\},\{\mathscr{T}\}, \{\varphi\}]$ and $\eta^{'}=[ \{\vec{X}\},\{\mathscr{T}\}, \{\varphi^{'}\}]$. Attempt probability for the move is $\omega(\eta \rightarrow \eta^{'})=\omega(\eta^{'} \rightarrow \eta)=(2\Delta \theta N)^{-1}$.
 As noted before, the self-interaction between the nematic field is dependent on the membrane curvature, which in turn implicitly couples the membrane geometry to the texture of the nematic field. In appendix ~\ref{sec:implicit-coup}, we describe how the equilibrium shape of an otherwise spherical membrane is changed in the presence of a polar and nematic field. The explicit interaction of the protein with the membrane was introduced by Ramakrishnan et al. ~\cite{Ramakrishnan:2011cc,Ramakrishnan:2010hk} through an additional energy term ($\mathscr{H}_{\rm anis}$) that depends on the directional spontaneous curvature induced by the protein and the curvature of the membrane surface; this form of $\mathscr{H}_{\rm anis}$ is described in the section below.

\subsection{Anisotropic bending energy}\label{sec:aniso-energy}
The elastic behavior of the membrane becomes anisotropic when the in-plane  field $\hat{m}$ couples to its curvature tensor $\underline{{K}}$. As comprehensively discussed in~~\cite{Piran:2003,Helfrich:1988p3004,Nelson:1992p9}, such an interaction is described by an energy functional containing all possible gauge-invariant scalars constructed out of $\hat{m}$ and $\underline{K}$. The full form of the energy will have  contributions from terms like ($\hat{m}\underline{K}\hat{m}$), ($\hat{m}_{\perp}\underline{K}\hat{m}_{\perp}$), $(\hat{m}\underline{K}\hat{m}_{\perp})$, $(\hat{m}_{\perp}\underline{K}\hat{m})$, ($\hat{m}\underline{K}\underline{K}\hat{m}$), ($\hat{m}_{\perp}\underline{K}\underline{K}\hat{m}_{\perp}$),  $(\hat{m}\underline{K}\hat{m})^{2}$, ($\hat{m}_{\perp}\underline{K}\hat{m}_{\perp})^{2}$, $(\hat{m}\underline{K}\hat{m}_{\perp})^{2}$, and $(\hat{m}_{\perp}\underline{K}\hat{m})^{2}$; in addition, one may also consider gradients of $\hat{m}$. $\hat{m}_{\perp}$ is the in-plane field orientation perpendicular to $\hat{m}$, defined as $m_{\perp}^{a}=g^{ac}\gamma_{cb}m^{b}$, with $\gamma_{ab}$ being the antisymmetric tensor. For computational purposes, following the definition of $\hat{m}$ in section~~\ref{sec:inplane-meth}, we approximate the perpendicular nematic orientation as $\hat{m}^{\perp}(v)=b\hat{t}_{1}(v)-a\hat{t}_{2}(v)$, and it can be verified that $\hat{m} \cdot \hat{m}^{\perp}=0$. 

 To keep the complexity of the problem tractable, we choose the in-plane field to be an achiral nematic director of unit magnitude. The terms of the interaction energy are chosen such that the membrane is invariant under $\hat{m} \rightarrow -\hat{m}$ and $\underline{K} \rightarrow -\underline{K}$. To lowest order in $\hat{m}$, the explicit interaction of the nematic field with the membrane has the form~~\cite{Frank:2008p1047}:

\begin{equation}
\label{eq:Hnesur}
\mathscr{H}_{\rm anis}=\underbrace{\frac{1}{2}\int_{S}d{\mathbf{S}}\,\kappa_{\parallel}\left[m^{a}K_{ab}m^{b}-C_{0}^{\parallel}\right]^{2}}_{\mathscr{H}_{\rm anis}^{\parallel}}+\underbrace{\frac{1}{2}\int_{S}d{\mathbf{S}}\,\kappa_{\perp}\left[m_{\perp}^{a}K_{ab}m_{\perp}^{b}-C_{0}^{\perp}\right]^{2}}_{\mathscr{H}_{\rm anis}^{\perp}}. \end{equation}

\noindent Here, the Einstein summation convention over repeated indices is implied. It has to be noted that higher order terms like ($\hat{m}\underline{K}\underline{K}\hat{m}$), which possess all the symmetries listed above, have been neglected, and contributions from second order gradients are zero due to the constraint $m^{i}m_{i}=1$~~\cite{Frank:2008p1047}.

The directional rigidities $\kappa_{\parallel} \ge 0$ and $\kappa_{\perp} \ge 0$ are, respectively, the additional stiffness along and perpendicular to the nematic field orientation $\hat{m}$.  If the nematic field represents the curvature due to BAR domain proteins, $\kappa_{\parallel}$ and $\kappa_{\perp}$ can be thought of as a measure of the modified rigidities due to the protein structure and the molecular-level  interaction strength  with the lipids on the bilayer membrane. Similarly, the ensuing curvatures resulting from the interaction of the protein with the membrane are approximated by $C_{0}^{\parallel}$ and $C_{0}^{\perp}$, which are the directional spontaneous curvatures imposed by the in-plane field on the membrane in its vicinity. Their values can be positive, negative, or zero, depending on the type of proteins they represent. For example, $C_{0}^{\parallel}>0$ for a protein with a F-BAR domain, and $C_{0}^{\parallel}<0$ when it contains an I-BAR domain~~\cite{Frost:2009p157}.

\noindent On a triangulated surface, the anisotropic bending energy can be computed as,
\begin{equation}
\label{eq:Hnesur-disc}
\mathscr{H}_{\rm anis}= \frac{1}{2}\sum_{v=1}^{N} A_{v}\left\{ \kappa_{\parallel}\left[c^{\parallel}(v)-C_{0}^{\parallel}\right]^{2}+\kappa_{\perp}\left[c^{\perp}(v)-C_{0}^{\perp}\right]^{2}  \right \}. \end{equation}
\noindent
Here, $c^{\parallel}(v)=c_{1}\cos^{2}\varphi(v)+c_{2}\sin^{2}\varphi(v)$ and $c^{\perp}(v)=c_{2}\cos^{2}\varphi(v)+c_{1}\sin^{2}\varphi(v)$ are the directional curvatures on the membrane parallel and perpendicular to the nematic orientation calculated using Euler's theorem~~\cite{doCarmo:1976}.

\subsection{Properties of nematic membranes}
The equilibrium shapes of the nematic membrane, with total energy $\mathscr{H}_{\rm tot}=\mathscr{H}_{\rm sur}+\mathscr{H}_{\rm 2-atic}+\mathscr{H}_{\rm anis}$, were determined by Ramakrishnan et al. ~\cite{Ramakrishnan:2012dk} using Monte Carlo techniques for surfaces with in-plane field, defined in section~~\ref{sec:field-mcs}. As discussed in appendix ~\ref{sec:implicit-coup}, the nematic in-plane field can affect membrane shapes only when they are in the nematic phase; this can be achieved by setting $J_2=3k_{B}T$, at which the 3-dimensional system is above the critical value for the isotropic-nematic transition. In this case, the thermal fluctuations in the orientational field do not change the qualitative behavior, and the nematic texture only allows the presence of minimum number of defects; see appendix ~\ref{sec:implicit-coup} for a detailed discussion on defects in field texture and surface topology. Any additional defect generation is due to the large deformations of the underlying membrane.  The predictions of the mean field model are reproduced when the bending stiffness $\kappa$ is set to large values. The thermally excited shapes of a membrane with a 1-dimensional nematic field ($\kappa_{\perp}=0$) have been studied by Ramakrishnan et al. ~\cite{Ramakrishnan:2012dk}, and we discuss this below. 

\subsubsection{Conformational phase diagram for $\kappa_{\perp}=0$} \label{sec:confphasespace}
\begin{figure}[H]
\centering
\includegraphics[width=15cm,clip]{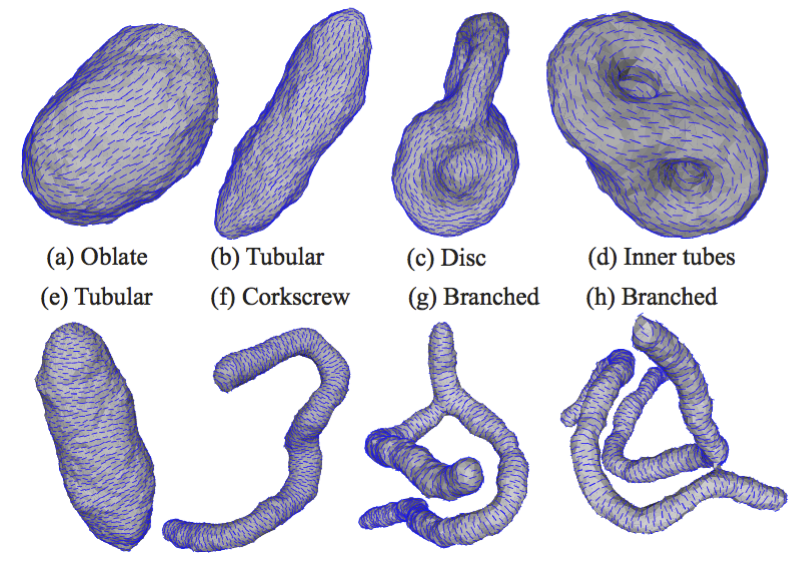}
\caption{\label{fig:conf-kperp0} Classes of shapes of a nematic membrane for $C_{0}^{\parallel}$ = 0.0 {\bf(a)}, $-$0.3 {\bf(b)}, $-$0.4 {\bf(c)}, $-$0.6 {\bf(d)}, 0.2 {\bf(e)}, 0.4 {\bf(f)}, 0.5 {\bf(g)} and 0.6 {\bf(h)}, with $\kappa=10k_{B}T$, $\kappa_{\parallel}=5k_{B}T$, $\kappa_{\perp}=0$ and $J_{2}=3k_{B}T$ {\sf(Reprinted figure with permission from  N. Ramakrishnan, John H. Ipsen, P. B. Sunil Kumar,  Soft Matter, {\bf 8}(11), 3058  and 2012. Copyright (2012) by the Royal Society of Chemistry)}.}
\end{figure}

 The equilibrium shapes of the nematic membrane as a function of $C_{0}^{\parallel}$ are given in Fig. ~\ref{fig:conf-kperp0} for $\kappa_{\parallel}=5k_{B}T$ ~\cite{Ramakrishnan:2012dk}.
The directional deformation induced by the in-plane nematic field augments the equilibrium shapes of the model membrane. Non-zero values of the directional spontaneous curvature ($C_{0}^{\parallel}$) stabilize many shapes ubiquitous in the biological cell: the shapes include {\it tubes} ($-$0.3 to 0.3), {\it corkscrews} (0.35 to 0.5), {\it branched shapes} ($>0.5$), {\it discs} ($-$0.35 to $-$0.55), and {\it inner tubes or caveola like shapes} ($<-0.55$). Recall that a fluid surface containing a nematic field with no anisotropic interactions deforms in a manner such that the four $+1/2$ disclinations are localized to the vertices of a tetrahedron. The presence of anisotropic interactions alters the free-energy landscape and allows for the presence of other defect textures. The prominent features distinguishing a class of shapes from another, for a membrane of constant surface area, are as follows:
\begin{enumerate}
\item
{\it Tubes } are cylindrical structures with exactly four $+1/2$ disclinations, with the tube radius dependent on the value of $C_{0}^{\parallel}$, $\kappa_{\parallel}$, and $\kappa$, as in Fig. ~\ref{fig:conf-kperp0}(a,b,e). Two individual disclinations pair up at the end cap of the tube, resulting in two defect pairs each of net charge +1. The average nematic orientation also responds to change in the directional spontaneous curvature, with $\langle \varphi \rangle \sim \pi/2$ for $C_{0}^{\parallel} \leq 0$ and $\langle \varphi \rangle \sim 0$ for $C_{0}^{\parallel}>0$.

\item
{\it The canal surface}, a snapshot of which is shown in Fig. ~\ref{fig:conf-kperp0}(f), closely resembles a tube  spiralling around its long axis.

\item
{\it Branched membranes } are  seen for large $C_{0}^{\parallel}$, where multiple tubes originate from a common region called an intersection or a neck.  Due to these large deformations, the number of $+1/2$ disclinations in  a branched membrane exceeds four but still preserves the Euler characteristic of the surface. For example, the nematic field on the branched shape shown in Fig. ~\ref{fig:conf-kperp0}(g), contains six $+1/2$  and two $-1/2$ disclinations. A pair of negative disclinations, each of charge $Q_{\rm neck}$, are localized to the region of intersection. This neck is finite sized (fig. ~\ref{fig:conf-kperp0}(g)) with $Q_{\rm neck}=-1/2$ and has exactly three tubular protrusions. On the other hand, the neck is said to be narrow (Fig. ~\ref{fig:conf-kperp0}(h)) when $Q_{\rm neck}=-1$ and bridges two membrane tubes. A similar organization has also been seen in the simulation of N-BAR domains on tubulated membranes ~\cite{Lyman:2011p10430}.

\item
{\it Discs:}  The highly curved rim along with the low curvature planar regions characterize a disc. The four $+1/2$ defects are  far separated from each other and localize to regions of negative principal curvatures as in Fig. ~\ref{fig:conf-kperp0}(c). 

\item 
{\it Inner tubes:} In the context of a biological cell, inner tubes are the invaginations in the plasma membrane  into the cytoplasmic side. These caveola like shapes, named after the membrane deformations caused by the caveolin proteins, are common in neuron cells and T-tubules. Inner tubes, shown in Fig. ~\ref{fig:conf-kperp0}(d) and Fig. ~\ref{fig:innertube}, are analogous to branched membranes in terms of the additional topological defects they posses. Unlike in the latter, the lengths of the tubes are dependent on the volume and self avoidance constraints of the membrane. 
\begin{figure}[H]
\centering
\includegraphics[height=2.0in]{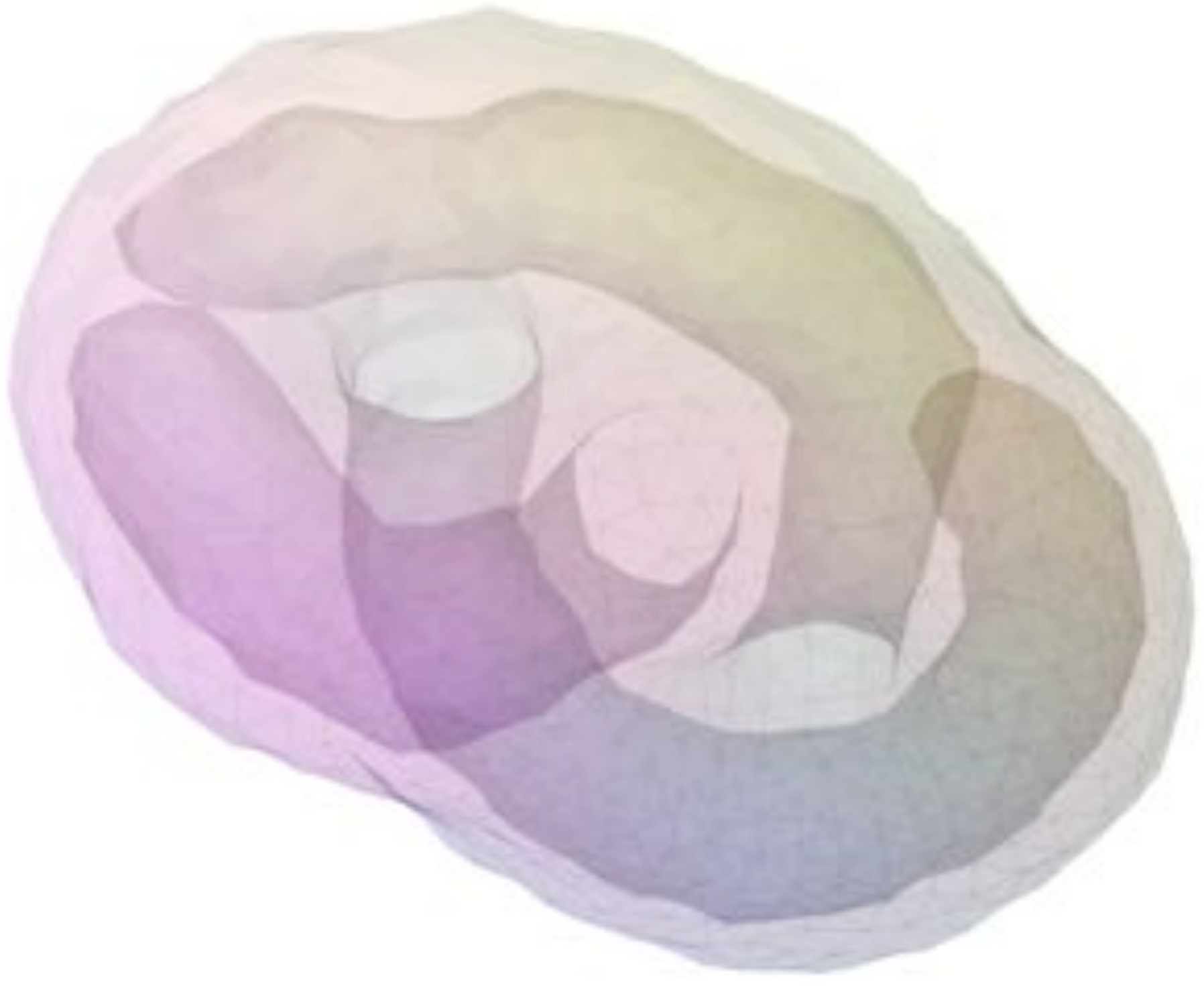}
\includegraphics[height=2.0in]{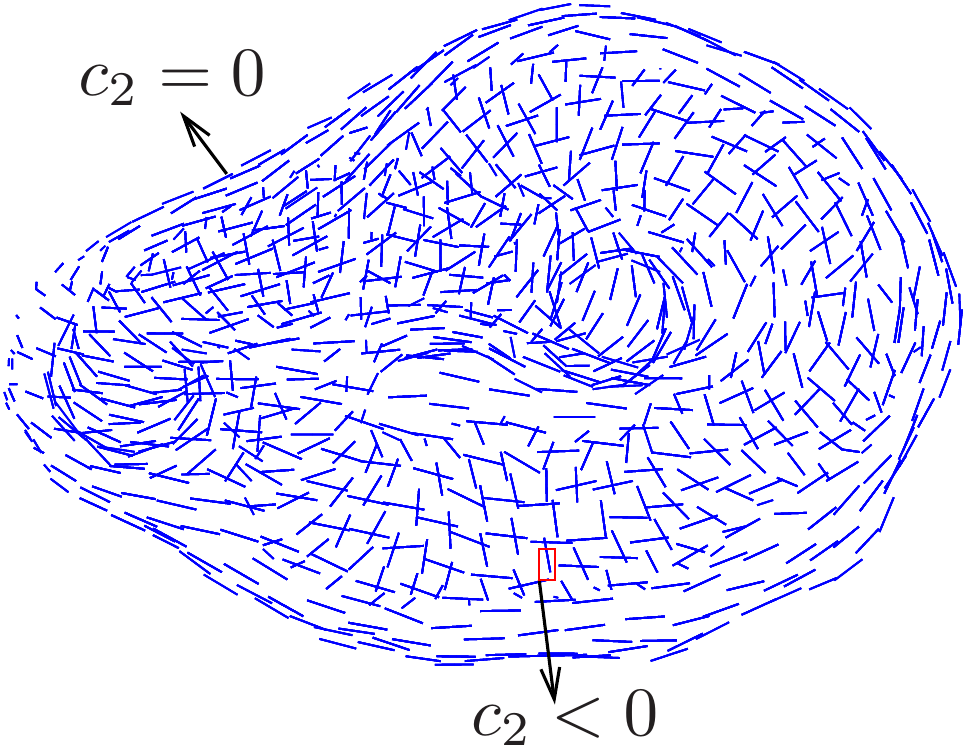}
\caption{\label{fig:innertube} Shown are two different aspects of the inner tubes.  On the left, the transparent view  reveals the tubes grown into the interior of the membrane. The nematic field orientation on the exterior surface and also on the inner tubes are shown in the right panel; for $C_{0}^{\parallel}<0$, the nematic field always orients along the minimal principal curvature direction, with this direction corresponding to $c_{2}=0$ on the outer surface and $c_{2}<0$ on the inner tubes.}
\end{figure} 
\end{enumerate}
It is helpful to consider the predominant sign of the principal curvatures, $c_{1}$ and $c_{2}$, for these shapes. The possible values of $c_{1}$ and $c_{2}$, for ideal geometries that belong to the above mentioned classes,  are listed in Table ~\ref{tab:curvlist}. 
\begin{table}[!h]
\centering
\begin{tabular}{lcccc}
\hline \hline
 {\quad\quad Class of shape }& \quad\quad & {$c_{1}$} & \quad \quad& {$c_{2}$ \quad\quad} \\
 \hline \hline
  & & & & \\
  {\quad\quad1. Tubular}& \quad \quad& $>0$ & \quad\quad & $=0$\quad\quad\\ [2ex]
  {\quad\quad2. Canal surface }&\quad\quad & $>0$ & \quad\quad &$=0$\quad\quad\\ [2ex]
  {\quad\quad3. Branched }&\quad \quad& $>0$ &\quad \quad& $=0$\quad\quad\\[2ex]
  {\quad\quad4. Disc like }&\quad\quad &  & \quad\quad \quad\quad&\\[0.5ex]
    \hspace*{15pt}\quad\quad{\it a. On the rim }&\quad\quad & $>0$ & \quad\quad &$=0$\quad\quad \\[0.5ex]
    \hspace*{15pt}\quad\quad{\it b. Region enclosed by rim }& \quad \quad&$=0$ &\quad \quad& $=0$ \quad\quad\\ [2ex]
  {\quad\quad5. Inner tubes }&\quad\quad &  & \quad\quad &\\[0.5ex]
    \hspace*{15pt}\quad\quad{\it a. Exterior side}& \quad \quad&$>0$ & \quad\quad &$=0$\quad\quad \\[0.5ex]
    \hspace*{15pt}\quad\quad{\it b. Interior side }& \quad\quad &$=0$ & \quad\quad &$ <0$\quad\quad \\ [1ex]
  \hline
\end{tabular}
 \caption{ \label{tab:curvlist} Predominant values of the principal curvatures for various classes of shapes.}

\end{table}

All shapes stabilized by scanning the value of the directional bending stiffness, $\kappa_{\parallel}$, fall into  one of the five classes listed above. It should also be observed that the characteristic value of $C_{0}^{\parallel}$, for which different classes of shapes are stabilized, is also a function of $\kappa_{\parallel}$. The precise state boundaries delineating the different states can be identified by computing the relative free energies, which we discuss in section~~\ref{sec:freeener}. The distribution of directional and principal curvatures as well as the effect of changing the bending rigidity on the emergent morphological states of the membrane are discussed further in Ramakrishnan~\textit{et.~al.} ~\cite{Ramakrishnan:2012dk}.

\subsection{Pairing of defects: the role of Gaussian curvature} \label{sec:defectpairing}
Charge of a topological defect is analogous to an electric charge. Two like charged defects repel each other with  an  interaction energy  logarithmically dependent on their separation,  $\mathscr{H}_{\rm def} \propto -\ln(R_{\rm def})$; this has been shown to be true for defects on both planar~~\cite{gennesprost1993} and curved surfaces~~\cite{Lubensky:1992p531}. In order to understand the spatial organization of  topological defects, Ramakrishnan et al. ~\cite{Ramakrishnan:2012dk}  computed the geodesic distance ($\xi$) between these defects on the triangulated surface, using the Dijkstra algorithm~~\cite{Dijkstra:1959p269}.
\begin{figure}[H]
\centering
\includegraphics[width=5in]{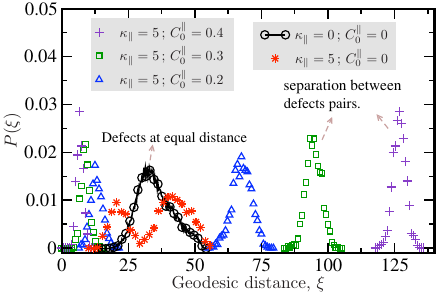}
\caption{\label{fig:geodesic-dist-plot} $P(\xi)$ is the distribution of the geodesic distance between a pair of defect cores, for shapes corresponding to different values of $C_{0}^{\parallel}$ and $\kappa_{\parallel}$ {\sf(Reprinted figure with permission from  N. Ramakrishnan, John H. Ipsen, P. B. Sunil Kumar,  Soft Matter, {\bf 8}(11), 3058  and 2012. Copyright (2012) by the Royal Society of Chemistry)}. }
\end{figure} 
The distribution of $\xi$  on a nematic membrane, for various set of parameters, is shown in Fig. ~\ref{fig:geodesic-dist-plot}. The analysis has been performed on tubular membrane shapes with exactly four $+1/2$ disclinations, but the results hold for other shapes too. When $\kappa_{\parallel} \neq 0$, two $+1/2$ disclinations come close to each other on a region of positive Gaussian curvature, resulting in defect pairs, each of strength +1, at either ends of the tube.  In Fig. ~\ref{fig:geodesic-dist-plot}, this pairing is reflected as two distinct peaks in $P(\xi)$  for $\kappa_{\parallel} \ne 0$, which is expressly different from the broad distribution seen when $\kappa_{\parallel}=0$. The peak at small $\xi$ represents the geodesic connecting defects within a pair and shifts to the left with increasing $C_{0}^{\parallel}$. The curvature dependence of defect localization is in good agreement with theoretical predictions of this phenomenon ~~\cite{Bowick:2009p955,Vitelli:2006p1462}.

Coupling of nematic defects to the curvature of the membrane and the resulting dynamics play an important role in the tubulation mechanisms of nematic membranes. At temperatures where the nematic order is susceptible to thermal fluctuations, proliferation and annihilation of additional defect pairs control the resulting tubular morphologies. The different mechanisms leading to stabilization of tubular and branched membranes have been discussed in ~\cite{Ramakrishnan:2012dk}. Further, nematic membranes display long-wavelength thermal undulations in the form of canal-surface-like structures that are intermediate to the tubular and branched morphologies ~\cite{Ramakrishnan:2012dk}. The qualitative behavior of a nematic membrane with both its anisotropic stiffnesses being non-zero  ($\kappa_{\parallel} \neq 0$ and $\kappa_{\perp} \neq 0$) is the same as described above, and hence is not discussed here (see ~\cite{Ramakrishnan:2012wm} for details).

\subsection{Implication of defect structures in biological membranes}
\begin{figure}[H]
\centering
\subfigure[]{
\begin{minipage}[]{0.45\textwidth}
\begin{center}
\label{fig:shibata1}
\includegraphics[height=1.35in]{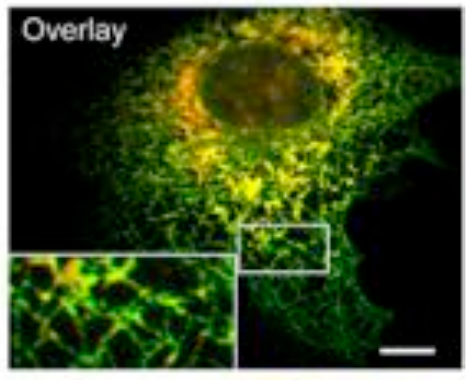}

\noindent \textsf{ \footnotesize Localization of calreticulons to ER edges. $\alpha$-Calreticulon proteins have been colored green.}
\vspace*{5pt}
\end{center}
\end{minipage}
}
\hspace*{0.15cm}
\subfigure[]{
\begin{minipage}[]{0.475\textwidth}
\begin{center}
\label{fig:shibata2}
\includegraphics[height=1.35in]{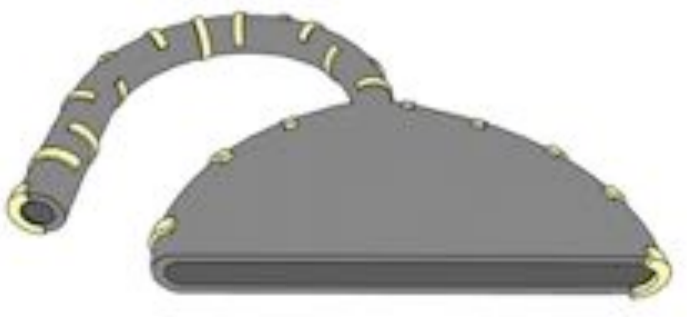}

\noindent\textsf{ \footnotesize An illustration of reticulon proteins on ER sheets.}
\vspace*{20pt}
\end{center}
\end{minipage}
}
\caption{ {\bf (a)} Immunofluorescence micrographs of a ER sheet with $\alpha$ - Calreticulon proteins (marked green with fluorescent markers) and {\bf (b)} an illustration showing the pattern of Dp1/Yop1p and reticulon proteins on an ER sheet and tubules \textsf{(Reprinted from {\it Cell} , {\bf 143} /5,  Y. Shibata et. al., Mechanisms Determining the Morphology of the Peripheral ER, 774--788, Copyright (2010), with permission from Elsevier).}} 
\end{figure}

 Even though the nematic ordering and the presence of defect structures in biological membranes has not been directly verified experimentally, Ramakrishnan et al. ~\cite{Ramakrishnan:2013gl} extended the anisotropic curvature model to investigate the behavior of partially decorated, single-, and multi-component nematic membranes. In the case of a partially decorated membrane, it has been shown that  the presence of an anisotropic ordering interaction promotes the aggregation of proteins with similar curvature properties into spatial domains~\cite{Ramakrishnan:2013gl}. Such a response leads to many interesting behaviors and also enriches the conformational phase space of the vesicular membrane with the protein field.
\begin{figure}[H]
\centering
\includegraphics[width=5in]{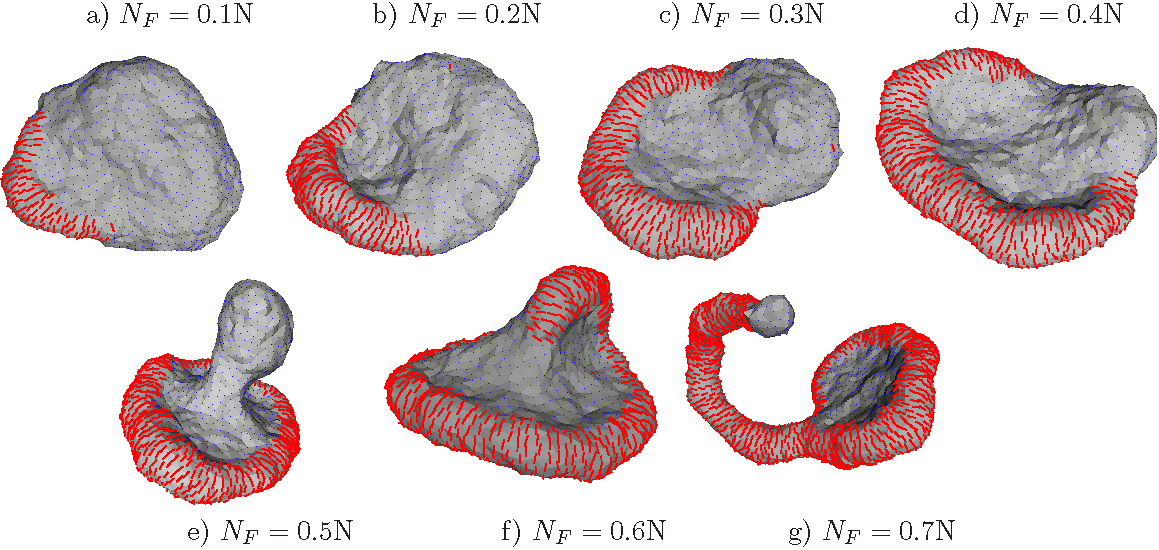}
\caption{\label{fig:funcconc} Equilibrium shapes of a nematic membrane with varying concentration of the nematic field ($N_{F}=0.1N-0.7N$), for a  prescribed value of the directional curvature, $C_{0}^{\parallel}=0.5$ {(\sf Reprinted from Biophysical Journal, {\bf 104}(5), N. Ramakrishnan, P. B. Sunil Kumar, John H. Ipsen, Membrane-Mediated Aggregation of Curvature-Inducing Nematogens and Membrane Tubulation, 1018--1028, Copyright (2013), with permission from Elsevier)}.}
\end{figure}
With $N_{F}$ being the relative concentration of the nematic field, Fig. ~\ref{fig:funcconc} shows the equilibrium shapes of a nematic membrane with $N_{F}$ in the range $0.1N-0.7N$, $\beta\kappa=10$, $\beta\kappa_{\parallel}=5$, and $C_{0}^{\parallel}=0.5$.  At low concentrations, the field localizes to the rim of a disc that becomes unstable at large concentrations ($N_{F}>0.6N$), resulting in membrane  tubulation. The equilibrium shapes of a partly decorated membrane as a function of the field concentration show striking similarities to the arrangement of reticulons and Dp1/Yop1 proteins in the peripheral ER~~\cite{Shibata:2010p2363}; the sheet like pattern (Fig. ~\ref{fig:funcconc}(c)) of the nematic membrane and the coexisting sheet and tubes (Fig. ~\ref{fig:funcconc}(g))  have both been observed in experiments on the ER. Green fluorescent regions in Fig. ~\ref{fig:shibata1}, reprinted from~~\cite{Shibata:2010p2363}, denote the spatial locations of the reticulon proteins on a sheet like ER and confirms that these proteins are confined to the rim of the sheet. Furthermore, at high concentrations of the nematic field, the observed conformations of the nematic membrane and the orientational pattern of the in-plane field  are in good agreement with the predictions of reference ~\cite{Shibata:2010p2363}, shown in Fig. ~\ref{fig:shibata2}. 
 
In this context, the conditions for tubulation; estimates for the tube radius, orientation angle, and their fluctuations; and application of the two-component nematic model to the problem of tube constriction by ATP activated dynamin have been discussed in detail in ~\cite{Ramakrishnan:2013gl}. 

\subsection{Mapping the length scales}
How do the model parameters like the spontaneous curvature and the density of nematic inclusions compare with experiments ? The size of the tubes remodeled by curvature-inducing protein can range from a few nanometers to a few hundred nanometers. Hence the length scale that can be associated with the triangulated model is not generic but is specific only to a class of proteins. For comparison, one can use the well-studied system of dynamin-driven tubulation of membranes. Dynamin proteins tubulate a spherical liposome, due to the curvature-inducing properties of their $\gamma$-GTPase and PH domains, and the resulting shape resembles the branched membrane discussed in Fig. ~\ref{fig:conf-kperp0}. Electron microscopy studies have shown  that the radius of these tubes to be in the range of $10-40$ nm~~\cite{Sweitzer:1998p3632,Hinshaw:2000p3770,Praefcke:2004p3309,Roux:2010p4079}, depending on the state and concentration of dynamin proteins. These proteins are observed to form a helical coat, each ring of the helix being formed from approximately 20 units, and the pitch of the helix, which is the separation between successive rings,  has been observed to be $\sim$15nm. 

\begin{figure}[H]
\centering
\includegraphics[height=1.0in]{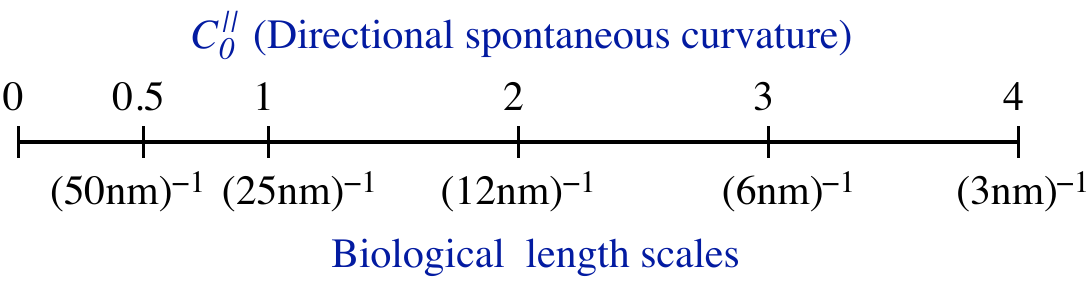}
\caption{\label{fig:lengthscale} A comparison of the coarse grained lengths to biological length scales for Dynamin.   }
\end{figure}

These experimentally observed network of tubes can be compared with the branched membrane shown in  Fig. ~\ref{fig:conf-kperp0}  for $C_0^{\parallel} \sim 1.0$. In this limit, five nematics (five membrane vertices) make up the circumference of the tube, which, when compared with the experimental values, yield the length of the tether to be $\approx 25$nm. Hence, $C_0^{\parallel}=1.0$ translates to a curvature of $\approx ({25\rm nm})^{-1}$, in real units, which is not far from the suggested value of the intrinsic curvature of dynamin. The five in-plane director  fields when mapped to the 20 dynamin units, making up a ring, leads us to the estimate that the in-plane nematic field at each vertex is the average orientation of approximately 4 dynamin proteins. A map comparing the coarse grained lengths, used in the simulation, with the biological scales is given in Fig. ~\ref{fig:lengthscale}. As mentioned earlier this mapping changes with the choice of the curvactant protein.


\section{Applications of  the particle-based EM2 model in the study of morphological transitions in membranes mediated by protein fields} \label{sec:em2}

The EM2 model can be employed as a competing approach to investigate the curvature-mediated morphological transitions induced by isotropic as well as anisotropic curvature-inducing proteins. In particular, it  has been used to understand the membrane remodeling behavior of N-BAR-domain and F-BAR-domain containing proteins ~\cite{Ayton:2009gw,Ayton:2007iw}. The EM2 model ~\cite{Ayton:2006ht}, introduced in section~~\ref{sec:membmodel-particlebased},  is a coarse grained simulation technique based on the elastic energy description of a lipid bilayer (eqn.~~\eqref{eqn:can-Helf}). In this model, the field theoretic continuum membrane is discretized into a set of quasiparticles that interact with each other through a bending potential, which is derived from eqn. ~\eqref{eqn:can-Helf-gen}. In spite of being a particle-based model, the EM2 approach allows one to access length and time scales corresponding to the macroscopic limit ($\mu m$, $ms$) and hence can be used to study the dynamics of protein-induced self-assembly of membranes. Formulation of this method involves a suite of techniques adopted from non-equilibrium molecular dynamics (NEMD), smoothed particle hydrodynamics (SPH) and smooth-particle applied mechanics (SPAM) and recasting the continuum elastic free energy in terms of pair potentials. Being able to express  the curvature energy in terms of a binary interaction potential gives one  a template for a general binary interaction potential.\footnotetext{We note here that in the spirit of reducing the Helfrich energy in terms of binary interaction potentials, a hybrid method that couples a continuum membrane to a particle-based fluid was introduced by Noguchi and Gompper~\cite{Noguchi:2004ky}, where the properties of the fluid particles are governed by multi-particle collision dynamics (MPCD). This model has been used to investigate the role of thermal fluctuations and viscosity in the shape transition of vesicles and red blood cells in shear and capillary flows ~\cite{Noguchi:2005p1410,Noguchi:2005br}. We do not discuss this method here because it has not been applied to membrane remodeling by curvature-inducing proteins.}

\subsection{Discretization to a particle-based model}  
In order to model the membrane interactions in terms of pair potentials, which reproduces the key thermodynamic properties, the continuum membrane should be discretized into a set of quasiparticles. The local number density of the quasiparticles ($\rho(\boldsymbol{r})$) can be expressed using a regularized delta function (see Appendix ~\ref{sec:regdelta} for details) as,

\begin{equation}
\rho(\boldsymbol{r})=\underset{j}{\sum}\delta_{h}(\boldsymbol{r}-\boldsymbol{r}_{j})\label{eq:rho_def}.
\end{equation}

For a quasiparticle $j$, located at position $\boldsymbol{r}_{j}$, the regularized delta function $\delta_{h}(\boldsymbol{r}-\boldsymbol{r}_{j})=(hL_{D}^{2})^{-1}$ for all values of $\boldsymbol{r}$ satisfying $\left|\boldsymbol{r}-\boldsymbol{r}_{j}  \right |\leq \boldsymbol{r}_{c}$ and $\delta_{h}(\boldsymbol{r}-\boldsymbol{r}_{j})=0$ if otherwise.  Here, the quasiparticle corresponds to a cuboid of height $h$ and sides $L_{D}$, which defines the length scale of coarse graining. \\

For any field variable $a(\boldsymbol{r})$, we have $a(\boldsymbol{r})\rho(\boldsymbol{r})=\underset{j}{\sum}a_{j}\delta_{h}(\boldsymbol{r}-\boldsymbol{r}_{j})$,
with $a_{j}$ given as ${\displaystyle \int}d\boldsymbol{r}a(\boldsymbol{r})\delta_{h}(\boldsymbol{r}-\boldsymbol{r}_{j})$. If the initially flat membrane patch, with surface area $A$, is discretized into $N$ quasiparticles, then the average density of the membrane is given by $\rho_{A}=N/A$, and the lipid density per quasiparticle is given by $\rho^{*}=\rho_{A}L_{D}^{2}$. \\

A quasiparticle $i$ with position $\boldsymbol{r}_{i}$ and outward unit normal $\boldsymbol{n}_{i}$ interacts with another quasiparticle $j$ with position $\boldsymbol{r}_{j}$ and outward unit normal $\boldsymbol{n}_{j}$ through a discretized bending potential \footnotetext{See reference ~\cite{Ayton:2006ht} for the complete derivation of the potential.} given  by,

\begin{equation}
{\cal H}_{\rm eff}=\underbrace{\dfrac{1}{2}\underset{i=1}{\overset{N}{\sum}}\underset{j\ne i}{\overset{N_{c,i}}{\sum}}\Delta {\cal U}^{\rm bend}_{ij}}_{{\cal H}_{\kappa,{\rm eff}}}+ \underbrace{\dfrac{1}{2}\underset{i=1}{\overset{N}{\sum}}\underset{j\ne i}{\overset{N_{c,i}}{\sum}}\Delta {\cal U}^{\rm stretch}_{ij}}_{{\cal H}_{\sigma,{\rm eff}}}.
\label{eq:em2bending}
\end{equation}

$N_{c,i}$ is the number of quasiparticles around particle $i$, within a  cutoff distance of $r_{c}$, and the bending and stretching potentials are given by, \\
\begin{equation}
\Delta {\cal U}^{\rm bend}_{ij}=\begin{cases}
\dfrac{8\kappa}{\rho_{A}N_{c,i}} \Phi_{ij} & \textrm{for}\,\,\, r_{ij}\le r_{c,i}\\
\,\,\,\,\,\,\,\,\,0 & \textrm{if otherwise}.
\end{cases}
\end{equation}

The curvature function is given by  $\Phi_{ij}= \left(\dfrac{\boldsymbol{n}_{i}\cdot\hat{\boldsymbol{r}}_{ij}}{r_{ij}}\right)^{2}+\left(\dfrac{\boldsymbol{n}_{j}\cdot\hat{\boldsymbol{r}}_{ij}}{r_{ij}}\right)^{2} $, and if $\lambda$ is the bulk modulus, the energy contribution due to stretching  is given by,
\begin{equation}
\Delta {\cal U}^{\rm stretch}_{ij}=\begin{cases}
\dfrac{2 \pi\lambda h}{N_{c,i}^{2}}\left[2\left(r_{ij}-r_{ij}^{0}\right)\right]^{2} & \textrm{for}\,\,\, r_{ij}\le r_{c,i}\\
\,\,\,\,\,\,\,\,\,0 & \textrm{if otherwise}.
\end{cases}
\end{equation}

\noindent $\boldsymbol{r}_{ij}$ and $\boldsymbol{r}_{ij}^{0}$ are the vectors connecting particles $i$ and $j$ in the deformed and undeformed states, respectively, with $r_{ij}=|\boldsymbol{r}_{ij}|$ and $r_{ij}^{0}=|\boldsymbol{r}_{ij}^{0}|$. The bending and stretching potentials have been chosen to ensure that the membrane has zero energy in the undeformed state (here $\boldsymbol{n}_{i}$ and $\boldsymbol{n}_{j}$ are perpendicular to $\boldsymbol{r}_{ij}$). The EM2 particles self-assemble into a membrane both in the absence and presence of an explicit solvent; however the membrane displays slightly altered dynamics  in the presence of a solvent. Two explicit solvent models\textemdash namely, the WCA solvent, and BLOBS solvent\textemdash have been used to model the interaction of the EM2 particle with the surrounding fluid. Details of the membrane and solvent models have been described in reference ~\cite{Ayton:2006ht}. 

The topology of the self-assembled structures formed by the quasi-particles is a dynamic variable in the EM2 model. Since topological changes involve energy changes, the topology-dependent Gaussian curvature term, which was not accounted for, but could be easily incorporated, in the triangulated surface model (section~~\ref{sec:dtmc}), should also be accounted for. The contribution from the  Gaussian curvature term  can be expressed as,
\begin{equation}
{\cal H}_{\kappa_{G},{\rm eff}}=\dfrac{\kappa_{G}}{\rho_{A}}\sum_{i=1}{N}c_{1,i}^{2}.
\end{equation}
Here $\kappa_{G}$ is the Gaussian modulus introduced in eqn.~\eqref{eqn:can-Helf-gen}. The total elastic contribution to the EM2 model has three contributions and is given by,
\begin{equation}
{\cal H}_{\rm eff}={\cal H}_{\kappa,{\rm eff}}+{\cal H}_{\sigma,{\rm eff}}+{\cal H}_{\kappa_{G},{\rm eff}}.
\end{equation}

\subsection{Isotropic and anisotropic protein fields in the EM2 model}
The quasiparticle discretization of the elastic energy function, given in eqn.~\eqref{eq:em2bending}, can be extended to accommodate the effect of both isotropic and anisotropic curvature fields that were  discussed earlier in section~~\ref{cellmemb-curvact}. The presence of membrane curving proteins enters into the model through the curvature function  $\Phi_{ij}$, as  in reference ~\cite{Ayton:2007iw},

\begin{equation}
\Phi_{ij}=
\left(\dfrac{\boldsymbol{n}_{i}\cdot\hat{\boldsymbol{r}}_{ij}}{r_{ij}}-\gamma \right)^{2}+\left(\dfrac{\boldsymbol{n}_{j}\cdot\hat{\boldsymbol{r}}_{ij}}{r_{ij}}+\gamma\right)^{2}.
\label{eq:EM2-iso}
\end{equation}

Here $\gamma$ is the protein-dependent spontaneous curvature function which for isotropically curving proteins, with spontaneous curvature $C_{0}$, has the form $\gamma=C_{0}/2$. Alternately, for anisotropically curving proteins, $\gamma$ is a more complex function that depends both on the  particle position and  orientation. Anisotropy is introduced into the EM2 model through a unit vector field $\boldsymbol{m}$ whose orientation along the tangent plane defines the direction of anisotropy. For an anisotropic protein, with maximum spontaneous curvature $C_{0}$,
\begin{equation}
\gamma(\boldsymbol{r}_{ij},\boldsymbol{m}_{i},\boldsymbol{m}_{j})=\dfrac{C_{0}}{2} \left[ \left(\hat{\boldsymbol{m}}^{\parallel}_{i}.\hat{\boldsymbol{r}}_{ij} \right)^{2}+\left(\hat{\boldsymbol{m}}^{\parallel}_{j}.\hat{\boldsymbol{r}}_{ij} \right)^{2} \right ].
\label{eq:EM2-aniso}
\end{equation}

If $\mathbbm{P}_{i}$ be the tangent plane projection operator at quasiparticle $i$, then $\boldsymbol{m}^{\parallel}_{i}=\mathbbm{P}_{i}\boldsymbol{m}_{i}$. 

A more sophisticated version of the EM2 model has been proposed by Ayton et al. ~\cite{Ayton:2009gw}, where coupling between the density of the protein field ( given by $\phi_{\alpha}$ for quasiparticle $\alpha$ ) and membrane composition has been explicitly considered. In the proposed model, for a pair of quasiparticles $i$ and $j$, the following quantities are defined:

\begin{enumerate}
\item{\textit{Membrane coupling to protein density}}
\begin{enumerate}
\item The bending modulus  depends on protein field density as $\kappa_{ij}=\kappa-\eta \kappa (\phi_{i}+\phi_{j})/2$.
\item The spontaneous curvature is given by $C_{0,ij}=C_{0}f(\phi_{i},\phi_{j})$, with $f(\phi_{i},\phi_{j})=-(\phi_{i}+\phi_{j})/2$ if $(\phi_{i}+\phi_{j})<0$ and zero otherwise.
\end{enumerate}
\item{ \textit{Protein density coupling to membrane composition and geometry :}
Curvature-inducing proteins diffuse on the membrane, and hence, their density can respond to heterogeneities on the membrane surface. Two key factors driving change in protein density\textemdash curvature sensing and  sensitivity to lipid charge distribution\textemdash have been considered. The compositional coupling is represented by an additional energy contribution, ${\cal H}_{S,M}$.}
\begin{enumerate}
\item Intrinsic coupling (IC): In this formulation, the interaction between the membrane and the protein field is sensitive to the background membrane curvature.
\item Compositional coupling (CC): The protein field density is dependent on the density of negatively charged lipids.
\end{enumerate}
\item{Free energy contributions arising from spatially varying densities of the protein field and membrane lipid composition have been accounted for in this model}
\item{The effect of protein oligomerization has been included through an explicit term that captures the energy costs due to protein oligomerization.}
\end{enumerate}

Technical details of the model along with the expression for various energy contributions can be found in ~\cite{Ayton:2009gw}. In the following section we will highlight the key findings on membrane remodeling studied using the EM2 model for two BAR domain containing proteins namely N-BAR and F-BAR. The atomistic to coarse grained mapping for the protein field and membrane quasiparticles has been described in detail in ~\cite{Lyman:2011iu}.\\

\subsection{Membrane remodeling by N-BAR and F-BAR proteins}
In the studies of Ayton et al.  ~\cite{Ayton:2009gw,Ayton:2007iw}, the protein is represented as a curvature field with curvature profiles given by eqns.~\eqref{eq:EM2-iso} and ~\eqref{eq:EM2-aniso}. In this multi-scale model, the value of the spontaneous curvature $C_{0}$ is determined from molecular simulations of a single N-BAR domain interacting with the membrane of interest ~\cite{Blood:2006uc}. If $\rho$ is the density of the N-BAR proteins, $\delta A$ a small area patch, and $H_{\rm N-BAR}$ the spontaneous curvature induced by a single protein, then
\begin{equation}
C_{0}=\rho \delta AH_{\rm N-BAR}.
\end{equation}

$C_{0}$ has a maximum value of $H_{\rm N-BAR} \sim 0.15{\rm nm}^{-1}$, which corresponds to the maximum packing of N-BAR proteins on the membrane. In this framework, the density of the protein field is set by choosing an appropriate value of $C_{0}$. For example $C_{0}=0.15{\rm nm}^{-1}$ and $C_{0}=0.05{\rm nm}^{-1}$ correspond to protein densities of $\rho=1$ and $\rho=0.3333$, respectively .

\begin{figure}[H]
\centering
\includegraphics[width=7.5cm,clip]{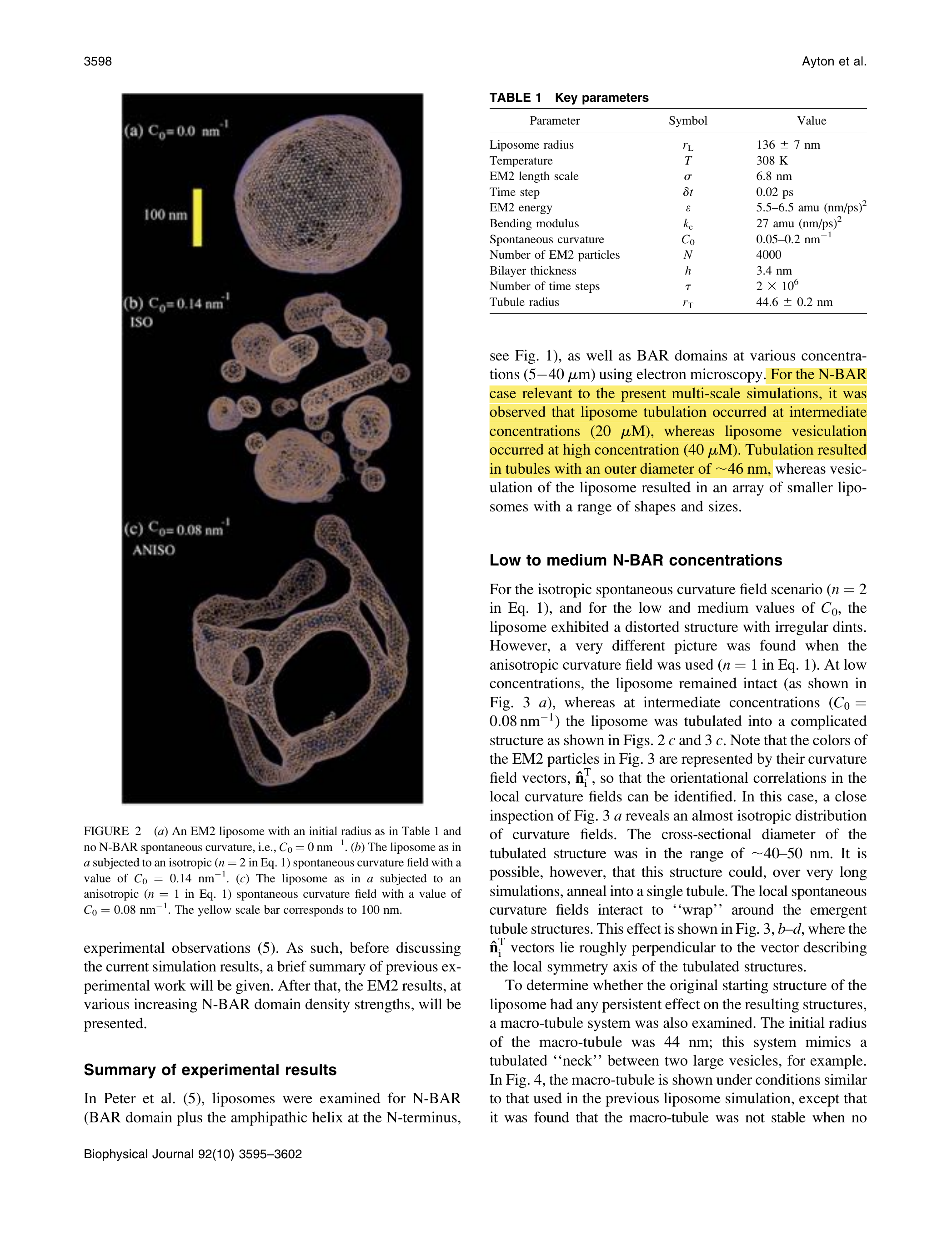}
\caption{\label{fig:ayton-2007-Iso}Membrane remodeling behavior studied using the EM2 model with the N-BAR domain modelled as an isotropic homogeneous curvature field. {\bf (a)} An initially vesicular membrane remains spherical when $C_{0}=0.0\,{\rm nm}^{-1}$ , and {\bf (b)} the vesicular structure breaks up into a collection of smaller vesicles at higher values of $C_{0}$; data has been shown for $C_{0}=0.14 \,{\rm nm}^{-1}$ {\sf (Reprinted from Biophys. J, {\bf 92} (10), G. Ayton, P. D. Blood, and G. A Voth, Membrane Remodeling from N-BAR Domain Interactions: Insights from Multi-Scale Simulation, 3595--3602, Copyright (2007), with permission from Elsevier)}.}
\end{figure}

Figs.~\ref{fig:ayton-2007-Iso}(a) and (b) show the membrane remodeling behavior of N-BAR domains using the isotropic curvature model (eqn.~\eqref{eq:EM2-iso}). At low N-BAR densities (small values of $C_{0}$), the initially vesicular membrane remains nearly spherical; a representative shape obtained for $C_{0}=0.0 {\rm nm}^{-1}$ has been shown in  Fig.~\ref{fig:ayton-2007-Iso}(a). However, with increase in the spontaneous curvature (increasing protein density), the spherical shape becomes unstable and breaks up into an ensemble of smaller vesicles; this phenomenon has been shown in Fig.~\ref{fig:ayton-2007-Iso}(b) for $C_{0}=0.14 {\rm nm}^{-1}$. The radius of the smaller vesicles is set by the imposed spontaneous curvature and can be approximated to be $(C_{0})^{-1}$. These results obtained from the EM2 model are consistent with those obtained from theoretical analysis and also from computations using dynamical triangulation Monte Carlo  techniques. 

\begin{figure}[H]
\centering
\includegraphics[width=7.5cm,clip]{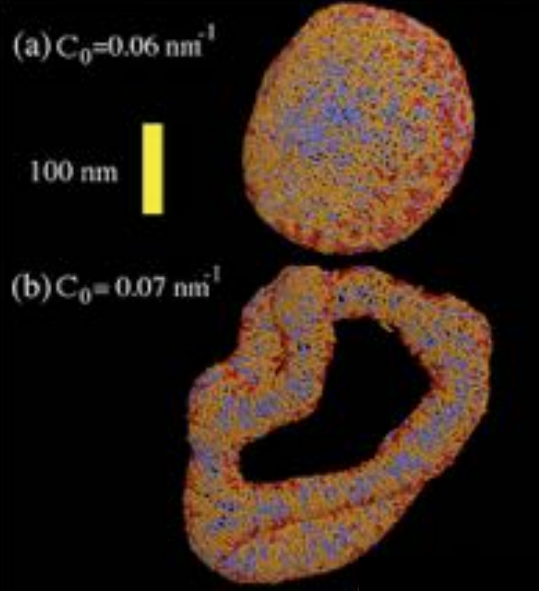}
\caption{\label{fig:ayton-2007-Aniso} Steady state shapes of a vesicular membrane evolved using the EM2 model with anisotropic curvature field. Spherical shapes which are stable for $C_{0} \leq 0.06\,{\rm nm}^{-1}$ {\bf (a)} evolve into a network of tubules when $C_{0}=0.07\,{\rm nm}^{-1}$  {\bf (b)} {\sf (Reprinted from Biophys. J, {\bf 92} (10), G. Ayton, P. D. Blood, and G. A Voth, Membrane Remodeling from N-BAR Domain Interactions: Insights from Multi-Scale Simulation, 3595-�3602, Copyright (2007), with permission from Elsevier).} }
\end{figure}

The membrane starts to show interesting remodeling dynamics when the  curvature field due to the BAR domains is treated as an anisotropic curvature field, whose curvature profile follows eqn.~\eqref{eq:EM2-aniso}. Fig.~\ref{fig:ayton-2007-Aniso} shows the steady-state self-assembled shapes of the EM2 particles for four different spontaneous curvatures. When $C_{0}<=0.06\,{\rm nm}^{-1}$, the protein-induced curvature field is too weak to promote membrane remodeling, and hence, the quasi particles assemble into vesicular shapes as seen in Fig.~\ref{fig:ayton-2007-Aniso}(a).   
With further increase in $C_{0}$, the membrane is driven into a network of tubular shapes. The topology of this tubular network depends on the values of $C_{0}$ and also on the strength of $\kappa_{G}$. The transition from spherical to tubular shapes, with increasing $C_{0}$, is consistent with that seen in Fig.~\ref{fig:conf-kperp0}.
\begin{figure}[H]
\centering
\includegraphics[width=12.5cm,clip]{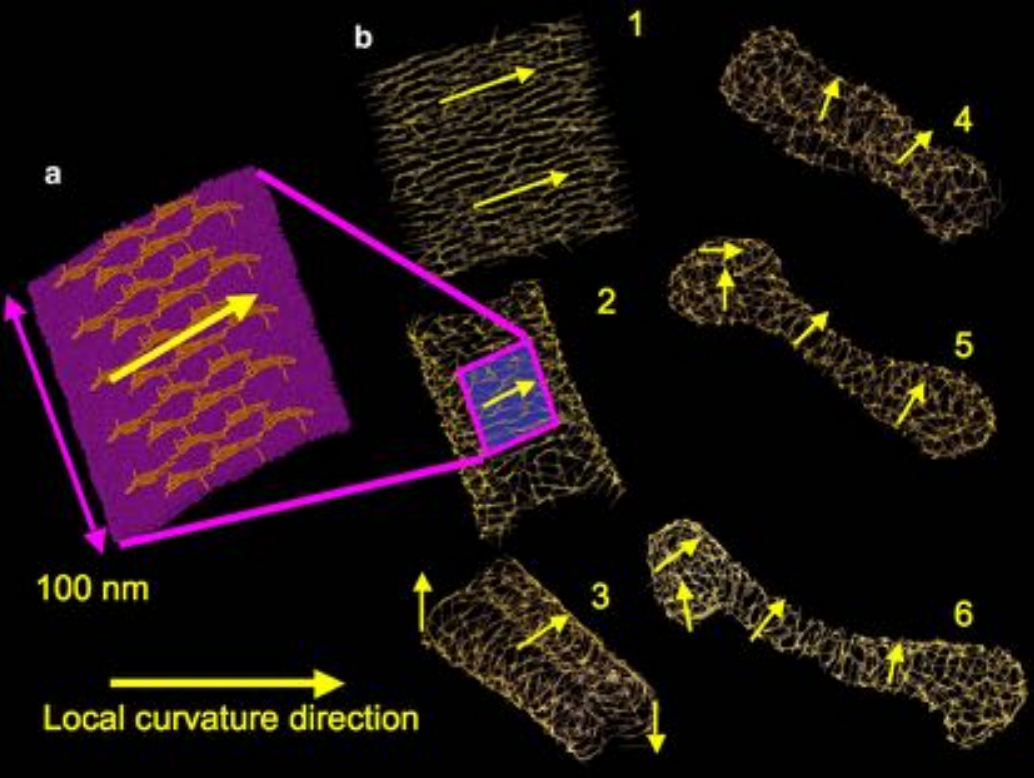}
\caption{\label{fig:ayton-2009-Seq}{\bf (a)} An atomistic representation of N-BAR proteins oligomerized on the membrane surface. Solid arrows show the local spontaneous curvature direction on the membrane. {\bf (b)} Evolution (1-6) of an initially flat 250 nm$^{2}$ membrane patch (1) into a tubule configuration (6) {\sf (Reprinted from Biophys. J, {\bf 97} (6), G. Ayton, E. Lyman, V. Krishna, R. D. Swenson, C. Mim, V. M. Unger and G. A Voth, New Insights into BAR Domain-Induced Membrane Remodeling, 1616--1625, Copyright (2009), with permission from Elsevier)}.}
\end{figure}
The dynamics of the EM2 particles and of the in-plane orientation field has been shown in Fig.~\ref{fig:ayton-2009-Seq}(b), where an initially planar membrane remodels into a tubule. Locally averaged in-plane orientations  are shown as solid lines while the orientations of the local membrane curvature are represented as diffuse lines in the background. For positive values of $C_{0}$, the orientation of the in-plane field is highly correlated with the maximum curvature direction on the membrane, and the steady state is reached when the tubule radius approaches $(C_{0})^{-1}$. This finding is consistent with that observed in the  case of the nematic membrane model. \\ 

The EM2 model provides an excellent simulation technique to study the dynamics of the membrane-protein system with explicit hydrodynamics  in the meso- and continuum scales. The parameters for the protein field are determined from an all atom  or coarse grained molecular simulation, but the validity of the linear relation between the protein density and spontaneous curvature employed here warrants further investigation. In the EM2 model, the membrane is formed as a result of self-assembly of the quasiparticles and hence the model can accommodate morphologies with varying topologies. The discretization of the elastic energy Hamiltonian assumes small curvature gradients locally, and as a result, the application of the model to systems with extreme curvature requires further sensitivity analysis.


\section{Free energy methods for membrane energetics} \label{sec:freeener}
In the previous sections, we studied the conformational phase of a biological membrane, with and without externally induced curvature fields, using analytical and computational techniques. We have shown that the evolution of suitable order parameters, in parametric space, in finite-temperature simulations can provide valuable insight into the nature of the morphological transitions sustained in specific model systems. However, in order to establish the stability of the different morphological states/phases, the free-energy landscape has to be delineated. Free-energy methods complement finite-temperature simulations by providing insight into how entropic contributions modulate the energy landscape associated with various morphologies. In this section, we describe methods  to numerically delineate the free-energy landscape for the membrane with isotropic and anisotropic curvature fields, using the method of thermodynamic integration.

\subsection{Thermodynamic integration (TI)}
In this approach, the free-energy difference between two membrane states $A$ and $B$ with energies $E_{A}$ and $E_{B}$, respectively, is computed by constructing a reversible path to transition from $A$ to $B$. The intermediate configurations of the membrane are controlled by the Kirkwood coupling parameter $0 \leq\lambda \leq 1$, such that the membrane is in state $A$ when $\lambda=0$ and state $B$ when $\lambda=1$. The energy of any given intermediate state can be written as $E(\lambda)= (1-\lambda)E_{A}+\lambda E_{B}$. Hence, the change in free-energy involved in a transition from state $A$ to $B$, calculated along the path $\cal{S}$, is given by ~\cite{Frenkel:2001},
\begin{equation}
\Delta F_{A\rightarrow B}=\int_{0}^{1} \left(\frac{\partial F}{\partial \lambda} \right) \,d\lambda .
\label{eqn:TIfree_energy}
\end{equation}

For a system whose energy depends on a coupling parameter, $\lambda$, the partition function can be written as, ~\cite{Frenkel:2001}:
\begin{equation}
\label{eq:par_fun}
Q(\lambda) = c\int{\exp\left[-\beta E(\lambda)\right]}\,dr^N,
\end{equation}
where $c$ is a constant. Since the Helmholtz free-energy $F(\lambda) = -k_BT\ln Q (\lambda)$, the derivative of the free-energy with respect to $\lambda$ can be written as:
\begin{equation}
\left(\frac{\partial F}{\partial \lambda}\right)_{N,V,T} = -\frac{1}{\beta}\frac{\partial }{\partial \lambda}\ln Q,
\end{equation}
yielding
\begin{equation}
\left(\frac{\partial F}{\partial \lambda}\right)_{N,V,T} = \left\langle \frac{\partial E}{\partial \lambda}\right\rangle _{\lambda}.
\label{eq:TI_lambda}
\end{equation}

Using eqn.~\eqref{eq:TI_lambda} in eqn.~\eqref{eqn:TIfree_energy} we can evaluate the change in free-energy without computing the absolute free energies for various membrane states along the path $\cal{S}$ as,
\begin{equation}
\Delta F_{A\rightarrow B}=\int_{0}^{1} \left \langle \frac{\partial E}{\partial \lambda} \right \rangle \,d\lambda.
\label{eqn:TIfree_energy_final}
\end{equation}
This method can be used to compute the free energies associated with the membrane shapes stabilized by isotropic and anisotropic curvactant fields.

\subsection{Free energy for a membrane in the Monge gauge subject to isotropic curvature fields}
To investigate this phenomenon from a free-energy perspective, we employ thermodynamic integration (TI). The elastic energy of a planar membrane with an external curvature field $C_{0}$, given by eqn.~\eqref{eqn:Monge-withscur}, has the form,
\begin{equation}
\mathscr{H}_{\rm sur}=\int \left \{ \frac{\kappa}{2} \left(\nabla^{2}h -C_{0}\right )^{2} 
		+ \left (\frac{\sigma}{2}+\frac{\kappa}{4}C_{0}^{2} \right )\left(\nabla h \right)^2 
\right \} dx dy.
\label{eq:Helfrich}
\end{equation} 

Agrawal and Radhakrishnan ~\cite{Agrawal:2009bt} demonstrated the applicability of TI to a model system of membrane deformations caused by a static (i.e. non-diffusing) heterogeneous curvature field. The spontaneous curvature field $C_{0}$ has a  radially symmetric profile  on the membrane over a localized region characterized by a linear extent $r_0$. The value of $C_0$ is taken to be zero in the membrane regions falling outside the localized region. Thus, the induced curvature field is described by:
\begin{equation}
C_0 = c_0\Gamma(r_0),
\label{eq:H0}
\end{equation}
where $\Gamma(r_0)$ is a function that is unity within a circular domain (centered at zero) of radius $r_0$ and zero otherwise, and $r_0$ is the linear extent (radius) of the curvature field projected on the $x$-$y$ plane. For the sake of illustration, we choose $c_0=0.04$ (nm)$^{-1}$. The free-energy change of the membrane is calculated as a function of the extent of the curvature field ($r_0$) as well as the magnitude of the curvature field ($c_0$). \\

\begin{figure}[t]
\centering
\includegraphics[width=7.5cm]{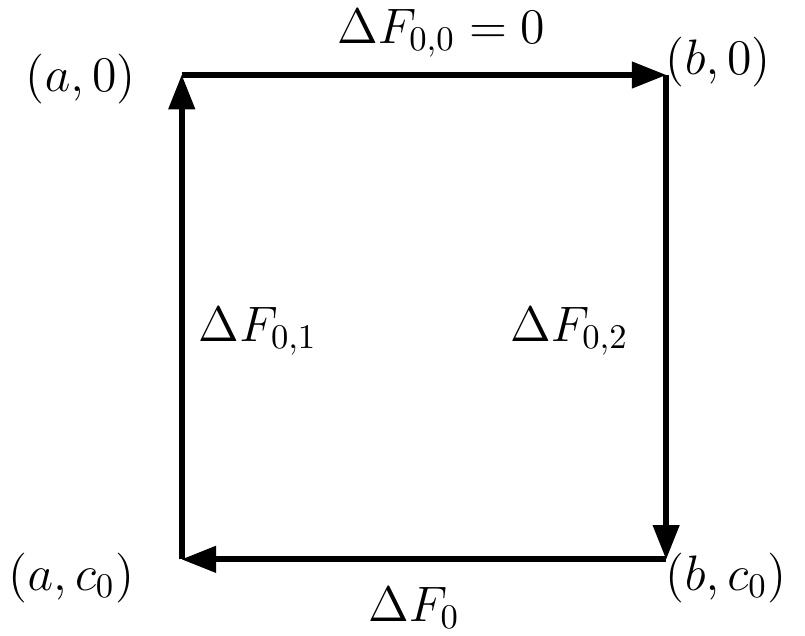}
\caption{Thermodynamic cycle to calculate $\Delta F_0$; $\Delta F_0 = -\Delta F_{0,1} + \Delta F_{0,0} + \Delta F_{0,2}$. $a$ and $b$ are the extent of the curvature-induced regions ($r_0$ values) while $c_0$ is the magnitude of the curvature. $\Delta F_{0,1}$ and $\Delta F_{0,2}$ are computed using eqn. ~\eqref{eq:TI}.}
\label{fig:thermo_cycle}
\end{figure}

In eqn. ~\eqref{eq:Helfrich} and eqn. ~\eqref{eq:H0}, when $c_0$ is set to zero, the planar state of the membrane is recovered, whereas for non-zero values of $c_0$, the desired state of the curvilinear membrane is obtained. We also note that the energy functional (eqn. ~\eqref{eq:Helfrich}) is differentiable with respect to $c_0$ but not differentiable with respect to $r_0$. Hence, to compute the free-energy changes, we choose $c_0$ as the thermodynamic integration variable (i.e. as the coupling parameter $\lambda$ in eqn. ~\eqref{eq:TI_lambda}) to obtain:
\begin{equation}
\frac{\partial F}{\partial c_0} = \left< \frac{\partial E}{\partial c_0}\right>_{c_0}.
\end{equation}
Using the expression for $E$ from eqn. ~\eqref{eq:Helfrich}, we obtain:
\begin{equation}
\frac{\partial F}{\partial c_0} = \left< \Gamma(r_0)\kappa\int\int{\left[-\left(\nabla^2h-c_0\Gamma(r_0)\right)+\left(\frac{c_0}{2}\right)\left(\nabla h\right)^2\right]}\,dxdy\right>_{c_0},
\end{equation}
Upon integration along $c_0$, this yields:
\begin{equation}
\label{eq:TI}
F(c_0,r_0)-F(0,r_0) = \int_0^{c_0}{\left< \Gamma(r_0)\kappa\int\int{\left[-\left(\nabla^2h-c_0\Gamma(r_0)\right)+\left(\frac{c_0}{2}\right)\left(\nabla h\right)^2\right]}\,dxdy\right>_{c_0}}\,dc_0.
\end{equation}
Here, $F(c_0,r_0)-F(0,r_0)$ is the free-energy change as derived from the partition function in eqn. ~\eqref{eq:par_fun}, where the energy is defined in Eq. ~\eqref{eq:Helfrich}. However, if we are interested in deformation free-energy, $F_0$, with reference to a state where $C_0 = 0$, we employ the relationship:
\begin{equation}
F_0 = F + \langle E_0\rangle - \langle E\rangle,
\label{eq:cycle}
\end{equation}
where $E_0$ is eqn.~\eqref{eq:Helfrich} with $C_{0}=0$. Thus, $\Delta F_0 = F_0(C_0,r_0)-F_0(0,r_0)$ gives the deformation free-energy change for a given extent of the localized region $r_0$ (such as size of the clathrin coat) when $C_0$ is varied. To calculate the free-energy as a function of $r_0$ for a fixed $C_0$, we employ a thermodynamic cycle defined in Fig.~~\ref{fig:thermo_cycle}. In this cycle, $\Delta F_{0,1}$ and $\Delta F_{0,2}$ required to deform a planar membrane to $C_0=c_0\Gamma(r_0=a)$ and $C_0=c_0\Gamma(r_0=b)$, respectively, are calculated through eqn. ~\eqref{eq:TI} and eqn. ~\eqref{eq:cycle}. \\

\begin{figure}[H]
\centering
\includegraphics[width=10cm,clip]{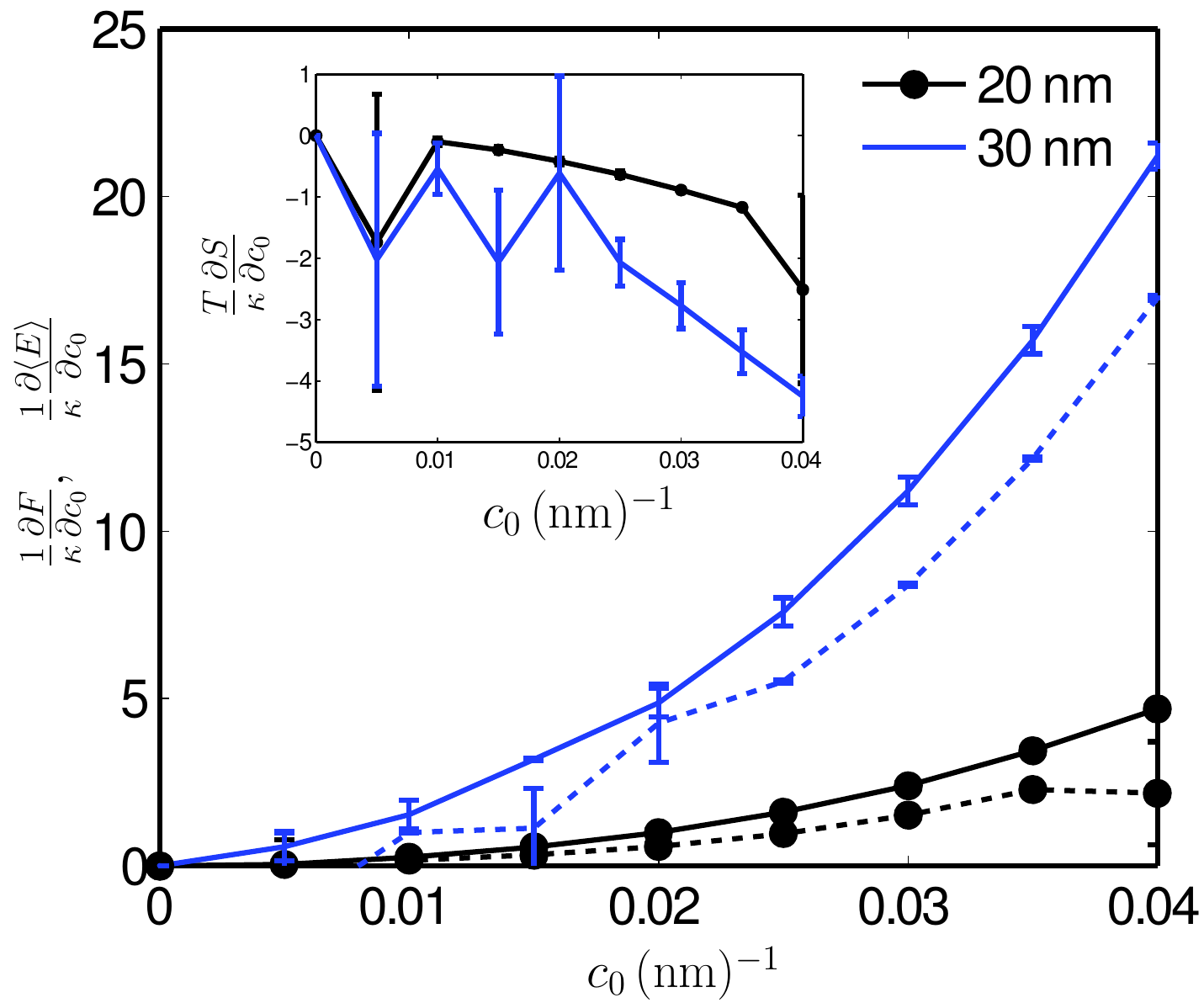}
\caption{$\dfrac{\partial F}{\partial C_0}$ (solid lines) and $\dfrac{\partial \langle E\rangle}{\partial C_0}$ (dotted lines) plotted for two values of $r_0$ = 20nm and 30 nm. The inset shows $T\dfrac{\partial S}{\partial C_0}$ as a function of $c_0$ {\sf(Reprinted figure with permission from  N. J. Agrawal, Radhakrishnan,  Phys. Rev. E, {\bf 80}, 011925  and 2009. Copyright (2009) by the American Physical Society)}.}
\label{fig:TI}
\end{figure}

The numerical results for the free-energy changes obtained by Agrawal and Radhakrishnan ~\cite{Agrawal:2009bt} using thermodynamic integration are shown in  Fig.~~\ref{fig:TI}. $\frac{\partial F}{\partial c_0}$ increases with increasing value of $c_0$ implying that the free-energy of the membrane, $F$, increases with increasing magnitude of $c_0$. Furthermore, for a larger extent $r_0$, the increase in free-energy is larger for the same change in $c_0$. Fig.~~\ref{fig:TI} shows the calculated values of $\left(\frac{\partial \langle E\rangle}{\partial C_0}\right)_{r_0}$ for different values of $c_0$ and $r_0$. The quantity $\frac{\partial \langle E\rangle}{\partial c_0} - \frac{\partial F}{\partial c_0}$ derived from these two plots yields the entropic contributions $T\frac{\partial S}{\partial c_0}$, which are plotted in the inset of Fig. ~\ref{fig:TI}.  As evident from these figures, the entropic contribution to the membrane free-energy decreases with $C_0$, with the decrease being more prominent for larger values of $r_0$. This provides support for the mechanism of entropy-mediated attraction between curvature-inducing bodies discussed in section~~\ref{sec:isotropic}.

\begin{figure}[H]
\centering
\includegraphics[width=10cm,clip]{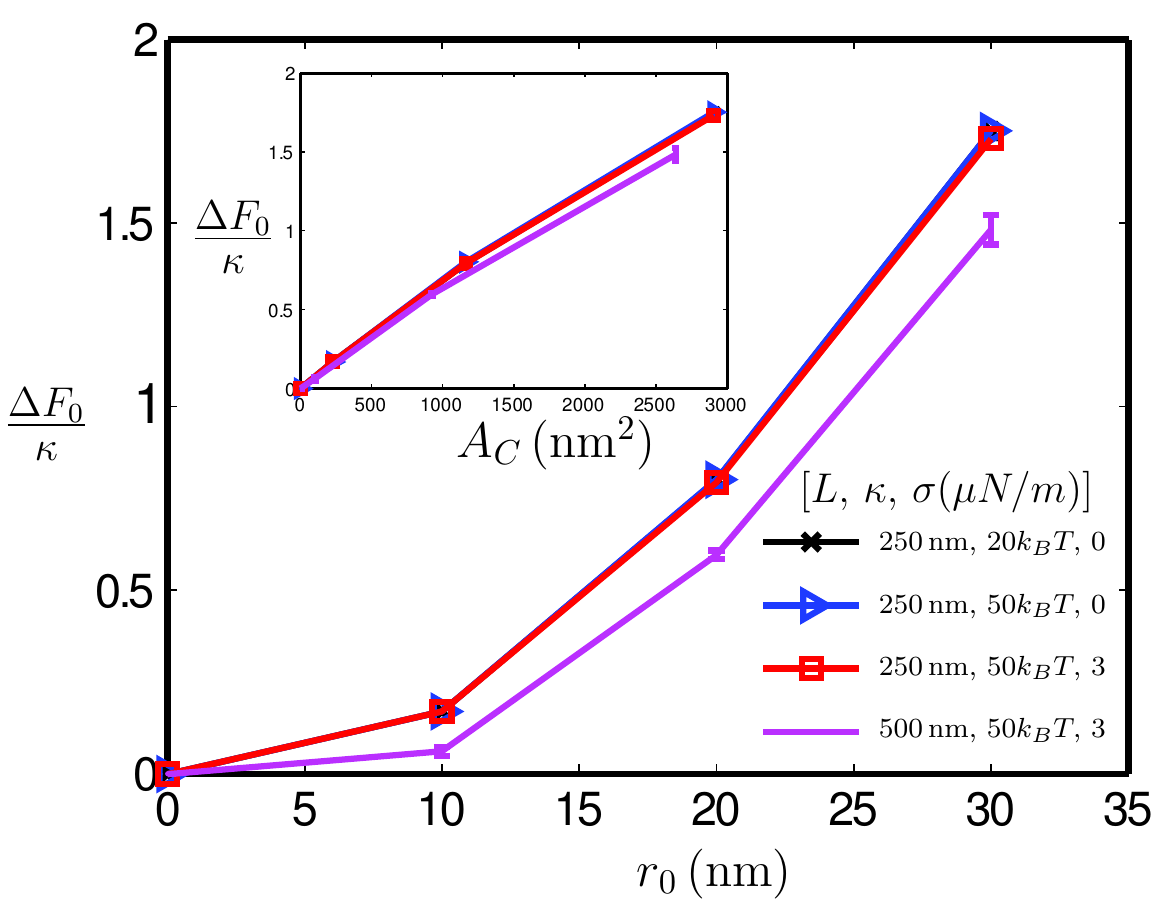}
\caption{Membrane energy change as a function of $r_0$ {\sf(Reprinted figure with permission from  N. J. Agrawal, Radhakrishnan,  Phys. Rev. E, {\bf 80}, 011925  and 2009. Copyright (2009) by the American Physical Society)}.}
\label{fig:free_energy}
\end{figure}

Using the thermodynamic cycle shown in Fig.~~\ref{fig:thermo_cycle}, Agrawal and Radhakrishnan ~\cite{Agrawal:2009bt} calculated the membrane deformation free-energy change as a function of the extent of $r_0$ (data not shown). Computing the deformation free-energy change  with respect to a planar membrane (i.e. $C_0 = 0$) gives the mean energy $\langle E_0\rangle$ with respect to the planar membrane. The change in the deformation free-energy ($F_0$) with respect to a planar membrane is plotted in Fig.~~\ref{fig:free_energy}, and the corresponding plots for change in deformation energy ($E_{0}$) and entropic energy ($TS$) can be found in reference ~\cite{Agrawal:2009bt}. Data are shown  for four different systems varying in length ($L$), bending rigidity ($\kappa$), and surface tension ($\sigma$).

The deformation free-energy of the membrane increases as the extent of the curvature field $r_0$ increases. Furthermore, changes in the non-dimensional deformation free-energy, $F_0/\kappa$, show similar trends for different systems of equal size, $L=250$nm. Thus, $\Delta F_0/\kappa$ depends only weakly on membrane bending rigidity $\kappa$ and membrane frame tension $\sigma$. The inset to Fig.~~\ref{fig:free_energy}  depicts the variation of $\Delta F_0/\kappa$ with area of the localized region subject to the curvature field, $A_c$, defined as:
\begin{equation}
A_C = \int\int{\Gamma(r_0)\left(1+\frac{1}{2}\left(\nabla h\right)^2\right)dxdy}.
\end{equation}
In this region, $\Delta F_0/\kappa$ scales almost linearly demonstrating that membrane free-energy is a linear function of $A_C$ for small deformations considered here. Interestingly, the increase in $\Delta F_0/\kappa$ is smaller for the larger membrane size. Noting that the difference in the entropy change for different sizes of membranes is small, the changes in $\Delta F_0/\kappa$ are a reflection of the changes in $\Delta E_0/\kappa$. 

\subsection{Thermodynamic integration methods for the anisotropic curvature model}
In the anisotropic curvature model, the membrane-curving nature of curvactant inclusions is captured in the anisotropic elastic term given by  eqn.~\eqref{eq:Hnesur}. Thermodynamic integration can also be used  to compute the change in free-energy  when a spherical homogeneous membrane undergoes a morphological change to the various phases listed in Figs.~\ref{fig:conf-kperp0} and ~\ref{fig:funcconc} due to its interaction with these membrane inclusions. In this model, the anisotropic bending rigidities can be chosen to be the Kirkwood coupling parameters, so that eqn.~\eqref{eq:Hnesur} can now be written as,
\begin{equation}
\label{eq:Hnesur-lambda}
\mathscr{H}_{\rm anis}(\lambda_{\parallel},\lambda_{\perp})=\frac{1}{2}\int_{S}d{\mathbf{S}}\,\left \{\lambda_{\parallel}\kappa_{\parallel}\left[m^{a}K_{ab}m^{b}-C_{0}^{\parallel}\right]^{2}\,+\,\lambda_{\perp}\kappa_{\perp}\left[m_{\perp}^{a}K_{ab}m_{\perp}^{b}-C_{0}^{\perp}\right]^{2} \right \}. 
\end{equation}
Here, two coupling parameters, $\lambda_{\parallel}$ and $\lambda_{\perp}$, that couple respectively to $\kappa_{\parallel}$ and $\kappa_{\perp}$ have been used to construct the two dimensional free-energy landscape associated with a nematic field coupled to the membrane. For simplicity, we will focus only on the effect of a nematic field with curvature profile given by $\kappa_{\parallel} \neq 0$ and $\kappa_{\perp}=0$. The total energy of the nematic field has three contributions given by $\mathscr{H}_{\rm tot}=\mathscr{H}_{\rm sur}+\mathscr{H}_{\rm 2-atic}+\mathscr{H}_{\rm anis}(\lambda_{\parallel})$. In the above-mentioned framework, the free-energy can be computed from thermodynamic integration using the relation,

\begin{equation}
\label{eqn:FE-expr-anisotropic}
\Delta F(\kappa, \kappa_{\parallel},C_{0}^{\parallel})=\int_{0}^{1} \left \langle \frac{\partial \mathscr{H}_{\rm tot}}{\partial \lambda_{\parallel}} \right \rangle d \lambda_{\parallel}.
\end{equation}

The dynamically triangulated Monte Carlo technique, introduced in section~~\ref{sec:field-mcs}, has been used to sample the phase space of the membrane.  So far, all our dicussions related to nematic membranes explored the effect of directional spontaneous curvatures with fixed isotropic and directional bending rigidities, $\kappa=20k_{B}T$ and $\kappa_{\parallel}=5k_{B}T$. To complement those results, we will construct the free-energy landscape as a function of $C_{0}^{\parallel}$ with the set of bending rigidities mentioned above. 

\begin{figure}[H]
\centering
\includegraphics[width=12.5cm,clip]{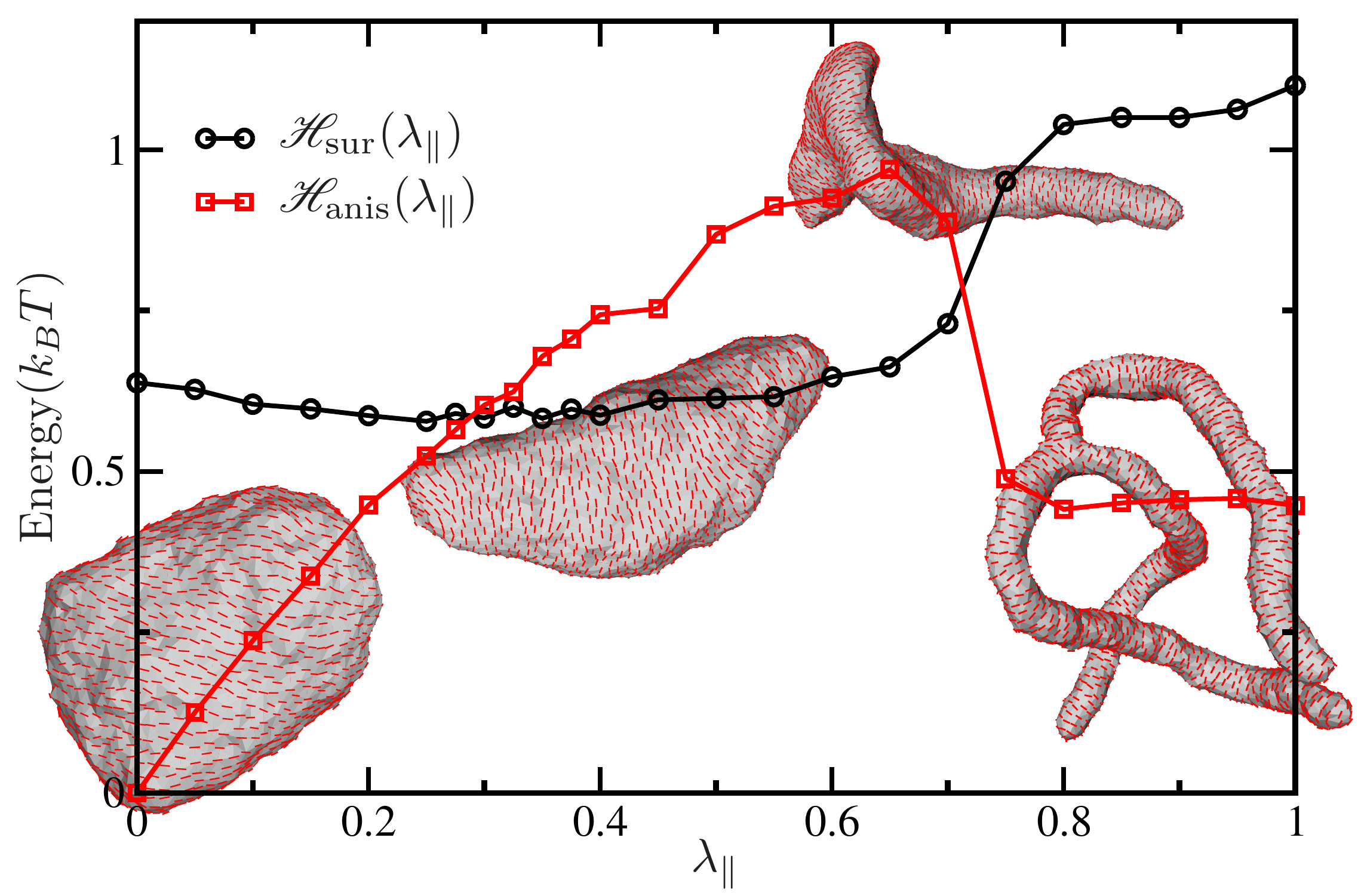}
\caption{\label{fig:kpar-energylambdapar}The elastic energy of a membrane ($\mathscr{H}_{\rm sur}$) and its interaction energy  with the nematic field ($\mathscr{H}_{\rm anis}(\lambda_{\parallel})$) as a function of $\lambda_{\parallel}$ for a nematic membrane, with $\kappa=20k_{B}T$, $C_{0}^{\parallel}=0.6$ and $\kappa_{\parallel}=5k_{B}T$. Shapes transitions are seen when the nematic coupling strength goes from $\kappa_{\parallel}=0 \rightarrow 5k_{B}T$. $\mathscr{H}_{\rm anis}$ starts to dictate the shape of the membrane when $\lambda_{\parallel} \simeq 0.7$ leading to tubular shapes.}
\end{figure}

Before proceeding to the discussion of the free-energy, we will recall the behavior of a membrane when subjected to a directional spontaneous curvature that promotes the formation of branched tubular shapes. The thermodynamic integration has been performed for $C_{0}^{\parallel}=0.6$, for which branched tubes were shown to be the equilibrium shapes in Fig.~\ref{fig:conf-kperp0}. The isotropic and anisotropic energy contributions, $\mathscr{H}_{\rm sur}$ and $\mathscr{H}_{\rm anis}$, for a triangulated membrane with $N=2030$ vertices, are shown along with the corresponding membrane conformations as a function of $\lambda_{\parallel}$ in Fig.~\ref{fig:kpar-energylambdapar}.

Whereas the elastic energy $\mathscr{H}_{\rm sur}$ prefers minimum curvature on the membrane surface, the anisotropic elastic term $\mathscr{H}_{\rm anis}(\lambda_{\parallel})$ would prefer membrane conformations with maximum principal curvature at every vertex,  equal to $C_{0}^{\parallel}$. As can be seen from Fig.~\ref{fig:kpar-energylambdapar}, when a spherical membrane transforms to a tube, $\mathscr{H}_{\rm anis}(\lambda_{\parallel})$ starts to minimize its contribution at the cost of $\mathscr{H}_{\rm sur}$ when $\lambda_{\parallel} \simeq 0.7$, which is a measure of the critical nematic membrane interaction strength required to tubulate a membrane. In the example above, this critical value can be shown to be equal to $\kappa_{\parallel}^{*}=0.7 \kappa_{\parallel}=3.5k_{B}T$, since we have fixed $\kappa_{\parallel}=5k_{B}T$ throughout our simulations. Such an analysis would be extremely helpful in protein design; say, in predicting the possible set of amino acids in a protein interacting with a membrane, provided a mapping scheme from the mesoscale to the atomistic scale exists.

\begin{figure}[H]
\centering
\includegraphics[width=12.5cm,clip]{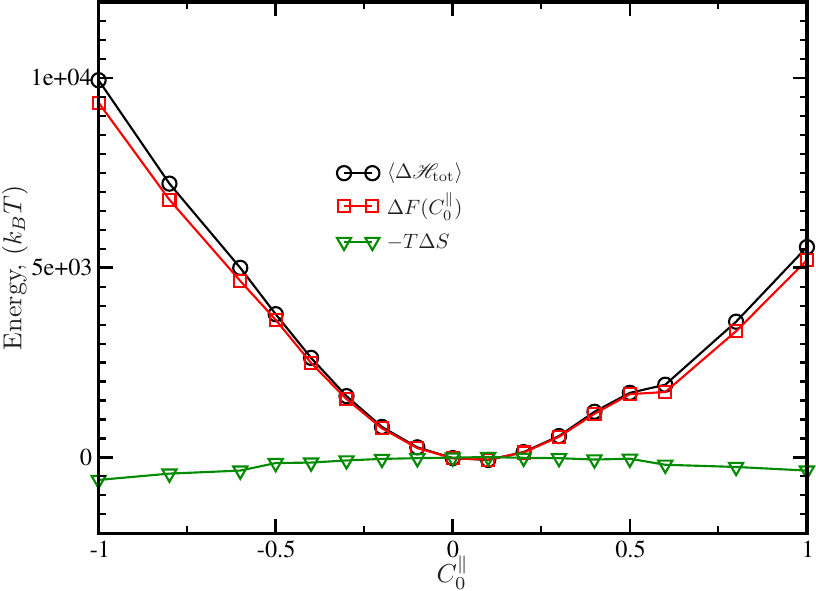}
\caption{\label{fig:kpar-freeenergy} The free-energy landscape connecting two membrane states with directional spontaneous curvatures 0 and $C_{0}^{\parallel}$, computed from the total energy $\mathscr{H}_{\rm tot}$ and free-energy $\Delta F(C_{0}^{\parallel}$). $\Delta F(C_{0}^{\parallel})$ approaches $\left \langle \Delta \mathscr{H}_{\rm tot} \right \rangle$ for smaller values of $C_{0}^{\parallel}$, whereas at large values of $C_{0}^{\parallel}$ the free-energy has contributions from the entropy, to the order of $600k_{B}T$; the sizeable entropic contribution stabilizes caveolae like morphologies (inward growing tubules) at large negative  $C_{0}^{\parallel}$ and tubular shapes at large positive  $C_{0}^{\parallel}$.}
\end{figure}

The free-energy change, $\Delta F(C_{0}^{\parallel})$,  for a nematic membrane with directional curvatures in the range $-1.0 \leq C_{0}^{\parallel}<1.0$, computed using eqn.~\eqref{eqn:FE-expr-anisotropic}, is shown in Fig.~\ref{fig:kpar-freeenergy}. Plotted alongside are the average energy difference $\left \langle \Delta \mathscr{H}_{\rm tot} \right \rangle$ and the entropic contribution to the free-energy, calculated  as $-T\Delta S=\Delta F(C_{0}^{\parallel})-\left \langle \Delta \mathscr{H}_{\rm tot} \right \rangle$. The energy difference in the thermodynamic state of the membrane for a given value of directional spontaneous curvature is defined as $\left \langle \Delta \mathscr{H}_{\rm tot} \right \rangle=\mathscr{H}_{\rm tot}(\lambda_{\parallel}=1)- \mathscr{H}_{\rm tot}(\lambda_{\parallel}=0)$.

As expected, the energy barrier given by the difference in the internal energies of the membrane  is always greater than or equal to the energy barrier computed through free-energy methods. $\left \langle \Delta \mathscr{H}_{\rm tot} \right \rangle \sim \Delta F$ when the membrane remains quasi-spherical at small magnitudes of $|C_{0}^{\parallel}|$. When the membrane transforms into tubes and inward tubes, for $|C_{0}^{\parallel}|>0.6$, the computed free-energy is lower than the  $\left \langle \Delta \mathscr{H}_{\rm tot} \right \rangle$ by around $600\,k_{B}T$ due to the large conformational entropy associated with these morphologies. The estimated  value of $-T\Delta S$ is $\sim 10\%$ of the total energy of the membrane, which agrees well with similar estimates for this value ~\cite{Agrawal:2009bt}. 

Other methods for computing free energies can be readily adapted within the continuum framework discussed above and using the machinery of Monte Carlo simulations. Recently, Tourdot et al. ~\cite{Tourdot:2014} introduced and compared three free-energy methods to delineate the free-energy landscape of curvature-inducing proteins on bilayer membranes. Specifically, they showed the utility of the Widom-test-particle/field insertion methodology in computing the excess chemical potential associated with curvature-inducing proteins on the membrane and in tracking the onset of morphological transitions on the membrane at elevated protein density. The authors validated their approach by comparing the results with those of thermodynamic integration and Bennett acceptance ratio methods.\\



\section{Conclusions} \label{sec:conclusion}

In summary, we have discussed various theoretical and modeling strategies along with specific applications utilized in the study of protein-induced curvature in membranes from equilibrium and dynamic/hydrodynamic perspectives. The future of such studies is very bright and ripe as we are beginning to unravel the true complexity of biological systems through highly quantitative experiments at the mesoscale. Given the central importance of protein-induced curvature mechanisms in cellular transport, an integration of these biophysical methods along with models of signal transduction is expected to provide unprecedented valuable insight into cell biology. With the advent of advances in algorithms for multiscale modeling as well as high-performance computing, future models are expected to mimic the underlying biological complexity, such as the signaling microenvironment, mechanotransduction machinery, and tissue-specific boundary conditions. Future extensions of such models to the non-equilibrium regime (which we have completely ignored in this article) are also expected to define as well as elucidate crucial cellular mechanisms that fall outside the purview of equilibrium or linear-response regimes.


\section{Acknowledgments}
The authors thank Dr. Neeraj Agrawal and Dr. John H. Ipsen for many insightful discussions. They also thank numerous colleagues at the University of Pennsylvania and the Indian Institute of Technology Madras for their various inputs. The authors are also very grateful to the anonymous referees who helped transform this article to its current form through their extensive and insightful feedback. The authors acknowledge funding from the US National Science Foundation (CBET-113267, CBET-1236514, DMR-1120901), the US National Institutes of Health (1R01EB006818 and U01-EB016027), and the European Commission (7th programme for research VPH-600841). Supercomputing resources were made available through XSEDE (extreme science and engineering discovery environment, MCB060006). Source files and example files for some of the codes discussed in this review are available online as supplementary material to this article.

\newpage
\section{Appendix}
\appendix
\renewcommand*{\thesection}{\Alph{section}}
\section{Differential geometry} \label{app:diffgeo}
In this appendix, we will focus on the general mathematical techniques used to compute the mean and Gaussian curvature associated with the continuum elastic energy given in eqn.~\eqref{eqn:can-Helf}. This material can be found elsewhere ~\cite{Piran:2003,doCarmo:1976,Cai:1995un}, but we present it here for completeness.

In the continuum approximation, a membrane is described by a two-dimensional surface, ${\cal S}$, embedded in a three-dimensional space, $\mathbb{R}^{3}$. The membrane is parameterized by a position vector $\vec{R}(x^{1},x^{2},x^{3})$, with the gauge defined by the orthonormal coordinates $x^{1}$, $x^{2}$, and $x^{3}$. If $x^{1}$ and $x^{2}$ are orthonormal vectors in the tangent plane of the membrane, the problem is simplified when one choses a suitable gauge for the surface such that $x^{3}=f(x^{1},x^{2})$, where $f$ is a function of $x^{1}$ and $x^{2}$. The tangent vector at any point on the membrane surface is defined as,
\begin{equation}
\vec{t}_{\alpha}=\partial_{\alpha}\vec{R}.
\end{equation}
Here, $\partial_{\alpha}=\partial/\partial x^{\alpha}$, with $\alpha=1, 2$. The unit surface normal can be determined from the tangents as,
\begin{equation}
\hat{n}=\frac{\vec{t}_{1} \times \vec{t}_{2}}{|\vec{t}_{1} \times \vec{t}_{2}|}.
\end{equation}
The metric of the gauge, defined by the metric tensor, is given by the  first fundamental form,
\begin{equation}
g_{\alpha,\beta}=\vec{t}_{\alpha} \cdot \vec{t}_{\beta}=\partial_{\alpha}\vec{R}\cdot\partial_{\beta}\vec{R},
\end{equation}
The metric tensor is a fundamental measure of length on the surface: the distance between any two points seperated by $d\vec{x}$ can be written in terms of the metric tensor as $ds^{2}=dx^{\alpha} dx^{\beta} g_{\alpha \beta}$, and the area of a surface patch is given by $dA=\sqrt{\det g}\,dx^{1}dx^{2}$.
The raising operator $g^{\alpha \beta}$, which is the inverse of the metric tensor,  is defined as,
\begin{equation}
g^{\alpha \beta}=\left(g_{\gamma \nu}^{-1}\right )_{\alpha \beta},
\end{equation}
and satisfies the following relation,
\begin{equation}
g^{\alpha \gamma}g_{\gamma \beta}=\delta^{\alpha}_{\beta}. 
\end{equation}
The curvature profile of a membrane is determined by the curvature tensor, given by the second fundamental form, as,
\begin{equation}
K_{\alpha \beta}=\partial_{\alpha\beta}\vec{R} \cdot \hat{n}. 
\end{equation}
A surface is characterized by the two curvature invariants determined from the curvature tensor\textemdash namely,
\begin{equation}
{\rm Mean \, Curvature :\quad} H=\frac{1}{2}K_{\alpha}^{\alpha}=\frac{1}{2}{\rm Tr}{(K_{\alpha \beta})} \quad {\rm and}
\end{equation}
\begin{equation}
{\rm Gaussian \, Curvature :\quad} G=\frac{\det{(K_{\alpha \beta})}}{{\det(g_{ab})}}.
\end{equation}

\section{Planar Monge gauge} \label{app:diffgeo-planar}
\noindent Small deformations of a planar membrane about the reference plane can be formulated and studied in the planar Monge gauge. The position vector in this gauge is a functional $\vec{R}(x,y,h(x,y))$, where $h(x,y)$ is the normal displacement of the membrane region $(x,y)$. If the position vector is $\vec{R}=\left\{x,y,h(x,y)\right\}$, various geometrical quantities, defined in ~\ref{app:diffgeo}, in this gauge are as follows: \\

\noindent {\em Tangent vector :} With $\alpha=x,y$, the unit tangent vectors at any point on the membrane surface are given by,
\beqn \vec{t}_{x}=\left\{\,1\,, 0\,,\partial_{x}h \right\}, \eeqn
and
\beqn \vec{t}_{y}=\left\{\,0, 1,\partial_{y}h \right\}.  \eeqn 

\noindent {\em Normal vector :} The unit normal vector to the surface is given by,
\beqn\hat{n}=\frac{\vec{t}_{x} \times \vec{t}_{y}}{\left|\vec{t}_{x} \times \vec{t}_{y} \right|}=\dfrac{ \left\{ -\partial_{x}h,\,-\partial_{y}h,\,1\right\}}{\sqrt{1+\left(\nabla h\right)^{2}}}. \eeqn

\noindent {\em Covariant metric tensor, $g_{\alpha \beta}$ :} The metric tensor in the Monge gauge takes the form,

\beqn g_{\alpha \beta}=\,\left [
\begin{array}{cc}
1+\left(\partial_{x}h\right)^{2} &\partial_{x}h\,\partial_{y}h \\
     & \\
\partial_{x}h\,\partial_{y}h & 1+\left(\partial_{y}h\right)^{2} 
\end{array}
\right ]\,.\nonumber\eeqn
The metric $g$ in the Monge gauge is,
\beqn g=\det(g_{\alpha \beta})=\left(1+\left(\nabla h\right)^{2} \right).\eeqn

\noindent {\em Contravariant metric tensor:}
The contravariant metric tensor $g^{\alpha \beta}$, the inverse of $g_{\alpha \beta}$, in the Monge gauge is,
\begin{equation}
g^{\alpha \beta}= \frac{1}{{1+(\nabla h)^{2}}} \left [ 
\begin{array}{cc}
1+\left(\partial_{y}h\right)^{2} & -\partial_{x}h\,\partial_{y}h \\
     & \\
-\partial_{x}h\,\partial_{y}h & 1+\left(\partial_{x}h\right)^{2} 
\end{array}
\right ].
\end{equation}

\noindent {\em Curvature tensor :}
The components of the curvature tensor are given by $K_{\alpha \beta}=\partial_{\alpha} \partial_{\beta} \vec{R}\,\cdot\, \hat{n}$ . In the Monge gauge,
\beqn
K= \dfrac{1}{\sqrt{1+\left(\nabla h\right)^{2}}}\,\,\,\left[
\begin{array}{cc}
\partial_{x}\partial_{x}h & \partial_{x}\partial_{y}h \\
 &  \\
\partial_{x}\partial_{y}h & \partial_{y}\partial_{y}h
\end{array}
\right].
\eeqn

\noindent {\em Curvature invariants :}
The gauge-invariant quantities that can be constructed from the curvature tensor are
\begin{itemize}
\item
Mean curvature \\
\beqn H=\frac{1}{2}{\rm Tr}(K)= \frac{1}{2} \dfrac{\nabla^{2}h}{\sqrt{1+\left(\nabla h\right)^{2}}}. \eeqn
\item Gaussian curvature
\beqn G=\frac{1}{2} \epsilon_{\alpha\beta} \epsilon_{\gamma \delta} K_{\alpha \gamma} K_{\beta \delta}= \det(K)= \frac{(
\partial_{x}\partial_{x}h)(\partial_{y}\partial_{y}h) - (\partial_{x}\partial_{y}h)^{2}}{\sqrt{1+\left(\nabla h\right)^{2}}}.
\eeqn
$\epsilon$ is the anti-symmetric Levi-Cevita tensor.
\end{itemize}
\subsection{Elastic energy of a planar membrane} \label{app:energy-planar}
Using the curvature invariants derived in appendix~~\ref{app:diffgeo-planar}, we obtain the expression for the elastic energy in the planar Monge gauge. In the Monge gauge, eqn.~\eqref{eqn:can-Helf} (and by including the contributions from surface tension) can be written as,
\begin{equation}
\mathscr{H}_{\rm sur}=\int \sqrt{1+(\nabla h)^{2}} \left \{ \frac{\kappa}{2} \left(\frac{\nabla^{2}h}{\sqrt{1+(\nabla h)^{2}}}-C_{0} \right )^{2} +\sigma \right \} dx dy.
\end{equation} 

In the limit of small deformations around the planar reference plane, where $|\nabla h| \ll 1$, expanding and retaining terms to fourth order gives the linearized form of the elastic energy truncated to quadratic order in curvature terms; it has the form:

\begin{equation}
\mathscr{H}_{\rm sur}=\int \left \{ \frac{\kappa}{2} \left(\nabla^{2}h -C_{0}\right )^{2} 
+ \sigma \right \} dx dy.
\label{eqn:Monge-withoutscur}
\end{equation} 

An alternate form, to quadratic order, is given by,

\begin{equation}
\mathscr{H}_{\rm sur}=\int \left \{ \frac{\kappa}{2} \left(\nabla^{2}h -C_{0}\right )^{2} 
+ \left (\frac{\sigma}{2}+\frac{\kappa}{4}C_{0}^{2} \right )  (\nabla h)^2
\right \} dx dy.
\label{eqn:Monge-withscur}
\end{equation} 
 
\section{Surface quantifiers on a triangulated surface}\label{geom-quant}
\begin{figure}[H]
\centering
\includegraphics[height=2.5in]{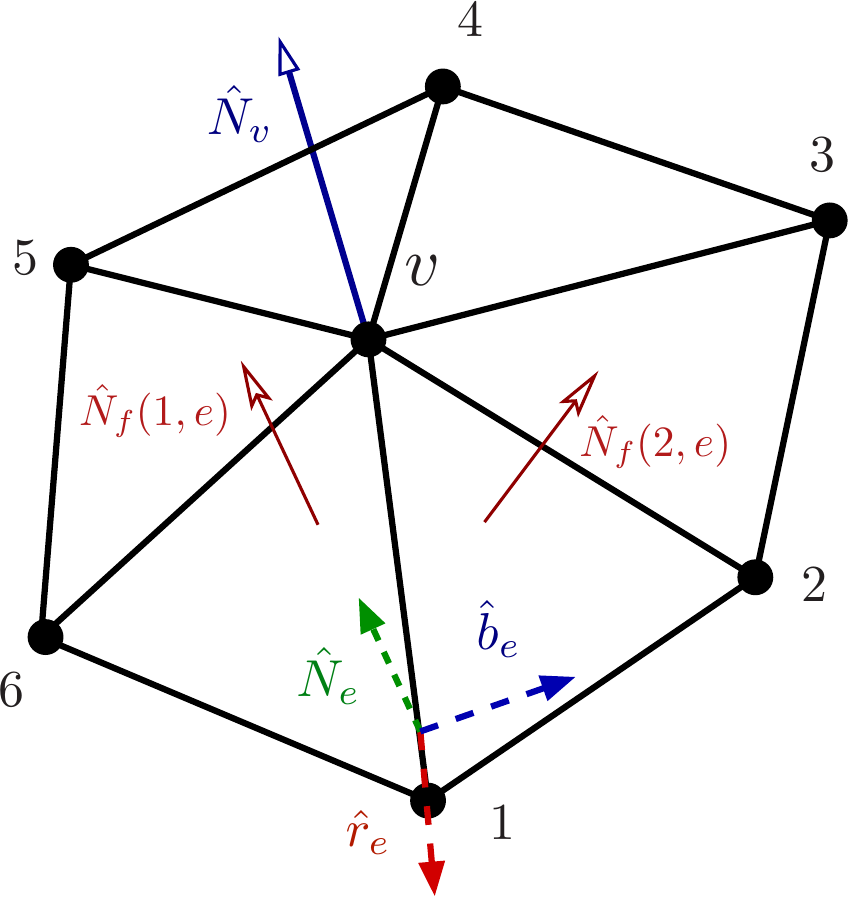}
\caption{\label{fig:patch-edgevec} $\vec{r}_{e}=\vec{x}(v)-\vec{x}(1)$ is the vector along the edge $e$, connecting vertices $v$ and $1$. $\hat{N}_{e}$ and $\hat{b}_{e}$ are respectively its normal and binormal. } 
\end{figure}

The principal curvatures and their corresponding directions are calculated at a vertex using its one-ring neighbourhood shown in Fig.~\ref{fig:patch-edgevec}. The approach is based on the construction of the discretized ``shape operator''  given by the differential form $-d\hat{N}$  in the plane of the surface, which  contains all information about the local surface topography.  Consider a local neighbourhood around a vertex  $v$  in Fig.~\ref{fig:patch-edgevec}.  $\vec{r}_{e}$ is the edge vector that links  $v$ to a neighbouring vertex. The set of edges linked to $v$ is $\{e\}_v$, and the oriented triangles or faces with $v$ as one of their vertices is $\{f\}_v$. The one-ring neighbourhood around vertex $v$ is well defined by $\{e\}_v$ and $\{f\}_v$. \\

\noindent {\it Edge and vertex normals:}
An edge $e$, tethering two vertices, is characterized by its length $|\vec{r}_{e}|$ and orientation $\hat{r}_{e}$.  As shown in Fig.~\ref{fig:patch-edgevec}, $\hat{r}_{e}$  along with the normal to the edge, $\hat{N}_{e}$, and binormal, $\hat{b}_{e}=\hat{N}_{e} \times \hat{r}_{e}$, define the Frenet frame on $e$. The edge normal is entirely dependent on the orientation of the set of faces, $\{f\}_e=[f_{1}(e), f_{2}(e)]$, sharing it. It is determined as, 

 \begin{equation}
 \hat{N}_{e}=\frac{\hat{N}_{f}(1,e)+\hat{N}_{f}(2,e)}{\left\| \hat{N}_{f}(1,e)+\hat{N}_{f}(2,e) \right\|}.
 \end{equation}
 \noindent
$\hat{N}_{f}(1,e)$ and $\hat{N}_{f}(2,e)$ are the unit normal vectors to faces $f_1(e)$ and $f_2(e)$, respectively, sharing the edge $e$. The term in the denominator is the normalization factor, and henceforth, the edge normal is a unit vector. Unlike in the case of an edge, there are ambiguities in the calculation of the normal at a vertex.

For instance, let $\cal{S}$ denote a continuous surface around vertex $v$ and let $\cal{C}$ be a closed contour on $\cal{S}$, enclosing $v$. The normal to the vertex $v$ is calculated by integrating the surface normal along the contour,
\begin{equation}
\label{eq:cont-vernormal}
\vec{N}_{v}=\int_{\cal{C}} \hat{N}_{\cal{S}}\left (\cal{C} \right ) d \cal{C}
\end{equation}
with $\hat{N}_{\cal{S}}(\cal{C})$ being the unit surface normal on the contour. When $\cal{S}$ is replaced by a triangulated patch, the value of $\hat{N}_{\cal{S}}(\cal{C})$  changes only at the interface between the faces\textemdash that is when the contour crosses an edge. Hence, the contributions from each face, in $\{f\}_{v}$, should be appropriately weighed when approximating the vertex normal using the discrete form of eqn.~\eqref{eq:cont-vernormal}. The above-mentioned ambiguity arises in the choice of this weight. We have approximated the normal at a vertex $v$ as,\begin{equation}
\label{eq:disc-vernormal}
 \hat{N}_{v}= \frac{\sum_{\{f\}_v} \Omega[A_{f}]\,\hat{N}_{f} }{\left\|\sum_{\{f\}_v}  \Omega[A_{f}]\,\hat{N}_{f}  \right\| },
\end{equation}
\noindent
with $A_{f}$ denoting the surface area of the face $f$. We have chosen the weight factor $\Omega[A_{f}]$ to be proportional to the area of the face, as in ~\cite{Taubin:1995,Hildebrandt:2004p216,Hildebrandt:2005p86}.

\noindent {\it Shape Operator at an edge:} The topographic details of the triangulated surface around a vertex are contained in the faces and edges. The shape operator, the discrete form of the curvature tensor, is constructed at a chosen vertex by superposing various measures that quantify the curvedness of the surface. One such measure that can be constructed from the geometry of the faces and edges is the shape operator at an edge, $\underline{\mathbf{S}_{\rm E}}(e)$.

As noted above, an object traversing a triangulated surface feels its curvature only when it crosses an edge. At any point $p$ on such an edge $e$, the curvedness can be quantified by the edge curvature, which is the gradient of the area vector of the triangles sharing $e$ ~\cite{Polthier_thesis_2002},
\begin{equation}
h(e)= \nabla_{p}({\rm area})\, \approx \, 2 \left \| \vec{r}_{e} \right \| \cos \left(\frac {\Phi(e)}{2} \right).
\end{equation}

$h(e)$ takes a non-zero value when the faces sharing an edge $e$ are non-planar. Note that $h(e)$ has the dimensions of length, and dividing it by an area makes it a curvature, an operation performed later.
\begin{figure}[H]
\centering
\includegraphics[height=1in]{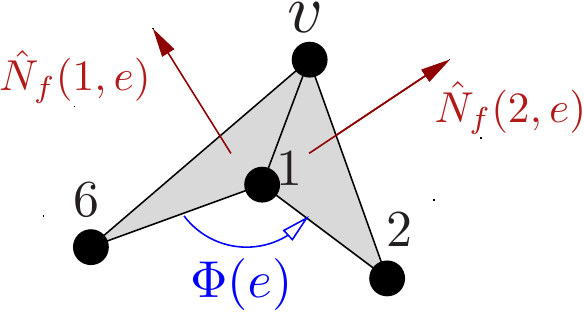}
\caption{\label{fig:dihedral-angle} An illustration of the signed dihedral angle between two faces sharing an edge $e$. } 
\end{figure}
$\Phi(e)$ is the signed dihedral angle between the faces $f_{1}(e)$ and  $f_{2}(e)$ (see Fig.~\ref{fig:dihedral-angle}) sharing edge $e$, calculated as
\begin{equation}
\Phi(e)  =  {\rm sign}\left[ \left\{\hat{N}_{f}(1,e)\times \hat{N}_{f}(2,e) \right\}\cdot  \vec{r}_{e}\right] 
 \arccos \left[\hat{N}_{f}(1,e) \cdot \hat{N}_{f}(2,e) \right] + \pi. 
\end{equation}
The edge curvature contribution at each edge can be used to construct the discretized ``edge shape operator'', which quantifies both the curvature and the orientation of $e$. This tensor has the form
\begin{equation}
\underline {\mathbf{S}_{\rm E}}(e) = h(e) \left [\hat{b}_{e} \otimes \hat{b}_{e}\right ].
\label{eq:shape_e}
\end{equation}
%
\noindent {\it Shape operator at a vertex:}
The complete description of a triangulated surface should contain details of its geometry and orientation. When defining the geometry of the surface the zero-dimensional vertices are primary since the higher-dimensional entities\textemdash namely, the edges and faces\textemdash are constructed from it. Hence it is natural to also define the orientational details of the surface on the vertices itself. This is done by constructing the shape operator at the tangent plane of a vertex, as was done for an edge. 
The shape operator on every edge originating from a vertex contributes to the shape operator at the vertex. Hence at a vertex $v$, the vertex operator $\underline{\mathbf{S}_{\rm V}}(v)$ is a superposition of all edge operators $\{\underline {\mathbf{S}_{\rm E}}(e)\}_{v}$ around it. Further, it is instructive to recall that $\underline {\mathbf{S}_{\rm E}}(e)$ was constructed in the direction of $\hat{b}_{e}$, which need not be along the tangent plane at $v$. The component of the edge shape operator along the tangent plane, at vertex $v$, is the relevant part  in the construction of the $\underline{\mathbf{S}_{\rm V}}(v)$. This contribution can be obtained by diagonalizing $\underline{\mathbf{S}_{\rm E}}(e)$ using the  projection operator at $v$~\cite{Hildebrandt:2004p216,Hildebrandt:2005p86}, 
\begin{equation}
\underline{{\bf P}_{v}} = \mathbbm{1} - \left [\hat{N}_{v}\otimes\hat{N}_{v} \right ],
\end{equation}

where $\mathbbm{1}$ is the unit diagonal matrix. The shape operator at $v$ is then a weighted sum of these contributions, given by 
\begin{equation}
\underline{\mathbf{S}_{\rm V}}(v) = \frac{1}{A_{v}}\,\,\sum_{\{e\}_v} W(e)\,\underline{{\bf P}_{v}}^{\dagger}\,\underline{\mathbf{S}_{\rm E}}(e)\,\underline{{\bf P}_{v}}.
\label{eq:curv_operator}
\end{equation}
$A_{v}=\sum_{\{f\}_v} A_{f}/3$ is the average surface area around $v$, whereas the weight factor for an edge is calculated as $W(e)=\hat{N}_{v} \cdot \hat{N}_{e}$. The shape operator, eqn.~\eqref{eq:curv_operator}, at the vertex $v$ is a rank 2 tensor expressed in the coordinates of the global reference system $[\hat{x},\hat{y},\hat{z}]$. The curvature tensor, in conventional differential geometry, is defined on the tangent plane of the vertex ~\cite{doCarmo:1976,Piran:2003}, with the principal directions $\hat{t}_{1}(v)$ and $\hat{t}_{2}(v)$ being the basis vectors. It should be observed that, the vertex normal $\hat{N}_{v}$ is orthogonal to the principal directions.  These three directions define the local frame of reference, $[\,\hat{t}_{1}(v),\, \hat{t}_{2}(v),\, \hat{N}_{v} ]$, also called the Darboux frame. If the curvature tensor is constructed in the Darboux frame, then the vertex normal $\hat{N}_{v}$ would be the eigendirection corresponding to an eigenvalue of zero. 

Since the eigenvalues of the curvature tensor are gauge invariant, we expect one eigenvalue of $\underline{\mathbf{S}_{\rm V}}(v)$ to be zero and the other two, namely $c_{1}$ and $c_{2}$, to be the principal curvatures. A numerical solution for the eigenspectrum of $\underline{\mathbf{S}_{\rm V}}(v)$ in the native cartesian frame of reference, is computationally expensive. Using a series of unitary transformations we can extract not only the principal curvatures but also the transformation matrix $\underline{\mathbf{T}_{\rm GL}}$, which allows us to switch forth and back between the global and Darboux frames. Knowledge of $\underline{\mathbf{T}_{\rm GL}}$ is crucial when studying membranes with an in-plane vector field. In the next part, techniques for the  transformations  mentioned are outlined. \\

\begin{figure}[H]
\centering
\includegraphics[height=1.5in,clip]{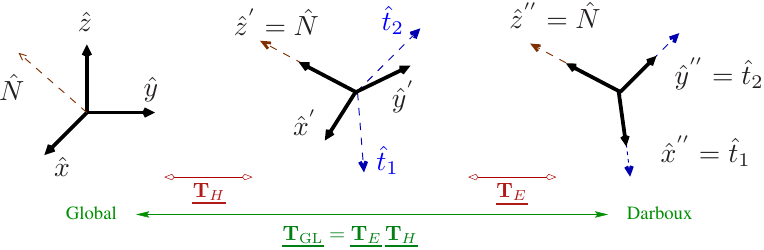}
\caption{Transformation from a global to local coordinate frame, using the Householder transformation.}
 \label{fig:hholder}
\end{figure}

\noindent {\it Transformation to the Darboux frame:}
Computation of the eigenvalues of $\underline{\mathbf{S}_{\rm V}}(v)$, the vertex shape operator constructed in the cartesian frame with basis vectors $[\hat{x},\hat{y},\hat{z}]$, is complicated by its non-zero off-diagonal elements. On the other hand, $\underline{\mathbf{S}_{\rm V}}(v)$ has a diagonal form when represented in the Darboux frame, $[\,\hat{t}_{1},\, \hat{t}_{2},\, \hat{N}]$, with diagonal elements $c_{1},\,c_{2}$,  and  zero, a representation more suitable for extracting the eigenspectrum of the shape operator. However, a direct transformation into the local frame is not feasible, since the principal directions $\hat{t}_{1}$ and $\hat{t}_{2}$ are not known a priori. In our approach, we use the Householder transformation (see Appendix~~\ref{app:householder}) technique to overcome this problem, as shown in Fig.~\ref{fig:hholder}.

The Householder matrix $\underline{\mathbf{T}_{H}}$ uses the vertex normal $\hat{N}_{v}$ to compute the tangent plane at vertex $v$, which need not be the principal directions themselves. Technical details pertaining to the construction of $\underline{\mathbf{T}_{H}}$ are outlined in Appendix~~\ref{app:householder}. We choose to rotate the global $\hat{z}$ direction into $\hat{N}_{v}$, whereas $\hat{x}$ and $\hat{y}$ are rotated into vectors ${\hat x}',\,{\hat y}'$ in the tangent plane at the vertex $v$.  The shape operator, at $v$,  in this frame $\underline {\mathbf{C}}(v)=\underline{\mathbf{T}_{H}^{\dagger}}(v)\,\underline {\mathbf{S}_{\rm V}}(v)\,\underline{\mathbf{T}_{H}}(v)$ is a 2x2 minor, with the two principal curvatures $c_{1}(v)$ and $c_{2}(v)$ as its eigenvalues. The principal curvatures and directions are computed from   $\underline{\mathbf{C}}(v)$, which is fairly simple. The eigenvector matrix, $\underline{\mathbf{T}_{E}}(v)$,  transforms $[\hat{x}^{'},\hat{y}^{'},\hat{N}(v)]$  into the Darboux frame at $v$. Any tensor in the global frame can now be transformed to this local frame by the transformation matrix $\underline{\mathbf{T}_{\rm GL}}=\underline{\mathbf{T}_{E}}(v)\,\underline{\mathbf{T}_{H}}(v)$. The curvature invariants\textemdash namely, the mean and Gaussian curvature\textemdash are $H(v)=(c_{1}(v)+c_{2}(v))/2$ and $R(v)=c_{1}(v)c_{2}(v)$, respectively. 
 
 There are also other methods that can be used for computing the principal surface properties on a triangulated surface ~\cite{Taubin:1995,Shimsoni:2003p626}. However, the results of the current algorithm for construction of the vertex shape operator are more accurate, especially when it comes to the prediction of principal directions.
\section{Householder transformation}\label{app:householder}
Consider two orthonormal frames of reference given by the coordinates $(\hat{x},\hat{y},\hat{z})$ and $(\hat{a},\hat{b},\hat{c})$.
The Householder matrix, $\underline{\mathbf{T}_{H}}$, can be used to rotate $\hat{z}$ in frame 1 to $\hat{c}$ in frame 2, such that $(\hat{x},\hat{y})$ now 
are some  arbitrary vectors in the plane formed by $(\hat{a},\hat{b})$. Define a vector,
\begin{equation}
W=\frac{\hat{x} \pm \hat{c}}{|\hat{x} \pm \hat{c}|}
\end{equation}
with a minus sign if $||\hat{x}-\hat{c}||\,>\,||\hat{x}+\hat{c}||$ and a plus if otherwise. The Householder matrix is then defined as,

\begin{equation}
\underline{\mathbf{T}_{H}}=\mathbbm{1}-2WW^\dagger
\end{equation}

\section{Parallel transport of a vector field on a triangulated patch}\label{sec:partpt}
 In order to compare the orientations of two distant in-plane vectors on a curved surface, it is necessary to perform a parallel transport of the vectors on the discretized surface~\cite{doCarmo:1976,Piran:2003}. In practice, we need only to define the parallel transport between neighbouring vertices, i.e. a transformation  $\hat{m} (v^{'}) \rightarrow \underline{\bf \Gamma}(v,\,v^{'}) \hat{m}(v)$, which brings $\hat{m}(v)$ correctly into the tangent plane of the vertex $v^{'}$, so that its angle with respect to the geodesic connecting $v$ and $v^{'}$ is preserved. 
 
\begin{figure}[H]
\centering
\includegraphics[height=1.5in]{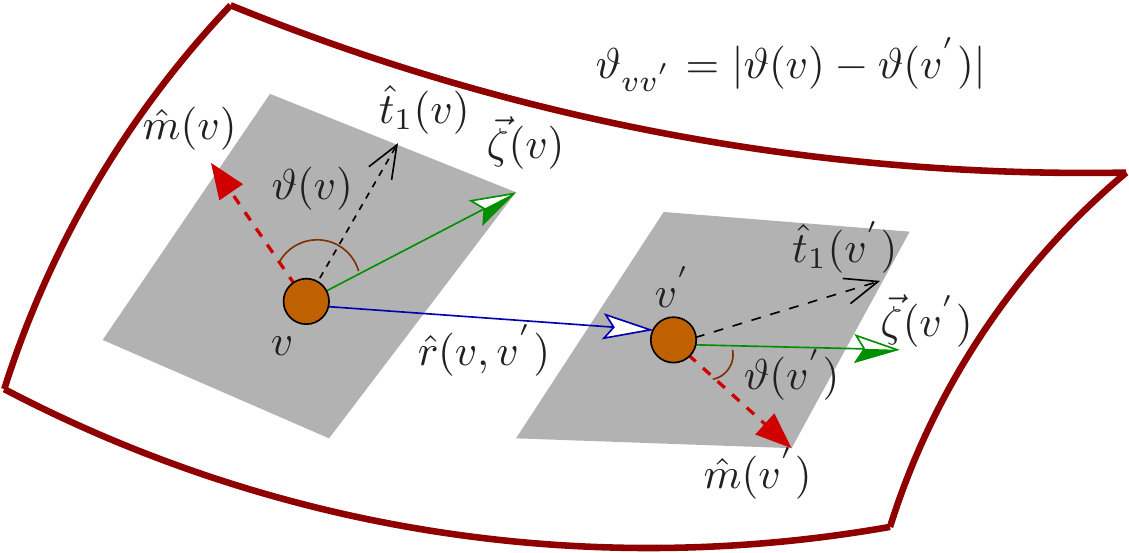}
\caption{\label{fig:partpt} Illustration of the  parallel transport of a vector between two neighbouring vertices $v$ and $v^{'}$ in a triangulated surface. }
\end{figure}

As shown in Fig.~\ref{fig:partpt}, if ${\hat r}(v,v^{'})$  is the unit vector  connecting   a vertex $v$ to its neighbor $v^{'}$ and ${\vec \zeta}(v)$=$\underline{{\bf P}_{v}}{\hat r}(v,v^{'})$  and ${\vec \zeta}(v^{'})$= $\underline{{\bf P}_{v^{'}}} {\hat  r}(v^{'},v)$ are its projections on to the tangent planes at $v$ and $v^{'}$, then  our best estimate for  the directions of the geodesic  connecting them are the unit vectors ${\hat  \zeta}(v)$ and   ${\hat \zeta}(v^{'})$. 

The decomposition of $\hat{m}(v)$ along the  orientation of the geodesic and its perpendicular in the tangent plane of $v$, is thus,
\begin{equation}
 \hat{m}(v) =  \left[\hat{m}(v) \cdot \hat{\zeta}(v)\right] \hat{\zeta}(v)+ 
                     \left[\hat{m}(v)\cdot (\hat{N}_{v} \times \hat{\zeta}(v))\right] \left(\hat{N}_{v} \times \hat{\zeta}(v)\right)
\end{equation} 
\noindent
Parallelism now demands that these  coordinates,  with respect to the geodesic orientation, are the same  in the tangent plane of $v^{'}$; therefore,
\begin{equation}
\underline{{\bf \Gamma}}(v,v^{'})\hat{m}(v)=  \left[\hat{m}(v) \cdot \hat{\zeta}(v)\right] \hat{\zeta}(v^{'})+
 \left\{\hat{m}(v) \cdot (\hat{N}_{v} \times \hat{\zeta}(v))\right\} \left[ \hat{N}_{v^{'}} \times \hat{\zeta}(v^{'})\right]
\end{equation} 
\noindent
This parallel transport operation allows us to define the angle $\theta_{vv^{'}}$ between vectors in the tangent plane at neighbouring vertices and, in turn, their cosine and sine as:
\begin{eqnarray}
 \cos (\theta_{vv^{'}})&= & \hat{m}(v^{'}) \cdot \,\underline{{\bf \Gamma}}(v,v^{'})\hat{m}(v) \\ 
 \sin (\theta_{vv^{'}})&=& \left[\hat{N}_{v^{'}}\times\hat{m}(v^{'}) \right] \cdot \,\underline{{\bf \Gamma}}(v,v^{'})\hat{m}(v) \nonumber
\end{eqnarray}

\section{Properties of a membrane with in-plane order}
\label{sec:implicit-coup}
The term implicit coupling indicates that the anisotropic constituents do not have any specific interaction with the membrane. The membrane exhibits anisotropic behavior, when it prefers a particular direction over others, which can be due to the presence of an ordering interaction, leading to the formation of spatially ordered patterns of its anisotropic constituents. The presence of such an order can limit the number of micro states accessible to the membrane, and hence, the membrane feels the presence of the anisotropic components through a reduction in the entropic contribution, which is implicit.  The equilibrium properties of the membrane significantly change when the surface vector field orients into an ordered state. This section describes how the morphology of the fluid membrane is remodeled due to the presence of an ordered phase of the in-plane field, using the Monte Carlo techniques described above. For this purpose, a nematic field ($p=2$) has been chosen to decorate the vertices of the triangulated surface, with 100\% surface coverage. The nematic field is ordered by choosing a non zero value for the exchange interaction $J_{2}$. In order to establish the importance of anisotropy the conformations of the field decorated membrane, with $\kappa=0$, are compared against the branched polymer phase of a pure fluid membrane of similar rigidity.
\subsection{Defect structure in polar and nematic fields}
\label{sec:defects}
\begin{figure}[H]
\centering
\subfigure[Frustrated Ising magnet]{
\begin{minipage}{0.4\textwidth}
\centering
\includegraphics[height=1.75in]{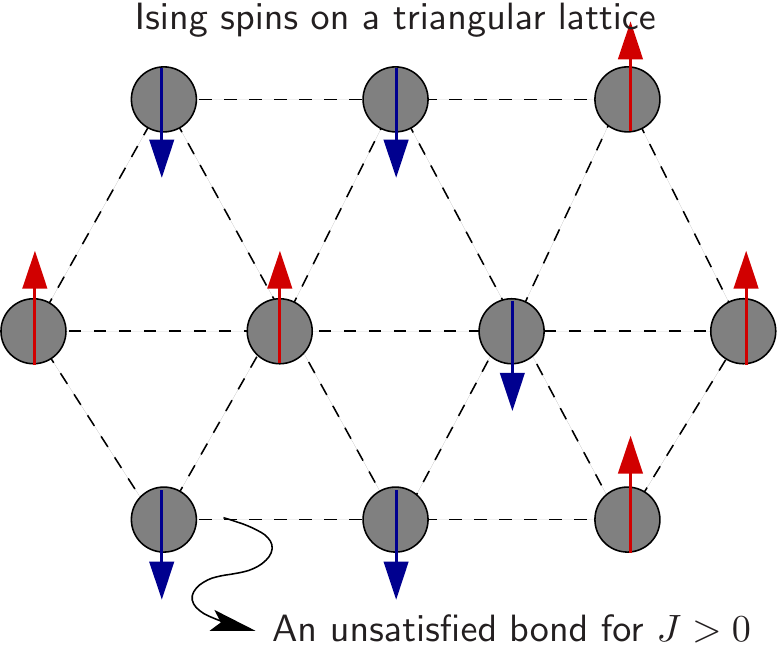}
\vspace*{10pt}
\label{fig:frustIsing}
\end{minipage}
}
\hspace*{20pt}
\subfigure[Disclinations on a tesellated sphere]{
\begin{minipage}{0.4\textwidth}
\centering
\includegraphics[height=1.75in]{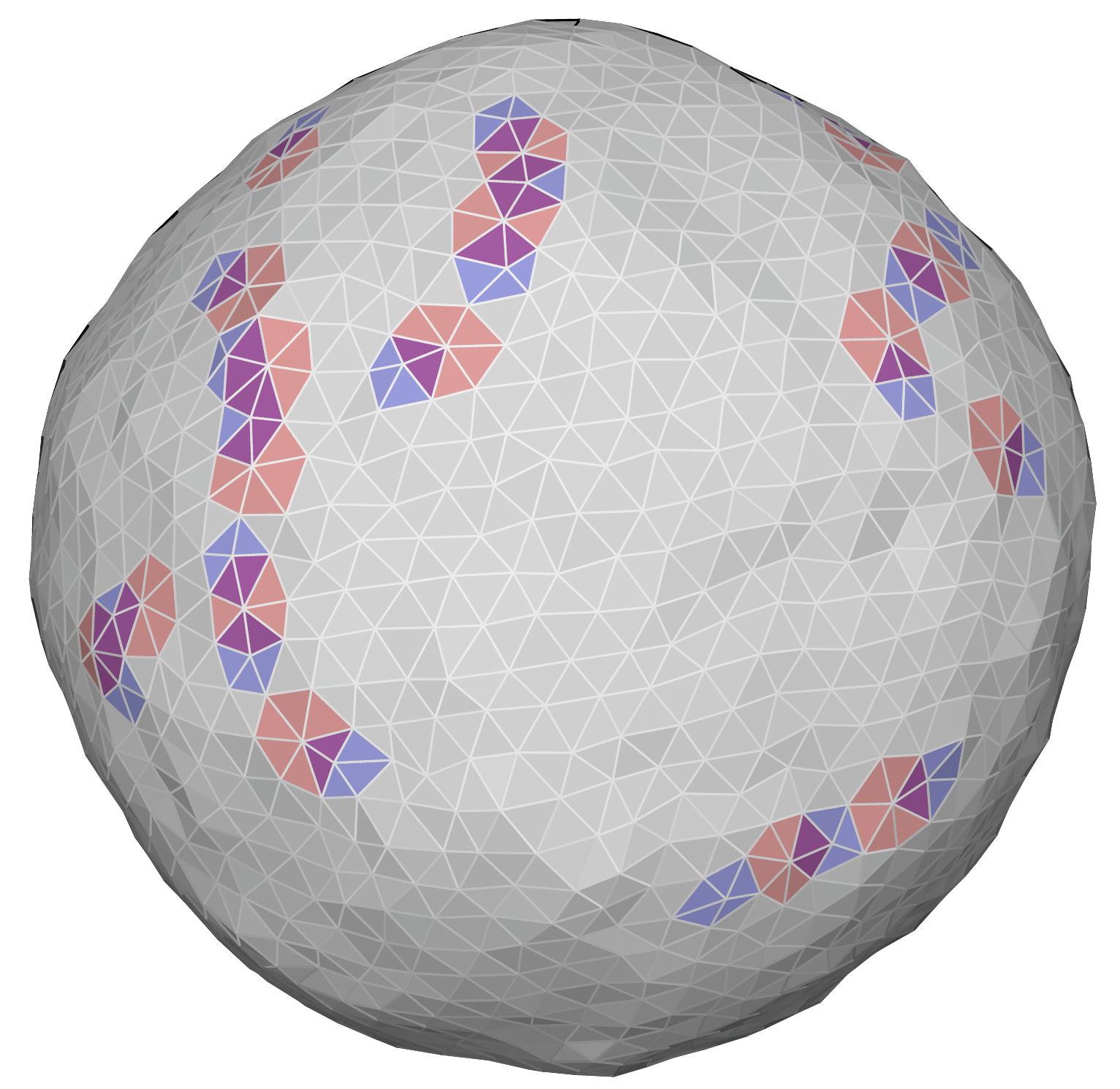}
\vspace*{10pt}
\label{fig:scarredmem}
\end{minipage}
}
\caption{\label{fig:geomfrust} Two examples of geometrically frustrated systems in nature. {\bf (a)} An Ising magnet with antiferromagnetic bonds on a triangular lattice; a bond with unfavorable spins orientations is marked with a wiggly arrow. {\bf (b)} Five fold and seven disclinations  are marked with blue and red shades on a tesellated sphere. Euler theorem requires the presence of at least 12 five fold disclinations to account for the non zero Gaussian curvature of the sphere.}
\end{figure}

Textures of a vector field on curved surfaces significantly differ from those on a plane surface. Unlike on  flat surfaces, spatial orientations of the field are frustrated by the topology of the curved surface. Geometric frustration is defined as the inability to impose the desired local order throughout the system, due to constraints that are purely geometric in nature. A well-known example is a system of Ising spins arranged on a triangular lattice with antiferromagnetic bonds; here, the ground state prefers anti-parallel spins at the ends of each bond. The triangular nature of the lattice, as illustrated in Fig.~\ref{fig:frustIsing}, disallows such a state leaving one unsatisfied bond per plaquette, with parallel spin orientations, as a result of which the minimal energy state differs from the ground state. Frustration is also observed in a wide range of physical and biological systems like water, spin ice, high -transition superconductors, buckled colloidal monolayers, etc. ~\cite{Han:2008p3701}. Further, the 5-fold and 7-fold disclinations in the tessellation of a sphere, giving rise to a scarred pattern, is another case of frustration in nature and is shown in Fig.~\ref{fig:scarredmem}.

Similarly, an in-plane surface vector field is also frustrated by the topology of the underlying surface that gives rise to orientational singularities, also called {\it topological defects}. In the case of an in-plane field on planar surfaces, the  defects are not part of the zero-temperature state and are seen only at high temperatures due to the thermally induced vortex unbinding.  However, for non-planar topologies, with non-zero total Gaussian curvature, a minimal number of defects, each of strength $1/p$ is part of the field texture at zero temperature. The number of defects is related to the surface topology through the Euler number $\chi$, introduced earlier in section~~\ref{sec:GBtheorem} as,
\begin{equation}
\chi=\sum_{D=1}^{N_{D}} q_{D},
\label{eq:eulcharge}
\end{equation}

\noindent where $q_{D}$ is the strength or the {\it topological charge} of defect $D$. The number of defects  $N_{D}$ and the charge of the defect $q_{D}$ depend on the symmetry of the in-plane field and topology of the membrane surface. The following section discusses the calculation of the defect charge from the winding number of the in-plane field.

\subsection{Defect core and its winding number} In the continuum description of the tangential field, a defect core  represents a mathematical singularity, {\it i.e.} spatial locations at which the gradient of the vector field diverges.~\cite{gennesprost1993,Chaikin:1995td}. This singular behavior is normally taken care of by setting the order parameter at the core, of radius $R_{\rm core}$, to zero and the texture of a field around a vortex is characterized by its vorticity or winding number.
\begin{figure}[H]
\centering
\subfigure[Continuum : W=$\int_{\cal{C}}\,d\theta \left (\cal{C} \right )=2\pi$]{
\begin{minipage}{0.4\textwidth}
\centering
\includegraphics[height=1.5in]{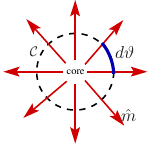}
\vspace*{20pt}
\label{fig:WNcont}
\end{minipage}
}
\subfigure[ Discrete : W(v)=$\sum_{v_{i}\in v}\vartheta_{v_{i}v_{i+1}}=2\pi$]{
\begin{minipage}{0.4\textwidth}
\centering
\includegraphics[height=1.5in]{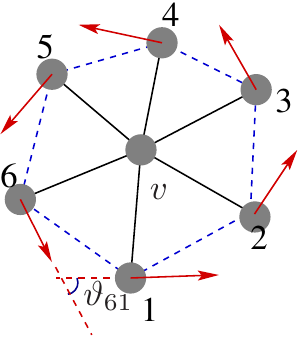}
\vspace*{20pt}
\label{fig:WNdisc}
\end{minipage}
}
\caption{\label{fig:WN}{\bf (a)} A curve $\cal{C}$ enclosing a defect core is shown, along with the vector field $\hat{m}$. {\bf (b)} Calculation of winding number around a vertex $v$ on a triangulated surface : $\theta_{61}$ is the angle the vector field at vertex $6$ subtends with the field at vertex $1$. }
\end{figure}

 As in Fig.~\ref{fig:WNcont}, if $\cal{C}$ is a curve enclosing a defect core and $d\vartheta(\cal{C})$ the change in field orientation along a small segment $d\cal{C}$, then the winding number is given by $W=\int_{\cal{C}}\,d\vartheta (\cal{C})$. Unlike in a continuum, the notion of a defect core has a slightly different meaning on a discrete surface since the in-plane field is well defined at all vertices of the triangulated membrane.
Hence, the singularities in the field orientation are identified using the winding number around each vertex, calculated as $W(v)=\sum_{v_{i}\in v}\vartheta_{v_{i}v_{i+1}}$. The relative orientation of the vector field at vertices $v_{i}$ and $v_{i+1}$, both in the one ring neighbourhood of vertex $v$, is calculated using the parallel transport defined in section~~\ref{sec:partpt}.

\subsection{Topological charge and its minimum} The  topological or defect charge is a non-dimensional measure of the vorticity of the defect defined from the winding number as
\begin{equation}
q_{D}=\frac{W}{2\pi}\,.
\end{equation}

\begin{figure}[H]
\centering
\includegraphics[width=9cm,clip]{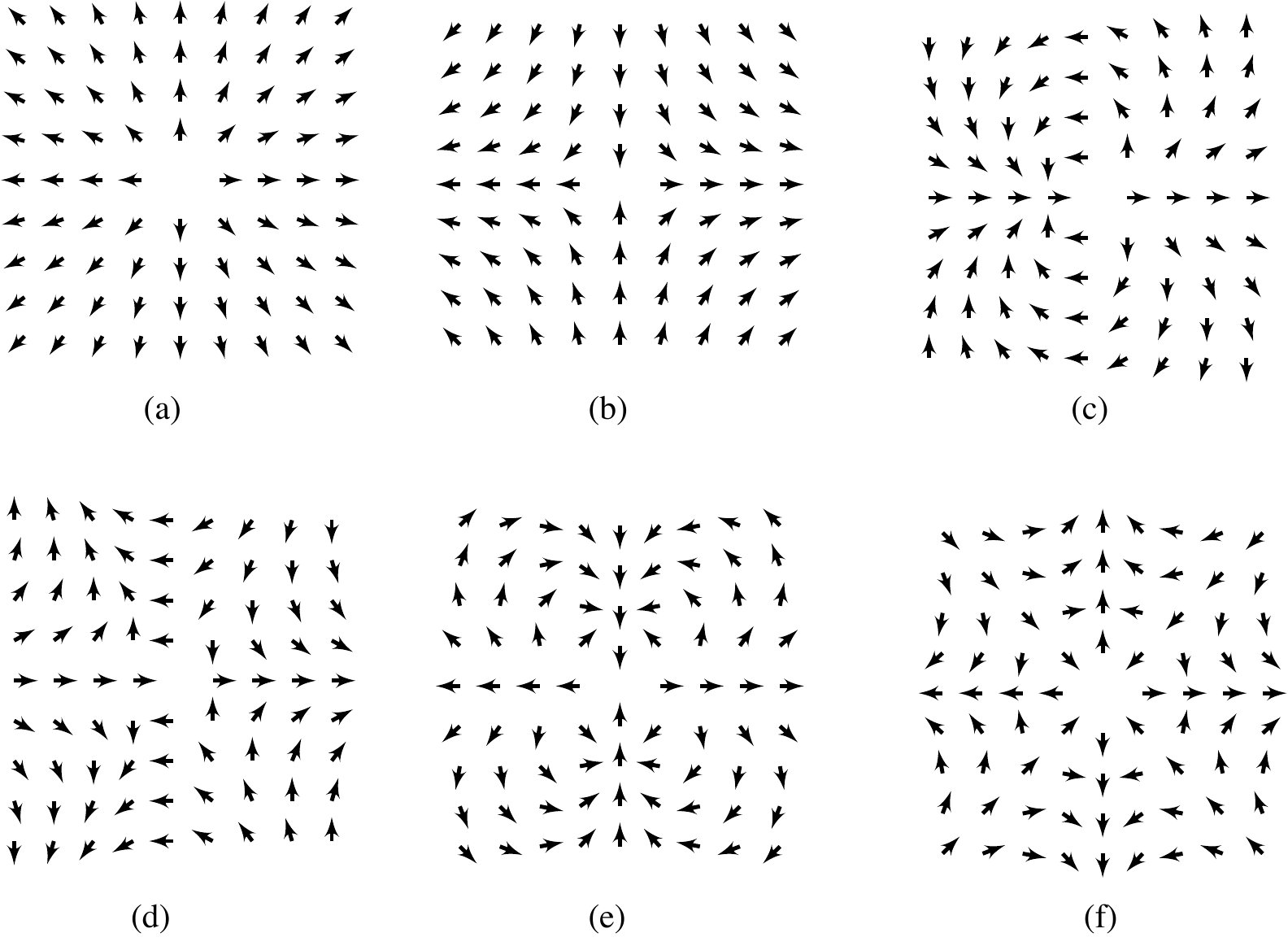}
\caption{ \label{fig:poldef}Polar field on a planar surface with its texture described by $\hat{m}(x,y)=\cos \phi \,\hat{x}+\sin \phi \,\hat{y}$, with $\phi(r,\theta)=q_{D}\theta$. Shown are patterns corresponding to {\bf (a)} $q_{D}=+1$, {\bf (b)} $q_{D}=-1$, {\bf (c)} $q_{D}=+2$, {\bf (d)} $q_{D}=-2$, {\bf (e)} $q_{D}=+3$, and {\bf (f)} $q_{D}=-3$.}
\end{figure}

\begin{figure}[H]
\centering
\includegraphics[height=3in]{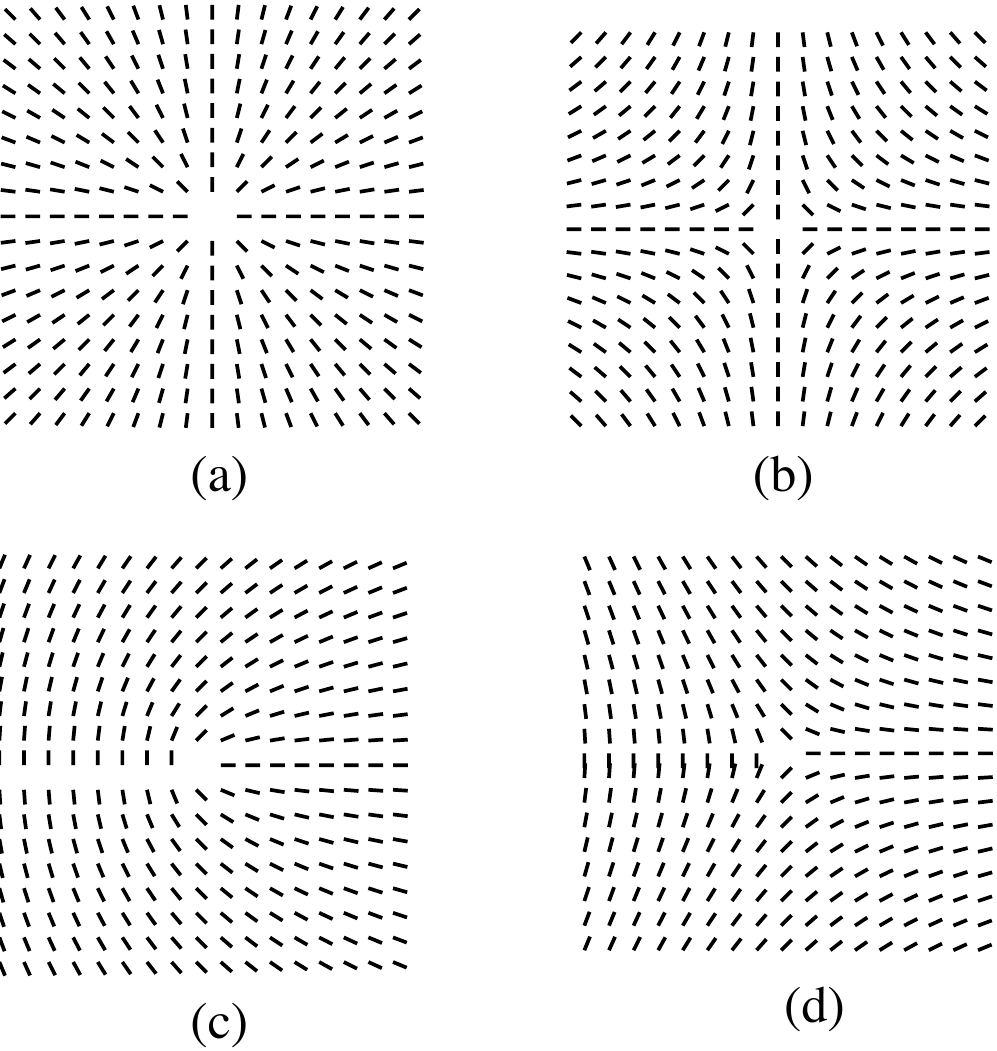}
\caption{\label{fig:nemdisc} Nematic field textures illustrating the various defect structures. Shown are the patterns with {\bf (a)} $q_{D}=+1$, {\bf (b)} $q_{D}=-1$, {\bf (c)} $q_{D}=+1/2$, and {\bf (d)} $q_{D}=-1/2$}
\end{figure}

Patterns of a planar vector field corresponding to varying defect strengths are shown in Fig.~\ref{fig:poldef} and Fig.~\ref{fig:nemdisc} for  polar and nematic fields, respectively.
 In general, all possible topological defects of a $p$-atic vector field can be represented by $n/p$, with $n=\pm1,\pm2,\dots$. 

\begin{figure}[H]
\centering
\subfigure[Polar field : $\chi p=2$, $q_{D}=+1$]{
\begin{minipage}{0.4\textwidth}
\centering
\includegraphics[height=2.0in]{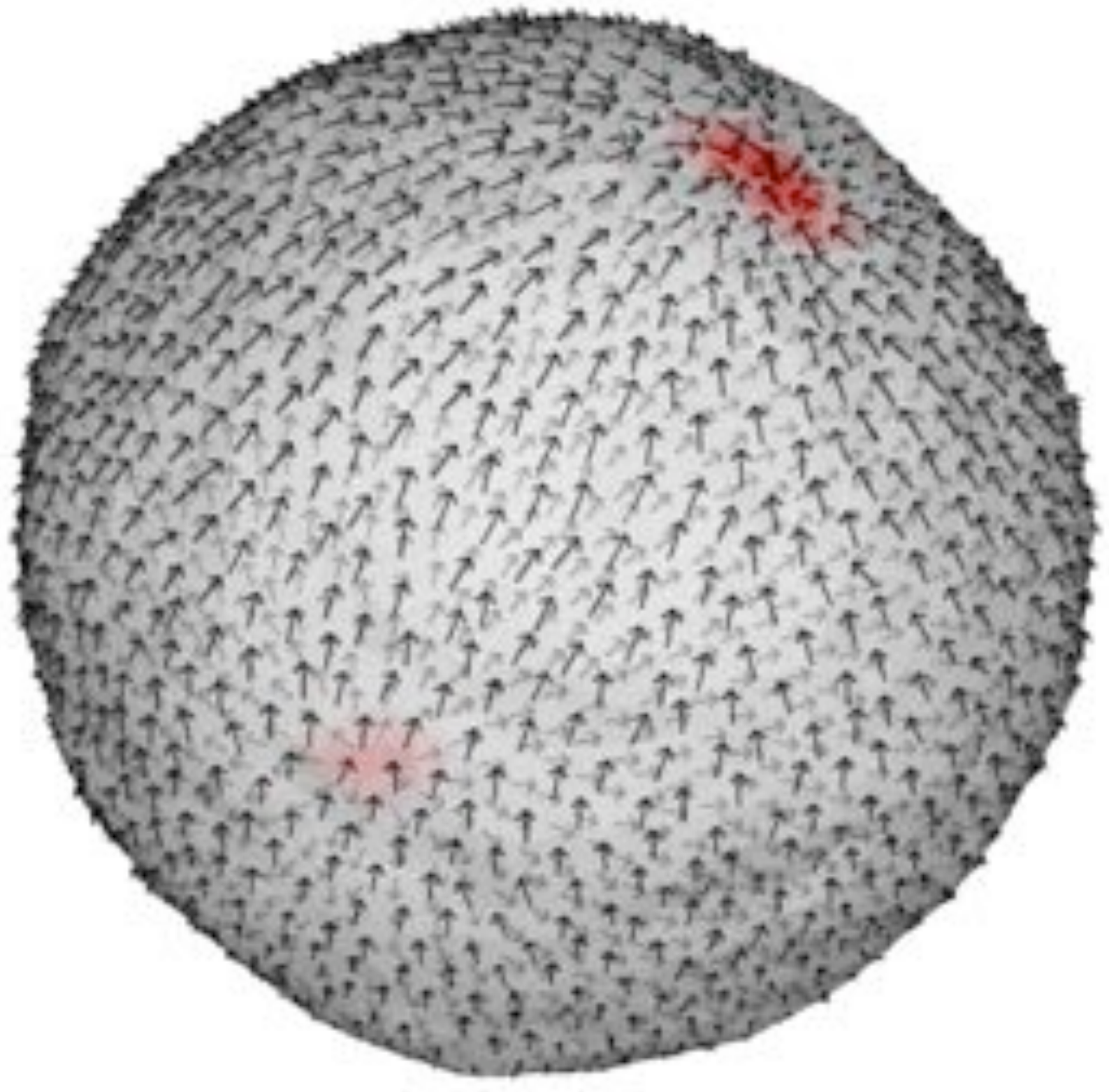}
\vspace*{10pt}
\label{fig:sphpol}
\end{minipage}
}
\hspace*{20pt}
\subfigure[Nematic field : $\chi p=4$, $q_{D}=+1/2$]{
\begin{minipage}{0.4\textwidth}
\centering
\includegraphics[height=2.0in]{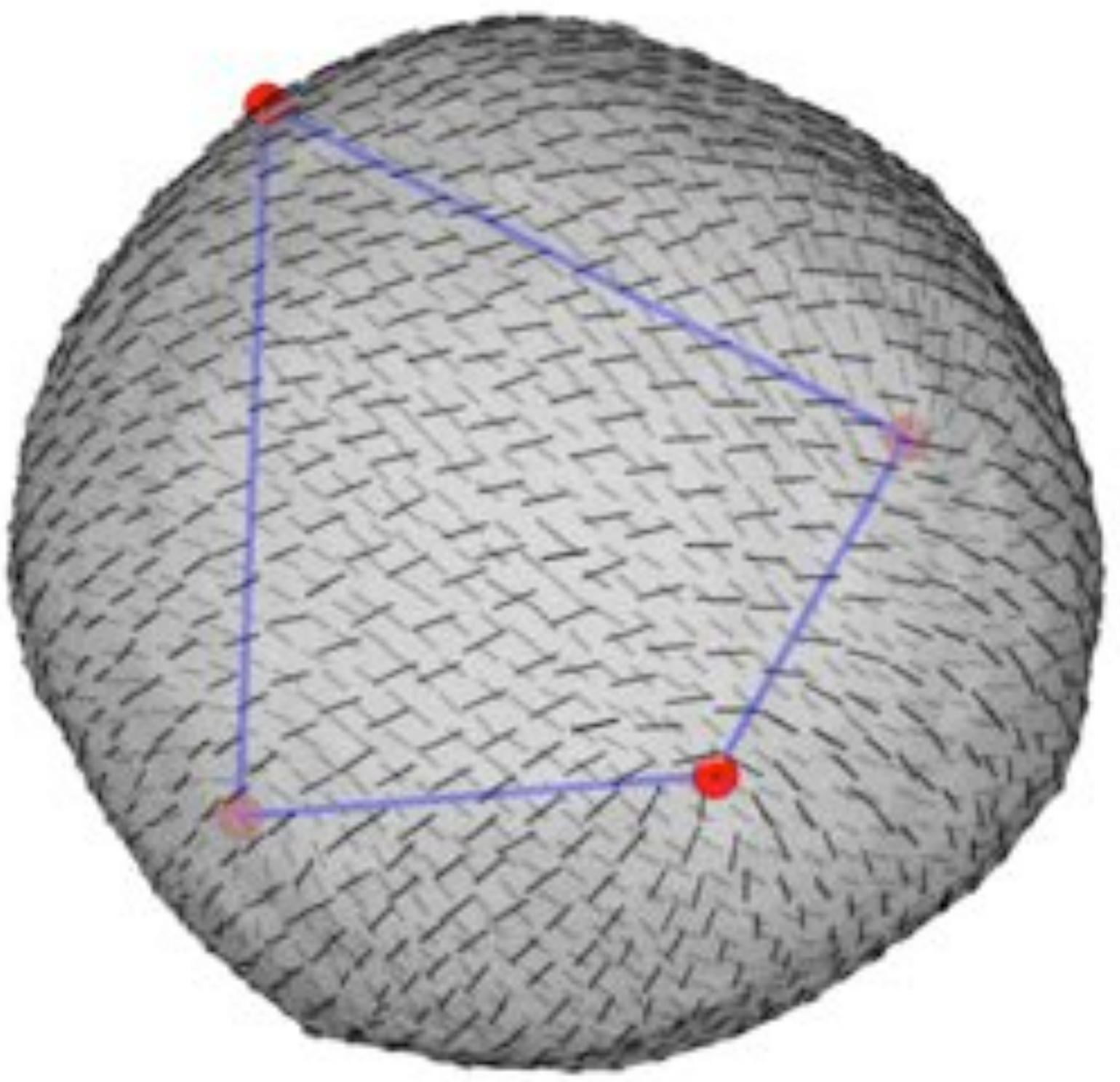}
\vspace*{10pt}
\label{fig:sphnem}
\end{minipage}
}
\caption{\label{fig:rigidsphere-def} Patterns of an in-plane field on a rigid spherical surface, with the shaded regions marking the defect core. {\bf (a)} A polar field with two +1 defects at the poles of the sphere. {\bf (b)} Four +1/2 disclinations of the nematic field arrange themselves on the vertices of a tetrahedron, resulting in the baseball texture.}
\end{figure}

The energy of a defect core  is a function of  its topological charge ($q_{D}$) and core radius ($R_{\rm core}$). In the case of a $p$-atic field, it scales quadratically with the charge and logarithmically with the core radius and has the form~\cite{gennesprost1993,Bowick:2009p955},
 \begin{equation}
 \mathscr{H}_{\rm core}=\pi J_{p} q_{D}^{2} \ln\left(R_{\rm core}/a\right),
 \end{equation}
  with $a$ being the short distance cutoff.  As a result of such a dependence, only defects of strength $1/p$, which minimize $\mathscr{H}_{\rm core}$, are accommodated in the low-temperature equilibrium texture of the $p$-atic field. From eqn.~\eqref{eq:eulcharge}, the minimal number of defects in the equilibrium texture of an in-plane field on a surface with Euler number $\chi$ can be shown to be equal to $\chi p$. The textures corresponding to polar ($p=1$) and nematic ($p=2$) fields on a surface of spherical topology ($\chi=2$) are shown in Fig.~\ref{fig:sphpol} and Fig.~\ref{fig:sphnem}, respectively.

\subsection{Organization of defects of non deformable spherical surfaces}
When the membrane surface is rigid, as expected, the polar field on  a spherical surface organizes such that the two +1 defects, arising due to topological frustration, are positioned at the poles of the sphere, giving rise to the familiar hairy ball pattern.  The nematic field texture, on the other hand, stabilizes four disclinations, each of strength +1/2, as shown in Fig.~\ref{fig:sphnem}. Earlier computational studies on packing nematic like objects on spherical surfaces have shown that the four +1/2 disclinations are located on a great circle inscribed on the surface of the sphere ~\cite{Shin:2008p174}. The pattern of the nematic field, in our system, instead displays the baseball texture ~\cite{Vitelli:2006p1462} with the defects positioned at the vertices of a tetrahedron,  as predicted in ~\cite{Lubensky:1992p531}. The observed difference has it origin in the one constant approximation of the Frank's free energy used here.

\begin{figure}[H]
\centering
\subfigure[$q_{D}=+1$]{
\begin{minipage}{0.2\textwidth}
\centering
\includegraphics[height=1.0in]{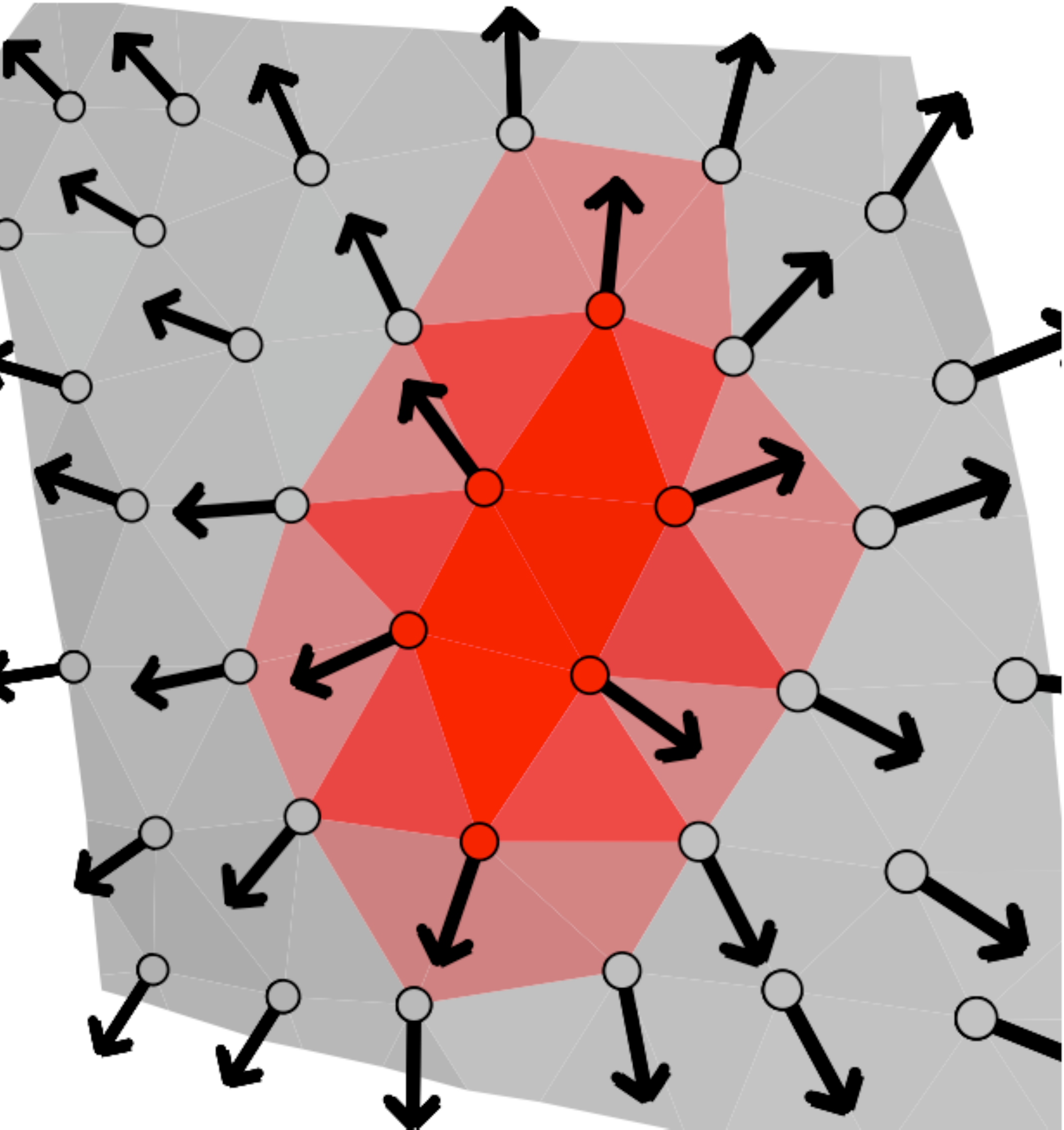}
\vspace*{10pt}
\label{fig:plusone}
\end{minipage}
}
\subfigure[$q_{D}=-1$]{
\begin{minipage}{0.2\textwidth}
\centering
\includegraphics[height=1.0in]{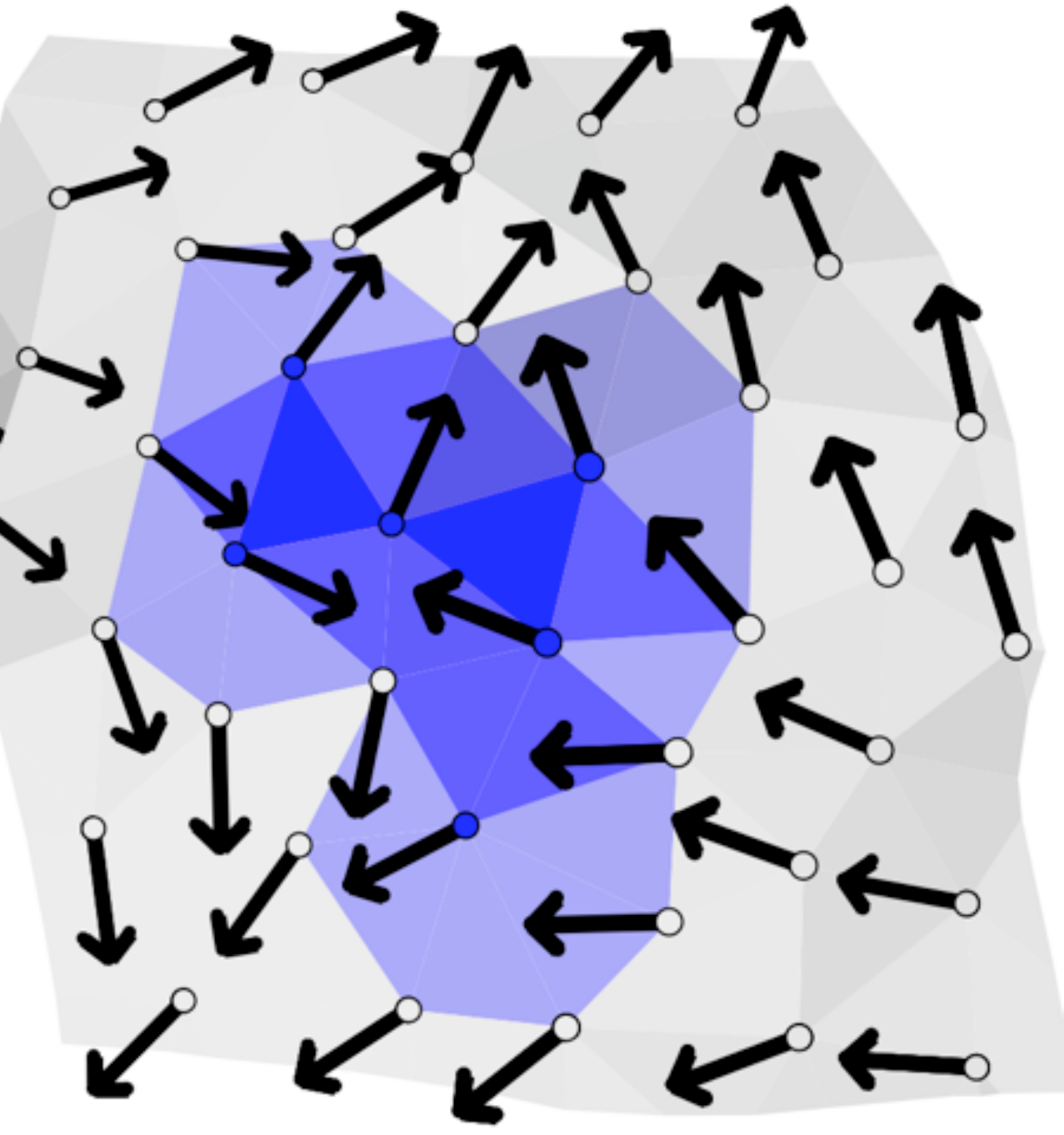}
\vspace*{10pt}
\label{fig:minusone}
\end{minipage}
}
\subfigure[$q_{D}=+1/2$]{
\begin{minipage}{0.2\textwidth}
\centering
\includegraphics[height=1.2in]{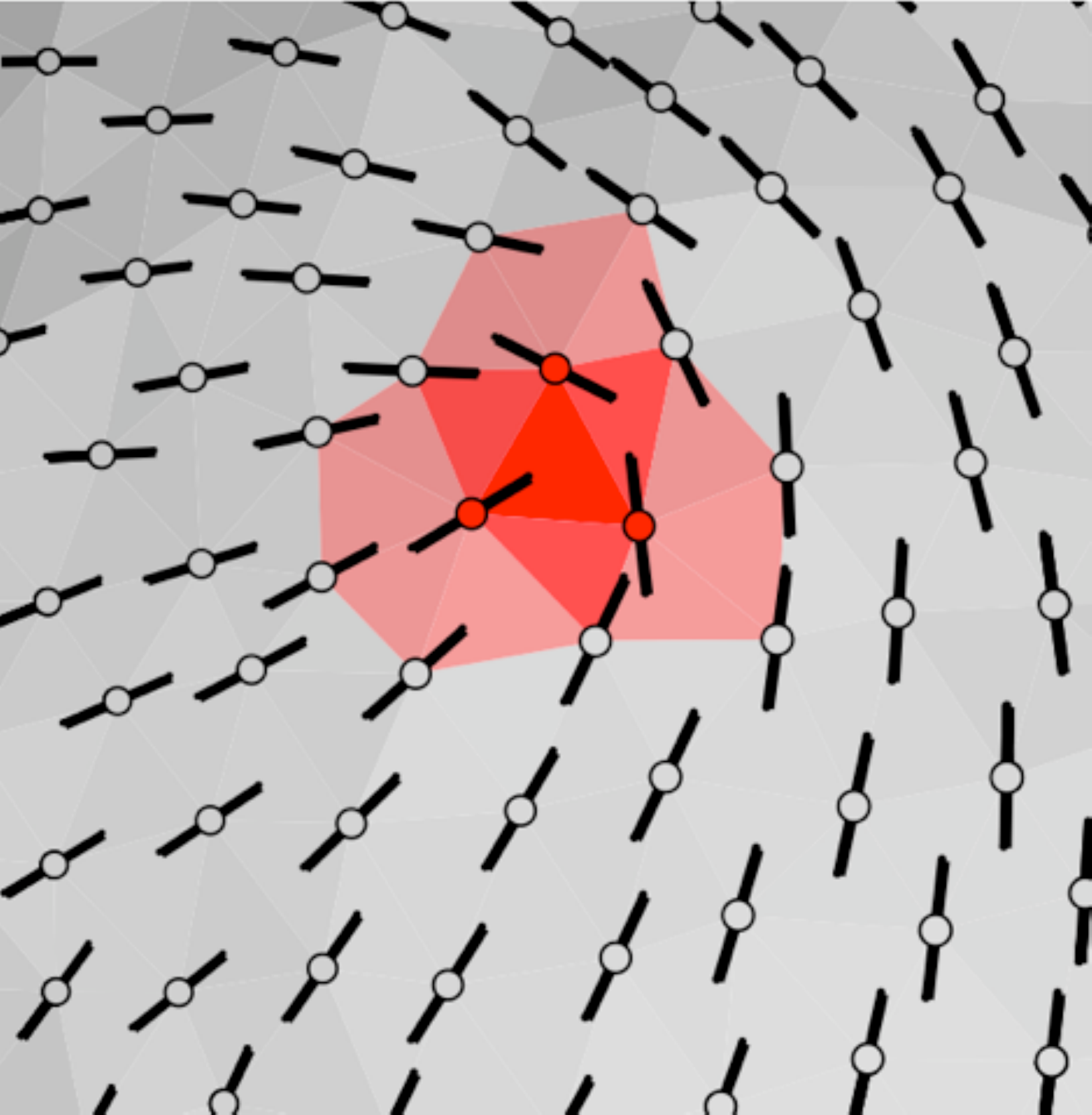}
\vspace*{10pt}
\label{fig:plushalf}
\end{minipage}
}
\subfigure[$q_{D}=+1/2$]{
\begin{minipage}{0.2\textwidth}
\centering
\includegraphics[height=1.15in]{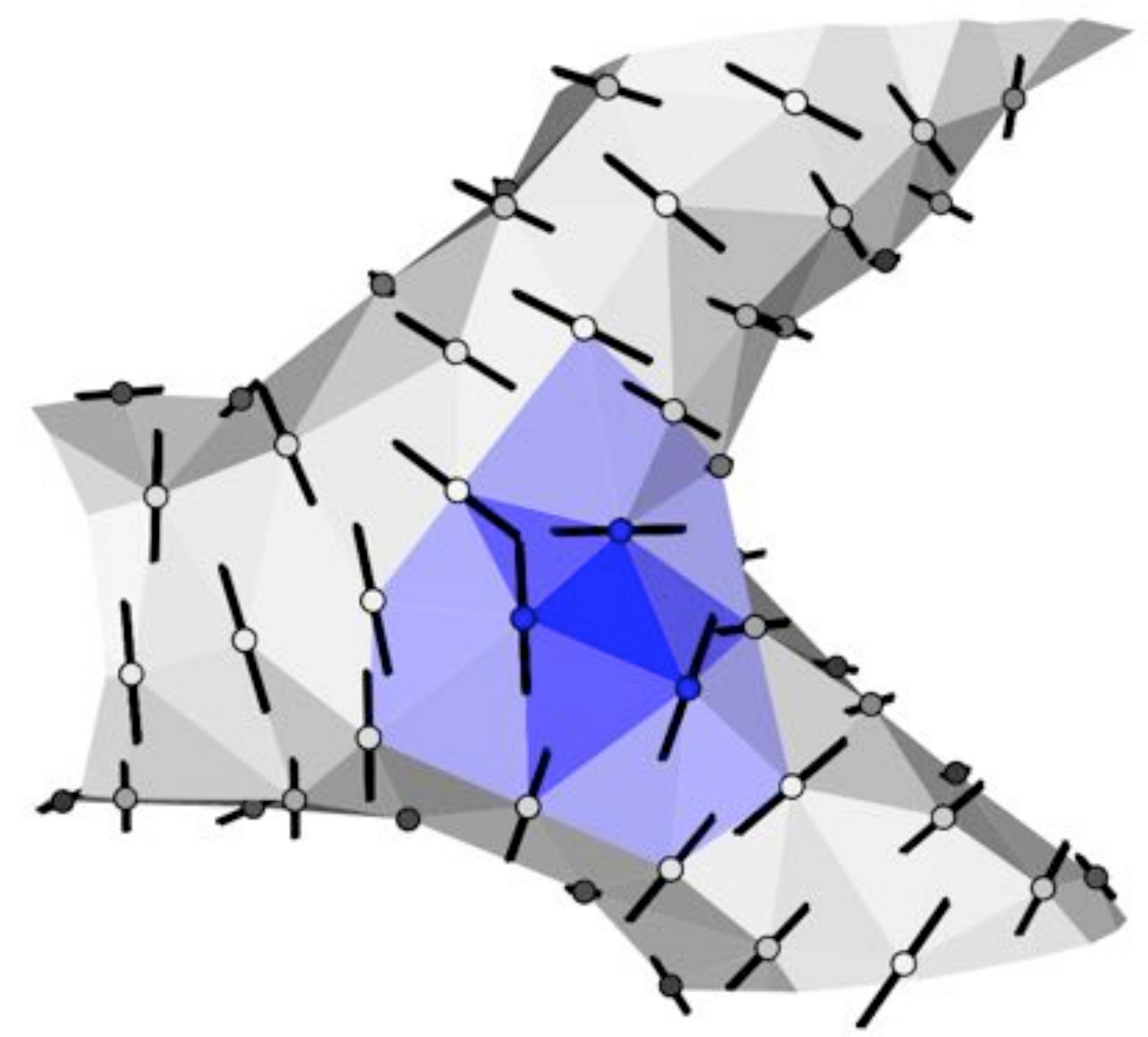}
\label{fig:minuhalf}
\end{minipage}
}
\caption{\label{fig:discsur-def} Mugshot of four different defect cores;   +1 {\bf (a)} and  -1 {\bf (b)} defects for a polar field; and  +1/2 {\bf (c)} and -1/2 {\bf (d)} defects for a nematic field.}
\end{figure}

The ground state defects of a polar and nematic field, as seen in the simulations, are presented in Fig.~\ref{fig:discsur-def}. So far, the behavior of the orientational field was investigated in the context of non-deformable surfaces. These properties hold even  when the underlying surface deforms into complicated morphologies. The role of an in-plane field in deforming membrane surfaces and the host of other interesting phenomena  arising due to it are investigated in the next section. 

 \subsection{Non-curvature-inducing in-plane field and deformable membranes}
 Though the interaction of the surface vector field with the membrane is  implicit, it tends to deform the membrane into a variety of shapes. Simple arguments based on energy minimization could shed some light on this interesting behavior. The total energy of a field-decorated membrane $\mathscr{H}_{\rm tot}$, previously defined in eqn.~\eqref{eq:Zin-plane}, contains contributions from the elastic energy $(\mathscr{H}_{\rm sur})$, due to surface deformations, and the self-interaction energy $(\mathscr{H}_{p-{\rm atic}})$, due to the field. In addition to these is the energy of the defect core, $\mathscr{H}_{\rm core}$. In the case of membrane conformations with minimal topological defects ($\chi p$) and smooth field texture, $\mathscr{H}_{\rm core}$ and $\mathscr{H}_{p-{\rm atic}}$ are nearly constants. The minimum of $\mathscr{H}_{\rm tot}$ would then be seen for membrane conformations with minimum elastic energy, which corresponds to a sphere. This indicates that, changes in the membrane shape are driven by  a competing interaction that is much stronger than $\mathscr{H}_{\rm sur}$.
  
  A known source of such an implicit term is the interaction between the topological defects, which are now a part of the membrane surface. Topological charges are analogous to electric charges and hence attract when unlike and repel when alike. This interaction, for defect charges on a planar surface, is known to depend logarithmically on the   separation between the defects. The arrangement of the nematic disclination on the vertices of a tetrahedron, in Fig.~\ref{fig:sphnem}, indicates that such a behavior holds for defects on a curved surface too. Hence, the defects of the vector field maximize their spatial separation by deforming the membrane at the cost of the much softer elastic energy.
  
   An alternate reasoning for the deformation of a membrane by its in-plane polar field is due to Mackintosh and Lubensky ~\cite{MacKintosh:1991p685}.  They argue that, as the field couples to the Gaussian curvature ($G$) of the surface, it tends to minimize $G$ everywhere by deforming the surface into a cylinder, except at the defect core located at the pole.

 \begin{figure}[H]
\centering
\subfigure[\,Polar field : Ellipsoidal membrane]{
\begin{minipage}{0.4\textwidth}
\centering
\includegraphics[height=2.25in]{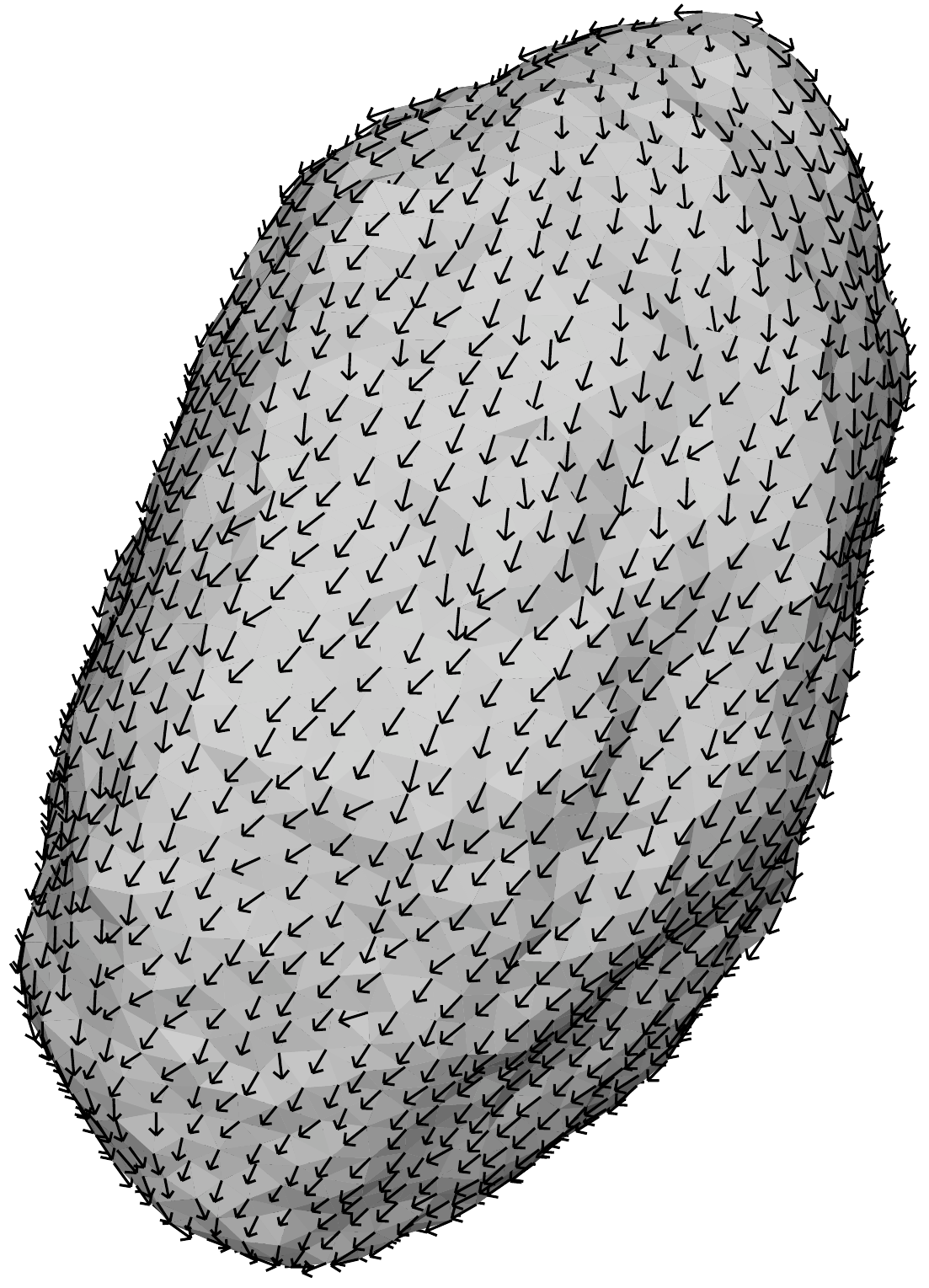}
\vspace*{20pt}
\label{fig:cylXY}
\end{minipage}
}
\subfigure[\,Nematic field : Tetrahedron]{
\begin{minipage}{0.4\textwidth}
\centering
\includegraphics[height=2.25in]{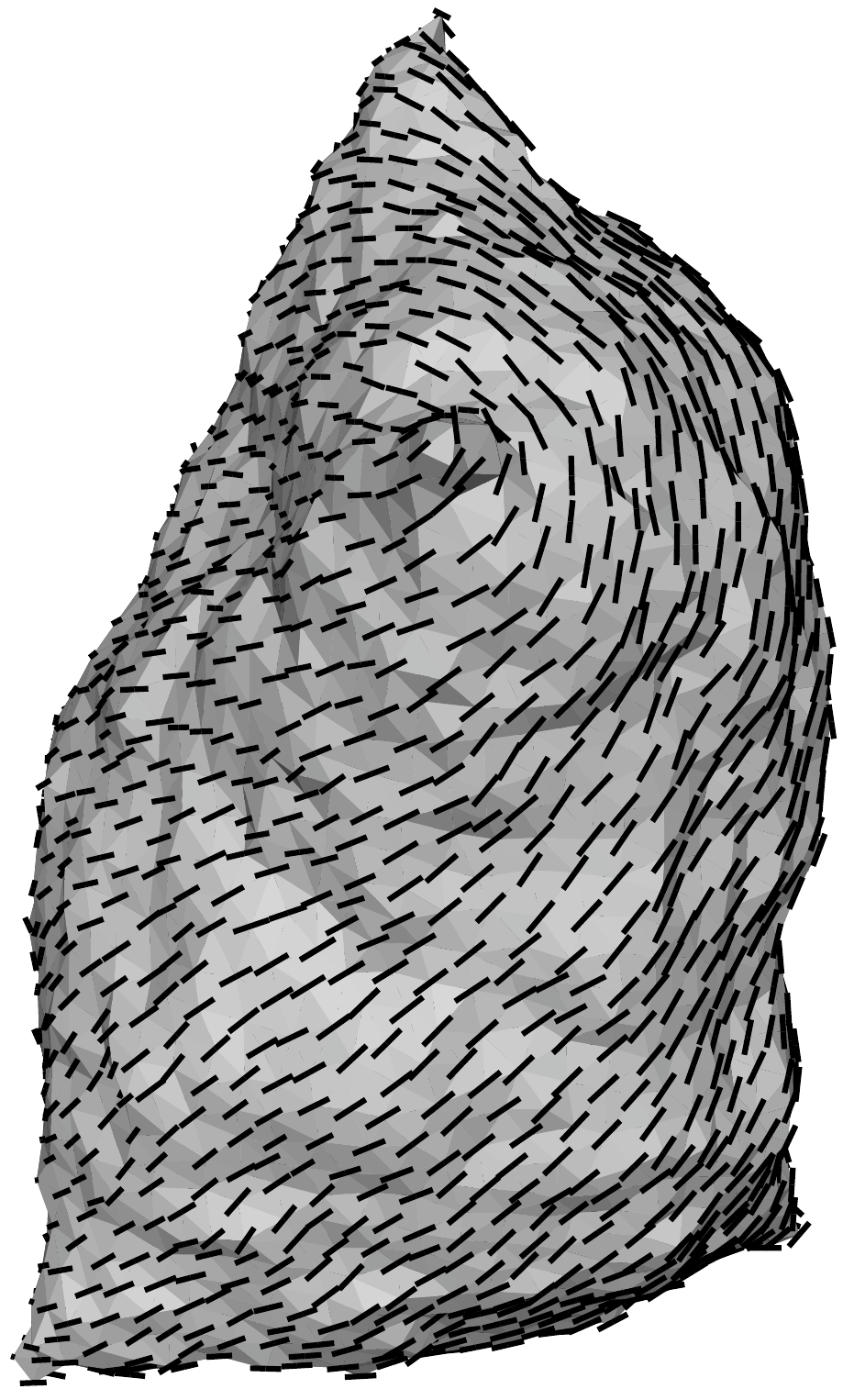}
\vspace*{20pt}
\label{fig:tetranem}
\end{minipage}
}
\caption{\label{fig:deformablesurf} Deformations of a stiff membrane ($\kappa=10$) due to an ordered in-plane field; {\bf (a)} A polar field stabilizes cylindrical shapes; and {\bf (b)} a  tetrahedron is stabilized when the field is nematic, with $J_{2}=5k_{B}T$.}
\end{figure}
Low-temperature equilibrium conformations of the membrane, in our simulations, are in excellent agreement with earlier theoretical predictions ~\cite{Lubensky:1992p531,MacKintosh:1991p685,Park:1992p946}. Figure ~\ref{fig:cylXY} shows the predicted ellipsoidal membrane for a polar field($p=1$), whereas the tetrahedral arrangement of defects, characteristic of a nematic field, is shown in Figure ~\ref{fig:tetranem}. In the case of polar fields, the two $+1$ antipodal vortices are positioned at the ellipsoidal caps, whereas the four +1/2 disclinations are at the vertices of the tetrahedron for the nematic field. Mean field membrane shapes associated with a $p$-atic vector field, with p=1 through 6, are given in ~\cite{Park:1992p946}.

\section{Regularized delta function}\label{sec:regdelta}
The discretization of a continuum membrane to  quasiparticles is obtained through the use of a regularized delta function ~\cite{Peskin:2002go}. These functions with compact support over a given interval satisfy some (most) of the properties of the Dirac delta in the distributional sense and are represented by $\delta_{h}(\boldsymbol{r})$, where $h$ would denote the spread of the function's support.

For $\boldsymbol{r}\in\mathbb{R}^{n}$, $\delta_{h}(\boldsymbol{r})=\dfrac{1}{\Delta V}\underset{i=1..n}{\prod}\phi\left(\dfrac{r_{i}}{h}\right)$,
$\Delta V$ is the elemental volume (discrete) and $\{r_{i}\}$ are
the components of the position vector. The function $\phi(x)$ satisfies
the following properties: 

\begin{eqnarray}
\phi(x) & = & \phi(-x)\nonumber \\
\underset{j}{\sum}\,\phi(x-x_j) & = & 1\,\quad\quad\forall x\label{eq:Regularized_Delta_Prop}\\
\underset{j}{\sum}\,x_j\phi(x-x_j) & = & x\,\quad \quad \forall x\nonumber 
\end{eqnarray}

To uniquely fix the form of $\phi(x)$, more conditions on its derivatives, support, and other features, will have to be imposed, the choice of which depends on the particular application. One common condition  is $\underset{j}{\sum}\left(\phi(x-j)\right)^{2}=C\,\forall x$, for a $C$  which is independent of $x$.


\begin{thebibliography}{332}
\expandafter\ifx\csname natexlab\endcsname\relax\def\natexlab#1{#1}\fi
\providecommand{\bibinfo}[2]{#2}
\ifx\xfnm\relax \def\xfnm[#1]{\unskip,\space#1}\fi
\bibitem[{Mouritsen(2005)}]{olemouritsen:2005}
\bibinfo{author}{O.~G. Mouritsen}, \bibinfo{title}{Life - As a matter of fat :
  The emerging science of lipidomics}, The Frontiers collection,
  \bibinfo{publisher}{Springer}, \bibinfo{address}{Germany},
  \bibinfo{year}{2005}.
\bibitem[{Guidotti(1972)}]{Guidotti:1972wd}
\bibinfo{author}{G.~Guidotti},
\newblock \bibinfo{title}{Membrane proteins},
\newblock \bibinfo{journal}{Ann. Rev. Biochem.} \bibinfo{volume}{41}
  (\bibinfo{year}{1972}) \bibinfo{pages}{731--752}.
\bibitem[{Voeltz and Prinz(2007)}]{Voeltz:2007p1399}
\bibinfo{author}{G.~K. Voeltz}, \bibinfo{author}{W.~A. Prinz},
\newblock \bibinfo{title}{Sheets, ribbons and tubules - how organelles get
  their shape},
\newblock \bibinfo{journal}{Nat. Rev. Mol. Cell Biol.} \bibinfo{volume}{8}
  (\bibinfo{year}{2007}) \bibinfo{pages}{258--264}.
\bibitem[{Joyce and Pollard(2008)}]{Joyce:2008kw}
\bibinfo{author}{J.~A. Joyce}, \bibinfo{author}{J.~W. Pollard},
\newblock \bibinfo{title}{Microenvironmental regulation of metastasis},
\newblock \bibinfo{journal}{Nat Rev Cancer} \bibinfo{volume}{9}
  (\bibinfo{year}{2008}) \bibinfo{pages}{239--252}.
\bibitem[{Escrib{\'a} et~al.(2008)Escrib{\'a}, Gonz{\'a}lez-Ros, Go{\~n}i,
  Kinnunen, Vigh, S{\'a}nchez-Magraner, Fern{\'a}ndez, Busquets, Horv{\'a}th,
  and Barcel{\'o}-Coblijn}]{Escriba:2008hb}
\bibinfo{author}{P.~V. Escrib{\'a}}, \bibinfo{author}{J.~M. Gonz{\'a}lez-Ros},
  \bibinfo{author}{F.~M. Go{\~n}i}, \bibinfo{author}{P.~K.~J. Kinnunen},
  \bibinfo{author}{L.~Vigh}, \bibinfo{author}{L.~S{\'a}nchez-Magraner},
  \bibinfo{author}{A.~M. Fern{\'a}ndez}, \bibinfo{author}{X.~Busquets},
  \bibinfo{author}{I.~Horv{\'a}th}, \bibinfo{author}{G.~Barcel{\'o}-Coblijn},
\newblock \bibinfo{title}{Membranes: a meeting point for lipids, proteins and
  therapies},
\newblock \bibinfo{journal}{J Cellular Mol Med} \bibinfo{volume}{12}
  (\bibinfo{year}{2008}) \bibinfo{pages}{829--875}.
\bibitem[{Cavey et~al.(2008)Cavey, Rauzi, Lenne, and Lecuit}]{Cavey:2008p751}
\bibinfo{author}{M.~Cavey}, \bibinfo{author}{M.~Rauzi}, \bibinfo{author}{P.-F.
  Lenne}, \bibinfo{author}{T.~Lecuit},
\newblock \bibinfo{title}{A two-tiered mechanism for stabilization and
  immobilization of e-cadherin},
\newblock \bibinfo{journal}{Nature} \bibinfo{volume}{453}
  (\bibinfo{year}{2008}) \bibinfo{pages}{751--756}.
\bibitem[{Jefferson et~al.(2004)Jefferson, Leung, and Liem}]{Jefferson:2004cf}
\bibinfo{author}{J.~J. Jefferson}, \bibinfo{author}{C.~L. Leung},
  \bibinfo{author}{R.~K.~H. Liem},
\newblock \bibinfo{title}{Plakins: Goliaths that link cell junctions and the
  cytoskeleton},
\newblock \bibinfo{journal}{Nature} \bibinfo{volume}{5} (\bibinfo{year}{2004})
  \bibinfo{pages}{542--553}.
\bibitem[{Dannhauser and Ungewickell(2012)}]{Dannhauser:2012gy}
\bibinfo{author}{P.~N. Dannhauser}, \bibinfo{author}{E.~J. Ungewickell},
\newblock \bibinfo{title}{Reconstitution of clathrin-coated bud and vesicle
  formation with minimal components},
\newblock \bibinfo{journal}{Nature Cell Biology} \bibinfo{volume}{14}
  (\bibinfo{year}{2012}) \bibinfo{pages}{634--639}.
\bibitem[{Wendland(2002)}]{Wendland:2002js}
\bibinfo{author}{B.~Wendland},
\newblock \bibinfo{title}{Epsins: adaptors in endocytosis?},
\newblock \bibinfo{journal}{Nat. Rev. Mol. Cell Biol.} \bibinfo{volume}{3}
  (\bibinfo{year}{2002}) \bibinfo{pages}{971--977}.
\bibitem[{Ford et~al.(2002)Ford, Mills, Peter, Vallis, Praefcke, Evans, and
  McMahon}]{Ford:2002ifb}
\bibinfo{author}{M.~G.~J. Ford}, \bibinfo{author}{I.~G. Mills},
  \bibinfo{author}{B.~J. Peter}, \bibinfo{author}{Y.~Vallis},
  \bibinfo{author}{G.~J.~K. Praefcke}, \bibinfo{author}{P.~R. Evans},
  \bibinfo{author}{H.~T. McMahon},
\newblock \bibinfo{title}{Curvature of clathrin-coated pits driven by epsin},
\newblock \bibinfo{journal}{Nature} \bibinfo{volume}{419}
  (\bibinfo{year}{2002}) \bibinfo{pages}{361--366}.
\bibitem[{Lai et~al.(2012)Lai, Jao, Lyman, Gallop, Peter, McMahon, Langen, and
  Voth}]{Lai:2012hk}
\bibinfo{author}{C.-L. Lai}, \bibinfo{author}{C.~C. Jao},
  \bibinfo{author}{E.~Lyman}, \bibinfo{author}{J.~L. Gallop},
  \bibinfo{author}{B.~J. Peter}, \bibinfo{author}{H.~T. McMahon},
  \bibinfo{author}{R.~Langen}, \bibinfo{author}{G.~A. Voth},
\newblock \bibinfo{title}{Membrane binding and self-association of the epsin
  n-terminal homology domain},
\newblock \bibinfo{journal}{Journal of Molecular Biology} \bibinfo{volume}{423}
  (\bibinfo{year}{2012}) \bibinfo{pages}{800--817}.
\bibitem[{Angst et~al.(2001)Angst, Marcozzi, and Magee}]{Angst:2001um}
\bibinfo{author}{B.~D. Angst}, \bibinfo{author}{C.~Marcozzi},
  \bibinfo{author}{A.~I. Magee},
\newblock \bibinfo{title}{The cadherin superfamily: diversity in form and
  function},
\newblock \bibinfo{journal}{Journal of Cell Science} \bibinfo{volume}{114}
  (\bibinfo{year}{2001}) \bibinfo{pages}{629--641}.
\bibitem[{Causeret et~al.(2005)Causeret, Taulet, Comunale, Favard, and
  Gauthier-Rouvi{\`e}re}]{Causeret:2005hb}
\bibinfo{author}{M.~Causeret}, \bibinfo{author}{N.~Taulet},
  \bibinfo{author}{F.~Comunale}, \bibinfo{author}{C.~Favard},
  \bibinfo{author}{C.~Gauthier-Rouvi{\`e}re},
\newblock \bibinfo{title}{N-cadherin association with lipid rafts regulates its
  dynamic assembly at cell-cell junctions in c2c12 myoblasts},
\newblock \bibinfo{journal}{Mol. Biol. Cell} \bibinfo{volume}{16}
  (\bibinfo{year}{2005}) \bibinfo{pages}{2168--2180}.
\bibitem[{M{\'a}rquez et~al.(2012)M{\'a}rquez, Favale, Nieto, Pescio, and
  Sterin-Speziale}]{Marquez:2012br}
\bibinfo{author}{M.~G. M{\'a}rquez}, \bibinfo{author}{N.~O. Favale},
  \bibinfo{author}{F.~L. Nieto}, \bibinfo{author}{L.~G. Pescio},
  \bibinfo{author}{N.~Sterin-Speziale},
\newblock \bibinfo{title}{Changes in membrane lipid composition cause
  alterations in epithelial cell-cell adhesion structures in renal papillary
  collecting duct cells},
\newblock \bibinfo{journal}{BBA - Biomembranes} \bibinfo{volume}{1818}
  (\bibinfo{year}{2012}) \bibinfo{pages}{491--501}.
\bibitem[{Alberts et~al.(1994)Alberts, Johnson, Lewis, Raff, Roberts, and
  Walter}]{Alberts:1994}
\bibinfo{author}{B.~Alberts}, \bibinfo{author}{A.~Johnson},
  \bibinfo{author}{J.~Lewis}, \bibinfo{author}{M.~Raff},
  \bibinfo{author}{K.~Roberts}, \bibinfo{author}{P.~Walter},
  \bibinfo{title}{Molecular Biology of the cell}, \bibinfo{publisher}{Garland
  Publishing}, \bibinfo{address}{Singapore}, \bibinfo{edition}{third} edition,
  \bibinfo{year}{1994}.
\bibitem[{van Meer et~al.(2008)van Meer, Voelker, and
  Feigenson}]{vanMeer:2008p3294}
\bibinfo{author}{G.~van Meer}, \bibinfo{author}{D.~R. Voelker},
  \bibinfo{author}{G.~W. Feigenson},
\newblock \bibinfo{title}{Membrane lipids: where they are and how they behave},
\newblock \bibinfo{journal}{Nat. Rev. Mol. Cell Biol.} \bibinfo{volume}{9}
  (\bibinfo{year}{2008}) \bibinfo{pages}{112--124}.
\bibitem[{Farese and Walther(2009)}]{Farese:2009p3290}
\bibinfo{author}{R.~V. Farese}, \bibinfo{author}{T.~C. Walther},
\newblock \bibinfo{title}{Lipid droplets finally get a little r-e-s-p-e-c-t},
\newblock \bibinfo{journal}{Cell} \bibinfo{volume}{139} (\bibinfo{year}{2009})
  \bibinfo{pages}{855--860}.
\bibitem[{Israelachvili et~al.(1977)Israelachvili, Mitchell, and
  Ninham}]{Israelachvili:1977ve}
\bibinfo{author}{J.~N. Israelachvili}, \bibinfo{author}{D.~J. Mitchell},
  \bibinfo{author}{B.~W. Ninham},
\newblock \bibinfo{title}{Theory of self-assembly of lipid bilayers and
  vesicles},
\newblock \bibinfo{journal}{Biochimica Biophysics Acta (BBA) - Biomembranes}
  \bibinfo{volume}{470} (\bibinfo{year}{1977}) \bibinfo{pages}{185--201}.
\bibitem[{Koynova and Caffrey(2002)}]{Koynova:2002gb}
\bibinfo{author}{R.~Koynova}, \bibinfo{author}{M.~Caffrey},
\newblock \bibinfo{title}{An index of lipid phase diagrams},
\newblock \bibinfo{journal}{Chem. Phys. Lipids} \bibinfo{volume}{115}
  (\bibinfo{year}{2002}) \bibinfo{pages}{107--219}.
\bibitem[{Singer and Nicolson(1972)}]{Singer:1972p2064}
\bibinfo{author}{S.~J. Singer}, \bibinfo{author}{G.~L. Nicolson},
\newblock \bibinfo{title}{The fluid mosaic model of the structure of cell
  membranes},
\newblock \bibinfo{journal}{Science} \bibinfo{volume}{175}
  (\bibinfo{year}{1972}) \bibinfo{pages}{720--731}.
\bibitem[{Spector and Yorek(1985)}]{Spector:1985vu}
\bibinfo{author}{A.~A. Spector}, \bibinfo{author}{M.~A. Yorek},
\newblock \bibinfo{title}{Membrane lipid-composition and cellular function},
\newblock \bibinfo{journal}{The Journal of Lipid Research} \bibinfo{volume}{26}
  (\bibinfo{year}{1985}) \bibinfo{pages}{1015--1035}.
\bibitem[{Paula et~al.(1996)Paula, Volkov, Van~Hoek, Haines, and
  Deamer}]{Paula:1996p339}
\bibinfo{author}{S.~Paula}, \bibinfo{author}{A.~G. Volkov},
  \bibinfo{author}{A.~N. Van~Hoek}, \bibinfo{author}{T.~H. Haines},
  \bibinfo{author}{D.~W. Deamer},
\newblock \bibinfo{title}{Permeation of protons, potassium ions, and small
  polar molecules through phospholipid bilayers as a function of membrane
  thickness},
\newblock \bibinfo{journal}{Biophys. J.} \bibinfo{volume}{70}
  (\bibinfo{year}{1996}) \bibinfo{pages}{339--348}.
\bibitem[{Finkelstein(1976)}]{Finkelstein:1976vo}
\bibinfo{author}{A.~Finkelstein},
\newblock \bibinfo{title}{Water and nonelectrolyte permeability of lipid
  bilayer membranes},
\newblock \bibinfo{journal}{J. Gen. Physiol.} \bibinfo{volume}{68}
  (\bibinfo{year}{1976}) \bibinfo{pages}{127}.
\bibitem[{Honerkamp-Smith et~al.(2009)Honerkamp-Smith, Veatch, and
  Keller}]{HonerkampSmith:2009fh}
\bibinfo{author}{A.~R. Honerkamp-Smith}, \bibinfo{author}{S.~L. Veatch},
  \bibinfo{author}{S.~L. Keller},
\newblock \bibinfo{title}{An introduction to critical points for biophysicists;
  observations of compositional heterogeneity in lipid membranes},
\newblock \bibinfo{journal}{Biochimica et Biophysica Acta (BBA) - Biomembranes}
  \bibinfo{volume}{1788} (\bibinfo{year}{2009}) \bibinfo{pages}{53--63}.
\bibitem[{Semrau and Schmidt(2009)}]{Semaru:2009p3174}
\bibinfo{author}{S.~Semrau}, \bibinfo{author}{T.~Schmidt},
\newblock \bibinfo{title}{Membrane heterogeneity - from lipid domains to
  curvature effects},
\newblock \bibinfo{journal}{Soft Matter} \bibinfo{volume}{5}
  (\bibinfo{year}{2009}) \bibinfo{pages}{3174}.
\bibitem[{Munro(2003)}]{Munro:2003p377}
\bibinfo{author}{S.~Munro},
\newblock \bibinfo{title}{Lipid rafts: Elusive or illusive?},
\newblock \bibinfo{journal}{Cell} \bibinfo{volume}{115} (\bibinfo{year}{2003})
  \bibinfo{pages}{377--388}.
\bibitem[{Edidin(2003)}]{Edidin:2003p1606}
\bibinfo{author}{M.~Edidin},
\newblock \bibinfo{title}{Lipids on the frontier: a century of cell-membrane
  bilayers},
\newblock \bibinfo{journal}{Nat. Rev. Mol. Cell Biol.} \bibinfo{volume}{4}
  (\bibinfo{year}{2003}) \bibinfo{pages}{414--418}.
\bibitem[{Kaiser et~al.(2009)Kaiser, Lingwood, Levental, Sampaio, Kalvodova,
  Rajendran, and Simons}]{Kaiser:2009p497}
\bibinfo{author}{H.-J. Kaiser}, \bibinfo{author}{D.~Lingwood},
  \bibinfo{author}{I.~Levental}, \bibinfo{author}{J.~L. Sampaio},
  \bibinfo{author}{L.~Kalvodova}, \bibinfo{author}{L.~Rajendran},
  \bibinfo{author}{K.~Simons},
\newblock \bibinfo{title}{Order of lipid phases in model and plasma membranes},
\newblock \bibinfo{journal}{Proc. Natl. Acad. Sci. USA.} \bibinfo{volume}{106}
  (\bibinfo{year}{2009}) \bibinfo{pages}{16645--16650}.
\bibitem[{Veatch and Keller(2002)}]{Veatch:2002ta}
\bibinfo{author}{S.~L. Veatch}, \bibinfo{author}{S.~L. Keller},
\newblock \bibinfo{title}{Organization in lipid membranes containing
  cholesterol},
\newblock \bibinfo{journal}{Phys. Rev. Lett.} \bibinfo{volume}{89}
  (\bibinfo{year}{2002}) \bibinfo{pages}{268101}.
\bibitem[{Bagatolli and Sunil~Kumar(2009)}]{Bagatolli:2009SM}
\bibinfo{author}{L.~Bagatolli}, \bibinfo{author}{P.~B. Sunil~Kumar},
\newblock \bibinfo{title}{Phase behavior of multicomponent membranes:
  Experimental and computational techniques},
\newblock \bibinfo{journal}{Soft Matter} \bibinfo{volume}{5}
  (\bibinfo{year}{2009}) \bibinfo{pages}{3234}.
\bibitem[{Veatch and Keller(2003)}]{Veatch:2003hn}
\bibinfo{author}{S.~L. Veatch}, \bibinfo{author}{S.~L. Keller},
\newblock \bibinfo{title}{Separation of liquid phases in giant vesicles of
  ternary mixtures of phospholipids and cholesterol},
\newblock \bibinfo{journal}{Biophys. J.} \bibinfo{volume}{85}
  (\bibinfo{year}{2003}) \bibinfo{pages}{3074--3083}.
\bibitem[{Go{\~n}i et~al.(2008)Go{\~n}i, Alonso, Bagatolli, Brown, Marsh,
  Prieto, and Thewalt}]{Goni:2008dt}
\bibinfo{author}{F.~M. Go{\~n}i}, \bibinfo{author}{A.~Alonso},
  \bibinfo{author}{L.~A. Bagatolli}, \bibinfo{author}{R.~E. Brown},
  \bibinfo{author}{D.~Marsh}, \bibinfo{author}{M.~Prieto},
  \bibinfo{author}{J.~L. Thewalt},
\newblock \bibinfo{title}{Phase diagrams of lipid mixtures relevant to the
  study of membrane rafts},
\newblock \bibinfo{journal}{Biochimica et Biophysica Acta (BBA) - Molecular and
  Cell Biology of Lipids} \bibinfo{volume}{1781} (\bibinfo{year}{2008})
  \bibinfo{pages}{665--684}.
\bibitem[{Hamada et~al.(2011)Hamada, Kishimoto, Nagasaki, and
  Takagi}]{Hamada:2011fq}
\bibinfo{author}{T.~Hamada}, \bibinfo{author}{Y.~Kishimoto},
  \bibinfo{author}{T.~Nagasaki}, \bibinfo{author}{M.~Takagi},
\newblock \bibinfo{title}{Lateral phase separation in tense membranes},
\newblock \bibinfo{journal}{Soft Matter} \bibinfo{volume}{7}
  (\bibinfo{year}{2011}) \bibinfo{pages}{9061}.
\bibitem[{Ursell et~al.(2009)Ursell, Klug, and Phillips}]{Ursell:2009p176}
\bibinfo{author}{T.~S. Ursell}, \bibinfo{author}{W.~S. Klug},
  \bibinfo{author}{R.~Phillips},
\newblock \bibinfo{title}{Morphology and interaction between lipid domains},
\newblock \bibinfo{journal}{Proc. Natl. Acad. Sci. USA.} \bibinfo{volume}{106}
  (\bibinfo{year}{2009}) \bibinfo{pages}{13301--13306}.
\bibitem[{Simons and Toomre(2000)}]{Simons:2000fa}
\bibinfo{author}{K.~Simons}, \bibinfo{author}{D.~Toomre},
\newblock \bibinfo{title}{Lipid rafts and signal transduction},
\newblock \bibinfo{journal}{Nat. Rev. Mol. Cell Biol.} \bibinfo{volume}{1}
  (\bibinfo{year}{2000}) \bibinfo{pages}{31--39}.
\bibitem[{Simons and Sampaio(2011)}]{Simons:tx}
\bibinfo{author}{K.~Simons}, \bibinfo{author}{J.~L. Sampaio},
\newblock \bibinfo{title}{Membrane organization and lipid rafts},
\newblock \bibinfo{journal}{Cold Spring Harb Perspect Biol}
  (\bibinfo{year}{2011}).
\bibitem[{Simons and Vaz(2004)}]{Simons:2004p187}
\bibinfo{author}{K.~Simons}, \bibinfo{author}{W.~L.~C. Vaz},
\newblock \bibinfo{title}{Model systems, lipid rafts, and cell membranes},
\newblock \bibinfo{journal}{Ann. Rev. Biophys. Biomol. Struct.}
  \bibinfo{volume}{33} (\bibinfo{year}{2004}) \bibinfo{pages}{269--295}.
\bibitem[{Simons and Ikonen(1997)}]{Simons:1997jq}
\bibinfo{author}{K.~Simons}, \bibinfo{author}{E.~Ikonen},
\newblock \bibinfo{title}{Functional rafts in cell membranes},
\newblock \bibinfo{journal}{Nature} \bibinfo{volume}{387}
  (\bibinfo{year}{1997}) \bibinfo{pages}{569--572}.
\bibitem[{K{\"a}s and Sackmann(1991)}]{Kas:1991fk}
\bibinfo{author}{J.~K{\"a}s}, \bibinfo{author}{E.~Sackmann},
\newblock \bibinfo{title}{Shape transitions and shape stability of giant
  phospholipid vesicles in pure water induced by area-to-volume changes},
\newblock \bibinfo{journal}{Biophys. J.} \bibinfo{volume}{60}
  (\bibinfo{year}{1991}) \bibinfo{pages}{825--844}.
\bibitem[{Hotani(1984)}]{Hotani:1984tz}
\bibinfo{author}{H.~Hotani},
\newblock \bibinfo{title}{Transformation pathways of liposomes},
\newblock \bibinfo{journal}{Journal of Molecular Biology} \bibinfo{volume}{178}
  (\bibinfo{year}{1984}) \bibinfo{pages}{113--120}.
\bibitem[{Lipowsky(1991)}]{Lipowsky:1991p1059}
\bibinfo{author}{R.~Lipowsky},
\newblock \bibinfo{title}{The conformation of membranes},
\newblock \bibinfo{journal}{Nature} \bibinfo{volume}{349}
  (\bibinfo{year}{1991}) \bibinfo{pages}{475}.
\bibitem[{Seifert(1997)}]{Seifert:1997p1058}
\bibinfo{author}{U.~Seifert},
\newblock \bibinfo{title}{Configurations of fluid membranes and vesicles},
\newblock \bibinfo{journal}{Adv. Phys.} \bibinfo{volume}{46}
  (\bibinfo{year}{1997}) \bibinfo{pages}{13}.
\bibitem[{Devaux(1991)}]{Devaux:1991ti}
\bibinfo{author}{P.~F. Devaux},
\newblock \bibinfo{title}{Static and dynamic lipid asymmetry in cell
  membranes},
\newblock \bibinfo{journal}{Biochemistry} \bibinfo{volume}{30}
  (\bibinfo{year}{1991}) \bibinfo{pages}{1163--1173}.
\bibitem[{Slochower et~al.(2014)Slochower, Wang, Tourdot, Radhakrishnan, and
  Janmey}]{Slochower2014}
\bibinfo{author}{D.~R. Slochower}, \bibinfo{author}{Y.-H. Wang},
  \bibinfo{author}{R.~W. Tourdot}, \bibinfo{author}{R.~Radhakrishnan},
  \bibinfo{author}{P.~A. Janmey},
\newblock \bibinfo{title}{Counterion-mediated pattern formation in membranes
  containing anionic lipids},
\newblock \bibinfo{journal}{Advances in Colloid and Interface Science}
  (\bibinfo{year}{2014}) \bibinfo{pages}{in press. DOI
  10.1016/j.cis.2014.01.016}.
\bibitem[{Canham(1970)}]{Canham:1970p61}
\bibinfo{author}{P.~B. Canham},
\newblock \bibinfo{title}{The minimum energy of bending as a possible
  explanation of the biconcave shape of the human red blood cell},
\newblock \bibinfo{journal}{J. Theor. Biol.} \bibinfo{volume}{26}
  (\bibinfo{year}{1970}) \bibinfo{pages}{61--81}.
\bibitem[{Helfrich(1973)}]{Helfrich:1973p693}
\bibinfo{author}{W.~Helfrich},
\newblock \bibinfo{title}{Elastic properties of lipid bilayers: theory and
  possible experiments},
\newblock \bibinfo{journal}{Z. Naturforsch. C} \bibinfo{volume}{28}
  (\bibinfo{year}{1973}) \bibinfo{pages}{1--12}.
\bibitem[{Landau and Lifshitz(1970)}]{Landau:Elasticity}
\bibinfo{author}{L.~D. Landau}, \bibinfo{author}{E.~M. Lifshitz},
  \bibinfo{title}{Theory of Elasticity}, \bibinfo{publisher}{Pergamon Press},
  \bibinfo{address}{Oxford}, \bibinfo{year}{1970}.
\bibitem[{Woodka et~al.(2012)Woodka, Butler, Porcar, Farago, and
  Nagao}]{Woodka:2012cf}
\bibinfo{author}{A.~C. Woodka}, \bibinfo{author}{P.~D. Butler},
  \bibinfo{author}{L.~Porcar}, \bibinfo{author}{B.~Farago},
  \bibinfo{author}{M.~Nagao},
\newblock \bibinfo{title}{Lipid bilayers and membrane dynamics: Insight into
  thickness fluctuations},
\newblock \bibinfo{journal}{Phys. Rev. Lett.} \bibinfo{volume}{109}
  (\bibinfo{year}{2012}) \bibinfo{pages}{058102}.
\bibitem[{J{\"a}hnig(1996)}]{Jahnig:1996cv}
\bibinfo{author}{F.~J{\"a}hnig},
\newblock \bibinfo{title}{What is the surface tension of a lipid bilayer
  membrane?},
\newblock \bibinfo{journal}{Biophys. J.} \bibinfo{volume}{71}
  (\bibinfo{year}{1996}) \bibinfo{pages}{1348--1349}.
\bibitem[{do~Carmo(1976)}]{doCarmo:1976}
\bibinfo{author}{M.~P. do~Carmo}, \bibinfo{title}{Differential geometry of
  curves and surfaces}, \bibinfo{publisher}{Prentice Hall},
  \bibinfo{address}{Engelwood Cliffs, New Jersey}, \bibinfo{year}{1976}.
\bibitem[{Helfrich(1985)}]{Helfrich:1985wi}
\bibinfo{author}{W.~Helfrich},
\newblock \bibinfo{title}{Effect of thermal undulations on the rigidity of
  fluid membranes and interfaces},
\newblock \bibinfo{journal}{J. Phys. France} \bibinfo{volume}{46}
  (\bibinfo{year}{1985}) \bibinfo{pages}{1263--1268}.
\bibitem[{Peliti and Leibler(1985)}]{Peliti:1985p1690}
\bibinfo{author}{L.~Peliti}, \bibinfo{author}{S.~Leibler},
\newblock \bibinfo{title}{Effects of thermal fluctuations on systems with small
  surface tension},
\newblock \bibinfo{journal}{Phys. Rev. Lett.} \bibinfo{volume}{54}
  (\bibinfo{year}{1985}) \bibinfo{pages}{1690--1693}.
\bibitem[{F{\"o}ster(1986)}]{Foster:1986jb}
\bibinfo{author}{D.~F{\"o}ster},
\newblock \bibinfo{title}{On the scale dependence, due to thermal fluctuations,
  of the elastic properties of membranes},
\newblock \bibinfo{journal}{Phys. Lett. A} \bibinfo{volume}{114}
  (\bibinfo{year}{1986}) \bibinfo{pages}{115--120}.
\bibitem[{Kleinert(1986)}]{Kleinert:1986p347}
\bibinfo{author}{H.~Kleinert},
\newblock \bibinfo{title}{Thermal softening of curvature elasticity in
  membranes},
\newblock \bibinfo{journal}{Phys. Lett.} \bibinfo{volume}{114A}
  (\bibinfo{year}{1986}) \bibinfo{pages}{263--268}.
\bibitem[{De~Gennes and Taupin(1982)}]{Gennes:1982p240}
\bibinfo{author}{P.~G. De~Gennes}, \bibinfo{author}{C.~Taupin},
\newblock \bibinfo{title}{Microemulsions and the flexibility of oil/water
  interfaces},
\newblock \bibinfo{journal}{J. Phys. Chem.} \bibinfo{volume}{86}
  (\bibinfo{year}{1982}) \bibinfo{pages}{2294--2304}.
\bibitem[{Helfrich(1986)}]{Helfrich:1986bx}
\bibinfo{author}{W.~Helfrich},
\newblock \bibinfo{title}{Size distributions of vesicles : the role of the
  effective rigidity of membranes},
\newblock \bibinfo{journal}{J. Phys. France} \bibinfo{volume}{47}
  (\bibinfo{year}{1986}) \bibinfo{pages}{321--329}.
\bibitem[{Helfrich(1987)}]{Helfrich:1987if}
\bibinfo{author}{W.~Helfrich},
\newblock \bibinfo{title}{Measures of integration in calculating the effective
  rigidity of fluid surfaces},
\newblock \bibinfo{journal}{J. Phys. France} \bibinfo{volume}{48}
  (\bibinfo{year}{1987}) \bibinfo{pages}{285--289}.
\bibitem[{Helfrich(1998)}]{Helfrich:1998dk}
\bibinfo{author}{W.~Helfrich},
\newblock \bibinfo{title}{Stiffening of fluid membranes and entropy loss of
  membrane closure: Two effects of thermal undulations},
\newblock \bibinfo{journal}{Eur. Phys. J. B} \bibinfo{volume}{1}
  (\bibinfo{year}{1998}) \bibinfo{pages}{481}.
\bibitem[{Pinnow and Helfrich(2000)}]{Pinnow:2000hr}
\bibinfo{author}{H.~A. Pinnow}, \bibinfo{author}{W.~Helfrich},
\newblock \bibinfo{title}{Effect of thermal undulations on the bending
  elasticity and spontaneous curvature of fluid membranes},
\newblock \bibinfo{journal}{Eur. Phys. J. E} \bibinfo{volume}{3}
  (\bibinfo{year}{2000}) \bibinfo{pages}{149--157}.
\bibitem[{Marsh(1997)}]{Marsh:1997p865}
\bibinfo{author}{D.~Marsh},
\newblock \bibinfo{title}{Renormalization of the tension and area expansion
  modulus in fluid membranes},
\newblock \bibinfo{journal}{Biophys. J.} \bibinfo{volume}{73}
  (\bibinfo{year}{1997}) \bibinfo{pages}{865--869}.
\bibitem[{Marsh(2006)}]{Marsh:2006ft}
\bibinfo{author}{D.~Marsh},
\newblock \bibinfo{title}{Elastic curvature constants of lipid monolayers and
  bilayers},
\newblock \bibinfo{journal}{Chemistry and Physics of Lipids}
  \bibinfo{volume}{144} (\bibinfo{year}{2006}) \bibinfo{pages}{146--159}.
\bibitem[{Cai and Lubensky(1994)}]{Cai:1994p410}
\bibinfo{author}{W.~Cai}, \bibinfo{author}{T.~Lubensky},
\newblock \bibinfo{title}{Covariant hydrodynamics of fluid membranes},
\newblock \bibinfo{journal}{Phys. Rev. Lett.} \bibinfo{volume}{73}
  (\bibinfo{year}{1994}) \bibinfo{pages}{1186--1189}.
\bibitem[{Cai and Lubensky(1995)}]{Cai:1995un}
\bibinfo{author}{W.~Cai}, \bibinfo{author}{T.~Lubensky},
\newblock \bibinfo{title}{Hydrodynamics and dynamic fluctuations of fluid
  membranes},
\newblock \bibinfo{journal}{Phys. Rev. E} \bibinfo{volume}{52}
  (\bibinfo{year}{1995}) \bibinfo{pages}{4251--4266}.
\bibitem[{Dill and Bromberg(2003)}]{Dill:2003vs}
\bibinfo{author}{K.~A. Dill}, \bibinfo{author}{S.~Bromberg},
  \bibinfo{title}{Molecular driving forces}, \bibinfo{publisher}{Garland
  Science}, \bibinfo{address}{New York}, \bibinfo{year}{2003}.
\bibitem[{David and Leibler(1991)}]{DAVID:1991p3533}
\bibinfo{author}{F.~David}, \bibinfo{author}{S.~Leibler},
\newblock \bibinfo{title}{Vanishing tension of fluctuating membranes},
\newblock \bibinfo{journal}{J. Phys. II France} \bibinfo{volume}{1}
  (\bibinfo{year}{1991}) \bibinfo{pages}{959--976}.
\bibitem[{Nelson et~al.(2003)Nelson, Piran, and Weinberg}]{Piran:2003}
\bibinfo{editor}{D.~Nelson}, \bibinfo{editor}{T.~Piran},
  \bibinfo{editor}{S.~Weinberg} (Eds.), \bibinfo{title}{Statistical Mechanics
  of Membranes and Surfaces}, \bibinfo{publisher}{World Scientific},
  \bibinfo{address}{Singapore}, \bibinfo{edition}{second} edition,
  \bibinfo{year}{2003}.
\bibitem[{Miao et~al.(1994)Miao, Seifert, Wortis, and
  D{\"o}bereiner}]{Miao:1994p112}
\bibinfo{author}{L.~Miao}, \bibinfo{author}{U.~Seifert},
  \bibinfo{author}{M.~Wortis}, \bibinfo{author}{H.-G. D{\"o}bereiner},
\newblock \bibinfo{title}{Budding transitions of fluid-bilayer vesicles - the
  effect of area-difference elasticity},
\newblock \bibinfo{journal}{Phys. Rev. E} \bibinfo{volume}{49}
  (\bibinfo{year}{1994}) \bibinfo{pages}{5389--5407}.
\bibitem[{Chaikin and Lubensky(1995)}]{Chaikin:1995td}
\bibinfo{author}{P.~M. Chaikin}, \bibinfo{author}{T.~C. Lubensky},
  \bibinfo{title}{Principles of condensed matter physics},
  \bibinfo{publisher}{Cambridge University Press}, \bibinfo{address}{Cambridge
  ; New York}, \bibinfo{year}{1995}.
\bibitem[{Reister-Gottfried et~al.(2007)Reister-Gottfried, Leitenberger, and
  Seifert}]{ReisterGottfried:2007ep}
\bibinfo{author}{E.~Reister-Gottfried}, \bibinfo{author}{S.~Leitenberger},
  \bibinfo{author}{U.~Seifert},
\newblock \bibinfo{title}{Hybrid simulations of lateral diffusion in
  fluctuating membranes},
\newblock \bibinfo{journal}{Phys. Rev. E} \bibinfo{volume}{75}
  (\bibinfo{year}{2007}) \bibinfo{pages}{011908}.
\bibitem[{Lin and Brown(2004)}]{Lin:2004p256001}
\bibinfo{author}{L.~C.-L. Lin}, \bibinfo{author}{F.~L. Brown},
\newblock \bibinfo{title}{Brownian dynamics in fourier space: Membrane
  simulations over long length and time scales},
\newblock \bibinfo{journal}{Phys. Rev. Lett.} \bibinfo{volume}{93}
  (\bibinfo{year}{2004}) \bibinfo{pages}{256001}.
\bibitem[{Brown(2008)}]{Brown:2008cf}
\bibinfo{author}{F.~L.~H. Brown},
\newblock \bibinfo{title}{Elastic modeling of biomembranes and lipid bilayers},
\newblock \bibinfo{journal}{Annu. Rev. Phys. Chem.} \bibinfo{volume}{59}
  (\bibinfo{year}{2008}) \bibinfo{pages}{685--712}.
\bibitem[{Evans and Rawicz(1990)}]{Evans:1990he}
\bibinfo{author}{E.~Evans}, \bibinfo{author}{W.~Rawicz},
\newblock \bibinfo{title}{Entropy-driven tension and bending elasticity in
  condensed-fluid membranes},
\newblock \bibinfo{journal}{Phys. Rev. Lett.} \bibinfo{volume}{64}
  (\bibinfo{year}{1990}) \bibinfo{pages}{2094--2097}.
\bibitem[{Evans et~al.(1996)Evans, Bowman, Leung, Needham, and
  Tirrell}]{Evans:1996fs}
\bibinfo{author}{E.~Evans}, \bibinfo{author}{H.~Bowman},
  \bibinfo{author}{A.~Leung}, \bibinfo{author}{D.~Needham},
  \bibinfo{author}{D.~Tirrell},
\newblock \bibinfo{title}{Biomembrane templates for nanoscale conduits and
  networks},
\newblock \bibinfo{journal}{Science} \bibinfo{volume}{273}
  (\bibinfo{year}{1996}) \bibinfo{pages}{933--935}.
\bibitem[{Der{\'e}nyi et~al.(2002)Der{\'e}nyi, J{\"u}licher, and
  Prost}]{Derenyi:2002kx}
\bibinfo{author}{I.~Der{\'e}nyi}, \bibinfo{author}{F.~J{\"u}licher},
  \bibinfo{author}{J.~Prost},
\newblock \bibinfo{title}{Formation and interaction of membrane tubes},
\newblock \bibinfo{journal}{Phys. Rev. Lett.} \bibinfo{volume}{88}
  (\bibinfo{year}{2002}) \bibinfo{pages}{238101}.
\bibitem[{Powers et~al.(2002)Powers, Huber, and Goldstein}]{Powers:2002jg}
\bibinfo{author}{T.~Powers}, \bibinfo{author}{G.~Huber},
  \bibinfo{author}{R.~Goldstein},
\newblock \bibinfo{title}{Fluid-membrane tethers: Minimal surfaces and elastic
  boundary layers},
\newblock \bibinfo{journal}{Phys. Rev. E} \bibinfo{volume}{65}
  (\bibinfo{year}{2002}) \bibinfo{pages}{041901}.
\bibitem[{Allain et~al.(2004)Allain, Storm, Roux, Amar, and
  Joanny}]{Allain:2004ix}
\bibinfo{author}{J.~M. Allain}, \bibinfo{author}{C.~Storm},
  \bibinfo{author}{A.~Roux}, \bibinfo{author}{M.~Amar}, \bibinfo{author}{J.~F.
  Joanny},
\newblock \bibinfo{title}{Fission of a multiphase membrane tube},
\newblock \bibinfo{journal}{Phys. Rev. Lett.} \bibinfo{volume}{93}
  (\bibinfo{year}{2004}) \bibinfo{pages}{158104}.
\bibitem[{Zhong-can and Helfrich(1989)}]{Zhongcan:1989ue}
\bibinfo{author}{O.~Zhong-can}, \bibinfo{author}{W.~Helfrich},
\newblock \bibinfo{title}{Bending energy of vesicle membranes: General
  expressions for the first, second, and third variation of the shape energy
  and applications to spheres and cylinders},
\newblock \bibinfo{journal}{Phys. Rev. A} \bibinfo{volume}{39}
  (\bibinfo{year}{1989}) \bibinfo{pages}{5280--5288}.
\bibitem[{J{\"u}licher and Seifert(1994)}]{Julicher:1994bk}
\bibinfo{author}{F.~J{\"u}licher}, \bibinfo{author}{U.~Seifert},
\newblock \bibinfo{title}{Shape equations for axisymmetric vesicles: A
  clarification},
\newblock \bibinfo{journal}{Phys. Rev. E} \bibinfo{volume}{49}
  (\bibinfo{year}{1994}) \bibinfo{pages}{4728--4731}.
\bibitem[{J{\"u}licher and Lipowsky(1996)}]{Julicher:1996co}
\bibinfo{author}{F.~J{\"u}licher}, \bibinfo{author}{R.~Lipowsky},
\newblock \bibinfo{title}{Shape transformations of vesicles with intramembrane
  domains},
\newblock \bibinfo{journal}{Phys. Rev. E} \bibinfo{volume}{53}
  (\bibinfo{year}{1996}) \bibinfo{pages}{2670--2683}.
\bibitem[{Seifert et~al.(1991)Seifert, Berndl, and Lipowsky}]{Seifert91}
\bibinfo{author}{U.~Seifert}, \bibinfo{author}{K.~Berndl},
  \bibinfo{author}{R.~Lipowsky},
\newblock \bibinfo{title}{Shape transformations of vesicles: Phase diagram for
  spontaneous- curvature and bilayer-coupling models},
\newblock \bibinfo{journal}{Phys. Rev. A} \bibinfo{volume}{44}
  (\bibinfo{year}{1991}) \bibinfo{pages}{1182}.
\bibitem[{Gao et~al.(2005)Gao, Shi, and Freund}]{Gao05}
\bibinfo{author}{H.~Gao}, \bibinfo{author}{W.~Shi}, \bibinfo{author}{L.~B.
  Freund},
\newblock \bibinfo{title}{Mechanics of receptor-mediated endocytosis},
\newblock \bibinfo{journal}{Proc. Natl. Acad. Sci. USA.} \bibinfo{volume}{102}
  (\bibinfo{year}{2005}) \bibinfo{pages}{9469--9474}.
\bibitem[{Liu et~al.(2006{\natexlab{a}})Liu, Yao, Jing, Li, and Sun}]{Liu06}
\bibinfo{author}{Z.-L. Liu}, \bibinfo{author}{K.-L. Yao},
  \bibinfo{author}{X.-B. Jing}, \bibinfo{author}{X.-A. Li},
  \bibinfo{author}{X.-Z. Sun},
\newblock \bibinfo{title}{Endocytic vesicle scission by lipid phase boundary
  forces},
\newblock \bibinfo{journal}{Proc. Natl. Acad. Sci. USA.} \bibinfo{volume}{103}
  (\bibinfo{year}{2006}{\natexlab{a}}) \bibinfo{pages}{10277--10282}.
\bibitem[{Liu et~al.(2006{\natexlab{b}})Liu, Kaksonen, Drubin, and
  Oster}]{Liu:2006fc}
\bibinfo{author}{J.~Liu}, \bibinfo{author}{M.~Kaksonen}, \bibinfo{author}{D.~G.
  Drubin}, \bibinfo{author}{G.~Oster},
\newblock \bibinfo{title}{Endocytic vesicle scission by lipid phase boundary
  forces},
\newblock \bibinfo{journal}{Proc. Natl. Acad. Sci. USA.} \bibinfo{volume}{103}
  (\bibinfo{year}{2006}{\natexlab{b}}) \bibinfo{pages}{10277--10282}.
\bibitem[{Agrawal et~al.(2010)Agrawal, Nukpezah, and
  Radhakrishnan}]{Agrawal:2010eu}
\bibinfo{author}{N.~J. Agrawal}, \bibinfo{author}{J.~Nukpezah},
  \bibinfo{author}{R.~Radhakrishnan},
\newblock \bibinfo{title}{Minimal mesoscale model for protein-mediated
  vesiculation in clathrin-dependent endocytosis},
\newblock \bibinfo{journal}{PLoS Comput. Biol.} \bibinfo{volume}{6}
  (\bibinfo{year}{2010}) \bibinfo{pages}{e1000926}.
\bibitem[{Brakke(1992)}]{Brakke:1992tn}
\bibinfo{author}{K.~A. Brakke},
\newblock \bibinfo{title}{The surface evolver},
\newblock \bibinfo{journal}{Experimental mathematics} \bibinfo{volume}{1}
  (\bibinfo{year}{1992}) \bibinfo{pages}{141--165}.
\bibitem[{Dasgupta et~al.(2013)Dasgupta, Auth, and Gompper}]{Dasgupta:2013iz}
\bibinfo{author}{S.~Dasgupta}, \bibinfo{author}{T.~Auth},
  \bibinfo{author}{G.~Gompper},
\newblock \bibinfo{title}{Wrapping of ellipsoidal nano-particles by fluid
  membranes},
\newblock \bibinfo{journal}{Soft Matter} \bibinfo{volume}{9}
  (\bibinfo{year}{2013}) \bibinfo{pages}{5473}.
\bibitem[{Dasgupta et~al.(2014)Dasgupta, Auth, and Gompper}]{Dasgupta:2014hr}
\bibinfo{author}{S.~Dasgupta}, \bibinfo{author}{T.~Auth},
  \bibinfo{author}{G.~Gompper},
\newblock \bibinfo{title}{Shape and orientation matter for the cellular uptake
  of nonspherical particles},
\newblock \bibinfo{journal}{Nano Lett.} \bibinfo{volume}{14}
  (\bibinfo{year}{2014}) \bibinfo{pages}{687--693}.
\bibitem[{Lin and Brown(2004)}]{Lin:2004eg}
\bibinfo{author}{L.~C.-L. Lin}, \bibinfo{author}{F.~Brown},
\newblock \bibinfo{title}{Brownian dynamics in fourier space: Membrane
  simulations over long length and time scales},
\newblock \bibinfo{journal}{Phys. Rev. Lett.} \bibinfo{volume}{93}
  (\bibinfo{year}{2004}) \bibinfo{pages}{256001}.
\bibitem[{Lin and Brown(2005)}]{Lin:2005iv}
\bibinfo{author}{L.~C.-L. Lin}, \bibinfo{author}{F.~Brown},
\newblock \bibinfo{title}{Dynamic simulations of membranes with cytoskeletal
  interactions},
\newblock \bibinfo{journal}{Phys. Rev. E} \bibinfo{volume}{72}
  (\bibinfo{year}{2005}) \bibinfo{pages}{011910}.
\bibitem[{Lin and Brown(2006)}]{Lin:2006ic}
\bibinfo{author}{L.~C.-L. Lin}, \bibinfo{author}{F.~L.~H. Brown},
\newblock \bibinfo{title}{Simulating membrane dynamics in nonhomogeneous
  hydrodynamic environments},
\newblock \bibinfo{journal}{J. Chem. Theory Comput.} \bibinfo{volume}{2}
  (\bibinfo{year}{2006}) \bibinfo{pages}{472--483}.
\bibitem[{Brown(2007)}]{Brown:2007ip}
\bibinfo{author}{F.~L.~H. Brown},
\newblock \bibinfo{title}{Simple models for biomembrane structure and
  dynamics},
\newblock \bibinfo{journal}{Computer Physics Communications}
  \bibinfo{volume}{177} (\bibinfo{year}{2007}) \bibinfo{pages}{172--175}.
\bibitem[{Sigurdsson et~al.(2013)Sigurdsson, Brown, and
  Atzberger}]{Sigurdsson:2013cx}
\bibinfo{author}{J.~K. Sigurdsson}, \bibinfo{author}{F.~L.~H. Brown},
  \bibinfo{author}{P.~J. Atzberger},
\newblock \bibinfo{title}{Hybrid continuum-particle method for fluctuating
  lipid bilayer membranes with diffusing protein inclusions},
\newblock \bibinfo{journal}{J. Comp. Phys.} \bibinfo{volume}{252}
  (\bibinfo{year}{2013}) \bibinfo{pages}{65--85}.
\bibitem[{Polyakov(1981)}]{Polyakov:1981p207}
\bibinfo{author}{A.~M. Polyakov},
\newblock \bibinfo{title}{Quantum geometry of fermionic strings},
\newblock \bibinfo{journal}{Phys. Lett. B} \bibinfo{volume}{103}
  (\bibinfo{year}{1981}) \bibinfo{pages}{207 -- 210}.
\bibitem[{David(1985)}]{David:1985p303}
\bibinfo{author}{F.~David},
\newblock \bibinfo{title}{Randomly triangulated surfaces in - 2 dimensions},
\newblock \bibinfo{journal}{Phys. Lett. B} \bibinfo{volume}{159}
  (\bibinfo{year}{1985}) \bibinfo{pages}{303 -- 306}.
\bibitem[{Kazakov et~al.(1985)Kazakov, Kostov, and Migdahl}]{Kazakov:1985p295}
\bibinfo{author}{V.~A. Kazakov}, \bibinfo{author}{I.~K. Kostov},
  \bibinfo{author}{A.~A. Migdahl},
\newblock \bibinfo{title}{Critical properties of randomly triangulated planar
  random surfaces},
\newblock \bibinfo{journal}{Phys. Lett. B} \bibinfo{volume}{157}
  (\bibinfo{year}{1985}) \bibinfo{pages}{295--300}.
\bibitem[{Maritan and Stella(1987)}]{Stella:1987p561}
\bibinfo{author}{A.~Maritan}, \bibinfo{author}{A.~L. Stella},
\newblock \bibinfo{title}{Some exact results from self-avoiding random
  surfaces},
\newblock \bibinfo{journal}{Nucl. Phys. B} \bibinfo{volume}{280}
  (\bibinfo{year}{1987}) \bibinfo{pages}{561 -- 575}.
\bibitem[{Kantor et~al.(1986)Kantor, Kardar, and Nelson}]{Kantor:1986p3356}
\bibinfo{author}{Y.~Kantor}, \bibinfo{author}{M.~Kardar},
  \bibinfo{author}{D.~R. Nelson},
\newblock \bibinfo{title}{Statistical mechanics of tethered surfaces},
\newblock \bibinfo{journal}{Phys. Rev. Lett.} \bibinfo{volume}{57}
  (\bibinfo{year}{1986}) \bibinfo{pages}{791}.
\bibitem[{Kantor and Nelson(1987)}]{Kantor:1987p3357}
\bibinfo{author}{Y.~Kantor}, \bibinfo{author}{D.~R. Nelson},
\newblock \bibinfo{title}{Phase transitions in flexible polymeric surfaces},
\newblock \bibinfo{journal}{Phys. Rev. A} \bibinfo{volume}{36}
  (\bibinfo{year}{1987}) \bibinfo{pages}{4020}.
\bibitem[{Ho and Baumg{\"a}rtner(1990{\natexlab{a}})}]{Ho:1990p295}
\bibinfo{author}{J.-S. Ho}, \bibinfo{author}{A.~Baumg{\"a}rtner},
\newblock \bibinfo{title}{Simulations of fluid self-avoiding membranes},
\newblock \bibinfo{journal}{Europhys. Lett.} \bibinfo{volume}{12}
  (\bibinfo{year}{1990}{\natexlab{a}}) \bibinfo{pages}{295}.
\bibitem[{Ho and Baumg{\"a}rtner(1990{\natexlab{b}})}]{Ho:1990p5747}
\bibinfo{author}{J.-S. Ho}, \bibinfo{author}{A.~Baumg{\"a}rtner},
\newblock \bibinfo{title}{Crumpling of fluid vesicles},
\newblock \bibinfo{journal}{Phys. Rev. A} \bibinfo{volume}{41}
  (\bibinfo{year}{1990}{\natexlab{b}}) \bibinfo{pages}{5747 -- 5750}.
\bibitem[{Ramakrishnan et~al.(2010)Ramakrishnan, Sunil~Kumar, and
  Ipsen}]{Ramakrishnan:2010hk}
\bibinfo{author}{N.~Ramakrishnan}, \bibinfo{author}{P.~B. Sunil~Kumar},
  \bibinfo{author}{J.~H. Ipsen},
\newblock \bibinfo{title}{Monte carlo simulations of fluid vesicles with
  in-plane orientational ordering},
\newblock \bibinfo{journal}{Phys. Rev. E} \bibinfo{volume}{81}
  (\bibinfo{year}{2010}) \bibinfo{pages}{041922}.
\bibitem[{Frenkel and Smit(2001)}]{Frenkel:2001}
\bibinfo{author}{D.~Frenkel}, \bibinfo{author}{B.~Smit},
  \bibinfo{title}{Understanding Molecular Simulation : From Algorithms to
  Applications}, \bibinfo{publisher}{Academic Press}, \bibinfo{edition}{2}
  edition, \bibinfo{year}{2001}.
\bibitem[{Metropolis et~al.(1953)Metropolis, Rosenbluth, Rosenbluth, Teller,
  and Teller}]{metropolis:1953p1087}
\bibinfo{author}{N.~Metropolis}, \bibinfo{author}{A.~W. Rosenbluth},
  \bibinfo{author}{M.~N. Rosenbluth}, \bibinfo{author}{A.~H. Teller},
  \bibinfo{author}{E.~Teller},
\newblock \bibinfo{title}{Equation of state calculations by fast computing
  machines},
\newblock \bibinfo{journal}{J. Chem. Phys.} \bibinfo{volume}{21}
  (\bibinfo{year}{1953}) \bibinfo{pages}{1087--1092}.
\bibitem[{Kroll and Gompper(1992)}]{Kroll:1992jb}
\bibinfo{author}{D.~Kroll}, \bibinfo{author}{G.~Gompper},
\newblock \bibinfo{title}{Scaling behavior of randomly triangulated
  self-avoiding surfaces},
\newblock \bibinfo{journal}{Phys. Rev. A} \bibinfo{volume}{46}
  (\bibinfo{year}{1992}) \bibinfo{pages}{3119--3122}.
\bibitem[{Gompper and Kroll(1995)}]{Gompper:1995fga}
\bibinfo{author}{G.~Gompper}, \bibinfo{author}{D.~Kroll},
\newblock \bibinfo{title}{Phase diagram and scaling behavior of fluid
  vesicles},
\newblock \bibinfo{journal}{Phys. Rev. E} \bibinfo{volume}{51}
  (\bibinfo{year}{1995}) \bibinfo{pages}{514--525}.
\bibitem[{Paulose et~al.(2012)Paulose, Vliegenthart, Gompper, and
  Nelson}]{Paulose:2012dl}
\bibinfo{author}{J.~Paulose}, \bibinfo{author}{G.~A. Vliegenthart},
  \bibinfo{author}{G.~Gompper}, \bibinfo{author}{D.~R. Nelson},
\newblock \bibinfo{title}{Fluctuating shells under pressure},
\newblock \bibinfo{journal}{Proc. Natl. Acad. Sci. USA.} \bibinfo{volume}{109}
  (\bibinfo{year}{2012}) \bibinfo{pages}{19551--19556}.
\bibitem[{Drouffe et~al.(1991)Drouffe, Maggs, and Leibler}]{Drouffe:1991}
\bibinfo{author}{J.~M. Drouffe}, \bibinfo{author}{A.~C. Maggs},
  \bibinfo{author}{S.~Leibler},
\newblock \bibinfo{title}{Computer simulations of self-assembled membranes},
\newblock \bibinfo{journal}{Science} \bibinfo{volume}{254}
  (\bibinfo{year}{1991}) \bibinfo{pages}{1353--1356}.
\bibitem[{Ayton and Voth(2002)}]{Ayton:2002p3357}
\bibinfo{author}{G.~Ayton}, \bibinfo{author}{G.~Voth},
\newblock \bibinfo{title}{Bridging microscopic and mesoscopic simulations of
  lipid bilayers},
\newblock \bibinfo{journal}{Biophys. J.} \bibinfo{volume}{83}
  (\bibinfo{year}{2002}) \bibinfo{pages}{3357--3370}.
\bibitem[{Noguchi and Takasu(2001)}]{Noguchi:2001bn}
\bibinfo{author}{H.~Noguchi}, \bibinfo{author}{M.~Takasu},
\newblock \bibinfo{title}{Self-assembly of amphiphiles into vesicles: A
  brownian dynamics simulation},
\newblock \bibinfo{journal}{Phys. Rev. E} \bibinfo{volume}{64}
  (\bibinfo{year}{2001}) \bibinfo{pages}{041913}.
\bibitem[{Noguchi and Takasu(2002)}]{Noguchi:2002iv}
\bibinfo{author}{H.~Noguchi}, \bibinfo{author}{M.~Takasu},
\newblock \bibinfo{title}{Adhesion of nanoparticles to vesicles: a brownian
  dynamics simulation},
\newblock \bibinfo{journal}{Biophys. J.} \bibinfo{volume}{83}
  (\bibinfo{year}{2002}) \bibinfo{pages}{299--308}.
\bibitem[{Farago(2003)}]{Farago:2003jf}
\bibinfo{author}{O.~Farago},
\newblock \bibinfo{title}{``water-free'' computer model for fluid bilayer
  membranes},
\newblock \bibinfo{journal}{J. Chem. Phys.} \bibinfo{volume}{119}
  (\bibinfo{year}{2003}) \bibinfo{pages}{596}.
\bibitem[{Cooke and Deserno(2005)}]{Cooke:2005p4710}
\bibinfo{author}{I.~R. Cooke}, \bibinfo{author}{M.~Deserno},
\newblock \bibinfo{title}{Solvent-free model for self-assembling fluid bilayer
  membranes: Stabilization of the fluid phase based on broad attractive tail
  potentials},
\newblock \bibinfo{journal}{J. Chem. Phys.} \bibinfo{volume}{123}
  (\bibinfo{year}{2005}) \bibinfo{pages}{4710}.
\bibitem[{Brannigan and Brown(2004)}]{Brannigan:2004fr}
\bibinfo{author}{G.~Brannigan}, \bibinfo{author}{F.~L.~H. Brown},
\newblock \bibinfo{title}{Solvent-free simulations of fluid membrane bilayers},
\newblock \bibinfo{journal}{J. Chem. Phys.} \bibinfo{volume}{120}
  (\bibinfo{year}{2004}) \bibinfo{pages}{1059}.
\bibitem[{Ayton et~al.(2006)Ayton, McWhirter, and Voth}]{Ayton:2006ht}
\bibinfo{author}{G.~S. Ayton}, \bibinfo{author}{J.~L. McWhirter},
  \bibinfo{author}{G.~A. Voth},
\newblock \bibinfo{title}{A second generation mesoscopic lipid bilayer model:
  Connections to field-theory descriptions of membranes and nonlocal
  hydrodynamics},
\newblock \bibinfo{journal}{J. Chem. Phys.} \bibinfo{volume}{124}
  (\bibinfo{year}{2006}) \bibinfo{pages}{064906}.
\bibitem[{Noguchi and Gompper(2006)}]{Noguchi:2006pre}
\bibinfo{author}{H.~Noguchi}, \bibinfo{author}{G.~Gompper},
\newblock \bibinfo{title}{Meshless membrane model based on the moving
  least-squares method},
\newblock \bibinfo{journal}{Phys. Rev. E} \bibinfo{volume}{73}
  (\bibinfo{year}{2006}) \bibinfo{pages}{21903}.
\bibitem[{Kohyama(2009)}]{Kohyama:2009p3334}
\bibinfo{author}{T.~Kohyama},
\newblock \bibinfo{title}{Simulations of flexible membranes using a
  coarse-grained particle-based model with spontaneous curvature variables},
\newblock \bibinfo{journal}{Physica A} \bibinfo{volume}{388}
  (\bibinfo{year}{2009}) \bibinfo{pages}{3334--3344}.
\bibitem[{Yuan et~al.(2010)Yuan, Huang, Li, Lykotrafitis, and
  Zhang}]{Yuan:2010ww}
\bibinfo{author}{H.~Yuan}, \bibinfo{author}{C.~Huang}, \bibinfo{author}{J.~Li},
  \bibinfo{author}{G.~Lykotrafitis}, \bibinfo{author}{S.~Zhang},
\newblock \bibinfo{title}{One-particle-thick, solvent-free, coarse-grained
  model for biological and biomimetic fluid membranes},
\newblock \bibinfo{journal}{Phys. Rev. E} \bibinfo{volume}{82}
  (\bibinfo{year}{2010}) \bibinfo{pages}{011905}.
\bibitem[{Ayton et~al.(2007)Ayton, Blood, and Voth}]{Ayton:2007p3485}
\bibinfo{author}{G.~S. Ayton}, \bibinfo{author}{P.~D. Blood},
  \bibinfo{author}{G.~A. Voth},
\newblock \bibinfo{title}{Membrane remodeling from n-bar domain interactions:
  Insights from multi-scale simulation},
\newblock \bibinfo{journal}{Biophys. J.} \bibinfo{volume}{92}
  (\bibinfo{year}{2007}) \bibinfo{pages}{3595--3602}.
\bibitem[{Ayton et~al.(2009)Ayton, Lyman, Krishna, Swenson, Mim, Unger, and
  Voth}]{Ayton:2009gw}
\bibinfo{author}{G.~S. Ayton}, \bibinfo{author}{E.~Lyman},
  \bibinfo{author}{V.~Krishna}, \bibinfo{author}{R.~D. Swenson},
  \bibinfo{author}{C.~Mim}, \bibinfo{author}{V.~M. Unger},
  \bibinfo{author}{G.~A. Voth},
\newblock \bibinfo{title}{New insights into bar domain-induced membrane
  remodeling},
\newblock \bibinfo{journal}{Biophys. J.} \bibinfo{volume}{97}
  (\bibinfo{year}{2009}) \bibinfo{pages}{1616--1625}.
\bibitem[{Cui et~al.(2009)Cui, Ayton, and Voth}]{Cui:2009p2746}
\bibinfo{author}{H.~Cui}, \bibinfo{author}{G.~S. Ayton}, \bibinfo{author}{G.~A.
  Voth},
\newblock \bibinfo{title}{Membrane binding by the endophilin n-bar domain},
\newblock \bibinfo{journal}{Biophys. J.} \bibinfo{volume}{97}
  (\bibinfo{year}{2009}) \bibinfo{pages}{2746--2753}.
\bibitem[{Cui et~al.(2011)Cui, Lyman, and Voth}]{Cui:2011p1271}
\bibinfo{author}{H.~Cui}, \bibinfo{author}{E.~Lyman}, \bibinfo{author}{G.~A.
  Voth},
\newblock \bibinfo{title}{Mechanism of membrane curvature sensing by
  amphipathic helix containing proteins},
\newblock \bibinfo{journal}{Biophys. J.} \bibinfo{volume}{100}
  (\bibinfo{year}{2011}) \bibinfo{pages}{1271--1279}.
\bibitem[{Lyman et~al.(2011)Lyman, Cui, and Voth}]{Lyman:2011p10430}
\bibinfo{author}{E.~Lyman}, \bibinfo{author}{H.~Cui}, \bibinfo{author}{G.~A.
  Voth},
\newblock \bibinfo{title}{Reconstructing protein remodeled membranes in
  molecular detail from mesoscopic models},
\newblock \bibinfo{journal}{Phys. Chem. Chem. Phys.} \bibinfo{volume}{13}
  (\bibinfo{year}{2011}) \bibinfo{pages}{10430--10436}.
\bibitem[{Ayton et~al.(2004)Ayton, Tepper, Mirijanian, and Voth}]{Ayton:2004jr}
\bibinfo{author}{G.~S. Ayton}, \bibinfo{author}{H.~L. Tepper},
  \bibinfo{author}{D.~T. Mirijanian}, \bibinfo{author}{G.~A. Voth},
\newblock \bibinfo{title}{A new perspective on the coarse-grained dynamics of
  fluids},
\newblock \bibinfo{journal}{J. Chem. Phys.} \bibinfo{volume}{120}
  (\bibinfo{year}{2004}) \bibinfo{pages}{4074}.
\bibitem[{Tobias et~al.(1997)Tobias, Tu, and Klein}]{Tobias:1997gw}
\bibinfo{author}{D.~J. Tobias}, \bibinfo{author}{K.~Tu}, \bibinfo{author}{M.~L.
  Klein},
\newblock \bibinfo{title}{Atomic-scale molecular dynamics simulations of lipid
  membranes},
\newblock \bibinfo{journal}{Current Opinion in Colloid {\&} Interface Science}
  \bibinfo{volume}{2} (\bibinfo{year}{1997}) \bibinfo{pages}{15--26}.
\bibitem[{Tieleman et~al.(1997)Tieleman, Marrink, and
  Berendsen}]{Tieleman:1997ve}
\bibinfo{author}{D.~P. Tieleman}, \bibinfo{author}{S.-J. Marrink},
  \bibinfo{author}{H.~J. Berendsen},
\newblock \bibinfo{title}{A computer perspective of membranes: molecular
  dynamics studies of lipid bilayer systems},
\newblock \bibinfo{journal}{Biochimica Biophysics Acta (BBA) - Reviews on
  Biomembranes} \bibinfo{volume}{1331} (\bibinfo{year}{1997})
  \bibinfo{pages}{235--270}.
\bibitem[{Feller(2000)}]{Feller:2000bi}
\bibinfo{author}{S.~E. Feller},
\newblock \bibinfo{title}{Molecular dynamics simulations of lipid bilayers},
\newblock \bibinfo{journal}{Curr Opin Colloid in} \bibinfo{volume}{5}
  (\bibinfo{year}{2000}) \bibinfo{pages}{217--223}.
\bibitem[{Marrink et~al.(2009)Marrink, de~Vries, and Tieleman}]{Marrink:2009uo}
\bibinfo{author}{S.-J. Marrink}, \bibinfo{author}{A.~H. de~Vries},
  \bibinfo{author}{D.~P. Tieleman},
\newblock \bibinfo{title}{Lipids on the move: simulations of membrane pores,
  domains, stalks and curves},
\newblock \bibinfo{journal}{Biochimica Biophysics Acta (BBA) - Biomembranes}
  \bibinfo{volume}{1788} (\bibinfo{year}{2009}) \bibinfo{pages}{149--168}.
\bibitem[{van~der Ploeg(1982)}]{vanderPloeg:1982by}
\bibinfo{author}{P.~van~der Ploeg},
\newblock \bibinfo{title}{Molecular dynamics simulation of a bilayer membrane},
\newblock \bibinfo{journal}{J. Chem. Phys.} \bibinfo{volume}{76}
  (\bibinfo{year}{1982}) \bibinfo{pages}{3271--3276}.
\bibitem[{Heller et~al.(1993)Heller, Schaefer, and Schulten}]{Heller:1993wk}
\bibinfo{author}{H.~Heller}, \bibinfo{author}{M.~Schaefer},
  \bibinfo{author}{K.~Schulten},
\newblock \bibinfo{title}{Molecular dynamics simulation of a bilayer of 200
  lipids in the gel and in the liquid crystal phase},
\newblock \bibinfo{journal}{J. Phys. Chem.} \bibinfo{volume}{97}
  (\bibinfo{year}{1993}) \bibinfo{pages}{8343--8360}.
\bibitem[{Tu et~al.(1996)Tu, Tobias, Blasie, and Klein}]{Tu:1996wi}
\bibinfo{author}{K.~Tu}, \bibinfo{author}{D.~J. Tobias}, \bibinfo{author}{J.~K.
  Blasie}, \bibinfo{author}{M.~L. Klein},
\newblock \bibinfo{title}{Molecular dynamics investigation of the structure of
  a fully hydrated gel-phase dipalmitoylphosphatidylcholine bilayer},
\newblock \bibinfo{journal}{Biophys. J.}  (\bibinfo{year}{1996}).
\bibitem[{Tieleman and Berendsen(1996)}]{Tieleman:1996dz}
\bibinfo{author}{D.~P. Tieleman}, \bibinfo{author}{H.~J.~C. Berendsen},
\newblock \bibinfo{title}{Molecular dynamics simulations of a fully hydrated
  dipalmitoylphosphatidylcholine bilayer with different macroscopic boundary
  conditions and parameters},
\newblock \bibinfo{journal}{J. Chem. Phys.} \bibinfo{volume}{105}
  (\bibinfo{year}{1996}) \bibinfo{pages}{4871}.
\bibitem[{Berger et~al.(1997)Berger, Edholm, and J{\"a}hnig}]{Berger:1997ut}
\bibinfo{author}{O.~Berger}, \bibinfo{author}{O.~Edholm},
  \bibinfo{author}{F.~J{\"a}hnig},
\newblock \bibinfo{title}{Molecular dynamics simulations of a fluid bilayer of
  dipalmitoylphosphatidylcholine at full hydration, constant pressure, and
  constant temperature},
\newblock \bibinfo{journal}{Biophys. J.}  (\bibinfo{year}{1997}).
\bibitem[{Marrink et~al.(1998)Marrink, Berger, Tieleman, and
  J{\"a}hnig}]{Marrink:1998fi}
\bibinfo{author}{S.-J. Marrink}, \bibinfo{author}{O.~Berger},
  \bibinfo{author}{P.~Tieleman}, \bibinfo{author}{F.~J{\"a}hnig},
\newblock \bibinfo{title}{Adhesion forces of lipids in a phospholipid membrane
  studied by molecular dynamics simulations},
\newblock \bibinfo{journal}{Biophys. J.} \bibinfo{volume}{74}
  (\bibinfo{year}{1998}) \bibinfo{pages}{931--943}.
\bibitem[{Pasenkiewicz-Gierula et~al.(2000)Pasenkiewicz-Gierula, R{\'o}g,
  Kitamura, and Kusumi}]{PasenkiewiczGierula:2000jc}
\bibinfo{author}{M.~Pasenkiewicz-Gierula}, \bibinfo{author}{T.~R{\'o}g},
  \bibinfo{author}{K.~Kitamura}, \bibinfo{author}{A.~Kusumi},
\newblock \bibinfo{title}{Cholesterol effects on the phosphatidylcholine
  bilayer polar region: A molecular simulation study},
\newblock \bibinfo{journal}{Biophys. J.} \bibinfo{volume}{78}
  (\bibinfo{year}{2000}) \bibinfo{pages}{1376--1389}.
\bibitem[{Mashl et~al.(2001)Mashl, Scott, Subramaniam, and
  Jakobsson}]{Mashl:2001jm}
\bibinfo{author}{R.~J. Mashl}, \bibinfo{author}{H.~L. Scott},
  \bibinfo{author}{S.~Subramaniam}, \bibinfo{author}{E.~Jakobsson},
\newblock \bibinfo{title}{Molecular simulation of dioleoylphosphatidylcholine
  lipid bilayers at differing levels of hydration},
\newblock \bibinfo{journal}{Biophys. J.} \bibinfo{volume}{81}
  (\bibinfo{year}{2001}) \bibinfo{pages}{3005--3015}.
\bibitem[{Murzyn et~al.(2001)Murzyn, Rog, Jezierski, Takaoka, and
  Pasenkiewicz-Gierula}]{KMurzyn:2001uj}
\bibinfo{author}{K.~Murzyn}, \bibinfo{author}{T.~Rog},
  \bibinfo{author}{G.~Jezierski}, \bibinfo{author}{Y.~Takaoka},
  \bibinfo{author}{M.~Pasenkiewicz-Gierula},
\newblock \bibinfo{title}{Effects of phospholipid unsaturation on the
  membrane/water interface: a molecular simulation study},
\newblock \bibinfo{journal}{Biophys. J.} \bibinfo{volume}{81}
  (\bibinfo{year}{2001}) \bibinfo{pages}{170}.
\bibitem[{Marrink and Mark(2003)}]{Marrink:2003iu}
\bibinfo{author}{S.-J. Marrink}, \bibinfo{author}{A.~E. Mark},
\newblock \bibinfo{title}{Molecular dynamics simulation of the formation,
  structure, and dynamics of small phospholipid vesicles},
\newblock \bibinfo{journal}{J. Amer. Chem. Soc.} \bibinfo{volume}{125}
  (\bibinfo{year}{2003}) \bibinfo{pages}{15233--15242}.
\bibitem[{Hofs{\"a}{\ss} et~al.(2003)Hofs{\"a}{\ss}, Lindahl, and
  Edholm}]{ChristoferHofsass:2003ub}
\bibinfo{author}{C.~Hofs{\"a}{\ss}}, \bibinfo{author}{E.~Lindahl},
  \bibinfo{author}{O.~Edholm},
\newblock \bibinfo{title}{Molecular dynamics simulations of phospholipid
  bilayers with cholesterol},
\newblock \bibinfo{journal}{Biophys. J.} \bibinfo{volume}{84}
  (\bibinfo{year}{2003}) \bibinfo{pages}{2192}.
\bibitem[{de~Vries et~al.(2004)de~Vries, Mark, and Marrink}]{deVries:2004vc}
\bibinfo{author}{A.~H. de~Vries}, \bibinfo{author}{A.~E. Mark},
  \bibinfo{author}{S.-J. Marrink},
\newblock \bibinfo{title}{Molecular dynamics simulation of the spontaneous
  formation of a small dppc vesicle in water in atomistic detail},
\newblock \bibinfo{journal}{J. Amer. Chem. Soc.}  (\bibinfo{year}{2004}).
\bibitem[{Leontiadou et~al.(2004)Leontiadou, Mark, and
  Marrink}]{Leontiadou:2004ba}
\bibinfo{author}{H.~Leontiadou}, \bibinfo{author}{A.~E. Mark},
  \bibinfo{author}{S.-J. Marrink},
\newblock \bibinfo{title}{Molecular dynamics simulations of hydrophilic pores
  in lipid bilayers},
\newblock \bibinfo{journal}{Biophys. J.} \bibinfo{volume}{86}
  (\bibinfo{year}{2004}) \bibinfo{pages}{2156--2164}.
\bibitem[{Marrink and Mark(2003)}]{Marrink:2003wn}
\bibinfo{author}{S.-J. Marrink}, \bibinfo{author}{A.~E. Mark},
\newblock \bibinfo{title}{The mechanism of vesicle fusion as revealed by
  molecular dynamics simulations},
\newblock \bibinfo{journal}{J. Amer. Chem. Soc.}  (\bibinfo{year}{2003}).
\bibitem[{Sum et~al.(2003)Sum, Faller, and de~Pablo}]{Sum:2003bn}
\bibinfo{author}{A.~K. Sum}, \bibinfo{author}{R.~Faller},
  \bibinfo{author}{J.~J. de~Pablo},
\newblock \bibinfo{title}{Molecular simulation study of phospholipid bilayers
  and insights of the interactions with disaccharides},
\newblock \bibinfo{journal}{Biophys. J.} \bibinfo{volume}{85}
  (\bibinfo{year}{2003}) \bibinfo{pages}{2830--2844}.
\bibitem[{Edholm et~al.(1995)Edholm, Berger, and J{\"a}hnig}]{Edholm:1995ty}
\bibinfo{author}{O.~Edholm}, \bibinfo{author}{O.~Berger},
  \bibinfo{author}{F.~J{\"a}hnig},
\newblock \bibinfo{title}{Structure and fluctuations of bacteriorhodopsin in
  the purple membrane: a molecular dynamics study},
\newblock \bibinfo{journal}{J. Mol. Biol.}  (\bibinfo{year}{1995}).
\bibitem[{Pitman et~al.(2005)Pitman, Grossfield, Suits, and
  Feller}]{Pitman:2005gf}
\bibinfo{author}{M.~C. Pitman}, \bibinfo{author}{A.~Grossfield},
  \bibinfo{author}{F.~Suits}, \bibinfo{author}{S.~E. Feller},
\newblock \bibinfo{title}{Role of cholesterol and polyunsaturated chains in
  lipid−protein interactions: Molecular dynamics simulation of rhodopsin in a
  realistic membrane environment},
\newblock \bibinfo{journal}{J. Amer. Chem. Soc.} \bibinfo{volume}{127}
  (\bibinfo{year}{2005}) \bibinfo{pages}{4576--4577}.
\bibitem[{Sansom et~al.(2005)Sansom, Bond, Deol, Grottesi, Haider, and
  Sands}]{Sansom:2005en}
\bibinfo{author}{M.~S.~P. Sansom}, \bibinfo{author}{P.~J. Bond},
  \bibinfo{author}{S.~S. Deol}, \bibinfo{author}{A.~Grottesi},
  \bibinfo{author}{S.~Haider}, \bibinfo{author}{Z.~A. Sands},
\newblock \bibinfo{title}{Molecular simulations and lipid--protein
  interactions: potassium channels and other membrane proteins},
\newblock \bibinfo{journal}{Biochem Soc Trans} \bibinfo{volume}{33}
  (\bibinfo{year}{2005}) \bibinfo{pages}{916}.
\bibitem[{Lindahl and Sansom(2008)}]{Lindahl:2008fc}
\bibinfo{author}{E.~Lindahl}, \bibinfo{author}{M.~Sansom},
\newblock \bibinfo{title}{Membrane proteins: molecular dynamics simulations},
\newblock \bibinfo{journal}{Current Opinion in Structural Biology}
  \bibinfo{volume}{18} (\bibinfo{year}{2008}) \bibinfo{pages}{425--431}.
\bibitem[{de~Meyer et~al.(2008)de~Meyer, Venturoli, and Smit}]{Meyer:2008p1851}
\bibinfo{author}{F.~J.-M. de~Meyer}, \bibinfo{author}{M.~Venturoli},
  \bibinfo{author}{B.~Smit},
\newblock \bibinfo{title}{Molecular simulations of lipid-mediated
  protein-protein interactions},
\newblock \bibinfo{journal}{Biophys. J.} \bibinfo{volume}{95}
  (\bibinfo{year}{2008}) \bibinfo{pages}{1851--1865}.
\bibitem[{Lag{\"u}e et~al.(2005)Lag{\"u}e, Roux, and Pastor}]{Lague:2005ti}
\bibinfo{author}{P.~Lag{\"u}e}, \bibinfo{author}{B.~Roux},
  \bibinfo{author}{R.~W. Pastor},
\newblock \bibinfo{title}{Molecular dynamics simulations of the influenza
  hemagglutinin fusion peptide in micelles and bilayers: conformational
  analysis of peptide and lipids},
\newblock \bibinfo{journal}{J. Mol. Biol.}  (\bibinfo{year}{2005}).
\bibitem[{Doi and Edwards(1988)}]{Doi:1988ug}
\bibinfo{author}{M.~Doi}, \bibinfo{author}{S.~F. Edwards}, \bibinfo{title}{The
  theory of polymer dynamics}, \bibinfo{publisher}{Oxford University Press},
  \bibinfo{address}{Oxford, UK}, \bibinfo{year}{1988}.
\bibitem[{Goetz and Lipowsky(1998)}]{Goetz:1998p7397}
\bibinfo{author}{R.~Goetz}, \bibinfo{author}{R.~Lipowsky},
\newblock \bibinfo{title}{Computer simulations of bilayer membranes:
  Self-assembly and interfacial tension},
\newblock \bibinfo{journal}{J. Chem. Phys.} \bibinfo{volume}{108}
  (\bibinfo{year}{1998}) \bibinfo{pages}{7397--7409}.
\bibitem[{Stevens(2004)}]{Steven:2004p11942}
\bibinfo{author}{M.~J. Stevens},
\newblock \bibinfo{title}{Coarse-grained simulations of lipid bilayers},
\newblock \bibinfo{journal}{J. Chem. Phys.} \bibinfo{volume}{121}
  (\bibinfo{year}{2004}) \bibinfo{pages}{11942--11948}.
\bibitem[{Lyubartsev(2005)}]{Lyubartsev:2005de}
\bibinfo{author}{A.~P. Lyubartsev},
\newblock \bibinfo{title}{Multiscale modeling of lipids and lipid bilayers},
\newblock \bibinfo{journal}{Eur. Biophys. J.} \bibinfo{volume}{35}
  (\bibinfo{year}{2005}) \bibinfo{pages}{53--61}.
\bibitem[{Praprotnik et~al.(2008)Praprotnik, Delle~Site, and
  Kremer}]{Praprotnik:2008p646}
\bibinfo{author}{M.~Praprotnik}, \bibinfo{author}{L.~Delle~Site},
  \bibinfo{author}{K.~Kremer},
\newblock \bibinfo{title}{Multiscale simulation of soft matter: From scale
  bridging to adaptive resolution},
\newblock \bibinfo{journal}{Annu. Rev. Phys. Chem.} \bibinfo{volume}{59}
  (\bibinfo{year}{2008}) \bibinfo{pages}{545--571}.
\bibitem[{Peter and Kremer(2009{\natexlab{a}})}]{Peter:2009tx}
\bibinfo{author}{C.~Peter}, \bibinfo{author}{K.~Kremer},
\newblock \bibinfo{title}{Multiscale simulation of soft matter systems--from
  the atomistic to the coarse-grained level and back},
\newblock \bibinfo{journal}{Soft Matter}  (\bibinfo{year}{2009}{\natexlab{a}}).
\bibitem[{Peter and Kremer(2009{\natexlab{b}})}]{Peter:2010p65}
\bibinfo{author}{C.~Peter}, \bibinfo{author}{K.~Kremer},
\newblock \bibinfo{title}{Multiscale simulation of soft matter systems},
\newblock \bibinfo{journal}{Faraday Discuss.} \bibinfo{volume}{144}
  (\bibinfo{year}{2009}{\natexlab{b}}) \bibinfo{pages}{9--24}.
\bibitem[{Murtola et~al.(2004)Murtola, Falck, Patra, Karttunen, and
  Vattulainen}]{Murtola:2004ky}
\bibinfo{author}{T.~Murtola}, \bibinfo{author}{E.~Falck},
  \bibinfo{author}{M.~Patra}, \bibinfo{author}{M.~Karttunen},
  \bibinfo{author}{I.~Vattulainen},
\newblock \bibinfo{title}{Coarse-grained model for phospholipid/cholesterol
  bilayer},
\newblock \bibinfo{journal}{J. Chem. Phys.} \bibinfo{volume}{121}
  (\bibinfo{year}{2004}) \bibinfo{pages}{9156}.
\bibitem[{Murtola et~al.(2007)Murtola, Falck, Karttunen, and
  Vattulainen}]{Murtola:2007ja}
\bibinfo{author}{T.~Murtola}, \bibinfo{author}{E.~Falck},
  \bibinfo{author}{M.~Karttunen}, \bibinfo{author}{I.~Vattulainen},
\newblock \bibinfo{title}{Coarse-grained model for phospholipid/cholesterol
  bilayer employing inverse monte carlo with thermodynamic constraints},
\newblock \bibinfo{journal}{J. Chem. Phys.} \bibinfo{volume}{126}
  (\bibinfo{year}{2007}) \bibinfo{pages}{075101}.
\bibitem[{Shih et~al.(2006)Shih, Arkhipov, Freddolino, and
  Schulten}]{Shih:2006bi}
\bibinfo{author}{A.~Y. Shih}, \bibinfo{author}{A.~Arkhipov},
  \bibinfo{author}{P.~L. Freddolino}, \bibinfo{author}{K.~Schulten},
\newblock \bibinfo{title}{Coarse grained protein−lipid model with application
  to lipoprotein particles †},
\newblock \bibinfo{journal}{J. Phys. Chem. B} \bibinfo{volume}{110}
  (\bibinfo{year}{2006}) \bibinfo{pages}{3674--3684}.
\bibitem[{Arkhipov et~al.(2008)Arkhipov, Yin, and
  Schulten}]{Arkhipov:2008p3007}
\bibinfo{author}{A.~Arkhipov}, \bibinfo{author}{Y.~Yin},
  \bibinfo{author}{K.~Schulten},
\newblock \bibinfo{title}{Four-scale description of membrane sculpting by bar
  domains},
\newblock \bibinfo{journal}{Biophys. J.} \bibinfo{volume}{95}
  (\bibinfo{year}{2008}) \bibinfo{pages}{2806--2821}.
\bibitem[{Yin et~al.(2009)Yin, Arkhipov, and Schulten}]{Yin:2009p255}
\bibinfo{author}{Y.~Yin}, \bibinfo{author}{A.~Arkhipov},
  \bibinfo{author}{K.~Schulten},
\newblock \bibinfo{title}{Simulations of membrane tubulation by lattices of
  amphiphysin n-bar domains},
\newblock \bibinfo{journal}{Structure} \bibinfo{volume}{17}
  (\bibinfo{year}{2009}) \bibinfo{pages}{882--892}.
\bibitem[{Izvekov and Voth(2005)}]{Izvekov:2005iy}
\bibinfo{author}{S.~Izvekov}, \bibinfo{author}{G.~A. Voth},
\newblock \bibinfo{title}{Multiscale coarse graining of liquid-state systems},
\newblock \bibinfo{journal}{J. Chem. Phys.} \bibinfo{volume}{123}
  (\bibinfo{year}{2005}) \bibinfo{pages}{134105}.
\bibitem[{Noid et~al.(2008)Noid, Chu, Ayton, Krishna, Izvekov, Voth, Das, and
  Andersen}]{Noid:2008dc}
\bibinfo{author}{W.~G. Noid}, \bibinfo{author}{J.-W. Chu},
  \bibinfo{author}{G.~S. Ayton}, \bibinfo{author}{V.~Krishna},
  \bibinfo{author}{S.~Izvekov}, \bibinfo{author}{G.~A. Voth},
  \bibinfo{author}{A.~Das}, \bibinfo{author}{H.~C. Andersen},
\newblock \bibinfo{title}{The multiscale coarse-graining method. i. a rigorous
  bridge between atomistic and coarse-grained models},
\newblock \bibinfo{journal}{J. Chem. Phys.} \bibinfo{volume}{128}
  (\bibinfo{year}{2008}) \bibinfo{pages}{244114}.
\bibitem[{Marrink et~al.(2004)Marrink, de~Vries, and Mark}]{Marrink:2004p750}
\bibinfo{author}{S.-J. Marrink}, \bibinfo{author}{A.~H. de~Vries},
  \bibinfo{author}{A.~E. Mark},
\newblock \bibinfo{title}{Coarse grained model for semiquantitative lipid
  simulations},
\newblock \bibinfo{journal}{J. Phys. Chem. B} \bibinfo{volume}{108}
  (\bibinfo{year}{2004}) \bibinfo{pages}{750--760}.
\bibitem[{Marrink et~al.(2007)Marrink, Risselada, Yefimov, Tieleman, and
  de~Vries}]{Marrink:2007bw}
\bibinfo{author}{S.-J. Marrink}, \bibinfo{author}{H.~J. Risselada},
  \bibinfo{author}{S.~Yefimov}, \bibinfo{author}{D.~P. Tieleman},
  \bibinfo{author}{A.~H. de~Vries},
\newblock \bibinfo{title}{The martini force field: Coarse grained model for
  biomolecular simulations},
\newblock \bibinfo{journal}{J. Phys. Chem. B} \bibinfo{volume}{111}
  (\bibinfo{year}{2007}) \bibinfo{pages}{7812--7824}.
\bibitem[{Monticelli et~al.(2008)Monticelli, Kandasamy, Periole, Larson,
  Tieleman, and Marrink}]{Monticelli:2008ia}
\bibinfo{author}{L.~Monticelli}, \bibinfo{author}{S.~K. Kandasamy},
  \bibinfo{author}{X.~Periole}, \bibinfo{author}{R.~G. Larson},
  \bibinfo{author}{D.~P. Tieleman}, \bibinfo{author}{S.-J. Marrink},
\newblock \bibinfo{title}{The martini coarse-grained force field: Extension to
  proteins},
\newblock \bibinfo{journal}{J. Chem. Theory Comput.} \bibinfo{volume}{4}
  (\bibinfo{year}{2008}) \bibinfo{pages}{819--834}.
\bibitem[{Reynwar et~al.(2007)Reynwar, Illya, Harmandaris, Muller, Kremer, and
  Deserno}]{Reynwar07}
\bibinfo{author}{B.~J. Reynwar}, \bibinfo{author}{G.~Illya},
  \bibinfo{author}{V.~A. Harmandaris}, \bibinfo{author}{M.~M. Muller},
  \bibinfo{author}{K.~Kremer}, \bibinfo{author}{M.~Deserno},
\newblock \bibinfo{title}{Aggregation and vesiculation of membrane proteins by
  curvature-mediated interactions},
\newblock \bibinfo{journal}{Nature} \bibinfo{volume}{447}
  (\bibinfo{year}{2007}) \bibinfo{pages}{461--464}.
\bibitem[{Shillcock(2008)}]{Shillcock:2008ko}
\bibinfo{author}{J.~C. Shillcock},
\newblock \bibinfo{title}{Insight or illusion? seeing inside the cell with
  mesoscopic simulations},
\newblock \bibinfo{journal}{HFSP journal} \bibinfo{volume}{2}
  (\bibinfo{year}{2008}) \bibinfo{pages}{1--6}.
\bibitem[{Shillcock(2013)}]{Shillcock:2013fu}
\bibinfo{author}{J.~C. Shillcock},
\newblock \bibinfo{title}{Vesicles and vesicle fusion: coarse-grained
  simulations},
\newblock \bibinfo{journal}{Methods Mol. Biol.} \bibinfo{volume}{924}
  (\bibinfo{year}{2013}) \bibinfo{pages}{659--697}.
\bibitem[{V{\'a}cha et~al.(2012)V{\'a}cha, Martinez-Veracoechea, and
  Frenkel}]{Vacha:2012bd}
\bibinfo{author}{R.~V{\'a}cha}, \bibinfo{author}{F.~J. Martinez-Veracoechea},
  \bibinfo{author}{D.~Frenkel},
\newblock \bibinfo{title}{Intracellular release of endocytosed nanoparticles
  upon a change of ligand-receptor interaction},
\newblock \bibinfo{journal}{ACS Nano} \bibinfo{volume}{6}
  (\bibinfo{year}{2012}) \bibinfo{pages}{10598--10605}.
\bibitem[{Klein and Shinoda(2008)}]{klein-etal-08}
\bibinfo{author}{M.~Klein}, \bibinfo{author}{W.~Shinoda},
\newblock \bibinfo{title}{Large-scale molecular dynamics simulations of
  self-assembling systems},
\newblock \bibinfo{journal}{Science} \bibinfo{volume}{321}
  (\bibinfo{year}{2008}) \bibinfo{pages}{798--800}.
\bibitem[{Henderson(1992)}]{Henderson:1992vp}
\bibinfo{author}{D.~Henderson}, \bibinfo{title}{Fundamentals of Inhomogeneous
  Fluids}, \bibinfo{publisher}{CRC Press}, \bibinfo{year}{1992}.
\bibitem[{Schofield and Henderson(1982)}]{schofield-82}
\bibinfo{author}{P.~Schofield}, \bibinfo{author}{J.~R. Henderson},
\newblock \bibinfo{title}{Statistical mechanics of inhomogeneous fluids},
\newblock \bibinfo{journal}{Proc. R. Soc. Lon. A} \bibinfo{volume}{379}
  (\bibinfo{year}{1982}) \bibinfo{pages}{231}.
\bibitem[{Walton and Gubbins(1985)}]{Walton:1985to}
\bibinfo{author}{J.~Walton}, \bibinfo{author}{K.~E. Gubbins},
\newblock \bibinfo{title}{The pressure tensor in an inhomogeneous fluid of
  non-spherical molecules},
\newblock \bibinfo{journal}{Molecular Physics}  (\bibinfo{year}{1985}).
\bibitem[{Rossi and Testa(2009)}]{Rossi:2009wh}
\bibinfo{author}{G.~Rossi}, \bibinfo{author}{M.~Testa},
\newblock \bibinfo{title}{The stress tensor in thermodynamics and statistical
  mechanics},
\newblock \bibinfo{journal}{J. Chem. Phys.}  (\bibinfo{year}{2009}).
\bibitem[{Irving and Kirkwood(1950)}]{irving-50}
\bibinfo{author}{J.~Irving}, \bibinfo{author}{J.~Kirkwood},
\newblock \bibinfo{title}{The statistical mechanical theory of transport
  processes. the equations of hydrodynamics},
\newblock \bibinfo{journal}{J. Chem. Phys.} \bibinfo{volume}{18}
  (\bibinfo{year}{1950}) \bibinfo{pages}{817}.
\bibitem[{Varnik et~al.(2000)Varnik, Baschnagel, and Binder}]{Varnik:2000bd}
\bibinfo{author}{F.~Varnik}, \bibinfo{author}{J.~Baschnagel},
  \bibinfo{author}{K.~Binder},
\newblock \bibinfo{title}{Molecular dynamics results on the pressure tensor of
  polymer films},
\newblock \bibinfo{journal}{J. Chem. Phys.} \bibinfo{volume}{113}
  (\bibinfo{year}{2000}) \bibinfo{pages}{4444}.
\bibitem[{Venturoli et~al.(2006)Venturoli, Maddalena~Sperotto, Kranenburg, and
  Smit}]{Venturoli:2006p2740}
\bibinfo{author}{M.~Venturoli}, \bibinfo{author}{M.~Maddalena~Sperotto},
  \bibinfo{author}{M.~Kranenburg}, \bibinfo{author}{B.~Smit},
\newblock \bibinfo{title}{Mesoscopic models of biological membranes},
\newblock \bibinfo{journal}{Phys. Reports} \bibinfo{volume}{437}
  (\bibinfo{year}{2006}) \bibinfo{pages}{1--54}.
\bibitem[{Ollila et~al.(2009)Ollila, Risselada, Louhivuori, Lindahl,
  Vattulainen, and Marrink}]{Ollila-2009}
\bibinfo{author}{O.~H.~S. Ollila}, \bibinfo{author}{H.~J. Risselada},
  \bibinfo{author}{M.~Louhivuori}, \bibinfo{author}{E.~Lindahl},
  \bibinfo{author}{I.~Vattulainen}, \bibinfo{author}{S.-J. Marrink},
\newblock \bibinfo{title}{3d pressure field in lipid membranes and
  membrane-protein complexes},
\newblock \bibinfo{journal}{Phys. Rev. Lett.} \bibinfo{volume}{102}
  (\bibinfo{year}{2009}) \bibinfo{pages}{078101}.
\bibitem[{Jakobsen et~al.(2005)Jakobsen, Mouritsen, and Besold}]{ask-besold-05}
\bibinfo{author}{A.~F. Jakobsen}, \bibinfo{author}{O.~G. Mouritsen},
  \bibinfo{author}{G.~Besold},
\newblock \bibinfo{title}{Artifacts in dynamical simulations of coarse-grained
  model lipid bilayers},
\newblock \bibinfo{journal}{J. Chem. Phys.} \bibinfo{volume}{122}
  (\bibinfo{year}{2005}) \bibinfo{pages}{204901}.
\bibitem[{Venturoli et~al.(2006)Venturoli, Sperotto, and
  Kranenburg}]{Venturoli:2006tv}
\bibinfo{author}{M.~Venturoli}, \bibinfo{author}{M.~M. Sperotto},
  \bibinfo{author}{M.~Kranenburg},
\newblock \bibinfo{title}{Mesoscopic models of biological membranes},
\newblock \bibinfo{journal}{Phys. Reports}  (\bibinfo{year}{2006}).
\bibitem[{Grafmuller et~al.(2007)Grafmuller, Shillcock, and
  Lipowsky}]{Grafmuller07}
\bibinfo{author}{A.~Grafmuller}, \bibinfo{author}{J.~Shillcock},
  \bibinfo{author}{R.~Lipowsky},
\newblock \bibinfo{title}{Pathway of membrane fusion with two tension dependent
  energy barriers},
\newblock \bibinfo{journal}{Phys. Rev. Lett.} \bibinfo{volume}{98}
  (\bibinfo{year}{2007}) \bibinfo{pages}{218107}.
\bibitem[{Safran(1999)}]{Safran:1999ty}
\bibinfo{author}{S.~A. Safran},
\newblock \bibinfo{title}{Curvature elasticity of thin films},
\newblock \bibinfo{journal}{Advances in Physics} \bibinfo{volume}{48}
  (\bibinfo{year}{1999}) \bibinfo{pages}{395--448}.
\bibitem[{Szleifer et~al.(1990)Szleifer, Kramer, Benshaul, Gelbart, and
  Safran}]{Szleifer:1990we}
\bibinfo{author}{I.~Szleifer}, \bibinfo{author}{D.~Kramer},
  \bibinfo{author}{A.~Benshaul}, \bibinfo{author}{W.~M. Gelbart},
  \bibinfo{author}{S.~A. Safran},
\newblock \bibinfo{title}{Molecular theory of curvature elasticity in
  surfactant films},
\newblock \bibinfo{journal}{J. Chem. Phys.} \bibinfo{volume}{92}
  (\bibinfo{year}{1990}) \bibinfo{pages}{6800--6817}.
\bibitem[{Goetz and Lipowsky(1998)}]{goetz-98}
\bibinfo{author}{R.~Goetz}, \bibinfo{author}{R.~Lipowsky},
\newblock \bibinfo{title}{Computer simulations of bilayer membranes:
  Self-assembly and interfacial tension},
\newblock \bibinfo{journal}{J. Chem. Phys.} \bibinfo{volume}{108}
  (\bibinfo{year}{1998}) \bibinfo{pages}{7397}.
\bibitem[{Farago and Pincus(2004)}]{Farago:2004vj}
\bibinfo{author}{O.~Farago}, \bibinfo{author}{P.~Pincus},
\newblock \bibinfo{title}{Statistical mechanics of bilayer membrane with a
  fixed projected area},
\newblock \bibinfo{journal}{J. Chem. Phys.} \bibinfo{volume}{120}
  (\bibinfo{year}{2004}) \bibinfo{pages}{2934--2950}.
\bibitem[{Hu et~al.(2012{\natexlab{a}})Hu, de~Jong, Marrink, and
  Deserno}]{Hu:2012et}
\bibinfo{author}{M.~Hu}, \bibinfo{author}{D.~H. de~Jong},
  \bibinfo{author}{S.-J. Marrink}, \bibinfo{author}{M.~Deserno},
\newblock \bibinfo{title}{Gaussian curvature elasticity determined from global
  shape transformations and local stress distributions: a comparative study
  using the martini model},
\newblock \bibinfo{journal}{Faraday Discuss.} \bibinfo{volume}{161}
  (\bibinfo{year}{2012}{\natexlab{a}}) \bibinfo{pages}{365--382}.
\bibitem[{Hu et~al.(2012{\natexlab{b}})Hu, Briguglio, and Deserno}]{Hu:2012bi}
\bibinfo{author}{M.~Hu}, \bibinfo{author}{J.~J. Briguglio},
  \bibinfo{author}{M.~Deserno},
\newblock \bibinfo{title}{Determining the gaussian curvature modulus of lipid
  membranes in simulations},
\newblock \bibinfo{journal}{Biophys. J.} \bibinfo{volume}{102}
  (\bibinfo{year}{2012}{\natexlab{b}}) \bibinfo{pages}{1403--1410}.
\bibitem[{Hu et~al.(2013)Hu, Diggins, and Deserno}]{Hu:2013gj}
\bibinfo{author}{M.~Hu}, \bibinfo{author}{P.~Diggins, IV},
  \bibinfo{author}{M.~Deserno},
\newblock \bibinfo{title}{Determining the bending modulus of a lipid membrane
  by simulating buckling},
\newblock \bibinfo{journal}{J. Chem. Phys.} \bibinfo{volume}{138}
  (\bibinfo{year}{2013}) \bibinfo{pages}{214110}.
\bibitem[{Shibata et~al.(2009)Shibata, Hu, Kozlov, and
  Rapoport}]{Shibata:2009p643}
\bibinfo{author}{Y.~Shibata}, \bibinfo{author}{J.~Hu}, \bibinfo{author}{M.~M.
  Kozlov}, \bibinfo{author}{T.~A. Rapoport},
\newblock \bibinfo{title}{Mechanisms shaping the membranes of cellular
  organelles},
\newblock \bibinfo{journal}{Ann. Rev. Cell Dev. Biol.} \bibinfo{volume}{25}
  (\bibinfo{year}{2009}) \bibinfo{pages}{329--354}.
\bibitem[{Kozlov(2010)}]{Kozlov:2010p301}
\bibinfo{author}{M.~M. Kozlov},
\newblock \bibinfo{title}{Biophysics: Joint effort bends membrane},
\newblock \bibinfo{journal}{Nature} \bibinfo{volume}{463}
  (\bibinfo{year}{2010}) \bibinfo{pages}{439--440}.
\bibitem[{Zimmerberg and Kozlov(2006)}]{Zimmerberg:2006p510}
\bibinfo{author}{J.~Zimmerberg}, \bibinfo{author}{M.~M. Kozlov},
\newblock \bibinfo{title}{How proteins produce cellular membrane curvature},
\newblock \bibinfo{journal}{Nat. Rev. Mol. Cell Biol.} \bibinfo{volume}{7}
  (\bibinfo{year}{2006}) \bibinfo{pages}{9--19}.
\bibitem[{Wiggins and Phillips(2004)}]{Wiggins:2004ir}
\bibinfo{author}{P.~Wiggins}, \bibinfo{author}{R.~Phillips},
\newblock \bibinfo{title}{Analytic models for mechanotransduction: gating a
  mechanosensitive channel},
\newblock \bibinfo{journal}{Proc. Natl. Acad. Sci. USA.} \bibinfo{volume}{101}
  (\bibinfo{year}{2004}) \bibinfo{pages}{4071--4076}.
\bibitem[{Aimon et~al.(2014)Aimon, Callan-Jones, Berthaud, Pinot, Toombes, and
  Bassereau}]{Aimon:2014if}
\bibinfo{author}{S.~Aimon}, \bibinfo{author}{A.~Callan-Jones},
  \bibinfo{author}{A.~Berthaud}, \bibinfo{author}{M.~Pinot},
  \bibinfo{author}{G.~E.~S. Toombes}, \bibinfo{author}{P.~Bassereau},
\newblock \bibinfo{title}{Membrane shape modulates transmembrane protein
  distribution},
\newblock \bibinfo{journal}{Developmental Cell} \bibinfo{volume}{28}
  (\bibinfo{year}{2014}) \bibinfo{pages}{212--218}.
\bibitem[{Praefcke and McMahon(2004)}]{Praefcke:2004p3309}
\bibinfo{author}{G.~J.~K. Praefcke}, \bibinfo{author}{H.~T. McMahon},
\newblock \bibinfo{title}{The dynamin superfamily: universal membrane
  tubulation and fission molecules?},
\newblock \bibinfo{journal}{Nat. Rev. Mol. Cell Biol.} \bibinfo{volume}{5}
  (\bibinfo{year}{2004}) \bibinfo{pages}{133--147}.
\bibitem[{Shibata et~al.(2008)Shibata, Voss, Rist, Hu, Rapoport, Prinz, and
  Voeltz}]{Shibata:2008p544}
\bibinfo{author}{Y.~Shibata}, \bibinfo{author}{C.~Voss}, \bibinfo{author}{J.~M.
  Rist}, \bibinfo{author}{J.~Hu}, \bibinfo{author}{T.~A. Rapoport},
  \bibinfo{author}{W.~A. Prinz}, \bibinfo{author}{G.~K. Voeltz},
\newblock \bibinfo{title}{The reticulon and dp1/yop1p proteins form immobile
  oligomers in the tubular endoplasmic reticulum},
\newblock \bibinfo{journal}{J. Biol. Chem.} \bibinfo{volume}{283}
  (\bibinfo{year}{2008}) \bibinfo{pages}{18892--18904}.
\bibitem[{Hu et~al.(2008)Hu, Shibata, Voss, Shemesh, Li, Coughlin, Kozlov,
  Rapoport, and Prinz}]{Hu:2008p3289}
\bibinfo{author}{J.~Hu}, \bibinfo{author}{Y.~Shibata},
  \bibinfo{author}{C.~Voss}, \bibinfo{author}{T.~Shemesh},
  \bibinfo{author}{Z.~Li}, \bibinfo{author}{M.~Coughlin},
  \bibinfo{author}{M.~M. Kozlov}, \bibinfo{author}{T.~A. Rapoport},
  \bibinfo{author}{W.~A. Prinz},
\newblock \bibinfo{title}{Membrane proteins of the endoplasmic reticulum induce
  high-curvature tubules},
\newblock \bibinfo{journal}{Science} \bibinfo{volume}{319}
  (\bibinfo{year}{2008}) \bibinfo{pages}{1247--1250}.
\bibitem[{Campelo et~al.(2008)Campelo, McMahon, and Kozlov}]{Campelo:2008p3288}
\bibinfo{author}{F.~Campelo}, \bibinfo{author}{H.~McMahon},
  \bibinfo{author}{M.~Kozlov},
\newblock \bibinfo{title}{The hydrophobic insertion mechanism of membrane
  curvature generation by proteins},
\newblock \bibinfo{journal}{Biophys. J.} \bibinfo{volume}{95}
  (\bibinfo{year}{2008}) \bibinfo{pages}{2325}.
\bibitem[{Dawson et~al.(2006)Dawson, Legg, and Machesky}]{Dawson:2006p113}
\bibinfo{author}{J.~C. Dawson}, \bibinfo{author}{J.~A. Legg},
  \bibinfo{author}{L.~M. Machesky},
\newblock \bibinfo{title}{Bar domain proteins: a role in tubulation, scission
  and actin assembly in clathrin-mediated endocytosis},
\newblock \bibinfo{journal}{Trends Cell Biol.} \bibinfo{volume}{16}
  (\bibinfo{year}{2006}) \bibinfo{pages}{493--498}.
\bibitem[{Farsad et~al.(2001)Farsad, Ringstad, Takei, Floyd, Rose, and
  Camilli}]{Farsad:2001p3596}
\bibinfo{author}{K.~Farsad}, \bibinfo{author}{N.~Ringstad},
  \bibinfo{author}{K.~Takei}, \bibinfo{author}{S.~R. Floyd},
  \bibinfo{author}{K.~Rose}, \bibinfo{author}{P.~D. Camilli},
\newblock \bibinfo{title}{Generation of high curvature membranes mediated by
  direct endophilin bilayer interactions},
\newblock \bibinfo{journal}{J. Cell Biol.} \bibinfo{volume}{155}
  (\bibinfo{year}{2001}) \bibinfo{pages}{193--200}.
\bibitem[{Peter et~al.(2004)Peter, Kent, Mills, Vallis, Butler, Evans, and
  McMahon}]{Peter:2004p3597}
\bibinfo{author}{B.~J. Peter}, \bibinfo{author}{H.~M. Kent},
  \bibinfo{author}{I.~G. Mills}, \bibinfo{author}{Y.~Vallis},
  \bibinfo{author}{P.~J.~G. Butler}, \bibinfo{author}{P.~R. Evans},
  \bibinfo{author}{H.~T. McMahon},
\newblock \bibinfo{title}{Bar domains as sensors of membrane curvature: the
  amphiphysin bar structure},
\newblock \bibinfo{journal}{Science} \bibinfo{volume}{303}
  (\bibinfo{year}{2004}) \bibinfo{pages}{495--499}.
\bibitem[{Habermann(2004)}]{Habermann:2004p3595}
\bibinfo{author}{B.~Habermann},
\newblock \bibinfo{title}{The bar-domain family of proteins: a case of bending
  and binding?},
\newblock \bibinfo{journal}{EMBO Rep.} \bibinfo{volume}{5}
  (\bibinfo{year}{2004}) \bibinfo{pages}{250--255}.
\bibitem[{Frost et~al.(2009)Frost, Unger, and de~Camilli}]{Frost:2009p157}
\bibinfo{author}{A.~Frost}, \bibinfo{author}{V.~M. Unger},
  \bibinfo{author}{P.~de~Camilli},
\newblock \bibinfo{title}{The bar domain superfamily: membrane-molding
  macromolecules},
\newblock \bibinfo{journal}{Cell} \bibinfo{volume}{137} (\bibinfo{year}{2009})
  \bibinfo{pages}{191--196}.
\bibitem[{Gallop et~al.(2006)Gallop, Jao, Kent, Butler, Evans, Langen, and
  McMahon}]{Gallop:2006p306}
\bibinfo{author}{J.~L. Gallop}, \bibinfo{author}{C.~C. Jao},
  \bibinfo{author}{H.~M. Kent}, \bibinfo{author}{P.~J.~G. Butler},
  \bibinfo{author}{P.~R. Evans}, \bibinfo{author}{R.~Langen},
  \bibinfo{author}{H.~T. McMahon},
\newblock \bibinfo{title}{Mechanism of endophilin n-bar domain-mediated
  membrane curvature},
\newblock \bibinfo{journal}{EMBO J.} \bibinfo{volume}{25}
  (\bibinfo{year}{2006}) \bibinfo{pages}{2898--2910}.
\bibitem[{Henne et~al.(2007)Henne, Kent, Ford, Hegde, Daumke, Butler, Mittal,
  Langen, Evans, and McMahon}]{Henne:2007p94}
\bibinfo{author}{W.~M. Henne}, \bibinfo{author}{H.~M. Kent},
  \bibinfo{author}{M.~G.~J. Ford}, \bibinfo{author}{B.~G. Hegde},
  \bibinfo{author}{O.~Daumke}, \bibinfo{author}{P.~J.~G. Butler},
  \bibinfo{author}{R.~Mittal}, \bibinfo{author}{R.~Langen},
  \bibinfo{author}{P.~R. Evans}, \bibinfo{author}{H.~T. McMahon},
\newblock \bibinfo{title}{Structure and analysis of fcho2 f-bar domain: a
  dimerizing and membrane recruitment module that effects membrane curvature},
\newblock \bibinfo{journal}{Structure} \bibinfo{volume}{15}
  (\bibinfo{year}{2007}) \bibinfo{pages}{839--852}.
\bibitem[{Frost et~al.(2007)Frost, de~Camilli, and Unger}]{Frost:2007p225}
\bibinfo{author}{A.~Frost}, \bibinfo{author}{P.~de~Camilli},
  \bibinfo{author}{V.~M. Unger},
\newblock \bibinfo{title}{F-bar proteins join the bar family fold},
\newblock \bibinfo{journal}{Structure} \bibinfo{volume}{15}
  (\bibinfo{year}{2007}) \bibinfo{pages}{751--753}.
\bibitem[{Frost et~al.(2008)Frost, Perera, Roux, Spasov, Destaing, Egelman,
  de~Camilli, and Unger}]{Frost:2008p6}
\bibinfo{author}{A.~Frost}, \bibinfo{author}{R.~Perera},
  \bibinfo{author}{A.~Roux}, \bibinfo{author}{K.~Spasov},
  \bibinfo{author}{O.~Destaing}, \bibinfo{author}{E.~H. Egelman},
  \bibinfo{author}{P.~de~Camilli}, \bibinfo{author}{V.~M. Unger},
\newblock \bibinfo{title}{Structural basis of membrane invagination by f-bar
  domains},
\newblock \bibinfo{journal}{Cell} \bibinfo{volume}{132} (\bibinfo{year}{2008})
  \bibinfo{pages}{807--817}.
\bibitem[{Ahmed et~al.(2010)Ahmed, Goh, and Bu}]{Ahmed:2010p3565}
\bibinfo{author}{S.~Ahmed}, \bibinfo{author}{W.~I. Goh},
  \bibinfo{author}{W.~Bu},
\newblock \bibinfo{title}{I-bar domains, irsp53 and filopodium formation},
\newblock \bibinfo{journal}{Semin. Cell Dev. Biol.} \bibinfo{volume}{21}
  (\bibinfo{year}{2010}) \bibinfo{pages}{350--356}.
\bibitem[{Weissenhorn(2005)}]{Weissenhorn:2005p3594}
\bibinfo{author}{W.~Weissenhorn},
\newblock \bibinfo{title}{Crystal structure of the endophilin-a1 bar domain},
\newblock \bibinfo{journal}{J. Mol. Biol.} \bibinfo{volume}{351}
  (\bibinfo{year}{2005}) \bibinfo{pages}{653--661}.
\bibitem[{Huttner and Schmidt(2002)}]{Huttner:2002p1214}
\bibinfo{author}{W.~Huttner}, \bibinfo{author}{A.~Schmidt},
\newblock \bibinfo{title}{Membrane curvature: a case of endofeelin'},
\newblock \bibinfo{journal}{Trends Cell Biol.} \bibinfo{volume}{12}
  (\bibinfo{year}{2002}) \bibinfo{pages}{155--158}.
\bibitem[{Masuda et~al.(2006)Masuda, Takeda, Sone, Ohki, Mori, Kamioka, and
  Mochizuki}]{Masuda:2006p363}
\bibinfo{author}{M.~Masuda}, \bibinfo{author}{S.~Takeda},
  \bibinfo{author}{M.~Sone}, \bibinfo{author}{T.~Ohki},
  \bibinfo{author}{H.~Mori}, \bibinfo{author}{Y.~Kamioka},
  \bibinfo{author}{N.~Mochizuki},
\newblock \bibinfo{title}{Endophilin bar domain drives membrane curvature by
  two newly identified structure-based mechanisms},
\newblock \bibinfo{journal}{EMBO J.} \bibinfo{volume}{25}
  (\bibinfo{year}{2006}) \bibinfo{pages}{2889--2897}.
\bibitem[{Zimmerberg and McLaughlin(2004)}]{Zimmerberg:2004p129}
\bibinfo{author}{J.~Zimmerberg}, \bibinfo{author}{S.~McLaughlin},
\newblock \bibinfo{title}{Membrane curvature: How bar domains bend bilayers},
\newblock \bibinfo{journal}{Current Biology} \bibinfo{volume}{14}
  (\bibinfo{year}{2004}) \bibinfo{pages}{R250--R252}.
\bibitem[{Blood and Voth(2006)}]{Blood:2006uc}
\bibinfo{author}{P.~D. Blood}, \bibinfo{author}{G.~A. Voth},
\newblock \bibinfo{title}{Direct observation of bin/amphiphysin/rvs (bar)
  domain-induced membrane curvature by means of molecular dynamics
  simulations},
\newblock \bibinfo{journal}{Proc. Natl. Acad. Sci. USA.} \bibinfo{volume}{103}
  (\bibinfo{year}{2006}) \bibinfo{pages}{15068--15072}.
\bibitem[{Arkhipov et~al.(2008)Arkhipov, Yin, and Schulten}]{Arkhipov:2008p257}
\bibinfo{author}{A.~Arkhipov}, \bibinfo{author}{Y.~Yin},
  \bibinfo{author}{K.~Schulten},
\newblock \bibinfo{title}{Four-scale description of membrane sculpting by bar
  domains},
\newblock \bibinfo{journal}{Biophys. J.} \bibinfo{volume}{95}
  (\bibinfo{year}{2008}) \bibinfo{pages}{2806--21}.
\bibitem[{Arkhipov et~al.(2009)Arkhipov, Yin, and
  Schulten}]{Arkhipov:2009p3009}
\bibinfo{author}{A.~Arkhipov}, \bibinfo{author}{Y.~Yin},
  \bibinfo{author}{K.~Schulten},
\newblock \bibinfo{title}{Membrane-bending mechanism of amphiphysin n-bar
  domains},
\newblock \bibinfo{journal}{Biophys. J.} \bibinfo{volume}{97}
  (\bibinfo{year}{2009}) \bibinfo{pages}{2727--2735}.
\bibitem[{Collins(2006)}]{Collins:2006p464}
\bibinfo{author}{R.~N. Collins},
\newblock \bibinfo{title}{How the er stays in shape},
\newblock \bibinfo{journal}{Cell} \bibinfo{volume}{124} (\bibinfo{year}{2006})
  \bibinfo{pages}{464--466}.
\bibitem[{McMahon and Gallop(2005)}]{McMahon:2005p274}
\bibinfo{author}{H.~T. McMahon}, \bibinfo{author}{J.~L. Gallop},
\newblock \bibinfo{title}{Membrane curvature and mechanisms of dynamic cell
  membrane remodelling},
\newblock \bibinfo{journal}{Nature} \bibinfo{volume}{438}
  (\bibinfo{year}{2005}) \bibinfo{pages}{590--6}.
\bibitem[{Shibata et~al.(2006)Shibata, Voeltz, and Rapoport}]{Shibata:2006p311}
\bibinfo{author}{Y.~Shibata}, \bibinfo{author}{G.~K. Voeltz},
  \bibinfo{author}{T.~A. Rapoport},
\newblock \bibinfo{title}{Rough sheets and smooth tubules},
\newblock \bibinfo{journal}{Cell} \bibinfo{volume}{126} (\bibinfo{year}{2006})
  \bibinfo{pages}{435--439}.
\bibitem[{Shibata et~al.(2010)Shibata, Shemesh, Prinz, Kozlov, and
  Rapoport}]{Shibata:2010p2363}
\bibinfo{author}{Y.~Shibata}, \bibinfo{author}{T.~Shemesh},
  \bibinfo{author}{W.~A. Prinz}, \bibinfo{author}{M.~M. Kozlov},
  \bibinfo{author}{T.~A. Rapoport},
\newblock \bibinfo{title}{Mechanisms determining the morphology of the
  peripheral er},
\newblock \bibinfo{journal}{Cell} \bibinfo{volume}{143} (\bibinfo{year}{2010})
  \bibinfo{pages}{774--788}.
\bibitem[{Graham and Kozlov(2010)}]{Graham:2010p2433}
\bibinfo{author}{T.~R. Graham}, \bibinfo{author}{M.~M. Kozlov},
\newblock \bibinfo{title}{Interplay of proteins and lipids in generating
  membrane curvature},
\newblock \bibinfo{journal}{Curr. Opin. Cell Biol.} \bibinfo{volume}{22}
  (\bibinfo{year}{2010}) \bibinfo{pages}{430--436}.
\bibitem[{Bahrami et~al.(2012)Bahrami, Lipowsky, and Weikl}]{Bahrami:2012gb}
\bibinfo{author}{A.~H. Bahrami}, \bibinfo{author}{R.~Lipowsky},
  \bibinfo{author}{T.~R. Weikl},
\newblock \bibinfo{title}{Tubulation and aggregation of spherical nanoparticles
  adsorbed on vesicles},
\newblock \bibinfo{journal}{Phys. Rev. Lett.} \bibinfo{volume}{109}
  (\bibinfo{year}{2012}) \bibinfo{pages}{188102}.
\bibitem[{{\v S}ari{\'c} and Cacciuto(2012)}]{Saric:2012hb}
\bibinfo{author}{A.~{\v S}ari{\'c}}, \bibinfo{author}{A.~Cacciuto},
\newblock \bibinfo{title}{Mechanism of membrane tube formation induced by
  adhesive nanocomponents},
\newblock \bibinfo{journal}{Phys. Rev. Lett.} \bibinfo{volume}{109}
  (\bibinfo{year}{2012}) \bibinfo{pages}{188101}.
\bibitem[{Zhang et~al.(2012)Zhang, Nelson, and Beales}]{Zhang:2012ko}
\bibinfo{author}{S.~Zhang}, \bibinfo{author}{A.~Nelson}, \bibinfo{author}{P.~A.
  Beales},
\newblock \bibinfo{title}{Freezing or wrapping: The role of particle size in
  the mechanism of nanoparticle--biomembrane interaction},
\newblock \bibinfo{journal}{Langmuir} \bibinfo{volume}{28}
  (\bibinfo{year}{2012}) \bibinfo{pages}{12831--12837}.
\bibitem[{Zhang and Hinshaw(2001)}]{Zhang:2001p3767}
\bibinfo{author}{P.~Zhang}, \bibinfo{author}{J.~E. Hinshaw},
\newblock \bibinfo{title}{Three-dimensional reconstruction of dynamin in the
  constricted state},
\newblock \bibinfo{journal}{Nat. Cell Biol.} \bibinfo{volume}{3}
  (\bibinfo{year}{2001}) \bibinfo{pages}{922--926}.
\bibitem[{Roux et~al.(2010)Roux, Koster, Lenz, Sorre, Manneville, Nassoy, and
  Bassereau}]{Roux:2010p4141}
\bibinfo{author}{A.~Roux}, \bibinfo{author}{G.~Koster},
  \bibinfo{author}{M.~Lenz}, \bibinfo{author}{B.~Sorre}, \bibinfo{author}{J.-B.
  Manneville}, \bibinfo{author}{P.~Nassoy}, \bibinfo{author}{P.~Bassereau},
\newblock \bibinfo{title}{Membrane curvature controls dynamin polymerization},
\newblock \bibinfo{journal}{Proc. Natl. Acad. Sci. USA.} \bibinfo{volume}{107}
  (\bibinfo{year}{2010}) \bibinfo{pages}{4141--4146}.
\bibitem[{Morlot et~al.(2010)Morlot, Lenz, Prost, Joanny, and
  Roux}]{Morlot:2010hr}
\bibinfo{author}{S.~Morlot}, \bibinfo{author}{M.~Lenz},
  \bibinfo{author}{J.~Prost}, \bibinfo{author}{J.-F. Joanny},
  \bibinfo{author}{A.~Roux},
\newblock \bibinfo{title}{Deformation of dynamin helices damped by membrane
  friction},
\newblock \bibinfo{journal}{Biophys. J.} \bibinfo{volume}{99}
  (\bibinfo{year}{2010}) \bibinfo{pages}{3580--3588}.
\bibitem[{Blood et~al.(2008)Blood, Swenson, and Voth}]{Blood:2008p1866}
\bibinfo{author}{P.~Blood}, \bibinfo{author}{R.~Swenson},
  \bibinfo{author}{G.~Voth},
\newblock \bibinfo{title}{Factors influencing local membrane curvature
  induction by n-bar domains as revealed by molecular dynamics simulations},
\newblock \bibinfo{journal}{Biophys. J.} \bibinfo{volume}{95}
  (\bibinfo{year}{2008}) \bibinfo{pages}{1866--1876}.
\bibitem[{Kranenburg and Smit(2005)}]{Kranenburg:2005kv}
\bibinfo{author}{M.~Kranenburg}, \bibinfo{author}{B.~Smit},
\newblock \bibinfo{title}{Phase behavior of model lipid bilayers},
\newblock \bibinfo{journal}{J. Phys. Chem. B} \bibinfo{volume}{109}
  (\bibinfo{year}{2005}) \bibinfo{pages}{6553--6563}.
\bibitem[{Smith et~al.(1988)Smith, Sirota, Safinya, and
  Clark}]{Smith:1988p3535}
\bibinfo{author}{G.~S. Smith}, \bibinfo{author}{E.~B. Sirota},
  \bibinfo{author}{C.~R. Safinya}, \bibinfo{author}{N.~A. Clark},
\newblock \bibinfo{title}{Structure of the l$\beta$ phases in a hydrated
  phosphatidylcholine multimembrane},
\newblock \bibinfo{journal}{Phys. Rev. Lett.} \bibinfo{volume}{60}
  (\bibinfo{year}{1988}) \bibinfo{pages}{813}.
\bibitem[{Smith et~al.(1990)Smith, Sirota, Safinya, Plano, and
  Clark}]{Smith:1990p3537}
\bibinfo{author}{G.~S. Smith}, \bibinfo{author}{E.~B. Sirota},
  \bibinfo{author}{C.~R. Safinya}, \bibinfo{author}{R.~J. Plano},
  \bibinfo{author}{N.~A. Clark},
\newblock \bibinfo{title}{X-ray structural studies of freely suspended ordered
  hydrated dmpc multimembrane films},
\newblock \bibinfo{journal}{J. Chem. Phys.} \bibinfo{volume}{92}
  (\bibinfo{year}{1990}) \bibinfo{pages}{4519--4529}.
\bibitem[{Tardieu et~al.(1973)Tardieu, Luzzati, and Reman}]{Tardieu:1973p711}
\bibinfo{author}{A.~Tardieu}, \bibinfo{author}{V.~Luzzati},
  \bibinfo{author}{F.~C. Reman},
\newblock \bibinfo{title}{Structure and polymorphism of the hydrocarbon chains
  of lipids: A study of lecithin-water phases},
\newblock \bibinfo{journal}{J. Mol. Bio.} \bibinfo{volume}{75}
  (\bibinfo{year}{1973}) \bibinfo{pages}{711 -- 718}.
\bibitem[{Selinger et~al.(2001)Selinger, Spector, and
  Schnur}]{Selinger:2001p284}
\bibinfo{author}{J.~Selinger}, \bibinfo{author}{M.~Spector},
  \bibinfo{author}{J.~Schnur},
\newblock \bibinfo{title}{Theory of self-assembled tubules and helical
  ribbons},
\newblock \bibinfo{journal}{J. Phys. Chem. B} \bibinfo{volume}{105}
  (\bibinfo{year}{2001}) \bibinfo{pages}{7157--7169}.
\bibitem[{Sarasij et~al.(2007)Sarasij, Mayor, and Rao}]{Sarasij:2007p3509}
\bibinfo{author}{R.~Sarasij}, \bibinfo{author}{S.~Mayor},
  \bibinfo{author}{M.~Rao},
\newblock \bibinfo{title}{Chirality-induced budding: A raft-mediated mechanism
  for endocytosis and morphology of caveolae?},
\newblock \bibinfo{journal}{Biophys. J.} \bibinfo{volume}{92}
  (\bibinfo{year}{2007}) \bibinfo{pages}{3140}.
\bibitem[{Harris et~al.(1999)Harris, Kamien, and Lubensky}]{Harris:1999p3510}
\bibinfo{author}{A.~B. Harris}, \bibinfo{author}{R.~D. Kamien},
  \bibinfo{author}{T.~C. Lubensky},
\newblock \bibinfo{title}{Molecular chirality and chiral parameters},
\newblock \bibinfo{journal}{Rev. Mod. Phys.} \bibinfo{volume}{71}
  (\bibinfo{year}{1999}) \bibinfo{pages}{1745}.
\bibitem[{Helfrich and Prost(1988)}]{Helfrich:1988p3004}
\bibinfo{author}{W.~Helfrich}, \bibinfo{author}{J.~Prost},
\newblock \bibinfo{title}{Intrinsic bending force in anisotropic membranes made
  of chiral molecules},
\newblock \bibinfo{journal}{Phys. Rev. A} \bibinfo{volume}{38}
  (\bibinfo{year}{1988}) \bibinfo{pages}{3065}.
\bibitem[{Nelson and Powers(1992)}]{Nelson:1992p9}
\bibinfo{author}{P.~Nelson}, \bibinfo{author}{T.~Powers},
\newblock \bibinfo{title}{Rigid chiral membranes},
\newblock \bibinfo{journal}{Phys. Rev. Lett.} \bibinfo{volume}{69}
  (\bibinfo{year}{1992}) \bibinfo{pages}{3409--3412}.
\bibitem[{Selinger and Schnur(1993)}]{Selinger:1993p108}
\bibinfo{author}{J.~V. Selinger}, \bibinfo{author}{J.~M. Schnur},
\newblock \bibinfo{title}{Theory of chiral lipid tubules},
\newblock \bibinfo{journal}{Phys. Rev. Lett.} \bibinfo{volume}{71}
  (\bibinfo{year}{1993}) \bibinfo{pages}{4091}.
\bibitem[{Schnur(1993)}]{Schnur:1993p1277}
\bibinfo{author}{J.~M. Schnur},
\newblock \bibinfo{title}{Lipid tubules: a paradigm for molecularly engineered
  structures},
\newblock \bibinfo{journal}{Science} \bibinfo{volume}{262}
  (\bibinfo{year}{1993}) \bibinfo{pages}{1669--1676}.
\bibitem[{Kralj-Iglic et~al.(2002)Kralj-Iglic, Iglic, Gomiscek, Sevsek,
  Arrigler, and H{\"a}gerstrand}]{KraljIglic:2002p1533}
\bibinfo{author}{V.~Kralj-Iglic}, \bibinfo{author}{A.~Iglic},
  \bibinfo{author}{G.~Gomiscek}, \bibinfo{author}{F.~Sevsek},
  \bibinfo{author}{V.~Arrigler}, \bibinfo{author}{H.~H{\"a}gerstrand},
\newblock \bibinfo{title}{Microtubes and nanotubes of a phospholipid bilayer
  membrane},
\newblock \bibinfo{journal}{J. Phys. A} \bibinfo{volume}{35}
  (\bibinfo{year}{2002}) \bibinfo{pages}{1533--1549}.
\bibitem[{Oda et~al.(1997)Oda, Huc, and Candau}]{Oda:1997p124}
\bibinfo{author}{R.~Oda}, \bibinfo{author}{I.~Huc},
  \bibinfo{author}{S.~Candau},
\newblock \bibinfo{title}{Gemini surfactants, the effect of hydrophobic chain
  length and dissymmetry},
\newblock \bibinfo{journal}{Chem Commun}  (\bibinfo{year}{1997})
  \bibinfo{pages}{2105--2106}.
\bibitem[{Groves(2007)}]{Groves07}
\bibinfo{author}{J.~T. Groves},
\newblock \bibinfo{title}{Bending mechanics and molecular organization in
  biological membranes},
\newblock \bibinfo{journal}{Ann. Rev. Phys. Chem.} \bibinfo{volume}{58}
  (\bibinfo{year}{2007}) \bibinfo{pages}{697--717}.
\bibitem[{Neto et~al.(2006)Neto, Agero, Gazzinelli, and Mesquita}]{Neto06}
\bibinfo{author}{J.~C. Neto}, \bibinfo{author}{U.~Agero},
  \bibinfo{author}{R.~T. Gazzinelli}, \bibinfo{author}{O.~N. Mesquita},
\newblock \bibinfo{title}{Measuring optical and mechanical properties of a
  living cell with defocusing microscopy},
\newblock \bibinfo{journal}{Biophys. J.} \bibinfo{volume}{91}
  (\bibinfo{year}{2006}) \bibinfo{pages}{1108--1115}.
\bibitem[{Marguet et~al.(2006)Marguet, Lenne, Rigneault, and He}]{Marguet06}
\bibinfo{author}{D.~Marguet}, \bibinfo{author}{P.~F. Lenne},
  \bibinfo{author}{H.~Rigneault}, \bibinfo{author}{H.~T. He},
\newblock \bibinfo{title}{Dynamics in the plasma membrane: how to combine
  fluidity and order},
\newblock \bibinfo{journal}{Embo J.} \bibinfo{volume}{25}
  (\bibinfo{year}{2006}) \bibinfo{pages}{3446--3457}.
\bibitem[{Sorre et~al.(2009)Sorre, Callan-Jones, Manneville, Nassoy, Joanny,
  Prost, Goud, and Bassereau}]{Sorre:2009do}
\bibinfo{author}{B.~Sorre}, \bibinfo{author}{A.~Callan-Jones},
  \bibinfo{author}{J.-B. Manneville}, \bibinfo{author}{P.~Nassoy},
  \bibinfo{author}{J.-F. Joanny}, \bibinfo{author}{J.~Prost},
  \bibinfo{author}{B.~Goud}, \bibinfo{author}{P.~Bassereau},
\newblock \bibinfo{title}{Curvature-driven lipid sorting needs proximity to a
  demixing point and is aided by proteins},
\newblock \bibinfo{journal}{Proc. Natl. Acad. Sci. USA.} \bibinfo{volume}{106}
  (\bibinfo{year}{2009}) \bibinfo{pages}{5622--5626}.
\bibitem[{Sorre et~al.(2012)Sorre, Callan-Jones, Manzi, Goud, Prost, Bassereau,
  and Roux}]{Sorre:2012if}
\bibinfo{author}{B.~Sorre}, \bibinfo{author}{A.~Callan-Jones},
  \bibinfo{author}{J.~Manzi}, \bibinfo{author}{B.~Goud},
  \bibinfo{author}{J.~Prost}, \bibinfo{author}{P.~Bassereau},
  \bibinfo{author}{A.~Roux},
\newblock \bibinfo{title}{Nature of curvature coupling of amphiphysin with
  membranes depends on its bound density},
\newblock \bibinfo{journal}{Proc. Natl. Acad. Sci. USA.} \bibinfo{volume}{109}
  (\bibinfo{year}{2012}) \bibinfo{pages}{173--178}.
\bibitem[{Tian and Baumgart(2009)}]{Tian:2009fu}
\bibinfo{author}{A.~Tian}, \bibinfo{author}{T.~Baumgart},
\newblock \bibinfo{title}{Sorting of lipids and proteins in membrane curvature
  gradients},
\newblock \bibinfo{journal}{Biophys. J.} \bibinfo{volume}{96}
  (\bibinfo{year}{2009}) \bibinfo{pages}{2676--2688}.
\bibitem[{Tian et~al.(2009)Tian, Capraro, Esposito, and Baumgart}]{Tian:2009hx}
\bibinfo{author}{A.~Tian}, \bibinfo{author}{B.~R. Capraro},
  \bibinfo{author}{C.~Esposito}, \bibinfo{author}{T.~Baumgart},
\newblock \bibinfo{title}{Bending stiffness depends on curvature of ternary
  lipid mixture tubular membranes},
\newblock \bibinfo{journal}{Biophys. J.} \bibinfo{volume}{97}
  (\bibinfo{year}{2009}) \bibinfo{pages}{1636--1646}.
\bibitem[{Heinrich et~al.(2010)Heinrich, Tian, Esposito, and
  Baumgart}]{Heinrich:2010ej}
\bibinfo{author}{M.~Heinrich}, \bibinfo{author}{A.~Tian},
  \bibinfo{author}{C.~Esposito}, \bibinfo{author}{T.~Baumgart},
\newblock \bibinfo{title}{Dynamic sorting of lipids and proteins in membrane
  tubes with a moving phase boundary},
\newblock \bibinfo{journal}{Proc. Natl. Acad. Sci. USA.} \bibinfo{volume}{107}
  (\bibinfo{year}{2010}) \bibinfo{pages}{7208--7213}.
\bibitem[{Seifert(1993)}]{Seifert:1993bz}
\bibinfo{author}{U.~Seifert},
\newblock \bibinfo{title}{Curvature-induced lateral phase segregation in
  two-component vesicles},
\newblock \bibinfo{journal}{Phys. Rev. Lett.} \bibinfo{volume}{70}
  (\bibinfo{year}{1993}) \bibinfo{pages}{1335--1338}.
\bibitem[{Capraro et~al.(2010)Capraro, Yoon, Cho, and
  Baumgart}]{Capraro:2010jo}
\bibinfo{author}{B.~R. Capraro}, \bibinfo{author}{Y.~Yoon},
  \bibinfo{author}{W.~Cho}, \bibinfo{author}{T.~Baumgart},
\newblock \bibinfo{title}{Curvature sensing by the epsin n-terminal homology
  domain measured on cylindrical lipid membrane tethers},
\newblock \bibinfo{journal}{J. Amer. Chem. Soc.} \bibinfo{volume}{132}
  (\bibinfo{year}{2010}) \bibinfo{pages}{1200--1201}.
\bibitem[{Mim et~al.(2012)Mim, Cui, Gawronski-Salerno, Frost, Lyman, Voth, and
  Unger}]{Mim:2012je}
\bibinfo{author}{C.~Mim}, \bibinfo{author}{H.~Cui}, \bibinfo{author}{J.~A.
  Gawronski-Salerno}, \bibinfo{author}{A.~Frost}, \bibinfo{author}{E.~Lyman},
  \bibinfo{author}{G.~A. Voth}, \bibinfo{author}{V.~M. Unger},
\newblock \bibinfo{title}{Structural basis of membrane bending by the n-bar
  protein endophilin},
\newblock \bibinfo{journal}{Cell} \bibinfo{volume}{149} (\bibinfo{year}{2012})
  \bibinfo{pages}{137--145}.
\bibitem[{Marino et~al.(2005)Marino, Moon, and Hinshaw}]{Marino:2005gl}
\bibinfo{author}{M.~Marino}, \bibinfo{author}{K.-H. Moon},
  \bibinfo{author}{J.~E. Hinshaw},
\newblock \bibinfo{title}{The role of dynamin in membrane constriction revealed
  by cryo-em},
\newblock \bibinfo{journal}{Microscopy and Microanalysis} \bibinfo{volume}{11}
  (\bibinfo{year}{2005}).
\bibitem[{Baumgart et~al.(2011)Baumgart, Capraro, Zhu, and
  Das}]{Baumgart:2011en}
\bibinfo{author}{T.~Baumgart}, \bibinfo{author}{B.~R. Capraro},
  \bibinfo{author}{C.~Zhu}, \bibinfo{author}{S.~L. Das},
\newblock \bibinfo{title}{Thermodynamics and mechanics of membrane curvature
  generation and sensing by proteins and lipids},
\newblock \bibinfo{journal}{Annu. Rev. Phys. Chem.} \bibinfo{volume}{62}
  (\bibinfo{year}{2011}) \bibinfo{pages}{483--506}.
\bibitem[{Jiang and Powers(2008)}]{Jiang:2008je}
\bibinfo{author}{H.~Jiang}, \bibinfo{author}{T.~Powers},
\newblock \bibinfo{title}{Curvature-driven lipid sorting in a membrane tubule},
\newblock \bibinfo{journal}{Phys. Rev. Lett.} \bibinfo{volume}{101}
  (\bibinfo{year}{2008}) \bibinfo{pages}{018103}.
\bibitem[{Singh et~al.(2012)Singh, Mahata, Baumgart, and Das}]{Singh:2012gb}
\bibinfo{author}{P.~Singh}, \bibinfo{author}{P.~Mahata},
  \bibinfo{author}{T.~Baumgart}, \bibinfo{author}{S.~L. Das},
\newblock \bibinfo{title}{Curvature sorting of proteins on a cylindrical lipid
  membrane tether connected to a reservoir},
\newblock \bibinfo{journal}{Phys. Rev. E} \bibinfo{volume}{85}
  (\bibinfo{year}{2012}) \bibinfo{pages}{051906}.
\bibitem[{Zhu et~al.(2012)Zhu, Das, and Baumgart}]{Zhu:2012eg}
\bibinfo{author}{C.~Zhu}, \bibinfo{author}{S.~L. Das},
  \bibinfo{author}{T.~Baumgart},
\newblock \bibinfo{title}{Nonlinear sorting, curvature generation, and crowding
  of endophilin n-bar on tubular membranes},
\newblock \bibinfo{journal}{Biophys. J.} \bibinfo{volume}{102}
  (\bibinfo{year}{2012}) \bibinfo{pages}{1837--1845}.
\bibitem[{Sunil~Kumar et~al.(1999)Sunil~Kumar, Gompper, and
  Lipowsky}]{SunilKumar:1999bf}
\bibinfo{author}{P.~B. Sunil~Kumar}, \bibinfo{author}{G.~Gompper},
  \bibinfo{author}{R.~Lipowsky},
\newblock \bibinfo{title}{Modulated phases in multicomponent fluid membranes},
\newblock \bibinfo{journal}{Phys. Rev. E} \bibinfo{volume}{60}
  (\bibinfo{year}{1999}) \bibinfo{pages}{4610--4618}.
\bibitem[{Sunil~Kumar et~al.(2001)Sunil~Kumar, Gompper, and
  Lipowsky}]{SunilKumar:2001cr}
\bibinfo{author}{P.~B. Sunil~Kumar}, \bibinfo{author}{G.~Gompper},
  \bibinfo{author}{R.~Lipowsky},
\newblock \bibinfo{title}{Budding dynamics of multicomponent membranes},
\newblock \bibinfo{journal}{Phys. Rev. Lett.} \bibinfo{volume}{86}
  (\bibinfo{year}{2001}) \bibinfo{pages}{3911--3914}.
\bibitem[{Ramakrishnan et~al.(2013)Ramakrishnan, Sunil~Kumar, and
  Ipsen}]{Ramakrishnan:2013gl}
\bibinfo{author}{N.~Ramakrishnan}, \bibinfo{author}{P.~B. Sunil~Kumar},
  \bibinfo{author}{J.~H. Ipsen},
\newblock \bibinfo{title}{Membrane-mediated aggregation of curvature-inducing
  nematogens and membrane tubulation},
\newblock \bibinfo{journal}{Biophys. J.} \bibinfo{volume}{104}
  (\bibinfo{year}{2013}) \bibinfo{pages}{1018--1028}.
\bibitem[{Ramakrishnan et~al.(2012)Ramakrishnan, Ipsen, and
  Sunil~Kumar}]{Ramakrishnan:2012dk}
\bibinfo{author}{N.~Ramakrishnan}, \bibinfo{author}{J.~H. Ipsen},
  \bibinfo{author}{P.~B. Sunil~Kumar},
\newblock \bibinfo{title}{Role of disclinations in determining the morphology
  of deformable fluid interfaces},
\newblock \bibinfo{journal}{Soft Matter} \bibinfo{volume}{8}
  (\bibinfo{year}{2012}) \bibinfo{pages}{3058}.
\bibitem[{Liu et~al.(2012)Liu, Tourdot, Ramanan, Agrawal, and
  Radhakrishanan}]{Liu:2012es}
\bibinfo{author}{J.~Liu}, \bibinfo{author}{R.~Tourdot},
  \bibinfo{author}{V.~Ramanan}, \bibinfo{author}{N.~J. Agrawal},
  \bibinfo{author}{R.~Radhakrishanan},
\newblock \bibinfo{title}{Mesoscale simulations of curvature-inducing protein
  partitioning on lipid bilayer membranes in the presence of mean curvature
  fields},
\newblock \bibinfo{journal}{Molecular Physics} \bibinfo{volume}{110}
  (\bibinfo{year}{2012}) \bibinfo{pages}{1127--1137}.
\bibitem[{Ramanan et~al.(2011)Ramanan, Agrawal, Liu, Engles, Toy, and
  Radhakrishnan}]{Ramanan:2011ds}
\bibinfo{author}{V.~Ramanan}, \bibinfo{author}{N.~J. Agrawal},
  \bibinfo{author}{J.~Liu}, \bibinfo{author}{S.~Engles},
  \bibinfo{author}{R.~Toy}, \bibinfo{author}{R.~Radhakrishnan},
\newblock \bibinfo{title}{Systems biology and physical biology of
  clathrin-mediated endocytosis},
\newblock \bibinfo{journal}{Integr. Biol.} \bibinfo{volume}{3}
  (\bibinfo{year}{2011}) \bibinfo{pages}{803}.
\bibitem[{Gov(2006)}]{Gov06}
\bibinfo{author}{N.~Gov},
\newblock \bibinfo{title}{Diffusion in curved fluid membranes},
\newblock \bibinfo{journal}{Phys. Rev. E} \bibinfo{volume}{73}
  (\bibinfo{year}{2006}) \bibinfo{pages}{041918}.
\bibitem[{Divet et~al.(2005)Divet, Danker, and Misbah}]{Divet05}
\bibinfo{author}{F.~Divet}, \bibinfo{author}{G.~Danker},
  \bibinfo{author}{C.~Misbah},
\newblock \bibinfo{title}{Fluctuations and instability of a biological membrane
  induced by interaction with macromolecules},
\newblock \bibinfo{journal}{Phys. Rev. E} \bibinfo{volume}{72}
  (\bibinfo{year}{2005}) \bibinfo{pages}{041901}.
\bibitem[{Naji and Brown(2007)}]{Naji07}
\bibinfo{author}{A.~Naji}, \bibinfo{author}{F.~Brown},
\newblock \bibinfo{title}{Diffusion on ruffled membrane surfaces},
\newblock \bibinfo{journal}{J. Chem. Phys.} \bibinfo{volume}{126}
  (\bibinfo{year}{2007}) \bibinfo{pages}{235103--16}.
\bibitem[{Reister-Gottfried et~al.(2007)Reister-Gottfried, Leitenberger, and
  Seifert}]{Reister07}
\bibinfo{author}{E.~Reister-Gottfried}, \bibinfo{author}{S.~M. Leitenberger},
  \bibinfo{author}{U.~Seifert},
\newblock \bibinfo{title}{Hybrid simulations of lateral diffusion in
  fluctuating membranes},
\newblock \bibinfo{journal}{Phys. Rev. E} \bibinfo{volume}{75}
  (\bibinfo{year}{2007}) \bibinfo{pages}{011908--11}.
\bibitem[{Atilgan and Sun(2007)}]{Atilgan07}
\bibinfo{author}{E.~Atilgan}, \bibinfo{author}{S.~X. Sun},
\newblock \bibinfo{title}{Shape transitions in lipid membranes and protein
  mediated vesicle fusion and fission},
\newblock \bibinfo{journal}{J. Chem. Phys.} \bibinfo{volume}{126}
  (\bibinfo{year}{2007}) \bibinfo{pages}{095102}.
\bibitem[{Weinstein and Radhakrishnan(2006)}]{Weinstein06}
\bibinfo{author}{J.~Weinstein}, \bibinfo{author}{R.~Radhakrishnan},
\newblock \bibinfo{title}{A coarse-grained methodology for simulating
  interfacial dynamics in complex fluids: application to protein mediated
  membrane processes},
\newblock \bibinfo{journal}{Mol. Phys.} \bibinfo{volume}{104}
  (\bibinfo{year}{2006}) \bibinfo{pages}{3653--3666}.
\bibitem[{Agrawal et~al.(2008)Agrawal, Weinstein, and
  Radhakrishnan}]{Agrawal:2008ff}
\bibinfo{author}{N.~J. Agrawal}, \bibinfo{author}{J.~Weinstein},
  \bibinfo{author}{R.~Radhakrishnan},
\newblock \bibinfo{title}{Landscape of finite-temperature equilibrium behaviour
  of curvature-inducing proteins on a bilayer membrane explored using a
  linearized elastic free energy model},
\newblock \bibinfo{journal}{Molecular Physics} \bibinfo{volume}{106}
  (\bibinfo{year}{2008}) \bibinfo{pages}{1913--1923}.
\bibitem[{Chou et~al.(2001)Chou, Kim, and Oster}]{Chou:2001bm}
\bibinfo{author}{T.~Chou}, \bibinfo{author}{K.~S. Kim},
  \bibinfo{author}{G.~Oster},
\newblock \bibinfo{title}{Statistical thermodynamics of membrane
  bending-mediated protein--protein attractions},
\newblock \bibinfo{journal}{Biophys. J.} \bibinfo{volume}{80}
  (\bibinfo{year}{2001}) \bibinfo{pages}{1075--1087}.
\bibitem[{Grabe et~al.(2003)Grabe, Neu, Oster, and Nollert}]{Grabe03}
\bibinfo{author}{M.~Grabe}, \bibinfo{author}{J.~Neu},
  \bibinfo{author}{G.~Oster}, \bibinfo{author}{P.~Nollert},
\newblock \bibinfo{title}{Protein interactions and membrane geometry},
\newblock \bibinfo{journal}{Biophys. J.} \bibinfo{volume}{84}
  (\bibinfo{year}{2003}) \bibinfo{pages}{854--868}.
\bibitem[{Kim et~al.(1998)Kim, Neu, and Oster}]{Kim98}
\bibinfo{author}{K.~S. Kim}, \bibinfo{author}{J.~Neu},
  \bibinfo{author}{G.~Oster},
\newblock \bibinfo{title}{Curvature-mediated interactions between membrane
  proteins},
\newblock \bibinfo{journal}{Biophys J} \bibinfo{volume}{75}
  (\bibinfo{year}{1998}) \bibinfo{pages}{2274--2291}.
\bibitem[{Dan et~al.(1994)Dan, Derman, Pincus, and Safran}]{Dan94}
\bibinfo{author}{N.~Dan}, \bibinfo{author}{A.~Derman},
  \bibinfo{author}{P.~Pincus}, \bibinfo{author}{S.~Safran},
\newblock \bibinfo{title}{Membrane-induced interactions between inclusions},
\newblock \bibinfo{journal}{J. Phys. II} \bibinfo{volume}{4}
  (\bibinfo{year}{1994}) \bibinfo{pages}{1713--1725}.
\bibitem[{Wallace et~al.(2005)Wallace, Hooper, and Olmsted}]{Wallace05}
\bibinfo{author}{E.~J. Wallace}, \bibinfo{author}{N.~M. Hooper},
  \bibinfo{author}{P.~D. Olmsted},
\newblock \bibinfo{title}{The kinetics of phase separation in asymmetric
  membranes},
\newblock \bibinfo{journal}{Biophys. J.} \bibinfo{volume}{88}
  (\bibinfo{year}{2005}) \bibinfo{pages}{4072--4083}.
\bibitem[{Aranda-Espinoza et~al.(1996)Aranda-Espinoza, Berman, Dan, Pincus, and
  Safran}]{ArandaEspinoza:1996ux}
\bibinfo{author}{H.~Aranda-Espinoza}, \bibinfo{author}{A.~Berman},
  \bibinfo{author}{N.~Dan}, \bibinfo{author}{P.~Pincus},
  \bibinfo{author}{S.~Safran},
\newblock \bibinfo{title}{Interaction between inclusions embedded in
  membranes},
\newblock \bibinfo{journal}{Biophys. J.} \bibinfo{volume}{71}
  (\bibinfo{year}{1996}) \bibinfo{pages}{648--656}.
\bibitem[{Berman et~al.(1994)Berman, Pincus, and Safran}]{Berman:1994uv}
\bibinfo{author}{A.~Berman}, \bibinfo{author}{P.~Pincus},
  \bibinfo{author}{S.~A. Safran},
\newblock \bibinfo{title}{Membrane-induced interactions between inclusions},
\newblock \bibinfo{journal}{J. de Physique II} \bibinfo{volume}{4}
  (\bibinfo{year}{1994}) \bibinfo{pages}{1713--1725}.
\bibitem[{Dan et~al.(1993)Dan, Pincus, and Safran}]{Dan:1993uv}
\bibinfo{author}{N.~Dan}, \bibinfo{author}{P.~Pincus}, \bibinfo{author}{S.~A.
  Safran},
\newblock \bibinfo{title}{Membrane-induced interactions between inclusions},
\newblock \bibinfo{journal}{Langmuir} \bibinfo{volume}{9}
  (\bibinfo{year}{1993}) \bibinfo{pages}{2768--2771}.
\bibitem[{Goulian et~al.(2007)Goulian, Bruinsma, and Pincus}]{Goulian:2007wk}
\bibinfo{author}{M.~Goulian}, \bibinfo{author}{R.~Bruinsma},
  \bibinfo{author}{P.~Pincus},
\newblock \bibinfo{title}{Long-range forces in heterogeneous fluid membranes},
\newblock \bibinfo{journal}{Europhys. Lett.} \bibinfo{volume}{22}
  (\bibinfo{year}{2007}) \bibinfo{pages}{145--150}.
\bibitem[{Golestanian et~al.(1996)Golestanian, Goulian, and
  Kardar}]{Golestanian:1996wj}
\bibinfo{author}{R.~Golestanian}, \bibinfo{author}{M.~Goulian},
  \bibinfo{author}{M.~Kardar},
\newblock \bibinfo{title}{Fluctuation-induced interactions between rods on a
  membrane},
\newblock \bibinfo{journal}{Phys. Rev. E} \bibinfo{volume}{54}
  (\bibinfo{year}{1996}) \bibinfo{pages}{6725--6734}.
\bibitem[{Goulian(1996)}]{Goulian:1996td}
\bibinfo{author}{M.~Goulian},
\newblock \bibinfo{title}{Inclusions in membranes},
\newblock \bibinfo{journal}{Curr. Opinion in Coll.{\&} Interface Sci.}
  \bibinfo{volume}{1} (\bibinfo{year}{1996}) \bibinfo{pages}{358}.
\bibitem[{Goulian and Libchaber(1996)}]{Goulian:1996wu}
\bibinfo{author}{M.~Goulian}, \bibinfo{author}{A.~Libchaber},
\newblock \bibinfo{title}{A new technique for probing inter-membrane
  interactions},
\newblock \bibinfo{journal}{The J. of General Physiology.}
  \bibinfo{volume}{107} (\bibinfo{year}{1996}) \bibinfo{pages}{311--312}.
\bibitem[{Zhao et~al.(2013)Zhao, Liu, Yang, Capraro, Baumgart, Bradley,
  Ramakrishnan, Xu, Radhakrishnan, Svitkina, and Guo}]{Zhao:2013hi}
\bibinfo{author}{Y.~Zhao}, \bibinfo{author}{J.~Liu}, \bibinfo{author}{C.~Yang},
  \bibinfo{author}{B.~R. Capraro}, \bibinfo{author}{T.~Baumgart},
  \bibinfo{author}{R.~P. Bradley}, \bibinfo{author}{N.~Ramakrishnan},
  \bibinfo{author}{X.~Xu}, \bibinfo{author}{R.~Radhakrishnan},
  \bibinfo{author}{T.~Svitkina}, \bibinfo{author}{W.~Guo},
\newblock \bibinfo{title}{Exo70 generates membrane curvature for morphogenesis
  and cell migration},
\newblock \bibinfo{journal}{Developmental Cell} \bibinfo{volume}{26}
  (\bibinfo{year}{2013}) \bibinfo{pages}{266--278}.
\bibitem[{Kozlov(2007)}]{Kozlov07}
\bibinfo{author}{M.~M. Kozlov},
\newblock \bibinfo{title}{Biophysics: Bending over to attract},
\newblock \bibinfo{journal}{Nature} \bibinfo{volume}{447}
  (\bibinfo{year}{2007}) \bibinfo{pages}{387--389}.
\bibitem[{Agrawal and Radhakrishnan(2009)}]{Agrawal:2009bt}
\bibinfo{author}{N.~Agrawal}, \bibinfo{author}{R.~Radhakrishnan},
\newblock \bibinfo{title}{Calculation of free energies in fluid membranes
  subject to heterogeneous curvature fields},
\newblock \bibinfo{journal}{Phys. Rev. E} \bibinfo{volume}{80}
  (\bibinfo{year}{2009}) \bibinfo{pages}{011925}.
\bibitem[{Weikl(2001)}]{Weikl01}
\bibinfo{author}{T.~R. Weikl},
\newblock \bibinfo{title}{Fluctuation-induced aggregation of rigid membrane
  inclusions},
\newblock \bibinfo{journal}{Europhys. Lett.} \bibinfo{volume}{54}
  (\bibinfo{year}{2001}) \bibinfo{pages}{547}.
\bibitem[{Lee et~al.(2002)Lee, Hori, Groves, Dustin, and Chakraborty}]{Lee02}
\bibinfo{author}{S.-J.~E. Lee}, \bibinfo{author}{Y.~Hori},
  \bibinfo{author}{J.~T. Groves}, \bibinfo{author}{M.~L. Dustin},
  \bibinfo{author}{A.~K. Chakraborty},
\newblock \bibinfo{title}{The synapse assembly model},
\newblock \bibinfo{journal}{Trends Immunol.} \bibinfo{volume}{23}
  (\bibinfo{year}{2002}) \bibinfo{pages}{500--502}.
\bibitem[{Liu et~al.(2012)Liu, Agrawal, Eckmann, Ayyaswamy, and
  Radhakrishnan}]{Liu:2012ww}
\bibinfo{author}{J.~Liu}, \bibinfo{author}{N.~Agrawal},
  \bibinfo{author}{D.~Eckmann}, \bibinfo{author}{P.~Ayyaswamy},
  \bibinfo{author}{R.~Radhakrishnan},
\newblock \bibinfo{title}{Top-down mesocale models and free energy calculations
  of multivalent protein-protein and protein-membrane interactions in
  nanocarrier adhesion and receptor trafficking},
\newblock in: \bibinfo{editor}{T.~Schlick} (Ed.),
  \bibinfo{booktitle}{Innovations in Biomolecular Modeling and Simulations},
  \bibinfo{publisher}{Royal Society of Chemistry},
  \bibinfo{address}{Cambridge~UK}, \bibinfo{year}{2012}, pp.
  \bibinfo{pages}{272--287}.
\bibitem[{Iglic et~al.(2007)Iglic, Babnik, Bohinc, Fosnaric, Hagerstrand, and
  Kralj-Iglic}]{Iglic:2007vq}
\bibinfo{author}{A.~Iglic}, \bibinfo{author}{B.~Babnik},
  \bibinfo{author}{K.~Bohinc}, \bibinfo{author}{M.~Fosnaric},
  \bibinfo{author}{H.~Hagerstrand}, \bibinfo{author}{V.~Kralj-Iglic},
\newblock \bibinfo{title}{On the role of anisotropy of membrane constituents in
  formation of a membrane neck during budding of a multicomponent membrane},
\newblock \bibinfo{journal}{J. Biomech.} \bibinfo{volume}{40}
  (\bibinfo{year}{2007}) \bibinfo{pages}{579--585}.
\bibitem[{Bohinc et~al.(2006)Bohinc, Lombardo, Kraljiglic, Fosnaric, May,
  Pernus, Hagerstrand, and Iglic}]{Bohinc:2006vr}
\bibinfo{author}{K.~Bohinc}, \bibinfo{author}{D.~Lombardo},
  \bibinfo{author}{V.~Kraljiglic}, \bibinfo{author}{M.~Fosnaric},
  \bibinfo{author}{S.~May}, \bibinfo{author}{F.~Pernus},
  \bibinfo{author}{H.~Hagerstrand}, \bibinfo{author}{A.~Iglic},
\newblock \bibinfo{title}{Shape variation of bilayer membrane daughter vesicles
  induced by anisotropic membrane inclusions},
\newblock \bibinfo{journal}{Cell Mol. Biol. Lett.} \bibinfo{volume}{11}
  (\bibinfo{year}{2006}) \bibinfo{pages}{90--101}.
\bibitem[{Leibler and Andelman(1987)}]{Leibler:1987kg}
\bibinfo{author}{S.~Leibler}, \bibinfo{author}{D.~Andelman},
\newblock \bibinfo{title}{Ordered and curved meso-structures in membranes and
  amphiphilic films},
\newblock \bibinfo{journal}{J. Phys. France} \bibinfo{volume}{48}
  (\bibinfo{year}{1987}) \bibinfo{pages}{2013--2018}.
\bibitem[{Leibler(1986)}]{Leibler:1986kq}
\bibinfo{author}{S.~Leibler},
\newblock \bibinfo{title}{Curvature instability in membranes},
\newblock \bibinfo{journal}{J. Phys. France} \bibinfo{volume}{47}
  (\bibinfo{year}{1986}) \bibinfo{pages}{507--516}.
\bibitem[{Lipowsky(1992)}]{Lipowsky:1992fh}
\bibinfo{author}{R.~Lipowsky},
\newblock \bibinfo{title}{Budding of membranes induced by intramembrane
  domains},
\newblock \bibinfo{journal}{J. de Physique II} \bibinfo{volume}{2}
  (\bibinfo{year}{1992}) \bibinfo{pages}{1825--1840}.
\bibitem[{J{\"u}licher and Lipowsky(1993)}]{Julicher:1993hg}
\bibinfo{author}{F.~J{\"u}licher}, \bibinfo{author}{R.~Lipowsky},
\newblock \bibinfo{title}{Domain-induced budding of vesicles},
\newblock \bibinfo{journal}{Phys. Rev. Lett.} \bibinfo{volume}{70}
  (\bibinfo{year}{1993}) \bibinfo{pages}{2964--2967}.
\bibitem[{Sunil~Kumar and Rao(1998)}]{Kumar:1998jt}
\bibinfo{author}{P.~B. Sunil~Kumar}, \bibinfo{author}{M.~Rao},
\newblock \bibinfo{title}{Shape instabilities in the dynamics of a
  two-component fluid membrane},
\newblock \bibinfo{journal}{Phys. Rev. Lett.} \bibinfo{volume}{80}
  (\bibinfo{year}{1998}) \bibinfo{pages}{2489--2492}.
\bibitem[{Kohyama et~al.(2003)Kohyama, Kroll, and Gompper}]{Kohyama:2003dg}
\bibinfo{author}{T.~Kohyama}, \bibinfo{author}{D.~M. Kroll},
  \bibinfo{author}{G.~Gompper},
\newblock \bibinfo{title}{Budding of crystalline domains in fluid membranes},
\newblock \bibinfo{journal}{Phys. Rev. E} \bibinfo{volume}{68}
  (\bibinfo{year}{2003}) \bibinfo{pages}{061905}.
\bibitem[{Hui and Sen(1989)}]{Hui:1989fh}
\bibinfo{author}{S.~W. Hui}, \bibinfo{author}{A.~Sen},
\newblock \bibinfo{title}{Effects of lipid packing on polymorphic phase
  behavior and membrane properties},
\newblock \bibinfo{journal}{Proc. Natl. Acad. Sci. USA.} \bibinfo{volume}{86}
  (\bibinfo{year}{1989}) \bibinfo{pages}{5825--5829}.
\bibitem[{D{\"o}bereiner et~al.(1999)D{\"o}bereiner, Selchow, and
  Lipowsky}]{Dobereiner:1999tm}
\bibinfo{author}{H.-G. D{\"o}bereiner}, \bibinfo{author}{O.~Selchow},
  \bibinfo{author}{R.~Lipowsky},
\newblock \bibinfo{title}{Spontaneous curvature of fluid vesicles induced by
  trans-bilayer sugar asymmetry},
\newblock \bibinfo{journal}{Eur. Biophys. J.} \bibinfo{volume}{28}
  (\bibinfo{year}{1999}) \bibinfo{pages}{174--178}.
\bibitem[{Markowitz and Singh(1991)}]{Markowitz:1991p3276}
\bibinfo{author}{M.~Markowitz}, \bibinfo{author}{A.~Singh},
\newblock \bibinfo{title}{Self-assembling properties of
  1,2-diacyl-sn-glycero-3-phosphohyroxyethanol - a headgroup modified
  diacetylenic phospholipd},
\newblock \bibinfo{journal}{Langmuir} \bibinfo{volume}{7}
  (\bibinfo{year}{1991}) \bibinfo{pages}{16--18}.
\bibitem[{Fournier(1996)}]{Fournier:1996p488}
\bibinfo{author}{J.-B. Fournier},
\newblock \bibinfo{title}{Nontopological saddle-splay and curvature
  instabilities from anisotropic membrane inclusions},
\newblock \bibinfo{journal}{Phys. Rev. Lett.} \bibinfo{volume}{76}
  (\bibinfo{year}{1996}) \bibinfo{pages}{4436--4439}.
\bibitem[{Fournier and Galatola(1998)}]{Fournier:1998p2999}
\bibinfo{author}{J.-B. Fournier}, \bibinfo{author}{P.~Galatola},
\newblock \bibinfo{title}{Bilayer membranes with 2d-nematic order of the
  surfactant polar heads},
\newblock \bibinfo{journal}{Braz. J. Phys.} \bibinfo{volume}{28}
  (\bibinfo{year}{1998}) \bibinfo{pages}{329}.
\bibitem[{Hinshaw(2000)}]{Hinshaw:2000fi}
\bibinfo{author}{J.~E. Hinshaw},
\newblock \bibinfo{title}{Dynamin and its role in membrane fission},
\newblock \bibinfo{journal}{Annu. Rev. Cell Dev. Biol.} \bibinfo{volume}{16}
  (\bibinfo{year}{2000}) \bibinfo{pages}{483--519}.
\bibitem[{Praefcke and McMahon(2004)}]{Praefcke:2004bi}
\bibinfo{author}{G.~J.~K. Praefcke}, \bibinfo{author}{H.~T. McMahon},
\newblock \bibinfo{title}{The dynamin superfamily: universal membrane
  tubulation and fission molecules?},
\newblock \bibinfo{journal}{Nature} \bibinfo{volume}{5} (\bibinfo{year}{2004})
  \bibinfo{pages}{133--147}.
\bibitem[{Voeltz et~al.(2006)Voeltz, Prinz, Shibata, Rist, and
  Rapoport}]{Voeltz:2006ca}
\bibinfo{author}{G.~K. Voeltz}, \bibinfo{author}{W.~A. Prinz},
  \bibinfo{author}{Y.~Shibata}, \bibinfo{author}{J.~M. Rist},
  \bibinfo{author}{T.~A. Rapoport},
\newblock \bibinfo{title}{A class of membrane proteins shaping the tubular
  endoplasmic reticulum},
\newblock \bibinfo{journal}{Cell} \bibinfo{volume}{124} (\bibinfo{year}{2006})
  \bibinfo{pages}{573--586}.
\bibitem[{Dawson et~al.(2006)Dawson, Legg, and Machesky}]{Dawson:2006fn}
\bibinfo{author}{J.~C. Dawson}, \bibinfo{author}{J.~A. Legg},
  \bibinfo{author}{L.~M. Machesky},
\newblock \bibinfo{title}{Bar domain proteins: a role in tubulation, scission
  and actin assembly in clathrin-mediated endocytosis},
\newblock \bibinfo{journal}{Trends in Cell Biology} \bibinfo{volume}{16}
  (\bibinfo{year}{2006}) \bibinfo{pages}{493--498}.
\bibitem[{Lopez-Leon et~al.(2011)Lopez-Leon, Koning, Devaiah, Vitelli, and
  Fernandez-Nieves}]{LopezLeon:2011p3971}
\bibinfo{author}{T.~Lopez-Leon}, \bibinfo{author}{V.~Koning},
  \bibinfo{author}{K.~B.~S. Devaiah}, \bibinfo{author}{V.~Vitelli},
  \bibinfo{author}{A.~Fernandez-Nieves},
\newblock \bibinfo{title}{Frustrated nematic order in spherical geometries},
\newblock \bibinfo{journal}{Nat. Phys.} \bibinfo{volume}{7}
  (\bibinfo{year}{2011}) \bibinfo{pages}{391}.
\bibitem[{Ramakrishnan et~al.(2011)Ramakrishnan, Sunil~Kumar, and
  Ipsen}]{Ramakrishnan:2011cc}
\bibinfo{author}{N.~Ramakrishnan}, \bibinfo{author}{P.~B. Sunil~Kumar},
  \bibinfo{author}{J.~H. Ipsen},
\newblock \bibinfo{title}{Modeling anisotropic elasticity of fluid membranes},
\newblock \bibinfo{journal}{Macromol. Theory Simul.} \bibinfo{volume}{20}
  (\bibinfo{year}{2011}) \bibinfo{pages}{446--450}.
\bibitem[{Lebwohl and Lasher(1972)}]{Lebwohl:1972p426}
\bibinfo{author}{P.~A. Lebwohl}, \bibinfo{author}{G.~Lasher},
\newblock \bibinfo{title}{Nematic-liquid-crystal order - monte-carlo
  calculation},
\newblock \bibinfo{journal}{Phys. Rev. A} \bibinfo{volume}{6}
  (\bibinfo{year}{1972}) \bibinfo{pages}{426--429}.
\bibitem[{Frank and Kardar(2008)}]{Frank:2008p1047}
\bibinfo{author}{J.~R. Frank}, \bibinfo{author}{M.~Kardar},
\newblock \bibinfo{title}{Defects in nematic membranes can buckle into
  pseudospheres},
\newblock \bibinfo{journal}{Phys. Rev. E} \bibinfo{volume}{77}
  (\bibinfo{year}{2008}) \bibinfo{pages}{41705}.
\bibitem[{De~Gennes and Prost(1993)}]{gennesprost1993}
\bibinfo{author}{P.~G. De~Gennes}, \bibinfo{author}{J.~Prost},
  \bibinfo{title}{The Physics of Liquid Crystals}, The International Series of
  Monographs on Physics, \bibinfo{publisher}{Clarendon Press},
  \bibinfo{address}{Oxford}, \bibinfo{year}{1993}.
\bibitem[{Lubensky and Prost(1992)}]{Lubensky:1992p531}
\bibinfo{author}{T.~Lubensky}, \bibinfo{author}{J.~Prost},
\newblock \bibinfo{title}{Orientational order and vesicle shape},
\newblock \bibinfo{journal}{J. Phys. II France} \bibinfo{volume}{2}
  (\bibinfo{year}{1992}) \bibinfo{pages}{371--382}.
\bibitem[{Dijkstra(1959)}]{Dijkstra:1959p269}
\bibinfo{author}{E.~W. Dijkstra},
\newblock \bibinfo{title}{A note on two problems in connexion with graphs},
\newblock \bibinfo{journal}{Numerische Mathematik} \bibinfo{volume}{1}
  (\bibinfo{year}{1959}) \bibinfo{pages}{269--271}.
\bibitem[{Bowick and Giomi(2009)}]{Bowick:2009p955}
\bibinfo{author}{M.~Bowick}, \bibinfo{author}{L.~Giomi},
\newblock \bibinfo{title}{Two-dimensional matter: order, curvature and
  defects},
\newblock \bibinfo{journal}{Adv. Phys.} \bibinfo{volume}{58}
  (\bibinfo{year}{2009}) \bibinfo{pages}{449}.
\bibitem[{Vitelli and Nelson(2006)}]{Vitelli:2006p1462}
\bibinfo{author}{V.~Vitelli}, \bibinfo{author}{D.~R. Nelson},
\newblock \bibinfo{title}{Nematic textures in spherical shells},
\newblock \bibinfo{journal}{Phys. Rev. E} \bibinfo{volume}{74}
  (\bibinfo{year}{2006}) \bibinfo{pages}{21711}.
\bibitem[{Ramakrishnan(2012)}]{Ramakrishnan:2012wm}
\bibinfo{author}{N.~Ramakrishnan}, \bibinfo{title}{Effect of in-plane order and
  activity on vesicular morphology}, Ph.D. thesis, Thesis submitted to Indian
  Institute of Technology Madras, \bibinfo{address}{Chennai},
  \bibinfo{year}{2012}.
\bibitem[{Sweitzer and Hinshaw(1998)}]{Sweitzer:1998p3632}
\bibinfo{author}{S.~M. Sweitzer}, \bibinfo{author}{J.~E. Hinshaw},
\newblock \bibinfo{title}{Dynamin undergoes a gtp-dependent conformational
  change causing vesiculation},
\newblock \bibinfo{journal}{Cell} \bibinfo{volume}{93} (\bibinfo{year}{1998})
  \bibinfo{pages}{1021--1029}.
\bibitem[{Hinshaw(2000)}]{Hinshaw:2000p3770}
\bibinfo{author}{J.~E. Hinshaw},
\newblock \bibinfo{title}{Dynamin and its role in membrane fission},
\newblock \bibinfo{journal}{Ann. Rev. Cell Dev. Biol.} \bibinfo{volume}{16}
  (\bibinfo{year}{2000}) \bibinfo{pages}{483--519}.
\bibitem[{Roux et~al.(2010)Roux, Koster, Lenz, Sorre, Manneville, Nassoy, and
  Bassereau}]{Roux:2010p4079}
\bibinfo{author}{A.~Roux}, \bibinfo{author}{G.~Koster},
  \bibinfo{author}{M.~Lenz}, \bibinfo{author}{B.~Sorre}, \bibinfo{author}{J.-B.
  Manneville}, \bibinfo{author}{P.~Nassoy}, \bibinfo{author}{P.~Bassereau},
\newblock \bibinfo{title}{Membrane curvature controls dynamin polymerization},
\newblock \bibinfo{journal}{Proc. Natl. Acad. Sci. USA} \bibinfo{volume}{107}
  (\bibinfo{year}{2010}) \bibinfo{pages}{4141--4146}.
\bibitem[{Ayton et~al.(2007)Ayton, Blood, and Voth}]{Ayton:2007iw}
\bibinfo{author}{G.~S. Ayton}, \bibinfo{author}{P.~D. Blood},
  \bibinfo{author}{G.~A. Voth},
\newblock \bibinfo{title}{Membrane remodeling from n-bar domain interactions:
  Insights from multi-scale simulation},
\newblock \bibinfo{journal}{Biophys. J.} \bibinfo{volume}{92}
  (\bibinfo{year}{2007}) \bibinfo{pages}{3595--3602}.
\bibitem[{Noguchi and Gompper(2004)}]{Noguchi:2004ky}
\bibinfo{author}{H.~Noguchi}, \bibinfo{author}{G.~Gompper},
\newblock \bibinfo{title}{Fluid vesicles with viscous membranes in shear flow},
\newblock \bibinfo{journal}{Phys. Rev. Lett.} \bibinfo{volume}{93}
  (\bibinfo{year}{2004}) \bibinfo{pages}{258102}.
\bibitem[{Noguchi and Gompper(2005{\natexlab{a}})}]{Noguchi:2005p1410}
\bibinfo{author}{H.~Noguchi}, \bibinfo{author}{G.~Gompper},
\newblock \bibinfo{title}{Vesicle dynamics in shear and capillary flows},
\newblock \bibinfo{journal}{J. Phys.-Condens. Mat.} \bibinfo{volume}{17}
  (\bibinfo{year}{2005}{\natexlab{a}}) \bibinfo{pages}{3439}.
\bibitem[{Noguchi and Gompper(2005{\natexlab{b}})}]{Noguchi:2005br}
\bibinfo{author}{H.~Noguchi}, \bibinfo{author}{G.~Gompper},
\newblock \bibinfo{title}{Shape transitions of fluid vesicles and red blood
  cells in capillary flows},
\newblock \bibinfo{journal}{Proc. Natl. Acad. Sci. USA.} \bibinfo{volume}{102}
  (\bibinfo{year}{2005}{\natexlab{b}}) \bibinfo{pages}{14159--14164}.
\bibitem[{Lyman et~al.(2011)Lyman, Cui, and Voth}]{Lyman:2011iu}
\bibinfo{author}{E.~Lyman}, \bibinfo{author}{H.~Cui}, \bibinfo{author}{G.~A.
  Voth},
\newblock \bibinfo{title}{Reconstructing protein remodeled membranes in
  molecular detail from mesoscopic models},
\newblock \bibinfo{journal}{Phys. Chem. Chem. Phys.} \bibinfo{volume}{13}
  (\bibinfo{year}{2011}) \bibinfo{pages}{10430}.
\bibitem[{Tourdot et~al.(2014)Tourdot, Ramakrishnan, and
  Radhakrishnan}]{Tourdot:2014}
\bibinfo{author}{R.~W. Tourdot}, \bibinfo{author}{N.~Ramakrishnan},
  \bibinfo{author}{R.~Radhakrishnan},
\newblock \bibinfo{title}{Defining the free energy landscape of curvature
  inducing proteins on membrane bilayers},
\newblock \bibinfo{journal}{Phys. Rev. E}  (\bibinfo{year}{2014})
  \bibinfo{pages}{at review}.
\bibitem[{G.Taubin(1995)}]{Taubin:1995}
\bibinfo{author}{G.Taubin},
\newblock \bibinfo{title}{Estimating the tensor of curvature of a surface from
  a polyhedral approximation},
\newblock \bibinfo{journal}{Proc. Int. Conf. Comp. Vision}
  (\bibinfo{year}{1995}).
\bibitem[{Hildebrandt and Polthier(2004)}]{Hildebrandt:2004p216}
\bibinfo{author}{K.~Hildebrandt}, \bibinfo{author}{K.~Polthier},
\newblock \bibinfo{title}{Anisotropic filtering of non-linear surface
  features},
\newblock \bibinfo{journal}{Eurographics} \bibinfo{volume}{23}
  (\bibinfo{year}{2004}) \bibinfo{pages}{1--10}.
\bibitem[{Hildebrandt et~al.(2005)Hildebrandt, Polthier, and
  Wardetzky}]{Hildebrandt:2005p86}
\bibinfo{author}{K.~Hildebrandt}, \bibinfo{author}{K.~Polthier},
  \bibinfo{author}{M.~Wardetzky},
\newblock \bibinfo{title}{Smooth feature lines on surface meshes},
\newblock \bibinfo{journal}{Eurographics Symposium on Geometry Processing}
  (\bibinfo{year}{2005}) \bibinfo{pages}{1--6}.
\bibitem[{Polthier(2002)}]{Polthier_thesis_2002}
\bibinfo{author}{K.~Polthier}, \bibinfo{title}{Polyhedral Surfaces of Constant
  Mean Curvature}, Ph.D. thesis, TU-Berlin, \bibinfo{year}{2002}.
\bibitem[{Hameiri and Shimsoni(2003)}]{Shimsoni:2003p626}
\bibinfo{author}{E.~Hameiri}, \bibinfo{author}{I.~Shimsoni},
\newblock \bibinfo{title}{Estimating the principal curvatures and the darboux
  frame from real 3-d range data},
\newblock \bibinfo{journal}{IEEE transactions on systems man and cybernetics :
  Cybernetics} \bibinfo{volume}{33} (\bibinfo{year}{2003})
  \bibinfo{pages}{626--633}.
\bibitem[{Han et~al.(2008)Han, Shokef, Alsayed, Yunker, Lubensky, and
  Yodh}]{Han:2008p3701}
\bibinfo{author}{Y.~Han}, \bibinfo{author}{Y.~Shokef}, \bibinfo{author}{A.~M.
  Alsayed}, \bibinfo{author}{P.~Yunker}, \bibinfo{author}{T.~C. Lubensky},
  \bibinfo{author}{A.~G. Yodh},
\newblock \bibinfo{title}{Geometric frustration in buckled colloidal
  monolayers},
\newblock \bibinfo{journal}{Nature} \bibinfo{volume}{456}
  (\bibinfo{year}{2008}) \bibinfo{pages}{898--903}.
\bibitem[{Shin et~al.(2008)Shin, Bowick, and Xing}]{Shin:2008p174}
\bibinfo{author}{H.~Shin}, \bibinfo{author}{M.~Bowick},
  \bibinfo{author}{X.~Xing},
\newblock \bibinfo{title}{Topological defects in spherical nematics},
\newblock \bibinfo{journal}{Phys. Rev. Lett.} \bibinfo{volume}{101}
  (\bibinfo{year}{2008}) \bibinfo{pages}{037802}.
\bibitem[{MacKintosh and Lubensky(1991)}]{MacKintosh:1991p685}
\bibinfo{author}{F.~MacKintosh}, \bibinfo{author}{T.~Lubensky},
\newblock \bibinfo{title}{Orientational order, topology, and vesicle shapes},
\newblock \bibinfo{journal}{Phys. Rev. Lett.} \bibinfo{volume}{67}
  (\bibinfo{year}{1991}) \bibinfo{pages}{1169--1172}.
\bibitem[{Park et~al.(1992)Park, Lubensky, and Mackintosh}]{Park:1992p946}
\bibinfo{author}{J.~Park}, \bibinfo{author}{T.~C. Lubensky},
  \bibinfo{author}{F.~C. Mackintosh},
\newblock \bibinfo{title}{n-atic order and continuous shape changes of
  deformable surfaces of genus zero},
\newblock \bibinfo{journal}{Europhys. Lett.} \bibinfo{volume}{20}
  (\bibinfo{year}{1992}) \bibinfo{pages}{279}.
\bibitem[{Peskin(2002)}]{Peskin:2002go}
\bibinfo{author}{C.~S. Peskin},
\newblock \bibinfo{title}{The immersed boundary method},
\newblock \bibinfo{journal}{ANU} \bibinfo{volume}{11} (\bibinfo{year}{2002})
  \bibinfo{pages}{479--517}.

\end{thebibliography}

\end{document}